\documentstyle[12pt,extras,epsfig]{article}

\newcommand{\ipar}{\hspace*{6mm}}
\newcommand{\ilskp}{\hspace*{27mm}}
\renewcommand{\thefigure}{\thesection.\arabic{figure}}

\newcommand{\hlf}{\mbox{$\frac{1}{2}$}}
\newcommand{\hlfsm}{\mbox{\small $\frac{1}{2}$}}
\newcommand{\dq}{\mbox{$\displaystyle\frac{d^4 q}{(2\pi)^4}$}}
\newcommand{\sgmsb}{\mbox{$\overline{\sigma}_S$}}
\newcommand{\sgmvb}{\mbox{$\overline{\sigma}_V$}}
\newcommand{\etal}{{\it et al.}}
\newcommand{\bcn}{\begin{center}}
\newcommand{\beq}{\begin{equation}}
\newcommand{\beqn}{\begin{eqnarray}}
\newcommand{\Aslash}{\mbox{$\not \!\! A$}}
\newcommand{\Dslash}{\mbox{$\not \!\! D$}}
\newcommand{\Pslash}{\mbox{$\not \! P$}}
\newcommand{\dslash}{\mbox{$\not \! \partial$}}
\newcommand{\dqbt}{\frac{d^{3}q}{(2\pi)^3}}
\newcommand{\dqbf}{\frac{d^{4}q}{(2\pi)^4}}
\newcommand{\pslash}{\mbox{$\not \! p$}}
\newcommand{\pslashsm}{\mbox{\footnotesize $\not \! p$}}
\newcommand{\qslash}{\mbox{$\not \! q$}}
\newcommand{\kslash}{\mbox{$\not \! k$}}

\newcommand{\ecn}{\end{center}}
\newcommand{\eeq}{\end{equation}}
\newcommand{\eeqn}{\end{eqnarray}}
\newcommand{\etab}{\mbox{$\overline{\eta}$}}

\newcommand{\mbar}{\mbox{$\overline{m}$}}
\newcommand{\omegab}{\mbox{$\overline{\omega}$}}
\newcommand{\psib}{\mbox{$\overline{\psi}$}}
\newcommand{\psibpsi}{\mbox{$\langle\overline{\psi}\psi\rangle$\ }}
\newcommand{\xib}{\mbox{$\overline{\xi}$}}
\newcommand{\qedd}{\mbox{QED$_{\rm d}$}\ }
\newcommand{\qedt}{\mbox{QED$_3$}\ }
\newcommand{\qedf}{\mbox{QED$_4$}\ }
\newcommand{\cndst}{\mbox{$\langle \overline{q} q \rangle$}}

\newcommand{\psq}{\mbox{$p^{2}$}}

\newcommand{\Eq}[1]{Eq.~(\ref{#1})}
\newcommand{\Eqs}[1]{Eqs.~(\ref{#1})}
\newcommand{\Fig}[1]{Fig.~\ref{#1}}
\newcommand{\Table}[1]{Table~\ref{#1}}
\newcommand{\sect}[1]{\section{{\sc #1}}
       \setcounter{table}{0}\setcounter{figure}{0}
       \setcounter{equation}{0}\vspace*{-\parskip}}
\newcommand{\subsect}[1]{\subsection{{\underline{\rm #1}}}\vspace*{-\parskip}} 
\newcommand{\subsubsect}[1]{\bigskip\underline{#1.}\ipar}
\def\lsim{\mathrel{\rlap{\lower4pt\hbox{\hskip1pt$\sim$}}
    \raise1pt\hbox{$<$}}}         
\def\gsim{\mathrel{\rlap{\lower4pt\hbox{\hskip1pt$\sim$}}
    \raise1pt\hbox{$>$}}}         

\def\overleftrightarrow#1{\vbox{\ialign{##\crcr
    $\leftrightarrow$\crcr
    \noalign{\kern 1pt\nointerlineskip}
    $\hfil\displaystyle{#1}\hfil$\crcr}}}
\def\mapright#1{\smash{
    \mathop{\longrightarrow}\limits_{#1}}}
\begin{document}
\rightline{Preprint numbers: ADP-93-225/T142}
\rightline{ANL-PHY-7668-TH-93}
\parbox{200mm}{
\bcn
DYSON-SCHWINGER EQUATIONS AND THEIR \\ APPLICATION TO HADRONIC PHYSICS
\vspace*{5mm}

Craig D. Roberts\footnotemark[2] and 
       Anthony G. Williams\footnotemark[3]\footnotemark[1]\\
\vspace*{5mm}

\footnotemark[2]
Physics Division,  Argonne National Laboratory, Argonne, IL 60439-4843, USA\\
\vspace*{2mm}
\smallskip
\footnotemark[3]
Department of Physics and Mathematical Physics, University of Adelaide,\\
S.A. 5005, Australia\\
\vspace*{2mm}
\footnotemark[1]
Department of Physics and the Supercomputer Computations Research
Institute,\\ Florida State University, Tallahassee, FL 32306, U.S.A.\\
\vspace*{2mm}
{\it e-mail: cdroberts@anl.gov, awilliam@physics.adelaide.edu.au}
\ecn
}\parbox{1mm}{$\rule{0mm}{91mm}$}\hspace*{15mm}

\ilskp ABSTRACT

\hspace*{-\parindent}
We review the current status of nonperturbative studies of gauge field theory
using the Dyson-Schwinger equation formalism and its application to hadronic
physics.  We begin with an introduction to the formalism and a discussion of
renormalisation in this approach.  We then review the current status of
studies of Abelian gauge theories [e.g., strong coupling quantum
electrodynamics] before turning our attention to the non-Abelian gauge theory
of the strong interaction, quantum chromodynamics.  We discuss confinement,
dynamical chiral symmetry breaking and the application and contribution of
these techniques to our understanding of the strong interactions.
\bigskip

\ilskp KEYWORDS

confinement of quarks and gluons; dynamical chiral symmetry breaking;
Dyson-Schwinger equations; hadrons; quantum electrodynamics; quantum
chromodynamics.

\vspace*{20mm}

\begin{center}{\it To appear in 
\underline{Progress in Particle and Nuclear Physics}}
\end{center}

\vfill\eject
\tableofcontents
\sect{Introduction}
\label{sect-Intro}
As computer technology continues to improve, lattice gauge theory [LGT] will
become an increasingly useful means of studying hadronic physics through
investigations of discretised quantum chromodynamics [QCD].  [For a recent
review of LGT with numerous references see Rothe (1992).]  However, it is
equally important to develop other complementary nonperturbative methods
based on continuum descriptions.  In particular, with the advent of new
accelerators such as CEBAF and RHIC, there is a need for the development of
approximation techniques and models which bridge the gap between
short-distance, perturbative QCD and the extensive amount of low- and
intermediate-energy phenomenology in a single covariant framework.
Cross-fertilisation between LGT studies and continuum techniques provides a
particularly useful means of developing a detailed understanding of
nonperturbative QCD.

One such continuum approach is based on the infinite tower of Dyson-Schwinger
equations [DSEs]. The DSEs are coupled integral equations which relate the
Green's functions of a field theory to each other.  Solving these equations
provides a solution of the theory; a field theory being completely defined
when all of its $n$-point Green's functions are known.  The DSEs include, for
example, the Bethe-Salpeter equation [BSE] which is needed for the
description of relativistic two-body scattering and bound states.
Quantitative studies of a field theory must be based on one [or more]
systematic approximation schemes: for example, perturbation theory for weak
coupling; lattice studies using increasing numbers of sites and
$\beta$-values in order to represent a larger volume of the
spacetime-continuum; large $N$ expansions for $SU(N)$ gauge theories; $\hbar$
expansions for semiclassical approximations; etc.  For studies based on DSEs
it is unavoidable that the infinite tower of coupled equations must be
truncated at some point.  This means that the tower of equations must be
limited to some $n$, where $n$ is the maximum number of legs on any Green's
function included in the self-consistent solution of the equations.  Because
of the complex issues involved and the computational effort required, current
efforts are necessarily modest; for example, we know of no attempts as yet to
study coupled DSEs for $n>3$.  The penalty incurred by truncation is the need
to employ an Ansatz for the omitted function(s).  However, much can be
achieved by exploiting the need to maintain certain properties of the theory,
including the various global and local symmetries, multiplicative
renormalisability, analyticity, known perturbative behaviour in the weak
coupling limit, etc.  These can provide stringent constraints on the
Ans\"{a}tze.  In addition, there is the hope that future lattice gauge theory
simulations will be able to provide additional insight into the form of these
higher $n$-point functions.

In Sec.~\ref{sect-DSE} we introduce the DSE formalism, beginning with a
discussion of Abelian gauge theories [quantum electrodynamics in three- and
four-dimensions, \qedt and \qedf] and then extend this discussion to QCD.  In
Secs.~\ref{sect-QED3} and \ref{sect-QED4} we review in some detail the
current status of DSE-based studies of \qedt and \qedf with particular
emphasis on gauge covariance and multiplicative renormalisability.  Such
studies are extremely useful as a guide to the more complicated case of QCD.
We devote Sec.~\ref{sect-QCD-gluon} to a study of the gauge boson sector of
QCD and focus on studies of the infrared behaviour of the gluon propagator,
since this is thought to be crucial to confinement in QCD.  In
Sec.~\ref{sect-QCD-quark} we examine the quark sector of QCD, discussing the
crucial issues of dynamical chiral symmetry breaking [DCSB] and quark
confinement.  In Sec.~\ref{sect-hadrons} we review applications of the ideas
discussed in the preceding sections to studies of hadronic structure.
Finally, in Sec.~\ref{sect-summary}, we summarise and discuss possible future
extensions of these studies.

\sect{Dyson-Schwinger Equation Formalism}
\label{sect-DSE}
It has been known for quite some time that, from the field equations of a
quantum field theory, one can derive a system of coupled integral equations
relating the Green's functions for the theory to each other (Dyson, 1949;
Schwinger, 1951).  This infinite tower of equations is sometimes referred to
as the complex of Dyson-Schwinger equations.  An introduction to the formal
details of DSEs, including their derivation etc., can be found in a number of
text books; Bjorken and Drell (1965, pp.  283-376) and Itzykson and Zuber
(1980, pp.  475-481) are two examples.  For completeness we will summarise
those aspects of the formalism which are relevant to the gauge field theories
of quantum electrodynamics [QED] and quantum chromodynamics [QCD].  These are
the areas where the DSE approach to the solution of a quantum field theory
[QFT] has been applied most widely.  Our exposition employs the functional
integral formulation of these field theories; useful introductions to which
can be found in Itzykson and Zuber (1980, pp. 425-474) and Rivers (1987).
The separate but related question of whether the functional integral
formulation of these field theories can be rigorously defined is as yet
unresolved. A discussion of this problem can be found, for example, in Seiler
(1982).  The Euclidean-space, discretised, lattice theory is well-defined,
but there the question is simply deferred and becomes a question of
rigorously establishing the existence of the continuum limit.  This limit is
a critical point of the lattice theory, since, in units of the lattice
spacing, physical correlation lengths must diverge.

\subsect{Quantum Electrodynamics [QED]}
\label{subsect-QED}
In the following we will discuss quantum electrodynamics in $D=2$ and $D=3$
{\it space} dimensions, referred to as \qedt and \qedf respectively.  We
denote the number of space-time dimensions as ${\rm d}=D+1$ and write the action
for \qedd in Minkowski metric, with signature $g_{00}=1$ and $g_{ii}=-1$ for
$i=1\ldots D$:
\beq
S[\psib,\psi,A_\mu] = \int\, d^{\rm d} x\; \left[\,\sum_{f=1}^{N_f}\,
\psib^f \left( i\dslash - m_0^f + e_0^f \Aslash \right)\psi^f
 - \frac{1}{4} F_{\mu\nu}F^{\mu\nu}
\right]~.
\label{Sqed}
\eeq
[We have set $\hbar = 1 = c$.]  In \Eq{Sqed} the superscript $f$ is a {\it
flavour} label; \mbox{$f= 1\ldots N_f$} for a theory with $N_f$ distinct types
or {\it flavours} of electrically active fermions, each represented by
the field
$\psi^f(x)$; $m_0^f$ and $e_0^f$ are, respectively, the bare mass and charge
of each of these fermions; $A_\mu(x)$ is the gauge-boson [photon] field; and
\beq
F_{\mu\nu} = \partial_\mu A_\nu - \partial_\nu A_\mu
\eeq
is the Abelian gauge-boson field strength tensor.  We use
standard notation and conventions, where $\Aslash = A_\mu\gamma^\mu =
g^{\mu\nu}A_\mu \gamma_\nu$, $\{\gamma^\mu,\gamma^\nu\}=2g^{\mu\nu}$,
$\gamma_5=i\gamma^0\gamma^1\gamma^2\gamma^3$, etc.
Note that for an electron we would have for the [physical] charge
$e^f=-e$, where by definition
$e=|e|$ is the magnitude of the electron charge.

The quantum field theory associated with \Eq{Sqed} is defined by the generating
functional 
\beq
{\cal Z}[\etab,\eta,J_\mu]= 
\int\,d\mu(\psib,\psi,A)\,
\exp\left( iS[\psib,\psi,A_\mu] 
+ i\int\,d^{\rm d} x\;\left[
\,\sum_f\left(\psib^f\eta^f + \etab^f\psi^f\right) + A_\mu J^\mu \right]
\right)
\label{Zqed}
\eeq
where $\etab^f$, $\eta^f$ and $J_\mu$ are, respectively, source fields for the
fermion, antifermion and gauge boson and where we have defined
\beq
d\mu(\psib,\psi,A) = 
\prod_f\,{\cal D}\psib^f{\cal D}\psi^f \prod_\mu\, {\cal D}A_\mu~.
\label{dmu}
\eeq
[A constant normalisation factor is understood which ensures that
\mbox{${\cal Z}[0,0,0]= 1$}.]  This generating functional is the field
theoretic analogue of the partition function of a statistical mechanical
system and serves the same purpose; i.e., all the physical quantities of the
theory can be obtained from \mbox{${\cal Z}[\etab,\eta,J_\mu]$}.

Operationally, the integral in \Eq{Zqed} represents ``an integral over all
possible values of each of the fields at all spacetime points''.  Detailed
discussions of the functional integral can be found in
Rivers (1987), Pascual and Tarrach (1984, pp. 198-226) and Seiler
(1982).  For our purposes, however, we may proceed upon noting that the fermion
fields $\{\psib,\psi\}$ and their sources $\{\etab,\eta\}$ are elements of a
Grassmann algebra with involution which entails that all of these fields
anticommute with each other [this ensures that Fermi-statistics are obeyed by
the fermion fields]; and that $A_\mu$ and its source $J_\mu$ are c-number
fields.

In order to complete the operational definition of \qedd we note that the
action
in \Eq{Sqed} is invariant under the local Abelian gauge transformation
\begin{eqnarray}
\psi(x) \rightarrow\; ^\lambda\!\psi(x)&=& {\rm e}^{-ie_0\lambda(x)}\psi(x)
\;\;,\;\;
\psib(x) \rightarrow\; ^\lambda\!\,\psib(x)= {\rm e}^{ie_0\lambda(x)}
\psib(x)\;,\nonumber \\
A_\mu(x)\rightarrow\; ^\lambda\! A_\mu(x) &=& A_\mu(x) - \partial_\mu
\lambda(x)
\label{Abelian_gauge_transf}
\end{eqnarray}
where $\lambda(x)$ is an arbitrary scalar function and, for now, we have
suppressed the flavor indices.  This entails that the generating functional
as it has been defined so far is meaningless [even assuming that our above
caveats have been properly taken into account].  This is because gauge
invariance ensures that for any field configuration,
\{$\psib(x),\psi(x),A_\mu(x)$\}, there are [uncountably] many related
configurations,
\{$^\lambda\!\psib(x),^\lambda\!\psi(x),^\lambda\!A_\mu(x)$\},
which have the same action; i.e.,
\beq
S[\psib,\psi,A_\mu] = S[^\lambda\!\psib,^\lambda\!\psi,^\lambda\!A_\mu]
\;.
\eeq
The Grassmann integration over $\psib$ and $\psi$ gives the same result
independent of $\lambda(x)$ since the corresponding Grassmannian Jacobian for
this transformation is unity.  Hence there is an overall volume divergence in
the functional integration over the gauge field $A_\mu$ in \Eq{Zqed}.  This
is analogous to the divergence one obtains in the spacetime integral of a
translationally invariant function.  The proper definition of the measure in
\Eq{dmu} must ensure that the gauge field integration extends only over
gauge-inequivalent configurations (Bailin and Love, 1986, pp. 116-119).

This problem can be resolved via the introduction of the Faddeev-Popov
determinant (Popov, 1983). [Historically, this first arose in connection with
the correct quantisation of non-Abelian gauge theories
(Faddeev and Popov, 1967).]  The net effect of this procedure in
\qedd is simply to introduce a gauge fixing term into the action
of \Eq{Zqed} (Rivers, 1987, pp. 182-185).  A common choice
for this is the covariant gauge fixing term, which we use here, and which 
gives
\beq
S[\psib,\psi,A_\mu] \rightarrow S_\xi[\psib,\psi,A_\mu] 
= S[\psib,\psi,A_\mu] 
-\frac{1}{2\xi_0}\int\, d^{\rm d}x\, \left(\partial_\mu A^\mu \right)^2~,
\label{Sqedxi}
\eeq
where $\xi_0$ is the bare gauge fixing parameter.  It should be noted that
the difficulty with gauge fixing involving Gribov ambiguities does not appear
in Abelian gauge theories, [unless one adopts unusual nonlinear gauge fixing
prescriptions].  It can be shown that physical observables are independent of
the [renormalised] gauge parameter, $\xi$, and, in fact, that they are
independent of the actual form of the gauge fixing term (Bailin and Love,
1986, pp.~330-334 and references therein).  The unrenormalised quantum field
theory of electrodynamics is then defined by \Eq{Zqed} with the action
$S_\xi$ of \Eq{Sqedxi} and we may now proceed to obtain the unrenormalised
DSEs.

\subsubsect{Unrenormalised Dyson-Schwinger equation for the Photon
Polarisation Tensor}
\label{sec-PPT-M-UR}
The example of the derivation of the DSE for the photon polarisation tensor
in \qedf is given in Itzykson and Zuber (1980, pp. 476-477) and we summarise
these arguments here to illustrate the technique and to 
emphasise the fact that the DSEs are simply the Euler-Lagrange equations of
quantum field theory.  It should also be noted that independent of this
functional analysis one can directly obtain these equations by simply
decomposing the infinite sums of Feynman diagrams for
Green's functions in terms of other proper and full Green's functions.
This Feynman diagram approach is explained in some detail
in Bjorken and Drell (1965, pp.~283-376).

\begin{figure}[tb] 
  \centering{\
     \epsfig{figure=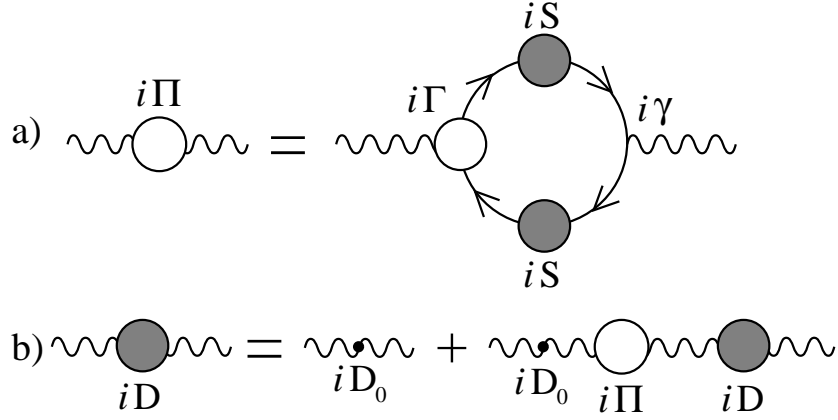,height=6cm}  }
\parbox{130mm}{\caption{
The Dyson-Schwinger equation for the photon propagator.
\label{photon_dse_fig} }}
\end{figure}

Consider the generating functional of \Eq{Zqed} with the action of
\Eq{Sqedxi}.  It can be shown [e.g., Itzykson and Zuber (1980, pp.~211-212)]
that the generating functional of {\it connected} Green's functions,
\mbox{${\cal G}[\etab,\eta,J_\mu]$}, is defined via
\beq
{\cal Z}[\etab,\eta,J_\mu] = \exp\left({\cal G}[\etab,\eta,J_\mu]\right)~.
\label{Gqed}
\eeq
In order to obtain the DSEs one must simply note that, in analogy with the
case of standard calculus, the functional integral of a total functional
derivative is zero given appropriate boundary conditions, [see, e.g.,
Collins (1984, pp.~13-15)].  Hence, for example,
\beqn
0 & = &\int d\mu(\psib,\psi,A)\, \frac{\delta}{\delta A_\mu(x)} 
        \exp\left\{i\left( S_\xi[\psib,\psi,A_\mu] + 
    \int\,d^{\rm d} x\;\left[ \psib^f\eta^f + \etab^f\psi^f + A_\mu J^\mu \right] 
                \right)\right\} \nonumber\\
& = &\int d\mu(\psib,\psi,A)\,
    \left\{ \frac{\delta S_\xi}{\delta A_\mu(x)} + J_\mu(x) \right\}
        \exp\left\{i\left( S_\xi[\psib,\psi,A_\mu] + 
    \int\,d^{\rm d} x\;\left[ \psib^f\eta^f + \etab^f\psi^f + A_\mu J^\mu \right] 
                \right)\right\} \nonumber\\
& = & \left\{ \frac{\delta S_\xi}{\delta A_\mu(x)}
        \left[\frac{\delta}{i\delta J }, 
        \frac{\delta}{i\delta\etab}, -\frac{\delta}{i\delta\eta}\right]
        + J_\mu(x) \right\} {\cal Z}[\etab,\eta,J_\mu] ~.
\label{ELeqn}
\eeqn
Differentiating \Eq{Sqedxi} immediately gives
\beq
\frac{\delta S_\xi}{\delta A_\mu(x)}
=\left[ \partial_\rho \partial^\rho g_{\mu\nu}
- \left( 1- \frac{1}{\xi_0}\right) \partial_\mu \partial_\nu\right] A^\nu
+ \sum_f e_0^f \, \psib^f \gamma_\mu \psi^f
\eeq
from which it follows that after dividing through by ${\cal Z}$
we can write \Eq{ELeqn} as
\beq
\left[ \partial_\rho \partial^\rho g_{\mu\nu}
- \left( 1- \frac{1}{\xi_0}\right) \partial_\mu \partial_\nu\right]
\frac{\delta {\cal G}}{i\delta J_\nu(x)} 
+  \sum_f e_0^f\,\left(  \frac{\delta{\cal G}}{\delta\eta^f(x)} 
        \gamma_\mu \frac{\delta {\cal G}}{\delta \etab^f(x)}
+ \frac{\delta}{\delta\eta^f(x)}
        \left[\gamma_\mu \frac{\delta {\cal G}}{\delta \etab^f(x)}
        \right]\right)
  =  - J_\mu(x)~.
\label{FldEqn}
\eeq
Equation.~(\ref{FldEqn}) represents a compact form of the nonperturbative
equivalent of Maxwell's equations.  To illustrate this we use it to obtain an
expression for the photon vacuum polarisation.  We can now perform a Legendre
transformation and introduce the generating functional for the connected,
{\it one-particle irreducible} [1-PI] Green's functions,
\mbox{$\Gamma[\psib, \psi, A_\mu]$}:
\beq
{\cal G}[\etab,\eta,J_\mu] \equiv 
i \Gamma[\psib, \psi, A_\mu] + 
i\int\,d^{\rm d} x\;\left[ \psib^f\eta^f + \etab^f\psi^f + A_\mu J^\mu \right]~.
\label{Legendre_transf}
\eeq
An explanation of the fact that $\Gamma[\psib,\psi,A_\mu]$ generates the 1-PI
Green's functions is given in Itzykson and Zuber (1980, pp.~289-294).  The
1-PI Green's functions are also frequently referred to as the {\it proper}
Green's functions.  It follows from the rules of Grassmannian integration
that ${\cal Z}[\etab,\eta,J_\mu]$ and hence ${\cal G}[\etab,\eta,J_\mu]$
depend only on powers of {\it pairs} of $\etab$ and $\eta$, which implies
that setting $\etab=\eta=0$ after differentiating ${\cal G}$ [or ${\cal Z}$]
will only give a nonzero result for equal numbers of $\etab$ and $\eta$
derivatives.  Similarly, in the absence of fermion derivatives it can be seen
that only even numbers of derivatives of ${\cal Z}$ and ${\cal G}$ with
respect to $J$ survive when we set $J=0$, which immediately leads to Furry's
theorem for QED, [see, e.g., Itzykson and Zuber (1980, pg. 276)].  Throughout
this work we diagrammatically represent proper Green's functions by unfilled
circles and full [connected] Green's functions by shaded circles.  See, e.g.,
\Fig{photon_dse_fig}, where the use of the proper electron-photon
vertex, $\Gamma^{f\mu}$, is necessary to avoid the double-counting of Feynman
diagrams using the arguments of Bjorken and Drell {\it op cit.}
{}From \Eq{Legendre_transf} the following relations immediately follow:
\beq
\begin{array}{ccc}
A_\mu(x) = \displaystyle \frac{\delta{\cal G}}{i\delta J^\mu(x)}, &
\psi^f(x) = \displaystyle\frac{\delta{\cal G}}{i\delta \etab^f(x)}, &
\psib^f(x) =\displaystyle -\,\frac{\delta{\cal G}}{i\delta \eta^f(x)}, \\
J_\mu(x) = \displaystyle -\,\frac{\delta \Gamma}{\delta A^\mu(x)}, &
\eta^f(x) = \displaystyle -\,\frac{\delta\Gamma}{\delta \psib^f(x)}, &
\etab^f(x) = \displaystyle\frac{\delta\Gamma}{\delta \psi^f(x)}~.
\end{array}
\label{Lrels}
\eeq
{}From \Eq{Lrels} we see that we now have expressions for $\psib, \psi$, and
$A_\mu$ in terms of $\etab, \eta$, and $J_\mu$ and vice versa; i.e.,
$\psib^f_\alpha(x) = \psib^f_\alpha[\etab,\eta,J_\mu(x)] =
i\delta{\cal G}[\etab,\eta,J_\mu]/\delta\eta^f_\alpha(x)$ with the spinor
index $\alpha$ now explicitly shown.
It is now easy to see that setting $\psib=\psi=0$ after differentiating
$\Gamma$ gives a nonzero result only when there are equal numbers
of $\psib$ and $\psi$ derivatives in analogy to the case for ${\cal G}$.
Using \Eq{Lrels} and the fact that $\delta\psib^f_\alpha/\delta\psib^g_\beta
=  \delta_{\alpha\beta}\delta_{fg}\delta^{\rm d}(x-y)$ we find that
\beq
\left. i \int d^{\rm d} z\, \frac{\delta^2 {\cal G}}{\delta \eta_\alpha^f(x)
      \etab_\gamma^h(z)}
       \, \frac{\delta^2\Gamma}{\delta \psi_\gamma^h(z) \psib_\beta^g(y)}
        \right|_{\begin{array}{c} \eta=\etab=0 \\
                                \psi=\overline{\psi}=0
                \end{array}} 
= \delta_{rs}\delta_{fg}\delta^{\rm d}(x-y)\;.
\label{inverse}
\eeq
Hence it follows that, when the fermion sources ($\etab,\eta$) vanish,
\Eq{FldEqn} can be written as
\beq
\left. \frac{\delta\Gamma}{\delta A^\mu(x)}\right|_{\psi=\overline{\psi}=0}
= \left[ \partial_\rho \partial^\rho g_{\mu\nu}
- \left( 1- \frac{1}{\xi_0}\right) \partial_\mu \partial_\nu\right]
A^\nu(x) -  i \sum_f e_0^f 
{\rm tr}\left[ \gamma_\mu S^f(x,x,[A_\mu]) \right] \;,
\label{FEppta}
\eeq
where we have made the identification 
\beq
S^f(x,y,[A_\mu]) = 
\left(\left.
\frac{\delta^2 \Gamma}{\delta\psib^f(x) \delta\psi^f(y)}
\right|_{\psi=\overline{\psi}=0}
\right)^{-1}\;,
\label{SfA}
\eeq
which is the propagator of the fermion of flavour $f$ in an external
electromagnetic field $A_\mu$.   This identification follows from
\Eq{inverse} and the fact that the second derivative of ${\cal G}$
is the full 2-point fermion Green's function; i.e., the fermion
propagator in the presence of a background field, $S^f(x,y,[A_\mu])$.
Clearly, the full fermion Green's function $S^f(x,y)$ follows from
setting $A_\mu=0$ in \Eq{SfA}.

To obtain the DSE for the photon polarisation tensor it remains only to act
with \mbox{$\delta/\delta A_\nu(y)$} on \Eq{FEppta} and
set \mbox{$J_\mu(x)=0$}.  It can be shown that 
\beq
 \left. e_0^f \, \Gamma^f_\mu(x;y,z) = \frac{\delta}{\delta A^\mu(x)}
        \frac{\delta^2 \Gamma}{\delta\psib^f(y) \delta\psi^f(z)}
\right|_{0=A_\mu=\psi=\overline{\psi}} ~,
\label{e0Gamma}
\eeq
is the proper [i.e., 1-PI] fermion--gauge--boson vertex [not to be confused
with the generating functional $\Gamma$].  Similarly to the fermion case it
can be shown that the second derivative of $\Gamma$ with respect to $A$ gives
the inverse photon propagator, $(D^{-1})^{\mu\nu}(x,y)$.  Thus from
\Eq{FEppta} we obtain the DSE for the inverse photon propagator
\beq
(D^{-1})^{\mu\nu}(x,y) =
\left.\frac{\delta^2 \Gamma}{\delta A^\mu(x) \delta A^\nu(y)}
\right|_{A_\mu=\psi=\overline{\psi}=0} 
= \left[ \partial_\rho \partial^\rho g_{\mu\nu}
- \left( 1- \frac{1}{\xi_0}\right) \partial_\mu \partial_\nu\right]
\delta^{\rm d}(x-y) + \Pi_{\mu\nu}(x,y)
\label{InvProp}
\eeq
where we have defined the photon polarisation tensor, $\Pi_{\mu\nu}$:
\beq
\Pi_{\mu\nu}(x,y) = i \sum_f (e_0^f)^2  \int\,d^{\rm d}z_1\,d^{\rm d}z_2\,
 {\rm tr}\left[ \gamma_\mu S^f(x,z_1)\Gamma^f_\nu(y;z_1,z_2)
        S^f(z_2,x)\right]\;.
\label{PPTeq}
\eeq
Using translational invariance we can write the photon propagator in
momentum space as
\beq
\label{DmnqQED}
D^{\mu\nu}(q)=\frac{-g^{\mu\nu}+(q^\mu q^\nu/(q^2+i\epsilon))}{q^2+i\epsilon}
\frac{1}{1+\Pi(q^2)} - \xi_0\frac{q^\mu q^\nu}{(q^2+i\epsilon)^2}\;,
\label{photon_propagator}
\eeq
where we have made use of the Ward-Takahashi identity [WTI] for the photon
propagator,
\beq
q_\mu\Pi^{\mu\nu}(q)=0~,
\label{WTI_photon}
\eeq
to define $\Pi(q^2)$ by $\Pi^{\mu\nu}(q)\equiv (-g^{\mu\nu}q^2+q^\mu
q^\nu)\Pi(q^2)$.  Note that $\Pi(q^2)$ is independent of the gauge
parameter $\xi_0$ in QED as a result of current conservation.
A discussion of WTIs is given, for example, in Itzykson and
Zuber (1980, pp.~407-411) and in Bjorken and Drell (1965, pp.~299-303).
There is a whole family of such identities, which follow from the fact that
gauge invariance leads to the conservation of the electromagnetic [e.m.]
current.  Because renormalisation does not affect e.m. current conservation
the WTIs also apply to the renormalised Green's functions of the theory
provided that the regularisation of infinities is performed in a
gauge-invariant way.  That $\Pi^{\mu\nu}$ is transverse leads to the fact
that the photon remains massless, [see, e.g., Itzykson and Zuber (1980, pp.
318-329].  The cases $\xi_0=0$ and 1 are referred to as Landau and Feynman
gauges respectively [and similarly for $\xi$ after renormalisation].  The
normalisations in these equations are such that at lowest order in
perturbation theory we have $\Pi=0$, as well as
\beq
\Gamma^f_\nu(y;z_1,z_2) = \gamma_\nu\, \delta^{\rm d}(y-z_1)\,\delta^{\rm d}(y-z_2)
\;\;\;\; {\rm and} \;\;\;\; 
(i\not \! \partial - m_0^f)\, S^f(x,y) = \delta^{\rm d}(x-y)~.
\eeq
It should be noted that, once $e_0^f$ is factored out, there is no {\it
explicit} flavour dependence in the proper vertex since the same sum of
Feynman diagrams will contribute in each case.  We will later see that once
we define the renormalised charge through appropriate boundary conditions a
flavour-dependence still arises for nondegenerate fermion masses.

We have seen that second derivatives of the generating functional,
$\Gamma[\psib,\psi,A_\mu]$, give the inverse fermion and photon propagators
and that the third derivative gave the proper photon-fermion vertex.  In
general, all derivatives of the generating functional,
$\Gamma[\psib,\psi,A_\mu]$, higher than two produce the corresponding proper
Green's functions, where the number and type of derivatives give the number
and type of proper Green's function legs.

The DSE for the photon propagator is represented diagrammatically in
\Fig{photon_dse_fig}. Part a) of the figure represents \Eq{PPTeq} for the
photon polarisation tensor and part b) shows the photon propagator,
$D^{\mu\nu}(x,y)$, defined in terms of $\Pi^{\mu\nu}(x,y)$ and the bare photon
propagator, $D_0^{\mu\nu}(x,y)$ [given by \Eq{photon_propagator} with $\Pi=0$]
and corresponds to the inverse of \Eq{InvProp}.
The momentum-space representation of the DSEs is readily obtained by either
Fourier transforming the coordinate-space form or, more readily, by
using the standard rules for Feynman diagrams based on the lowest-order
perturbative contribution to the nonperturbative quantities.  For example,
for the photon polarisation tensor we obtain
\beq
i\Pi_{\mu\nu}(q)=(-1)\sum_f (e_0^f)^2\int \frac{d^{\rm d}\ell}{(2\pi)^{\rm d}}
{\rm tr}[(i\gamma_\mu)(iS^f(\ell))(i\Gamma^f(\ell,\ell+q))(iS^f(\ell+q))]\;,
\eeq
where the factor of $(-1)$ arises from the fermion loop in the usual way.
We have elected to not explicitly show such factors of $(-1)$ in the
diagrammatic representation.
In momentum space, \Fig{photon_dse_fig}b) corresponds to
$iD^{\mu\nu}(q)=iD_0^{\mu\tau}(q)[\delta^\nu_\tau
+i\Pi_{\tau\rho}(q)iD^{\rho\nu}(q)]$, which can be obtained from the
Fourier transform of \Eq{photon_propagator}.  Throughout this work
we use figures like \Fig{photon_dse_fig} and the associated notation
so that the momentum space equations are easily written down by inspection.

\subsubsect{Unrenormalised Dyson-Schwinger equation for the Fermion Self
Energy} 
\label{FermDSE} Following a procedure similar to that above (Itzykson and
Zuber, 1980, pp. 478-479) one may derive an integral equation for the fermion
propagator starting from
\beqn
0 & = &\int d\mu(\psib,\psi,A)\, \frac{\delta}{\delta\psib(x)} 
        \exp\left\{i\left( S_\xi[\psib,\psi,A_\mu] + 
    \int\,d^{\rm d} x\;\left[ \psib^f\eta^f + \etab^f\psi^f + A_\mu J^\mu \right] 
                \right)\right\} \nonumber\\
& = & \left\{ \frac{\delta S_\xi}{\delta\psib(x)}
        \left[\frac{\delta}{i\delta J }, 
        \frac{\delta}{i\delta\etab}, -\frac{\delta}{i\delta\eta}\right]
        + \eta(x) \right\} {\cal Z}[\etab,\eta,J_\mu] ~.
\label{ELeqn_fermion}
\eeqn
After differentiating with respect to $\eta$ and setting all sources to zero
[$\etab=\eta=J=0$] we can rewrite \Eq{ELeqn_fermion} as
\beq
(i\not \! \partial - m_0^f)\, S^f(x,y) 
- i (e_0^f)^2 \int\, d^{\rm d} z_1 \, d^{\rm d} z_2\, d^{\rm d} z_3\,
        \gamma_\mu D^{\mu\nu}(x,z_1) S^f(x,z_2) \Gamma^f_\nu(z_1;z_2,z_3) S^f(z_3,y) =
\delta^{\rm d}(x-y)~,
\label{Fdsea}
\eeq
where \mbox{$D_{\mu\nu}(x,y)$} is the photon propagator which couples
\Eq{Fdsea} to \Eq{PPTeq}.  So, one sees that the equations for the 2-point
functions are coupled to each other and that both also depend on the
3-point function, $\Gamma^{f\mu}$.  This is the first indication of the
general rule that the DSE for an $n$-point function is coupled to other
functions of lesser and the same order and to functions of order (n+1) and
(n+2).
\begin{figure}[tb] 
  \centering{\ \epsfig{figure=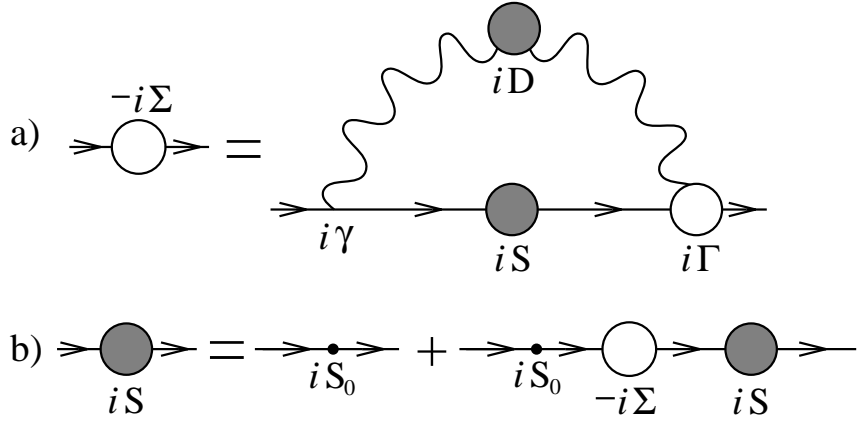,height=6cm} }
\parbox{130mm}{\caption{
The Dyson-Schwinger equation for the electron propagator.
\label{electron_dse_fig} }}
\end{figure}

The structure of \Eq{Fdsea} allows one to rewrite it in terms of the fermion
self energy, $-i\Sigma^f(x,y)$, defined such that
\beq
(i\not \! \partial - m_0^f)\, S^f(x,y) - \int d^{\rm d} z_1\,
\Sigma^f(x,z_1)\,S^f(z_1,y) = \delta^{\rm d}(x-y)~
\label{inv_electron_prop_eq}
\eeq
and hence satisfying
\beq
-i\Sigma^f(x,y) = (e_0^f)^2 \int\, d^{\rm d} z_1 \, d^{\rm d} z_2\, \gamma_\mu
D^{\mu\nu}(x,z_1) S^f(x,z_2) \Gamma^f_\nu(z_1;z_2,y)~.
\label{electron_dse_config_eq}
\eeq
The equation for the fermion self-energy is represented diagrammatically in
\Fig{electron_dse_fig}a) while part b) of this figure shows the definition of
the fermion self-energy ($-i\Sigma^f(p)$) in terms of the fermion propagator
$S^f(p)$ with $S^f_0(p)=1/(\pslash-m^f_0)$ the bare fermion propagator.
Again, the momentum-space form for the proper fermion self-energy
($-i\Sigma^f$) is easily obtained from \Fig{electron_dse_fig}a) using the
usual Feynman rules [or equivalently from Fourier transforming
\Eq{electron_dse_config_eq}] and can be written as
\beq
-i\Sigma^f(p)=(e_0^f)^2 \int\frac{d^{\rm d}\ell}{(2\pi)^{\rm d}}(i\gamma_\mu)
(iS^f(\ell))(iD^{\mu\nu}(p-\ell))(i\Gamma^f_\nu(\ell,p)) \;.
\label{electron_dse_eq}
\eeq
{}From \Fig{electron_dse_fig}b) or, equivalently, from
\Eq{inv_electron_prop_eq}, we can solve for $S^f(p)$ to give
$S^f(p)= 1/[(S^f_0)^{-1} - \Sigma^f(p)] = 1/[\pslash - m^f_0 - \Sigma^f(p)]$.

\begin{figure}[tb] 
  \centering{\ \epsfig{figure=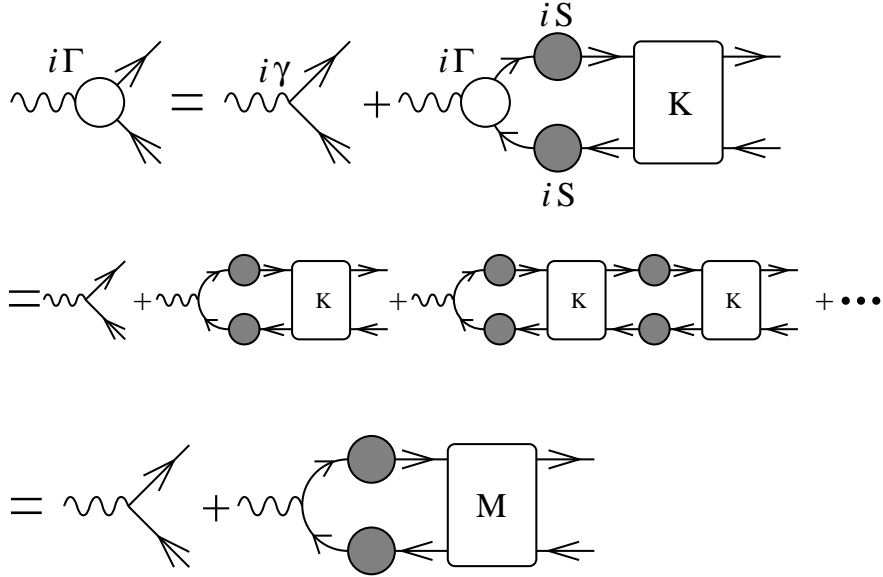,height=8.0cm} }
\parbox{130mm}{\caption{
The Dyson-Schwinger equation for the electron-photon proper vertex.
\label{el_ph_vertex_fig} }}
\end{figure}

\subsubsect{Unrenormalised Dyson-Schwinger equation for the Fermion-Photon
Vertex} \hspace*{-3pt}This equation can be derived in a simil\-ar way.  For
completeness, we present it here in momentum space where it is most concisely
written (Bjorken and Drell, 1965, pp. 291-293):
\beq
i\Gamma^f_\mu(p',p) = i\gamma_\mu + \sum_g \int \frac{d^{\rm d} \ell}{(2\pi)^{\rm d}}
(iS^g(p'+\ell)) (i\Gamma^g_\mu(p'+\ell,p+\ell))(iS^g(p+\ell))
K^{gf}(p+\ell,p'+\ell,\ell)~,
\label{Vtxa}
\eeq
where $K$ is referred to as the fermion-antifermion scattering kernel.  The
diagrammatic representation of this is given in \Fig{el_ph_vertex_fig}.
Clearly, $\Gamma^{\mu}$ is coupled to the fermion 2-point function [the
fermion propagator $S$] and also to a fermion-antifermion scattering
amplitude denoted by $M$, a 4-point function, which itself satisfies an
integral equation (Bjorken and Drell, 1965, pp.~293-298) and this illustrates
the general rule again.  The amplitude $M$ is 1-PI with respect to the
fermion lines and does not contain any fermion-antifermion annihilation
contributions [i.e., no intermediate single photon state] since these would
not be 1-PI contributions to $\Gamma^{f\mu}$ with respect to the photon line.
In an obvious shorthand notation we see from \Fig{el_ph_vertex_fig} that $M =
K+K(iS)^2K+K(iS)^2K(iS)^2K+\cdots = K+K(iS)^2M$.  Hence it follows that the
fermion-antifermion scattering kernel $K$ has no annihilation contributions
and is 2-PI with respect to the fermion-antifermion pair of lines.  The
latter requirement ensures that there is no double counting when $(iS)^2K$ is
iterated to form $M$.  It then follows that to lowest order in perturbation
theory $M^{fg}=K^{fg}=\delta^{fg}(e_0^f)^2(i\gamma_\mu)(iD^{\mu\nu}_0)
(i\gamma_\nu)$, where $D_0$ is the bare photon propagator.  The simplest
so-called {\it ladder approximation} consists of approximating $M$ by
iterating the lowest order perturbative contribution for the kernel $K$
together with the replacement of the fermion propagators by their
perturbative form; i.e., $S^f_0(p)=1/[\pslash-m^f_0]$.

Just as there was a WTI for photon vacuum polarisation [i.e.,
$q_\mu\Pi^{\mu\nu}(q)=0$] there is a WTI which relates the fermion propagator
to the proper fermion-photon vertex [see, for example, Itzykson and Zuber
(1980, pp.~407-411)]
\beq
k_\mu\Gamma^{f\mu}(p+k,p)=(S^f)^{-1}(p+k)-(S^f)^{-1}(p) \;.
\label{WTI_fermion}
\eeq

\subsubsect{Renormalisation of the Equations}  
\label{QEDRen}
Once the technique for deriving the unrenormalised equations is known then
the renormalised equations follow.  Essentially, one need only modify the
action of the theory to include the necessary counterterms and repeat the
above procedure.  For the purposes of this discussion we will temporarily
adopt a notation similar to that used in Bjorken and Drell (1965,
pp.~283-376), where the renormalised action and the quantities derived from
it are denoted by a tilde to distinguish these from the unrenormalised
quantities.

In $N_f$ flavour \qedd one has the following [infinite] renormalisation
constants (Dyson, 1949): the fermion wave function renormalisation, $Z_2$,
which relates the bare fermion field to the renormalised one; the photon wave
function renormalisation, $Z_3$; and the vertex renormalisation, $Z_1$, which
relates the bare and renormalised charge:
\beq
\begin{array}{ccc}
\psi^f_0 = \sqrt{Z^f_2} \psi^f~; &
A^\mu_0 = \sqrt{Z_3} A^\mu~; & e_0^f = \displaystyle\frac{Z^f_1}{Z^f_2
\sqrt{Z_3}} e^f~.
\end{array}
\label{Z_factors}
\eeq
A fundamental requirement and consequence of gauge invariant regularisation
and renormalisation of the theory is the Ward Identity (Ward, 1950;
Takahashi, 1957), which leads to
\beq
Z^f_1 = Z^f_2~.
\label{Ward_Z}
\eeq
This identity together with the multiplicative nature of renormalisation
ensures that the WTIs are preserved in the renormalised theory.  In addition
it ensures charge universality; i.e., that different species of fermion have
their charge renormalised by the same multiplicative factor.  The \qedd
renormalised action [i.e., including the counterterms] is obtained from
\Eq{Sqed} by identifying $\tilde S_\xi[\psib,\psi,A^\mu]\equiv
S_\xi[\psib_0,\psi_0,A^\mu_0]$ and making use of Eq.~(\ref{Z_factors}).
Hence the renormalised action is given by
\beqn
&&\tilde S_\xi[\psib,\psi,A_\mu]= \int\, d^{\rm d} x\; \left[
\,\sum_{f=1}^{N_f}\,
\left[ Z^f_2\psib^f \left( i\dslash - m_0^f \right)\psi^f
+Z^f_1e^f\psib^f\Aslash\psi^f\right] - \frac{Z_3}{4} F_{\mu\nu}F^{\mu\nu} -
\frac{Z_3}{2\xi_0}(\partial_\mu A^\mu)^2
\right] \nonumber \\
&&= \int\, d^{\rm d} x\; \left[\,\sum_{f=1}^{N_f}\,
\psib^f \left( i\dslash - m^f +e^f\Aslash\right)\psi^f
 - \frac{1}{4} F_{\mu\nu}F^{\mu\nu} - \frac{1}{2\xi}(\partial_\mu A^\mu)^2
\right. \nonumber\\ &&\left. + \sum_f\left[
(Z^f_2-1)\psib^f i\dslash\psi^f - \delta
m^f~\psib^f\psi^f +\delta e^f~\psib^f\Aslash\psi^f\right] - \frac{(Z_3-1)}{4}
F_{\mu\nu}F^{\mu\nu}
\right]~,
\label{Sqedxi_renorm}
\eeqn
where the four terms in the last line of this equation are called {\it
counterterms} and depend on both the regularisation parameter and the
renormalisation point.  We see that $\delta m^f \equiv m^f_B-m^f \equiv
Z^f_2m^f_0-m^f$ and $\delta e^f \equiv e^f_B-e^f \equiv (Z^f_1-1)e^f$. There
is no gauge-parameter counterterm since $\xi_0=Z_3\xi$, which follows from
\Eq{Z_factors} and from the WTI, $q_\mu\Pi^{\mu\nu}=0$, which ensures that
vacuum polarisation only affects the transverse part of the photon
propagator.  We will refer to $\psib^f_0, \psi^f_0, A^\mu_0, m^f_0, e^f_0,$
and $\xi_0$ as the {\it bare} quantities, although occasionally in the
literature this term is also used to describe $m^f_B$ and $e^f_B$.  This is
discussed for example in Collins (1984, pp.~10-11).  The renormalised [i.e.,
{\it physical}~] fermion masses and charges are $m^f$ and $e^f$,
respectively, and $\xi$ is the renormalised gauge parameter.

The renormalised action, $\tilde S_\xi$, will lead immediately to the
corresponding renormalised generating functionals $\tilde{\cal Z}$,
$\tilde{\cal G}$, and $\tilde\Gamma$, from Eqs.~(\ref{Zqed}), (\ref{Gqed}),
and (\ref{Legendre_transf}) respectively.  Using the same steps as before
these in turn lead to the renormalised Green's functions $\tilde D^{\mu\nu}$,
$\tilde S^f$, $\tilde\Gamma_\mu^f$, $\tilde K^{fg}$, etc.  The
renormalisation is carried out subject to the following boundary conditions
on renormalised quantities:
\beqn
 \left. (\tilde S^f)^{-1}(p)\right|_{p^2=(m^f)^2}=\pslash-m^f \;,\;\; \left.
\tilde\Gamma^{f\mu}(p,p)\right|_{\pslashsm =m^f}=\gamma^\mu \;,\;\;
\tilde\Pi(0)=0\;,
\label{QED_renorm_BCs}
\eeqn
together with the definition of the renormalised fermion and photon
propagators:
\beqn
(\tilde S^f)^{-1}(p)&=&\pslash-m^f-\tilde\Sigma^f(p) ~,\nonumber\\
\tilde D^{\mu\nu}(q)&=&
\frac{-g^{\mu\nu}+(q^\mu q^\nu/(q^2+i\epsilon))}{q^2+i\epsilon}
\frac{1}{1+\tilde\Pi(q^2)} - \xi\frac{q^\mu q^\nu}{(q^2+i\epsilon)^2}~.
\label{propagators}
\eeqn
The first condition in \Eq{QED_renorm_BCs} specifies that the renormalised
fermion propagator, $\tilde S^f(p)$, has a pole of residue one at the physical
fermion mass $m^f$; i.e., that $\tilde\Sigma^f(p)=0$ at $p^2=(m^f)^2$.  The
second condition ensures that when an on-shell fermion of flavour $f$ is
probed with a zero-momentum photon we measure the physical charge, $e^f$.  The
notation $\pslash=m^f$ means that all factors of $\pslash$ are replaced by
$m^f$, which is equivalent to having $\tilde\Gamma^{f\mu}$ act on a free
spinor at the fermion mass pole, $p^2=(m^f)^2$.  Note that for nondegenerate
fermion masses [i.e., $m^{f'}\ne m^f$] this boundary condition introduces a
subtle distinction between $\tilde\Gamma^f_\mu$ and $\tilde\Gamma^{f'}_\mu$,
since otherwise these vertices would be identical.  The last boundary
condition ensures that an on-shell photon has a pole at $p^2=0$ with unit
residue after vacuum polarisation corrections.

There are some subtle difficulties with these boundary conditions due to
infrared divergences but these are readily taken care of by introducing a
small photon mass as discussed, e.g., in Itzykson and Zuber (1980,
pp.~413-414).  It should be noted that this choice of boundary conditions
corresponds to the particular choice of the {\it on-shell} renormalisation
point, which is the usual choice for QED.  We know, however, from
renormalisation group arguments, that this choice is in fact arbitrary.  Since
QCD is believed to be confining, it has no corresponding natural choice of
renormalisation point.

Following these arguments gives, in place of \Eq{e0Gamma},
\beq
 \left. e^f \, \tilde\Gamma^f_\mu(x;y,z) = \frac{\delta}{\delta A^\mu(x)}
\frac{\delta^2 \tilde\Gamma}{\delta\psib^f(y) \delta\psi^f(z)}
\right|_{0=A_\mu=\psi=\overline{\psi}} ~.
\label{eGamma}
\eeq
Similarly, in place of Eqs.~(\ref{InvProp}) and (\ref{PPTeq}), we find
\beqn
(\tilde D^{-1})^{\mu\nu}(x,y) &=&
\left.\frac{\delta^2 \tilde\Gamma}{\delta A^\mu(x) \delta A^\nu(y)}
\right|_{A_\mu=\psi=\overline{\psi}=0} 
= Z_3\left[ \partial_\rho \partial^\rho g_{\mu\nu} - \left( 1-
\frac{1}{\xi_0}\right) \partial_\mu \partial_\nu\right]
\delta^{\rm d}(x-y) + \Pi'_{\mu\nu}(x,y) \nonumber\\
&=& Z_3\left[ \partial_\rho \partial^\rho g_{\mu\nu}
-\partial_\mu\partial_\nu\right]\delta^{\rm d}(x-y)
+\frac{1}{\xi}\partial_\mu\partial_\nu\delta^{\rm d}(x-y) + \Pi'_{\mu\nu}(x,y)
\label{InvProp_renorm}
\eeqn
where we have [{\it c.f.,} \Eq{PPTeq}]
\beq
\Pi'_{\mu\nu}(x,y)= i \sum_f Z_1^f (e^f)^2 \int\,d^{\rm d}z_1\,d^{\rm d}z_2\, 
{\rm tr}\left[ \gamma_\mu \tilde S^f(x,z_1)\tilde\Gamma_\nu(y;z_1,z_2)
\tilde S^f(z_2,x)\right]~.
\label{RPPTeq}
\eeq
In momentum space we then define $\Pi'_{\mu\nu}(q) \equiv
(-g_{\mu\nu}q^2+q_\mu q_\nu)\Pi'(q^2)$ and hence it follows from the momentum
space form of \Eq{InvProp_renorm} and from \Eq{propagators} that
$\tilde\Pi(q^2)=(Z_3-1)+\Pi'(q^2)$.  From the boundary condition
$\tilde\Pi(0)=0$ we find that
\beq
Z_3=1-\Pi'(0)
\label{Z_3}
\eeq
and so finally that
\beq
\label{PiTran}
i\tilde\Pi_{\mu\nu}(q)=i(-g_{\mu\nu}q^2+q_\mu q_\nu)\tilde\Pi(q^2)
=i(-g_{\mu\nu}q^2+q_\mu q_\nu)[\Pi'(q^2)-\Pi'(0)]~.
\eeq
Note that $q_\mu\Pi'^{\mu\nu}=0$ [assuming a covariant, gauge-invariant
regularisation scheme] just as before, and hence we have seen that the WTI
for the vacuum polarisation has survived renormalisation; i.e.,
$q_\mu\tilde\Pi^{\mu\nu}=0$, as expected.

Repeating the same arguments for the fermion self-energy we find
\Eq{Fdsea} becomes
\beq
Z_2^f(i\not \! \partial - m_0^f)\,\tilde S^f(x,y) -\int\, d^{\rm d}z_1
\Sigma^{f\prime}(x,z_1)\tilde S^f(z_1,y) = \delta^{\rm d}(x-y)~,
\label{RFdsea}
\eeq
where
\beq
-i\Sigma^{f\prime}(x,y) = Z_1^f (e^f)^2 \int\, d^{\rm d} z_1 \, d^{\rm d} z_2\,
\gamma_\mu \tilde D^{\mu\nu}(x,z_1) \tilde S^f(x,z_2)
\tilde\Gamma_\nu^f(z_1;z_2,y)~.
\label{Sigma_prime}
\eeq
In momentum space we have $(\tilde S^f)^{-1}(p) =
Z_2^f(\pslash-m^f_0)-\Sigma^{f\prime}(p) = \pslash-m^f-\tilde\Sigma^f(p)$.
Using Lorentz invariance allows us to decompose $\Sigma^{f\prime}$ into
Dirac-vector and -scalar pieces:
$\Sigma^{f\prime}(p)=\Sigma_d^{f\prime}(p^2)\pslash+\Sigma_s^{f\prime}(p^2)$;
and similarly for $\tilde\Sigma^f$.  Hence we find
$\tilde\Sigma_d^f(p^2)=\Sigma_d^{f\prime}(p^2)-(Z_2^f-1)$ and
$\tilde\Sigma_s^f(p^2)=\Sigma_s^{f\prime}(p^2)+(Z_2^fm_0^f-m^f)$.  Hence the
boundary condition $\tilde\Sigma^f(p)=0$ at $p^2=(m^f)^2$ gives
\beq
Z_2^f=1+\Sigma_d^{f\prime}((m^f)^2)\;\;,\;\;\;\;
m_0^f=\left[m^f-\Sigma_s^{f\prime}((m^f)^2)\right]/Z_2^f~.
\label{Z_2} 
\eeq
Then we have
$\tilde\Sigma^f(p)=\tilde\Sigma_d^f(p^2)\pslash+\tilde\Sigma_s^f(p^2)$, where
\beq
\tilde\Sigma_d^f(p^2)=\Sigma_d^{f\prime}(p^2)-\Sigma_d^{f\prime}((m^f)^2)
\;\;,\;\;\;\;
\tilde\Sigma_s^f(p^2)=\Sigma_s^{f\prime}(p^2)-\Sigma_s^{f\prime}((m^f)^2)~.
\label{Sigma_tilde}
\eeq
If we introduce the notation $(\tilde S^f)^{-1}(p)=A^f(p^2)\pslash-B^f(p^2)
=(Z^f)^{-1}(p^2)[\pslash-M^f(p^2)]$, then this leads to the renormalised DSE
for the fermion propagator in the form that we use in later sections:
\beq
A^f(p^2)=(Z^f)^{-1}(p^2)=1-\tilde\Sigma_d^f(p^2)\;\;,\;\;\;\;
B^f(p^2)=(Z^f)^{-1}(p^2)M^f(p^2)=m^f+\tilde\Sigma_s^f(p^2)~.
\label{DSE_AB}
\eeq 

For the fermion-photon proper vertex we have, in place of \Eq{Vtxa},
\beqn
\tilde\Gamma^f_\mu(p',p) &=& Z_1^f\gamma_\mu
- \int \frac{d^{\rm d} q}{(2\pi)^{\rm d}} \tilde S^f(p'+q)\tilde\Gamma^f_\mu(p'+q,p+q)
\tilde S^g(p+q) \tilde K^{fg}(p+q,p'+q,q)~\nonumber\\ &\equiv&
Z_1^f\gamma_\mu+\Lambda^{f\prime}_\mu(p',p)
\equiv \gamma_\mu+\tilde\Lambda^f_\mu(p',p)~,
\label{RVtxa}
\eeqn
which also defines both $\Lambda^{f\prime}_\mu$ and $\tilde\Lambda^f_\mu$.
Hence, we have
$\tilde\Lambda^f_\mu(p',p)=\Lambda^{f\prime}_\mu(p',p)+(Z_1^f-1)\gamma_\mu$
and from the boundary condition for the vertex in \Eq{QED_renorm_BCs} we find
\beq
\left. (Z_1^f-1)\gamma_\mu=
    -\Lambda^{f\prime}_\mu(p,p)\right|_{\pslashsm =m^f}~,
\label{Z_1}
\eeq
which gives finally that
\beq
\tilde\Lambda^f_\mu(p',p)=\Lambda^{f\prime}_\mu(p',p)
               -\Lambda^{f\prime}_\mu(p,p)\left|_{\pslashsm =m^f}\right.\;.
\label{Lambda_tilde}
\eeq
Since $(e^f\tilde\Gamma_\mu^f)$, $(\tilde D^{-1})_{\mu\nu}$, and $(\tilde
S^f)^{-1}$ are all obtained by differentiating
$\tilde\Gamma[\psib,\psi,A_\mu]$, then it follows, for example, from
Eqs.~(\ref{e0Gamma},\ref{Z_factors},\ref{Sqedxi_renorm},\ref{eGamma}) that
$(e^f_0\Gamma^f_\mu)=(Z^f_2\sqrt{Z_3})^{-1}(e^f\tilde\Gamma_\mu^f)$.
Similarly, $S^f=Z^f_2\tilde S^f$ and $D_{\mu\nu}=Z_3\tilde D_{\mu\nu}$.  Thus
from the unrenormalised WTI of \Eq{WTI_fermion} we find for the renormalised
quantities $(Z_2^f/Z_1^f)k_\mu\tilde\Gamma^{f\mu}(p+k,p)= (\tilde
S^f)^{-1}(p+k)-(\tilde S^f)^{-1}(p)$, which implies that since the theory is
renormalisable then $Z_1^f$ and $Z_2^f$ can only differ by a finite amount.
In fact from the boundary conditions in \Eq{QED_renorm_BCs} we see immediately that
we must have $Z_1^f=Z_2^f$, as was stated earlier.  The WTI implies the Ward
identity, which can be written as $\tilde \Gamma_\mu^f(p,p)=\partial (\tilde
S^f)^{-1}(p)/\partial p^\mu$.  [For ease of comparison with Bjorken and Drell
(1965), pp.~311-312, note that, e.g., $\tilde\Lambda$ and $\tilde\Pi$ used
here correspond to $\Lambda_c$ and $e^2\Pi_c$, respectively, in that work.]

In repeating the previous arguments and deriving the above renormalised
equations it soon becomes apparent that $Z$-factors are {\it only} introduced
when bare propagators or vertices appear in the DSE.  We have not discussed
details of the various regularisation schemes but it should be clear that all
of the regularisation-parameter dependence has been absorbed into these
factors.  Let $\Lambda$ denote a generic regularisation parameter such that
$\Lambda\to\infty$ as the regularisation is removed, e.g., $\Lambda$ might
denote a simple momentum cut-off or using dimensional regularisation, where
${\rm d}=4-\epsilon$, we would identify $\Lambda=1/\epsilon$, etc.  Then the
$Z$-factors are functions of the regularisation parameter and the
renormalisation point; i.e., $Z_2\equiv Z_2(\Lambda,\mu)$, etc.  For $m+n>3$,
where $m$ is the [even] number of external fermion legs and $n$ the number of
external photon legs, it can be shown that the unrenormalised Greens
functions can be decomposed into an infinite sum of {\it skeleton} diagrams
each of which is made up {\it only} of combinations of $S^f$, $D^{\mu\nu}$,
and $e_0\Gamma^f_\mu$.  However, we have seen that in the renormalised theory
the quantities $e^f_0\Gamma^f_\mu$, $D^{\mu\nu}$, and $S^f$ are simply
replaced by the renormalised quantities $e^f\tilde\Gamma^f_\mu$, $\tilde
D^{\mu\nu}$, and $\tilde S^f$.  Hence, for Green's functions with $m+n>3$ we
simply replace the skeleton expansion for the unrenormalised Greens function
by the same skeleton expansion with the renormalised quantities.  This is the
meaning of $\tilde K$ in \Eq{RVtxa}, for example.  This completes the
renormalisation program for QED.  A more detailed discussion of the skeleton
expansion and associated proof of renormalisability can be found in Bjorken
and Drell (1965).

\subsubsect{Review}
We have illustrated the general technique for obtaining the DSEs for a field
theory and have demonstrated by example the nature of the hierarchy of
equations; {\it i.e}, that the equations couple a given $n$-point function to
adjacent $n$-point functions in an infinite tower of coupled equations.  We
have also discussed in some detail the renormalisation of these equations.
As was already noted, choices of subtraction point [i.e., {\it
renormalisation point}] other than the on-shell subtraction point of
\Eq{QED_renorm_BCs} are possible.  From the previous discussion it is
clear that such a change can simply be absorbed as a finite rescaling of the
renormalised charges and masses whilst leaving the physical content of the
theory unchanged.  This observation is the basis of renormalisation group
studies of field theories.  We note in passing that with a suitably
unconventional choice of subtraction scheme one can find $Z_1^f\neq Z_2^f$,
although this is not the case in standard approaches [of course, this has no
physical significance].  The renormalised masses and charges in \qedf only
correspond to the physical ones when the on-shell renormalisation point is
used.  There is no corresponding on-shell renormalisation in QCD, since QCD
is believed to be a confining theory.  We have now introduced all of the
concepts that are necessary for the discussion of DSE studies of \qedd in
Secs.~\ref{sect-QED3} and \ref{sect-QED4}.  For our discussion of QCD in
Sec.~\ref{subsect-QCD} we simply point out the added complications due to the
non-Abelian nature of the theory.

\subsect{Quantum Chromodynamics [QCD]}
\label{subsect-QCD}

QCD is a non-Abelian gauge theory [based on the fundamental representation of
the group $SU(3)$], which leads to significantly different behaviour than that
shown by the Abelian theory of QED [based on the group $U(1)$].  In simple
terms this arises because the non-Abelian gauge fields are self-interacting.
The basic Lagrangian of QCD is given by
\beq
S[\psib,\psi,A_\mu] = \int\, d^{\rm d} x\; \left[\,\sum_{f=1}^{N_f}\,
\psib^f \left( i\Dslash - m_0^f \right)\psi^f
 - \frac{1}{4} F_{\mu\nu a}F^{\mu\nu}_a
\right]~,
\label{Sqcd}
\eeq
where
\beqn
F^{\mu\nu}_a &=& \partial^\mu A^\nu_a - \partial^\nu A^\mu_a
+g_0f_{abc}A^\mu_b A^\nu_c~,
\label{Fmunu_qcd}\\
D^\mu &=& \partial^\mu - ig_0A^\mu_a t_a = \partial^\mu-ig_0A^\mu~.
\label{cov_deriv_qcd}
\eeqn
and where $A^\mu\equiv A^\mu_a t_a$.  The fields $A^\mu_a$ are the gluon
fields and $\psi^f$ are the quark fields, where $f=1,\cdots,N_f$ is the quark
flavour index.  The indices $a=1,\cdots,8$ are colour indices and summation
over repeated indices is to be understood.  We use standard conventions
[e.g., Yndur\'ain (1993), Muta (1987)] where $t_a\equiv \lambda_a/2$ and the
hermitian [Gell-Mann] matrices $\lambda_a$ are the generators of $SU(3)$.
The $\lambda_a$ matrices satisfy ${\rm tr}(\lambda_a\lambda_b)=2\delta_{ab}$
and $[\lambda_a,\lambda_b]=2if_{abc}\lambda_c$, which leads to ${\rm tr}(t_a
t_b)=T_F\delta_{ab}=\delta_{ab}/2$ and $[t_a,t_b]=f_{abc}t_c\equiv
-C_{abc}t_c$.  [Note that Itzykson and Zuber (1980, pp. 563-567) define
instead antihermitian matrices $t_a\equiv i\lambda_a/2$].  We can write,
e.g., $(1/4)F^{\mu\nu}_aF_{\mu\nu a}=(1/2){\rm tr}(F^{\mu\nu}F_{\mu\nu})$,
where $F^{\mu\nu}\equiv t_a F^{\mu\nu}_a=
\partial^\mu A^\nu-\partial^\nu A^\mu
-ig_0[A^\mu,A^\nu] = (i/g_0)[D^\mu,D^\nu]$.  For $SU(N)$, the Casimir
invariants $C_A, C_F$, and $T_F$ take on the values $C_A=N$,
$C_F=(N^2-1)/2N$, and $T_F=1/2$.

As for QED, it is necessary to remove the infinite gauge-volume problem due
to integrations over gauge-equivalent field configurations.  This is remedied
by the prescription of Faddeev and Popov (1967) and has the added
complication that in order to maintain gauge invariance and unitarity in
covariant gauges it is necessary to introduce unphysical auxiliary fields
($\omegab_a,\omega_a$).  These are anticommuting spin-zero fields called {\it
ghost} fields and it is also necessary to integrate over these fields in
order to define the generation functional ${\cal Z}$ for QCD.  A technical
point, which we shall neglect henceforth, is the fact that the Faddeev-Popov
gauge-fixing procedure only uniquely specifies the gauge field configuration
with respect to infinitesimal local gauge transformations.  Under finite
transformations gauge-equivalent field configurations remain and are referred
to as Gribov copies [see, e.g., Itzykson and Zuber (1980, pp. 574-582) and
Gribov (1979)].  It is possible to choose ghost-free gauge-fixing
prescriptions such as the axial gauge, where $n\cdot A=0$ for some spacelike
four-vector $n^\mu$, ($n^2<0$).  The cases $n^2>0$ and $n^2=0$ are referred
to as the temporal and light-cone gauges, respectively.  The disadvantages of
such a choice are that Lorentz invariance is now broken explicitly and
unphysical singularities are present in the denominators of momentum-space
propagators; i.e., when $p\cdot n=0$.
 
The quantum field theory associated with \Eq{Sqcd} is defined for covariant
gauges by the generating functional
\beqn
\label{Zqcd}
\lefteqn{{\cal Z}[\etab,\eta,J_\mu,\xib,\xi] = }\\
&& 
\int\,d\mu(\psib,\psi,A,\omegab,\omega)\,
\exp\Big( iS_\xi[\psib,\psi,A_\mu,\omegab,\omega]
+ i\int\,d^{\rm d} x\;\left[ \psib^f\eta^f +
\etab^f\psi^f + A_{\mu a} J^\mu_a + \omegab_a\xi_a + \xib_a\omega_a \right]
\Big)\nonumber
\eeqn
where
\beq
S_\xi[\psib,\psi,A_\mu,\omegab,\omega] = S[\psib,\psi,A_\mu] + \int\,
d^{\rm d}x\,\left[
(\partial_\mu\omegab_a)(\delta_{ab}\partial^\mu-g_0f_{abc}A^\mu_c)\omega_b
-\frac{1}{2\xi_0} \left(\partial_\mu A^\mu \right)^2\right]~,
\label{Sqcdxi}
\eeq
where $\xi_0$ is the bare gauge fixing parameter.  The Faddeev-Popov technique
for non-Abelian gauge fields [as well as the gauge fixing procedure for QED]
can be understood in terms of the more general class of Becchi-Rouet-Stora
[BRS] transformations.  These are a generalisation of the local gauge
transformations to include the ghost and gauge-fixing terms.  The
Slavnov-Taylor identities [STIs] can be derived from the BRS invariance of
QCD and correspond to the WTIs of QED.  It can be shown that requiring BRS
invariance of a gauge theory automatically generates both the ghosts [where
appropriate] and the gauge fixing as well as ensuring the gauge-invariance of
physical observables, [see e.g., Pokorski (1987) pp.~64-85 and Pascual and
Tarrach (1984) and references therein].

Just as for QED we introduce appropriate counterterms into the QCD action in
\Eq{Sqcdxi} in order to define the QFT and then apply appropriate boundary
conditions at the chosen renormalisation point [a momentum scale that we
refer to as $\mu$].  Since the renormalisation point, $\mu$, is arbitrary,
physical observables must be independent of its choice.  In other words,
hadronic masses and cross-sections must be renormalisation-point independent
even though individual Green's functions [e.g., the quark and gluon
propagators] may not be.  In analogy with QED we identify the renormalised
action using $\tilde S_\xi[\psib,\psi,A^\mu,\omegab,\omega]\equiv
S_\xi[\psib_0,\psi_0,A^\mu_0,\omegab_0,\omega_0]$ together with the relations
[see, e.g., Yndur\'ain (1993) pp.~45-83]
\beq
\begin{array}{cccc}
\psi^f_0 = \sqrt{Z_F} \psi^f~; &
A^\mu_{a0} = \sqrt{Z_B} A^\mu_a~; &
\omega_{a0} = \sqrt{Z_\omega} \omega_a~;&
g_0 = \displaystyle Z_g g~.
\end{array}
\label{Z_factors_qcd}
\eeq
As a consequence of gauge invariant regularisation and renormalisation of the
theory the STIs are maintained, which leads to the fact that all quark {\it
colours} have the same $Z^f_F$ and that all gluons and ghosts have the same
$Z_B$ and $Z_\omega$ respectively.  In analogy with QED we see that
$\xi_0=Z_B \xi$, since only the transverse part of the gluon propagator is
modified by vacuum polarisation [i.e., $q_\mu\tilde\Pi^{\mu\nu}=0$].  Here,
$Z_g$ plays the role of the QED combination $Z_1^f/Z_2^f\sqrt{Z_3}$, so that
in place of $Z_1^f$ in QED we have $Z^f_\Gamma\equiv Z_g Z^f_F\sqrt{Z_B}$ in
QCD.  The STIs also imply that the same renormalisation constant $Z_g$
applies to the quark-gluon, ghost-gluon, three-gluon, and four-gluon
vertices.  The STIs are the reason that we do not need other independent
renormalisation constants for these couplings, [see, e.g., Itzykson and Zuber
(1980, pp.~593-594) and Muta (1987, pp.~158-179)].  We can define $\delta m^f
\equiv m^f_B-m^f \equiv Z^f_Fm^f_0-m^f \equiv (Z^f_FZ^f_m-1)m^f$ as for QED,
where the last result follows from the definition of $Z^f_m$ by $m^f_0\equiv
Z^f_mm^f$.

\begin{figure}[tb] 
  \centering{\ \epsfig{figure=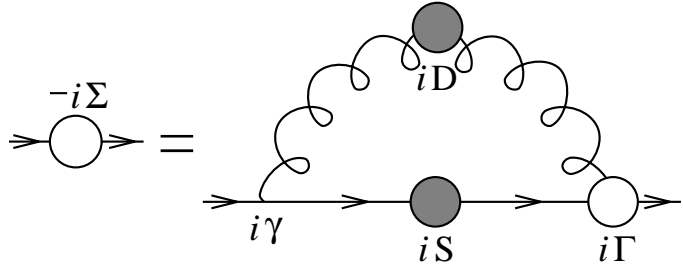,height=3.5cm} }
\parbox{130mm}{\caption{The Dyson-Schwinger equation for
the quark self-energy. 
\label{quark_dse_fig} }}
\end{figure}

\begin{figure}[tb] 
 \centering{\ \epsfig{figure=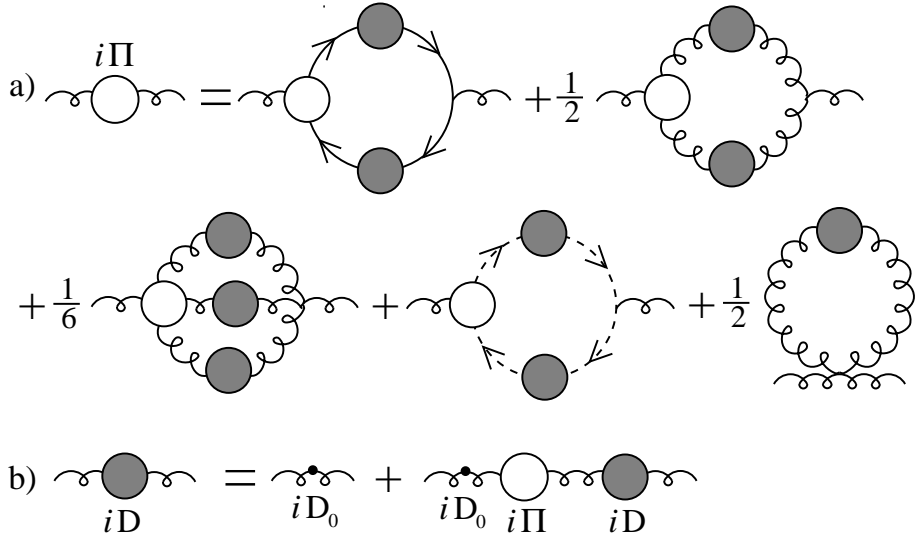,height=7.0cm} } \parbox{130mm}{
\caption{ The Dyson-Schwinger equation for the gluon propagator.
[Here and below the broken line represents the
propagator for the ghost field.]
\label{gluon_dse_fig}  }}
\end{figure}

\begin{figure}[tb] 
 \centering{\ \epsfig{figure=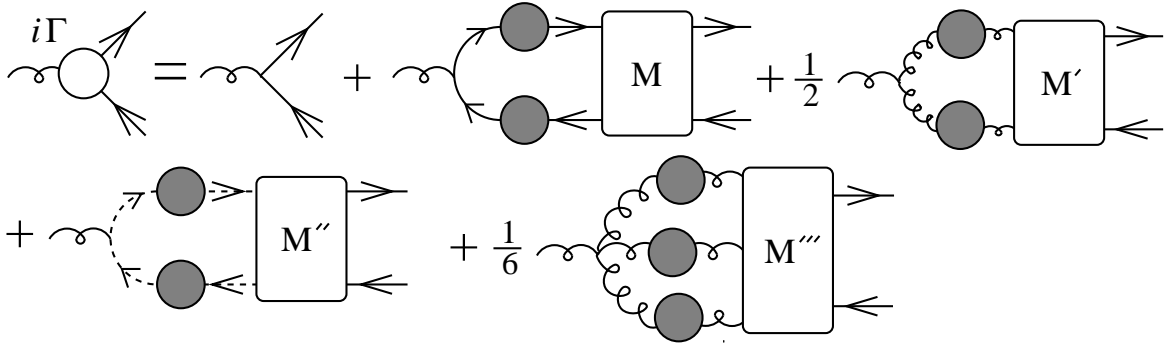,height=4.5cm} } \parbox{130mm}{
\caption{ The quark-gluon vertex Dyson-Schwinger equation.
\label{qu_gl_vertex_fig}  }}
\end{figure}

The derivation of the unrenormalised and renormalised DSEs proceeds in an
analogous way to that for QED and, as already stated, these are a direct
result of the BRS invariance of the theory.  In \Fig{quark_dse_fig} we show
the DSE for the quark self-energy.  The graphical representation of the quark
propagator DSE is the same as that for the electron given in
\Fig{electron_dse_fig}b).  Figure~\ref{gluon_dse_fig} specifies the
Dyson-Schwinger equation for the gluon propagator and can be compared with
the photon DSE in \Fig{photon_dse_fig}.  The symmetrisation factors of 1/2
and 1/6 arise from the usual Feynman rules, which also require a negative
sign [unshown] to be included for every fermion and ghost loop.  In
\Fig{qu_gl_vertex_fig} we show the quark-gluon proper vertex DSE.  The
notation is analogous to that used for the QED fermion-photon vertex in
\Fig{el_ph_vertex_fig}; i.e., the amplitudes $M, M', M''$, and $M'''$ are
1-PI with respect to all external legs and do not contain any single-gluon
intermediate states.


The difficult challenge is to find an Ansatz for the unknown renormalised
propagators, vertices, etc., of QCD which satisfy the DSEs and respect the
STIs of the theory.  Progress can be made by satisfying a subset of the DSEs
and STIs and then supplementing these with information gleaned from lattice
gauge theory, phenomenology, etc.  The test of such a scheme is to enlarge
the set of DSEs considered and look for a decreased dependence on the
phenomenological input.  Since the most significant difficulties in QCD arise
in the gluon sector this is the natural place to appeal to phenomenology
while attempting to construct explicit solutions and maintain important
symmetries in the quark sector.  Two symmetries of particular importance are
gauge invariance and chiral symmetry.

{\it Note:} Having completed the discussion of renormalisation, we now
dispense with the tilde notation for renormalised Green's functions,
self-energies, vacuum polarisations, etc.  In analogy with the QED case we
then denote the {\it renormalised} quark and gluon propagators [in momentum
space] by $S^f(p)$ and $D^{\mu\nu}_{ab}(p)\equiv\delta_{ab}D^{\mu\nu}(p)$
respectively, ($a,b=1,...,8$ are $SU(3)$ colour indices.  For a covariant
gauge the renormalised gluon propagator and quark propagators have forms
identical to those in \Eq{propagators}.  We can write for the proper
quark-gluon vertex $\Gamma^\mu_a(p',p)\equiv t_a\Gamma^\mu(p',p)$.

The renormalised quantities $S^f$, $\Sigma^f$, $D$, $\Pi$, $\Gamma^f$, $\xi$,
$g$, $m^f$, etc.  have a dependence on the both the renormalisation point
$\mu$ and the gauge parameter $\xi$ in general, but we do not explicitly
indicate these for notational brevity.  The renormalised coupling $g$ is
related to the running coupling constant $\alpha_s(Q^2)=g^2(Q^2)/4\pi$ by
\beq
g^2\equiv g^2(Q^2)\biggl|_{Q^2=\mu^2}
\equiv 4\pi\;\alpha_s(Q^2)\biggl|_{Q^2=\mu^2}\;,
\label{g2toalpha}
\eeq
where $q^2\equiv -Q^2 > 0$.  For $Q^2\gg\Lambda^2_{QCD}$ the running coupling
constant is given by the leading-log [i.e., one-loop] result
\beq
\alpha_s(Q^2)= {12\pi\over (33-2N_f)\ln (Q^2/\Lambda^2_{QCD})}\equiv
{d_M \pi\over \ln(Q^2/\Lambda^2_{QCD})}
\label{alpha_s}
\eeq
where $N_f$ is the number of quark flavors and $\Lambda_{QCD}$ is the scale
parameter of QCD.  We will see that $d_M\equiv 12/(33-2N_f)$ is the anomalous
dimension of the mass.  It was shown by Symanzik (1973) and Appelquist and
Carrazzone (1975) that when $Q^2\ll (m^f)^2$ these quarks can be neglected
since they only give rise to effects of ${\cal O}(Q^2/m^{f2})$.  This is
sometimes referred to as the decoupling theorem [Yndur\'ain (1993),
pp.~75-79].  Note that \Eq{alpha_s} is renormalisation-scheme and gauge
independent, as is its second-order [i.e., two-loop] extension.  The
[renormalised] gauge-parameter $\xi$ is chosen at the renormalisation point
and, in general, it also runs as the renormalisation scale $\mu$ is varied.

At the renormalisation point $\mu$ we impose the boundary conditions
\beqn
D^{\sigma\nu}(p)\biggl|_{p^2\simeq -\mu^2}\simeq\left[-g^{\sigma\nu} +
(1-\xi) {p^\sigma p^\nu\over p^2}\right]\;{1\over p^2}\;&,&\;\;
(S^f)^{-1}(p)\biggl|_{p^2\simeq-\mu^2}\simeq\rlap/p - m^f_\mu\;,\nonumber\\
\Gamma^{f\nu}(p',p)\biggl|_{p'^2\simeq p^2\simeq k^2\simeq -\mu^2}
&\simeq&\gamma^\nu~,
\label{QCD_renorm_BCs}
\eeqn
where $k^\mu\equiv (p'-p)^\mu$.  These are to be understood to imply that
$\Pi (p^2\simeq-\mu^2)\simeq 0$, $\Sigma^f (p^2\simeq-\mu^2)\simeq 0$,
$Z^f(p^2\simeq-\mu^2)\simeq 1$, and $M^f(p^2\simeq-\mu^2)\simeq m^f_\mu$
using the $(Z,M)$ representation of the fermion propagator in \Eq{DSE_AB}.
Here we have introduced the notation $m^f_\mu$ to denote the ``running'' mass
at the renormalisation point $\mu$.  In a nonconfining theory such as QED the
mass $m^f_\mu$ would only correspond to the {\em physical} mass when
$(m^f_\mu)^2=-\mu^2$ such as when the on-shell renormalisation scheme is
used.  The boundary condition on $\Lambda^{f\nu}(p',p)$ should obviously be
chosen so as not to violate the STI relating the quark propagator and the
quark-gluon vertex [see later] but since QCD is an asymptotically free theory
then the boundary condition given above is certainly approximately correct
for $\mu$ sufficiently large.  $\Lambda^{f\nu}(p',p)\simeq 0$ at $p'^2\simeq
p^2\simeq k^2\simeq -\mu^2$. Note that we have used approximation symbols in
specifying \Eq{QCD_renorm_BCs} since the exact boundary conditions depend on
the detailed choice of renormalisation scheme, e.g., MS or $\overline{\rm
MS}$ if we use dimensional regularisation.  In general there is some residual
renormalisation-scheme dependence as well as the obvious
renormalisation-point dependence for the renormalised quantities.  Within a
particular renormalisation scheme the renormalisation-point independence of
physical observables gives rise to the renormalisation group equations, [see,
e.g., Muta (1987) and Yndur\'ain (1993)].

We will now also omit the explicit flavour index on the quarks and restore it
only when necessary.  The STI for the gluon propagator can be written as
$q_\mu\Pi^{\mu\nu}(q)=0$ where
$\Pi^{\mu\nu}_{ab}(q)=\delta_{ab}\Pi^{\mu\nu}(q)$ and
$\Pi^{\mu\nu}(q)=[-g^{\mu\nu}q^2+q^\mu q^\nu]\Pi(q^2)$, just as was the case
for the photon.  Equivalently, we can write $q_\mu q_\nu D^{\mu\nu}(q)=\xi$.
The STI for the quark-gluon vertex can be written as (Marciano and Pagels,
1978, p.~172)
\beq
k_\sigma\Gamma^\sigma (p',p)[1+b(k^2)] =[1-B(k,p)] S^{-1} (p') - S^{-1}(p)
[1-B(k,p)]
\label{quark_gluon_STI}
\eeq
where $p'\equiv p+k$.  The ghost self-energy and ghost-quark scattering
kernel are denoted $b(k^2)$ and $B(k,p)$, respectively.  Note that in Landau
gauge $B(k,p) \simeq 0$ at the renormalisation point ($p^{\prime 2}\simeq
p^2\simeq k^2\simeq -\mu^2$) to lowest order in perturbation theory.  Without
ghosts \Eq{quark_gluon_STI} is identical to the corresponding WTI of QED.  In
general, other symmetries will also give rise to STIs.  In the limit that
there are no explicit chiral symmetry breaking [ECSB] quark masses, QCD is a
chirally symmetric theory and has a STI for the proper vertex for an
isovector axial-vector current [e.g., a W-boson] coupling to a quark,
\beq
k_\sigma\vec\Gamma^{\;\sigma}_5(p',p) = {\vec\tau\over 2}\left[S^{-1}(p')
\gamma_5 + \gamma_5\;S^{-1}(p)\right]~.
\label{axial_vector_WTI}
\eeq
Since ghost terms do not contribute this actually has the form of a WTI.  In
the presence of ECSB quark masses the chiral symmetry WTI is replaced, for
flavour $f$, by [see, e.g., Itzykson and Zuber (1980, pg.~557)]
\beq
k_\sigma\Gamma^{\;f\sigma}_5(p',p) =
[(S^f)^{-1}(p')\gamma_5+\gamma_5(S^f)^{-1}(p)]+2im^f\;\Gamma^f_5(p',p)\;.
\label{axial_vector_WTI_exp}
\eeq
The additional term contains the ECSB quark mass $m^f$ and the quark
pseudoscalar isovector vertex $\vec\Gamma^f_5$.

QCD is well known to be an asymptotically free theory, which means that if
the characteristic momenta in some physical process are sufficiently large
and space-like then QCD behaves as a free theory with logarithmic
corrections.  Let $\Gamma_i(q,g;\mu)$ represent some renormalised Green's
function, where $g$ is the renormalised coupling at the renormalisation point
$\mu$, $q^\nu$ is a large spacelike four-momentum characteristic of the
momenta of the external legs, and where we define, as usual, $Q^2\equiv
-q^2>0$.  In order to use perturbation theory effectively we also need
$Q^2\simeq\mu^2\gg\Lambda_{QCD}^2$ and so, as stated earlier, $g$ is as
defined in \Eq{g2toalpha}.  If we scale all of the external momenta by the
factor $\lambda$ then the characteristic momentum will also scale; i.e.,
$q^\nu\to q'^\nu\equiv \lambda q^\nu$ and $Q^2\to Q'^2\equiv -q'^2=\lambda^2
Q^2$.  The meaning of asymptotic freedom is that to leading order in
$\alpha_s$ we have (for $Q'^2\gg Q^2\gg\Lambda_{QCD}^2$)
\beq
\Gamma_i(\lambda q,g;\mu) \simeq
(\lambda)^{n_i}\Gamma_i(q,g';\mu)({1\over 2}\ln\lambda^2)^{d_i}\;,
\label{Gamma_scale}
\eeq
where we have defined $g=g(\mu^2)$ and $g'=g(\lambda^2\mu^2)$ and where
$g^2(Q^2)$ was defined in \Eq{g2toalpha}.  The exponent $n_i$ is the naive
[canonical] dimension of $\Gamma_i$ and the exponent $d_i$ is defined as its
anomalous dimension, [see, e.g., Yndur\'ain (1993), pp.~66-75, and also
Altarelli (1982) and Reya (1981)].  For example, for $Q^2\equiv
-q^2\gg\Lambda^2_ {QCD}$ we have the leading-log result for the running mass
\beq
M(-Q^2) = {\hat m\over \left[{1\over 2}\ln(Q^2/\Lambda^2_{QCD}\right]^{d_M}}
\;,
\label{MQ2}
\eeq
where $\hat m$ is the renormalisation group invariant mass parameter and
$d_M\equiv 12/(33-2N_f)$ is the anomalous dimension of the mass.  Note that
the asymptotic behaviour of $M(p^2)$ when $m\neq 0$ is gauge independent.
The parameter $\hat m$ sets the scale of the ECSB in QCD and is analogous to
$\Lambda_{QCD}$.  To lowest order the renormalised ECSB mass, $m_\mu$, is
related to $\hat m$ using \Eq{MQ2} by
\beq
m_\mu\equiv M(-\mu^2) = \hat m\biggl/\left[{1\over
2}\ln(\mu^2/\Lambda^2_{QCD} )\right]^{d_M}\;,
\label{ECSBmass}
\eeq
where the renormalisation-point dependence is explicitly indicated.  For
$Q^2\equiv -q^2$ and $Q^2\gg\mu^2\gg\Lambda^2_{QCD}$ the inverse quark
propagator is, to lowest order,
\beq
S^{-1}(q)=\left[Z^{-1}(q^2)\rlap/q\right] = \rlap/q
\left[{1\over 2}\ln(Q^2/\Lambda_{QCD}^2)\right]^{d_S} \;,
\label{SQ2}
\eeq
which gives $Z(-Q^2) = 1/\left[{1\over
2}\ln(Q^2/\Lambda_{QCD}^2)\right]^{d_S}$, where $d_S = -2\xi/(33-2N_f)$ is
the [gauge-dependent] anomalous quark dimension.  Note that $Z(-Q^2) = 1$ to
leading order in Landau gauge [$\xi$= 0]. Similarly, the asymptotic behaviour
of the transverse part of the gluon propagator at leading order is
\beq
D^{\sigma\nu({\rm tr})}(q) = -\left[g^{\sigma\nu} - {q^\mu q^\nu\over
q^2}\right]{1\over q^2}\left[{1\over 2}\ln(Q^2/\Lambda_{QCD}^2)\right
]^{-d_D}\;,
\label{DQ2}
\eeq
where $d_D = (39-9\xi-4N_f)/\left[4(33-2N_f)\right]$ is the gluon anomalous
dimension.  The quark-gluon vertex for $Q^2\equiv -p^2$ and
$Q^2\gg\mu^2\gg\Lambda^2_{QCD}$ is given by
\beq
\Gamma^\sigma(p',p)\biggl|_{p^{\prime 2}=p^2} = \gamma^\sigma\left[{1\over 2}
\ln (Q^2/\Lambda_{QCD}^2)\right]^{d_\Gamma}\;,
\label{GQ2}
\eeq
where $d_\Gamma = -(27+25\xi)/\left[8(33-2N_f)\right]$.  It is
straightforward to verify that $2d_S - 2d_\Gamma + d_D = 1/2$ which, since
$Z_g\equiv Z_\Gamma/Z_F\sqrt{Z_B}$, implies, for example, that the
$\beta$-function is independent of the choice of gauge to leading order, as
already stated.  It is also possible to similarly define a running gauge
parameter $\xi(Q^2)$, where $\xi\equiv \xi(Q^2=\mu^2)$.  These results all
follow from asymptotic freedom and the renormalisation group equations.  It
should be noted that the asymptotic spacelike behaviour of the various
quantities [e.g., $S(p)$, $Z(p^2)$, $M(p^2)$, $D^{\nu\sigma}(q)$, $\Pi(q^2)$,
and $\Gamma^\nu(p',p)$] has no explicit dependence on the renormalisation
point $\mu$, although there is an implicit dependence through the
renormalisation-point dependence of the gauge parameter.

If we consider the electromagnetic couplings of the quarks, then it follows
from electromagnetic gauge invariance that the photon-quark vertex satisfies
\beq
k_\sigma\Gamma_{\rm e.m.}^\sigma(p',p) = S^{-1}(p') - S^{-1}(p)\;.
\label{QCDphotonWTI}
\eeq
This is the same as the corresponding WTI of QED and has no ghost
complications.

In momentum space the renormalised inverse quark propagator is
\beq
S^{-1}(p) = \pslash-m-\Sigma(p)\equiv Z^{-1}(p^2)[\pslash-M(p^2)]\equiv
A(p^2)\pslash-B(p^2)\;,\;\;
\Sigma(p)\equiv \Sigma'(p) -\Sigma'(p)\left|_{p^2=-\mu^2}\right.\;,
\label{Squark}
\eeq
where
\beq
-i\Sigma'(p) = {4\over 3}Z_\Gamma
\;g^2\int {d^{\rm d}\ell\over (2\pi)^{\rm d}}(i\gamma_\mu)
(iS(\ell))(iD^{\mu\nu}(p-\ell))(i\Gamma_\nu(\ell,p))\;.
\label{Sigmaprime_QCD}
\eeq
We have used $\sum\limits_a\;\lambda^a\lambda^a/4 = 4/3$.  The signal that
DCSB has occurred is that the renormalised self-energy develops a nonzero
Dirac-scalar part; i.e., when $\Sigma_s(p^2)\neq 0$.  This leads to a nonzero
value for the quark condensate and, in the limit of exact chiral symmetry,
leads to the pion becoming a massless Goldstone boson.

In general, $\Lambda_{QCD}$ and $\hat m$ are renormalisa\-tion-scheme
dependent and so take different values in, for example, the modified minimal
subtraction [$\overline{MS}$] scheme, the minimal subtraction [$MS$] scheme,
and the momentum subtraction [MOM] scheme.  However, the difference does not
appear in leading order but only at second-order and higher in $\alpha_s$.
If we retain only leading-order terms in the asymptotic region we do not need
to concern ourselves with this.

%
\subsect{Euclidean Space Formulation}
\label{subsect-Euclidean}

Up to this point we have discussed the DSEs in Minkowski space.  To an
audience unfamiliar with studies of lattice gauge theory and the numerical
solution of DSEs this would presumably seem natural.  However, both of these
approaches are typically formulated in Euclidean space.

Our Euclidean space conventions are as follows: 1) we use a non-negative
metric for Euclidean four-vectors
\beq
a\cdot b \, = \, \delta_{\mu\nu}\,a_\mu\,b_\nu = \sum_{i=1}^{4}\,a_i\,b_i
\eeq
where $\delta_{\mu\nu}$ is the Kronecker delta; 2) our Dirac matrices are
hermitian and satisfy the algebra
\beq
\{\gamma_\mu,\gamma_\nu\} = 2\,\delta_{\mu\nu}~;
\eeq
and we have
\beq
\gamma_5 = -\,\gamma_1\,\gamma_2\,\gamma_3\,\gamma_4
\eeq
so that
\beq
{\rm tr}[\gamma_5\,\gamma_\lambda\,\gamma_\mu\,\gamma_\nu\,\gamma_\rho] =
-\,4\,\epsilon_{\lambda\mu\nu\rho}
\eeq
where $\epsilon_{\lambda\mu\nu\rho}$ is the completely antisymmetric
Levi-Civita tensor in ${\rm d}=4$ dimensions.  One realisation of this algebra is
\beqn
\gamma_4^E = \gamma^0 &\;\;{\rm and} \;\; 
        & \gamma_j^E = -i\gamma^j~,\;\; j=1,2, 3,
\eeqn
where $\gamma^0$ and $\gamma^j$ can be any one of the commonly used Minkowski
space representations of the usual Dirac algebra.  We note that with these
conventions a spacelike four-vector, $p_\mu$, has \mbox{$p^2 > 0$}.

A straightforward transcription procedure can be employed to determine the
action for the Euclidean field theory that corresponds to one formulated in
Minkowski space:
\beqn
\int^M\,d^4 x^M &\to & - i\,\int^E\,d^4 x^E ~,\label{TMEa}\\
\not\!\partial &\to & i\gamma^E\cdot \partial^E~,\\
\not\!\!A &\to & -i\gamma^E\cdot A^E~,\\
A_\mu\,B^\mu &\to & - A^E\cdot B^E ~,\label{TMEd}
\eeqn
where, as usual, $\not\!\!A$ represents \mbox{$g_{\mu\nu}\gamma^\mu_M
A_M^\nu$}.  These transcription rules can be used as a blind implementation
of an analytic continuation in the time variable, $x^0$: \mbox{$x^0
\to -i\,x_4$} with \mbox{$\vec x^M\to \vec x^E$}, etc.

Employing these rules in Eqs.~(\ref{Sqed}) and (\ref{Zqed}) it is easy to
verify that the generating functional for Euclidean \qedd is
\beq
{\cal Z}^E[\etab^E,\eta^E,J_\mu^E]=
\int\,d\mu^E(\psib^E,\psi^E,A^E)\,
\exp\left( -S[\psib^E,\psi^E,A_\mu^E] 
+ \int\,d^{\rm d} x\;\left[ \psib^{fE}\eta^{fE} + \etab^{fE}\psi^{fE} + A_\mu^E
J^{\mu\,E} \right]
\right)
\label{EZqed}
\eeq
with the action
\beq
S^E[\psib^E,\psi^E,A_\mu^E] = \int\, d^{\rm d} x^E\; \left[
\,\sum_{f=1}^{N_f}\,
\psib^{fE} \left( \gamma^E\cdot \partial^E +  m_0^f 
+ i e_0^f \gamma^E\cdot A^E \right)\psi^{fE} + \frac{1}{4}
F_{\mu\nu}^EF^{\mu\nu\,E}
\right]~.
\label{ESqed}
\eeq
[The analogue of this result in QCD is obvious.]

Equations (\ref{EZqed}) and (\ref{ESqed}) are the foundation of Lagrangian
lattice studies of \qedd; as are their analogues for other field theories.
When formulated on a discrete lattice, the measure in Eq.~(\ref{EZqed}) is
non-negative and this makes a direct numerical simulation of the theory
possible using the standard tools of statistical mechanics, which is an
important way in which lattice gauge theory benefits from being formulated in
Euclidean space.

In addition, the discrete lattice formulation in Euclidean space has allowed
some progress to be made in attempting to answer existence questions for
interacting gauge field theories (Seiler, 1982).  In fact, it is possible to
view the Euclidean formulation of a field theory as definitive [see, for
example, Symanzik (1969)].  The moments of the Euclidean measure, which can be
obtained operationally via functional differentiation of Eq.~(\ref{EZqed}) and
setting the sources to zero, are the Schwinger functions:
\beq
{\cal S}^n(x^1,\ldots,x^n)~,
\eeq
which are sometimes called Euclidean space Green functions.  Given a measure
[i.e.,the piece of Eq.~(\ref{EZqed}) that does not involve source terms] and
given that it satisfies certain conditions [i.e.,the Wightman and
Haag-Kastler axioms], then it can be shown that the Wightman functions,
\mbox{${\cal W}^n(x_1,\ldots,x_n)$}, can be obtained from the Schwinger
functions by analytic continuation in each of the time coordinates:
\beq
\label{WTS}
{\cal W}^n(x_1,\ldots,x_n) =
\lim_{x_4^i\rightarrow 0}\, 
{\cal S}^n([\vec{x}^1,x_4^1+ix_1^0],\ldots,[\vec{x}^n,x_4^n+ix_n^0])
\eeq
with \mbox{$x_1^0 < x_2^0 <\ldots < x_n^0$}.  These Wightman functions are
simply the vacuum expectation values of products of field operators from
which the Green functions [i.e., the Minkowski space propagators] are
obtained through the inclusion of step [$\theta$] functions in order to
obtain the appropriate time ordering.  [This is described in some detail in
Streater and Wightman (1980, Appendix); Seiler (1982); Glimm and Jaffe,
(1987, Chap. 6).]  Thus the Schwinger functions contain all of the
information necessary to calculate physical observables; this notion is used
directly in obtaining masses and charge radii in lattice simulations of QCD.
The same notion can be employed in the DSE approach since the Euclidean space
DSEs in \qedd can all be obtained from \Eq{EZqed} [and the DSEs for other
theories obtained from their respective Euclidean generating functionals] and
the solutions of these equations are the Schwinger functions.

\subsubsect{Euclidean $\rightarrow$ Minkowski Continuation}
In studying DSEs it has been commonplace to obtain DSEs from the Minkowski
space generating functional and then use transcription rules in momentum
space [i.e., $k^0\to ik_4$ and $\vec k^M\to -\vec k^E$]
\beqn
\int^M\,d^4 k^M & \rightarrow & i\,\int^E\,d^4 k^E ~,\label{TMEpa}\\
\not\! k & \rightarrow &  i\,\gamma^E \cdot k^E~, \label{TMEpb}\\
k_\mu q^\mu &\rightarrow & -\,k^E\cdot q^E~,\label{TMEpc}\\
 k_\mu x^\mu & \rightarrow & k^E\cdot x^E~,
\eeqn
[the last of which appears in four-dimensional Fourier transforms
and follows since $x^0\to -ix_4$ and $\vec x^M\to\vec x^E$] 
which are analogues of Eqs.~(\ref{TMEa})-(\ref{TMEd}), to obtain equations in
Euclidean metric which are assumed to be the Euclidean space counterparts of
the original equations.  This approach is often argued to be connected with
the ``Wick Rotation'' (Wick, 1954) and is discussed in most text books in
association with dimensional regularisation.  [In the following we take
\mbox{$(k^E \rightarrow \,-\,k^E)$} so that the inverse of the free-fermion
propagator is \mbox{$(i\gamma\cdot k +m$)}.]  As remarked in Itzykson and
Zuber (1980, pg.  485), this procedure is easy to justify in perturbation
theory when one neglects the possibility of dynamically generated
singularities in the first and third quadrants of the complex \mbox{$p^0$}
plane, however, in connection with {\em nonperturbative} studies this simple
assumption is highly nontrivial and in model studies is often incorrect.

Indeed, in a study in Euclidean space of an approximate DSE for the electron
in \qedf (Atkinson and Blatt, 1979) it was found that, instead of a single
physical branch point on the timelike $p^2$ axis, the electron propagator had
complex conjugate branch points whose position depended on the gauge
coupling.  A straightforward application of transcription rules is invalid in
this case.  Subsequently there have been a number of additional studies of
approximate and/or model DSEs for the electron propagator in \qedf (Maris
1993) and the quark propagator in QCD (Stainsby and Cahill, 1990; Maris and
Holties, 1992; Burden \etal, 1992b; Stainsby and Cahill, 1992; Stainsby,
1993; Maris, 1993) which have addressed this question; i.e., whether
singularities are generated dynamically in the complex $p^0$ plane which
complicate or even prohibit the Wick Rotation.  In each of the cases studied
so far the straightforward transcription has been found to be invalid.
Indeed, in one of these studies (Burden \etal, 1992b) the
quark propagator had an essential singularity in $p^2$ at timelike infinity
which completely prohibits the rotation of the $p^0$ contour.

In each of these cases the solution of that Minkowski space equation which is
obtained via a straightforward transcription of the Euclidean space equation
has no relation to the analytic continuation of the solution of the Euclidean
space equation.  This raises the question of how one may proceed between
Euclidean and Minkowski space in nonperturbative studies.  A possible answer
has already been given above; i.e., one can define the [model] field theory
in Euclidean space and solve for the Schwinger functions.  Assuming then that
the field theory ``exists'', in the sense that all of the necessary axioms
are satisfied, the Wightman functions are obtained as the analytic
continuation of the Schwinger functions, \Eq{WTS}, from which one may obtain
the Green functions and physical observables.  It should be noted that the
observed invalidity of the straightforward transcription procedure in
momentum space DSEs may be entirely due to the limitations of the various
model systems that have been studied and that no such problem is manifest in
the true tower of equations.  At the present time there appears to be no
definitive answer to this question.

In the following we will adopt the point of view that nonperturbative studies
of field theories are most easily defined in Euclidean space with physical
observables being obtained from the Schwinger functions by analytic
continuation.  As we have described above, this approach is the same as that
taken by the practitioners of lattice gauge theory and we emphasise that, at
least in principle, all physical observables can be obtained in this way.

\sect{Three-dimensional Quantum Electrodynamics}
\label{sect-QED3}
It will be seen upon comparison of Secs.~\ref{subsect-QED} and
\ref{subsect-QCD} that the DSEs in Abelian gauge theories are much simpler
because of the decoupling and effective absence of ghost fields in all
[linear] gauge fixing schemes.  For the same reason the Ward-Identities
relating $n$-point functions take a relatively simple form.  This makes
Abelian gauge theories an ideal framework within which to learn and gain
experience in solving a field theory using the tower of DSEs.

In this section we will focus on \qedt which, in addition to being useful for
pedagogical reasons, is an interesting theory in its own right.  We will work
almost exclusively in Euclidean space and use the conventions of
Sec.~\ref{subsect-Euclidean} but we will not explicitly indicate Euclidean
space quantities with the superscript $E$: this is to be understood where
appropriate. As an illustration we note that in our conventions we then
have, for example,
\beqn
& & 
\begin{array}{cc}
\displaystyle S(p)=\frac{1}{i\gamma\cdot p + m +\Sigma(p)}\;,
&  \; i\Gamma_\mu(p',p)=i\gamma_\mu +i\Lambda_\mu(p',p) 
\end{array} \\
& &
   \;\;{\rm and}\;\;  
\displaystyle 
D_{\mu\nu}(q)=\left(\delta_{\mu\nu}-\frac{q_\mu q_\nu}{q^2}\right)
\,D(q) + \xi\,\frac{q_\mu q_\nu}{q^4}~, 
\label{DmunuE} \\
& & \;\; {\rm where} \;\; \displaystyle
\label{Dq}
D(q) = \frac{1}{q^2\,[1+\Pi(q^2)] }~.
\eeqn
[We discuss our \qedt Dirac matrix conventions after \Eq{EProp}.] 

For those concerned with QCD and strongly interacting theories \qedt is
interesting because, in the absence of fermion loop contributions to the
photon polarisation tensor; i.e., in quenched approximation, the lattice
version of the theory has been shown to be confining (G\"{o}pfert and Mack,
1982).  This result can be seen heuristically by considering the classical
potential:
\begin{eqnarray}
\label{Vclass}
V(\vec{x}) & \equiv &
\int_{-\infty}^{\infty}dx_{3}\,\int\frac{d^3q}{(2\pi)^3}
\mbox{e}^{i(\vec{q}\cdot\vec{x}+q_{3}x_{3})}e^{2}D(q)
 =  \int \frac{d^{2}q}{(2\pi)^2}\, \mbox{e}^{i\vec{q}\cdot\vec{x}}\,
e^{2}D(\vec{q})~.
\end{eqnarray}
Neglecting the vacuum polarisation; i.e., setting \mbox{$\Pi(q^2)=0$}, one
obtains
\begin{eqnarray}
V(r) & = & \frac{e^{2}}{2\pi} \ln\left( r e^{2}\right)
\end{eqnarray}
which exhibits logarithmic confinement.

It is also of interest to those concerned with the dynamical generation of
mass in grand unified theories.  This is because the coupling in \qedt,
$e^2$, has the dimensions of mass.  In addition, the theory is
super-renormalisable which means that there are no ultraviolet divergences
whose regularisation would introduce a new scale that could complicate the
relation between the scale of dynamical mass generation and the natural
length scale provided by the coupling, $e^2$.

Another application is the study of field theory at finite temperature .  It
has been argued that the high temperature behaviour of a given
four-dimensional gauge theory, which has a small coupling, at distances
greater than some screening length is precisely described by the
corresponding three-dimensional theory (Appelquist and Pisarski, 1981).  It
is worth noting that \qedt at finite temperature, with fermions coupling to
two Abelian gauge fields [i.e., the usual electromagnetic field plus a
``statistical'' gauge field] has been applied to a description of high
T$_{\rm c}$ superconductivity in the quasi-planar oxides La$_2$CuO$_4$ and
YBa$_2$Cu$_6$ (Dorey and Mavromatos, 1992).

In every one of these cases it is the dynamical generation of mass; i.e., the
dynamical breaking of chiral symmetry, and its relation to, and effect on,
the gauge boson propagator that is of interest.  These quantities are most
easily studied within the DSE approach.  Indeed, an order parameter for DCSB
is the fermion condensate, \psibpsi, which is obtained from the fermion
propagator, \mbox{$S_F(x)$}, via:
\mbox{$\psibpsi = -{\rm tr}\,S_F(x=0) $}, when both the Lagrangian, $m_0$,
and renormalised, $m$, bare masses are zero.  The DSE for the
renormalised fermion self
energy, which is discussed in Sec.~\ref{FermDSE}, is therefore a natural
equation to study in order to address the issue of DCSB and in Euclidean
space it can be written as:
\begin{eqnarray}
\Sigma (p) & = & i \gamma\cdot p \,(Z_1 - 1) - (m - Z_1\,m_0)
+ Z_1 e^{2}\int \dqbt
\Gamma_{\mu}(p,q) 
                D_{\mu\nu}(p-q;\xi)S_{F}(q)\gamma_{\nu}~,
\label{eq:ERDSE}
\end{eqnarray}
where the Ward Identity $Z_1 = Z_2$ has been used, and in Euclidean space the
general form of the solution can be written in the form
\beq
\label{EProp}
S(p) = -\,i\gamma\cdot\,p\,\sigma_V(p^2) + \sigma_S(p^2) \;
\equiv \; -\,i\gamma\cdot\,p\,\frac{A(p^2)}{p^2 A(p^2)^2 + B(p^2)^2}
        + \frac{B(p^2)}{p^2 A(p^2)^2 + B(p^2)^2}~.
\eeq
[Note that we are no longer using a tilde to denote these renormalised
quantities.]  In the absence of bare mass a non zero value of \psibpsi
signals DCSB and a plot relating \psibpsi to relevant dimensionless
parameters in a given model can be used to study the transition to the
chiral symmetry breaking phase.

In \mbox{${\rm d}=3$} dimensions there are two inequivalent \mbox{$2\times
2$} representations of the algebra \mbox{$\{\gamma_{\mu},\gamma_{\nu}\} =
2\delta_{\mu\nu}$}.  Hence, to describe spinorial representations of the
Lorentz group, two component spinors are sufficient.  However, in this case
any mass term, whether explicit or dynamically generated, has the undesirable
property that it is odd under parity transformations.  This can be avoided if
one employs four component spinors and a
\mbox{$4\times 4$} representation of the Euclidean Dirac algebra; for
example, the set
\mbox{$\gamma_1$, $\gamma_2$, $\gamma_4$}.  It is clear that since both
$\gamma_5\equiv -\gamma_1\,\gamma_2\,\gamma_4$ and $\gamma_3$ anticommute
with this set then the massless theory is invariant under two
transformations:
\beqn
\psi \rightarrow e^{i \alpha \gamma_3} \psi \; 
& \;\; {\rm and} \;\; & 
\psi \rightarrow e^{i \alpha \gamma_5} \psi \label{CHab}
\eeqn
which are analogous to the chiral transformations in ${\rm d}=4$ dimensions.  In
this case there are two types of mass term
\beqn
m_1 \overline{\psi} \psi & \;\;{\rm and} \;\; & m_2 \overline{\psi}
\mbox{\small$\frac{1}{2}$}[\gamma_3,\gamma_5] \psi~.
\eeqn
The first of these is invariant under parity transformations but not under
Eqs.~(\ref{CHab}) while the second is invariant under the transformations of
Eqs.~(\ref{CHab}) but not under parity.  The $m_1$ term is clearly the ${\rm
d}=3$ analogue of the mass term in four-dimensions and it is common practice
to use only this type of mass term in \qedt (Pisarski, 1984).

\subsect{Quenched Approximation}
In massless \qedt the quenched approximation, which corresponds to neglecting
fermion loop contributions to the vacuum polarisation; i.e., to setting
\beq
\label{QuenchedA}
\Pi(q^2)\,\equiv\,0
\eeq
in \Eq{Dq}, leads to infrared divergences in ordinary perturbation theory
based on an expansion in the coupling, $e^2$.  This divergence is evident in
the lowest order vertex correction:
\beqn
\Lambda_\rho (k,p) & \sim & e^2\,\int\dqbt\, \frac{1}{q^2} 
\left(\delta_{\mu\nu} - (1-\xi)\frac{q_\mu q_\nu}{q^2}\right)
\gamma_\mu \frac{1}{i\gamma\cdot (p-q)}\gamma_\rho
        \frac{1}{i\gamma\cdot (k-q)}\gamma_\nu~.
\eeqn
In the limit $p\rightarrow\,0$, $q\rightarrow\,0$ this is
\beq
\label{IRdiv}
 \propto \, \left(\xi\,-\,\mbox{\small$\frac{2}{3}$} \right)
\int\dqbt\,\frac{1}{q^4}
\eeq
which has a manifest infrared divergence.  [In renormalising \qedt this
divergence is incorporated into the vertex renormalisation constant, $Z_1$,
which is described in Sec.~\ref{QEDRen}.  For ${\rm d}=4$ the gauge dependent
factor is \mbox{$ \xi - 1$} and hence $Z_1 = 1$ in Landau gauge at {\cal
O}($e^2$) in Abelian theories.]

One commonly used remedy for this problem is to soften the infrared behaviour
of the gauge boson propagator by including the fermion loop contribution to
the vacuum polarisation (Appelquist \etal, 1986).  For a fermion of mass
$m$ the lowest order [in $e^2$] contribution is given by
\begin{equation}
\Pi_{\mu\nu}(k) = \,-\,e^2 \int \dqbt \, 
\mbox{tr}\left[ \gamma_\mu \,\frac{1}{i\gamma\cdot (q+\hlf k) + m}\,
        \gamma_\nu \,\frac{1}{i\gamma\cdot (q-\hlf k) + m} \right]~.
\label{eq:Pi}
\eeq
Evaluating this using a gauge invariant regularisation scheme, such as
dimensional regularisation, one obtains [recall \Eq{PiTran}]
\beq
\label{PimnE}
\Pi_{\mu\nu}(k) = 
\left(\delta_{\mu\nu} - \frac{k_\mu\,k_\nu}{k^2}\right)\,\Pi(k^2)
\eeq
where the polarisation scalar is
\beqn
\label{PiScal}
\Pi(k) & = & \frac{e^2}{4\pi k^2} 
\left[ 2m+\frac{k^2 - m^2}{k} \, 
{\rm arcsin}\left(\frac{k}{\sqrt{k^2+4 m^2}}\right) \right]~.
\eeqn
In a theory with $N$ massless fermions the polarisation scalar is therefore
\beq
\label{PPS}
\Pi(k)  =  \frac{\tilde{\alpha}}{k}~,
\eeq
where $\tilde{\alpha} = N_f e^2/8$, in which case the gauge boson propagator is
given by \Eq{DmunuE} with
\beq
\label{DqN}
D(q) = \frac{1}{q^2 + \tilde{\alpha} q}
\eeq
which behaves as $1/q$ for $q^2 \approx 0$; i.e., the infrared divergence is
softened significantly without altering the appealing ultraviolet properties.

Allowing such a contribution from massless fermions introduces a
dimensionless parameter, $N_f$, the number of fermions, into the fermion DSE.
This breaks the scale invariance of this equation and makes possible the
existence of a phase transition associated with DCSB; i.e., a situation where
\psibpsi is nonzero for some values of $N_f$ and zero otherwise.  This is only
possible when there is a dimensionless parameter in the equation.  [There
have been a number of studies of this issue and we will discuss them below.]

However, the price of allowing such a contribution from massless fermions is
to eliminate confinement.  It is a simple matter to calculate the classical
potential, \Eq{Vclass}, in this case and one obtains (Burden and Roberts,
1991)
\begin{eqnarray}
V(r) & = & -\frac{e^{2}}{4}\left[ \mbox{\bf H}_{0}(\tilde{\alpha}r) -
N_{0}(\tilde{\alpha}r)\right]~,
\label{eq:mp}
\end{eqnarray}
where \mbox{${\bf H}_{0}(x)$} is a Struve Function and
\mbox{$N_{0}(x)$} a Neumann Function (Gradshteyn and Ryzhik, 1980, Secs.
8.4, 8.5).  From \Eq{eq:mp} one finds easily that
\beqn
V(r) \sim \ln (\tilde{\alpha}r)\;\;{\rm at~small}\;\; r \;\; 
& \;\; {\rm and} \;\; &
V(r) \simeq -\frac{e^{2}}{2\pi}\frac{1}{\tilde{\alpha}r}
\;\;{\rm at~large}\;\; r
\eeqn
and thus observes that the theory is no longer confining.  This deconfinement
is akin to that which occurs in lattice simulations with dynamical fermions:
the fixed source charges dress themselves with a cloud of massless, charged
fermions which strongly screen the source charge.

\subsect{Rainbow Approximation}
In studying DCSB using \Eq{eq:ERDSE} a commonly used approximation is to
write
\beq
\label{RA}
\Gamma_{\mu}(p,q) = \gamma_\mu
\eeq
which is referred to as the ``rainbow'' or ``ladder'' approximation [it is
the DSE analogue of the ladder approximation in the BSE].  If this
approximation is combined with the quenched approximation,
\Eq{QuenchedA}, then \Eq{eq:ERDSE} decouples from the remaining DSEs and it
can be studied in isolation.  This is the merit of this approximation,
however, there are significant problems associated with rainbow
approximation; notably the loss of gauge covariance and a solution for the
fermion propagator which has an unphysical singularity structure.

A commonly used simplification of \Eq{eq:ERDSE} is to set $Z_1\,=\,1$.  As
remarked in connection with \Eq{IRdiv}, this is true in \qedf at O($e^2$) but
in \qedt it is just a convenient simplifying truncation whose validity can
only be justified {\it a posteriori}.  [The proponents of a $1/N_f$ expansion
in \qedt, where $N_f$ is the number of light fermions [see \Eq{eq:Pi} and the
associated discussion] may argue that $Z_1\,=\,1$ and rainbow approximation
form a consistent pair of approximations valid at leading order in $1/N_f$~.]

In quenched, massless, rainbow approximation \qedt one finds easily that
\beq
A(p^2) \equiv 1
\eeq
in Landau gauge, $\xi=\,0$ in \Eq{DmunuE}, and obtains the following
nonlinear integral equation for $B(p^2)$:
\begin{eqnarray}
B(p^2) & = & \frac{2+\xi}{4\pi^{2}p}\int_{0}^{\infty}\, dq\,
\ln\left|\frac{p+q}{p-q}\right|
\frac{qB(q^2)}{q^2 +B^{2}(q^2)}~.
\label{eq:SDSE}
\end{eqnarray}
In other gauges $A(p^2;\xi\neq\,0)$ quickly approaches $1$ as $p^2$ increases
and it is a good approximation to use $A(p^2)=1$ in most studies of quenched,
massless, rainbow approximation \qedt.

An operator product expansion in \qedt suggests that when propagating in the
presence of a condensate \mbox{$\psibpsi\neq 0$} the fermion propagator will
receive a self mass contribution of the form (Burden and Roberts, 1991):
\begin{equation}
-\, 2\pi^3\, \psibpsi \delta^{3}(q)\;.
\end{equation}
The factor $\delta^{3}(q)$ is present because, by definition, the condensate
does not exchange momentum with the fermion and the numerical factors are
simply to ensure appropriate normalisation.  Including this term as a
perturbative contribution to the fermion propagator one finds
\begin{eqnarray}
S(q) & = & \frac{1}{i\gamma\cdot q} - \frac{2+\xi}{4}\,\frac{\psibpsi}{q^4} +
\ldots
\label{eq:SOPE}
\end{eqnarray}
and the OPE analysis then predicts that as \mbox{$p^2 \rightarrow \infty$}
\begin{eqnarray}
\frac{4}{2+\xi}\, p^2\, \Sigma(p) & \rightarrow & - \psibpsi~.
\label{eq:OPE}
\end{eqnarray}
This provides a means of checking, in the large-$p^2$ domain, the
approximations and solution procedure used to analyse the \qedt fermion DSE.

The large-$p^2$ asymptotic behaviour of the solution of \Eq{eq:SDSE} can be
obtained by analysing an approximate differential equation that is valid in
this domain.  Approximating the kernel as follows:
\begin{eqnarray}
\ln\left|\frac{p+q}{p-q}\right| & \approx &
\frac{2q}{p}\theta(p-q) + \frac{2p}{q}\theta(q-p)\;,
\end{eqnarray}
which is a good approximation for \mbox{$p^2\ll q^2$} or
\mbox{$p^2\gg q^2$} (Roberts and McKellar, 1990), one obtains the following
differential equation:
\begin{eqnarray}
\frac{d}{dp}\left(p^{3}\frac{d}{dp}B(p)\right) + 
\frac{2+\xi}{\pi^{2}}\,B(p)
& = & 0~.
\label{eq:de}
\end{eqnarray}
A differential equation valid over a greater $p^2$-domain could be obtained
using the methods discussed by Munczek and McKay (1990), however, this is not
necessary for a comparison with \Eq{eq:OPE}.

The solution of \Eq{eq:de} that is consistent with the ultraviolet boundary
condition \mbox{$B(p^2) \rightarrow 0$} as \mbox{$p^2\rightarrow\infty$} is
\begin{eqnarray}
B(p) & = & \kappa \frac{1}{p}
J_{2}\left(\sqrt{\frac{4(2+\xi)}{\pi^{2}p}}\right)
\;\;  \approx  \;\; 
\kappa\, \frac{2+\xi}{2\pi^{2}} \, \frac{1}{p^2}~,
\mbox{\hspace*{10mm}$p^2\sim\infty$}~,
\label{eq:deuv}
\end{eqnarray}
where \mbox{$J_{2}(x)$} is a Bessel function of integer order and $\kappa$ is
a constant that cannot be determined by the differential equation.  It can,
however, be determined by comparing \Eq{eq:deuv} with \Eq{eq:OPE}:
\begin{eqnarray}
\label{kappav}
\kappa & = & \, - \,\frac{\pi^{2}}{2}\psibpsi\;.
\end{eqnarray}

Burden and Roberts (1991) tested these predictions [Eq.~(\ref{eq:OPE}) c.f.
Eqs.~(\ref{eq:deuv}) and (\ref{kappav})] with the solution of \Eq{eq:ERDSE}
in quenched, massless, rainbow approximation and with $Z_1=\,1$.  In these
studies $A(p^2$) was not neglected and the coupled, nonlinear integral
equations for $A(p^2)$ and $B(p^2)$ were solved by iteration.  In \qedt, with
a \mbox{$4\times 4$} representation of the Dirac Matrices,
\begin{eqnarray}
\psibpsi & = & -\frac{2}{\pi^{2}}\int_{0}^{\infty}\,dp\,
\frac{p^2 B(p^2)}{p^2 A^{2}(p^2) + B^{2}(p^2)}\;,
\end{eqnarray}
which is a convergent integral.  As will be seen in \Table{OPEcf} there is
excellent agreement between the predictions and the numerical results.  This
indicates that at large spacelike $p^2$ the quenched, massless, rainbow
approximation DSE with $Z_1=\,1$ is a good approximation.
\begin{table}[tbh]
\label{OPEcf}
\begin{center}
\parbox{130mm}{\caption{
Comparison of the asymptotic form of the fermion self mass with \psibpsi to
test the OPE prediction.  [Adapted from Burden and Roberts (1991).] }}
\begin{tabular}{|c|r|c|c|}\hline
$\xi$ & p & $\displaystyle \frac{4}{2+\xi}\,p^2\, B(p^2)$ & $-\psibpsi$ \\ &
units of $e^2$ & units of $10^{-3}\,e^4$ & units of $10^{-3}\,e^4$
\\ \hline
 -0.2 & 511 & 2.638 & 2.638 \\ \cline{2-3} & 1000 & 2.638 & \\ \hline 
0.5 & 511 & 1.775 & 1.775 \\ \cline{2-3} & 1000 & 1.775 & \\ \hline 
1.2 & 511 & 1.352 & 1.352 \\ \cline{2-3} & 1000 & 1.352 & \\ \hline
\end{tabular}
\end{center}
\end{table}

\subsect{Beyond rainbow approximation}
\label{beyond-rainbow-QED3}
The loss of gauge covariance when using the rainbow approximation is a direct
consequence of the violation of the Ward Identity:
\beq
(p-q)_\mu\,i\Gamma_\mu(p,q) = S^{-1}(p) - S^{-1}(q)~.
\eeq
It is clear that if the solution of the fermion DSE, \Eq{eq:ERDSE}, is
momentum dependent, as it will be if the kernel is, then this identity is not
satisfied by the bare vertex.  The correct form of the fermion--gauge-boson
vertex is therefore crucial in restoring gauge covariance to the DSE.

The structure of this vertex has been analysed to order $\alpha^2$ in Abelian
theories by Ball and Chiu (1980).  One important result of this study is that
the dressed vertex should be free of kinematic singularities; i.e., that
$\Gamma_{\mu}(p,q)$ should have a well defined limit as
$p^2\rightarrow\,q^2$.  With this in mind, quenched \qedt was studied by
Burden and Roberts (1991) with the following ``light-cone regular'' Ansatz
for the fermion--gauge-boson vertex:
\begin{eqnarray}
i\Gamma_{\mu}(p,q) & = & i\left[a A(p^2) +(1-a) A(q^2)\right]\gamma_{\mu}
\nonumber\\
& + & \frac{(p+q)_\mu}{p^2 - q^2}\,
\left\{ i\left[ A(p^2)-A(q^2)\right]
                           \left[ (1-a)\gamma\cdot p + a \gamma\cdot q\right]
+\left[ B(p^2) - B(q^2)\right]\right\}~.
\label{eq:TEVtx}
\end{eqnarray}
This vertex is a simple modification of that proposed by Ball and Chiu
(1980).

It is important to note that the constraint that the vertex be regular as
$p^2\rightarrow\,q^2$ entails that it cannot be purely longitudinal and hence
that the solution of the DSE will be sensitive to the Ansatz even in Landau
gauge; $\xi=\,0$ in \Eq{DmunuE}.  The parameter $a$ in
\Eq{eq:TEVtx} was included in order to allow, in a simple way, for a
variation of the transverse part of the vertex so that its effect on the
solution could be studied.

Burden and Roberts (1991) studied \Eq{eq:ERDSE} in the quenched, massless
limit, with the vertex Ansatz of \Eq{eq:TEVtx} and with the additional
approximation of setting $Z_1\,=\,1$.  In this case the fermion DSE reduces
to a pair of coupled, one-dimensional integral equations:
\begin{eqnarray}
\lefteqn{ p^2\left(A(p^2)-1\right)  =  \frac{1}{4\pi^2} \int_0^\infty
      \frac{q^2 \, dq}{q^2 A(q^2)^2 + B(q^2)^2} \times } \nonumber \\ & &
\left( \xi \left\{\frac{1}{p^2-q^2} \left[ \left\{q^2A(q^2) -
p^2A(p^2)\right\}A(q^2) - \left\{B(p^2)-B(q^2)\right\}B(q^2) \right]
\right.\right.  \nonumber \\ & & + \left. \frac{1}{2pq}\ln \left|
\frac{p+q}{p-q} \right| \left[ \left\{q^2A(q^2) + p^2A(p^2)\right\}A(q^2) -
\left\{B(p^2)-B(q^2)\right\}B(q^2) \right] \right\} \nonumber \\ & & +
\frac{2}{p^2-q^2} \left( 1- \frac{p^2+q^2}{2pq} \ln
\left|\frac{p+q}{p-q}\right|\right) \nonumber \\ & & \left. \times \rule{0
mm}{5 mm} \left[ \left\{ (1-a)p^2+aq^2 \right\} \left\{A(p^2)-A(q^2) \right\}
A(q^2) + \left\{B(p^2)-B(q^2) \right\} B(q^2) \right] \right) \label{a4}
\end{eqnarray}
and
\begin{eqnarray}
\lefteqn{ B(p^2)  =   \frac{1}{4\pi^2} \int_0^\infty
      \frac{q^2 \, dq}{q^2 A(q^2)^2 + B(q^2)^2} \times } \nonumber \\ & &
\left\{\frac{1}{p^2-q^2} \left[ (\xi-2) + \frac{p^2+q^2}{pq} \ln
\left|\frac{p+q}{p-q}\right| \right] \left\{ A(p^2)B(q^2) - A(q^2)B(p^2)
\right\} \right.  \nonumber \\ 
& & + \left.\frac{1}{2pq} \ln \left| \frac{p+q}{p-q}\right| 
\left[ 4(1-a)A(q^2)B(q^2) + \rule{0 mm}{5 mm}
(4a+\xi)A(p^2)B(q^2) + \xi A(q^2)B(p^2) \right] \right\}~.\label{b4}
\end{eqnarray}

These equations are scale invariant.  The mass scale is set by \mbox{$\mu =
e^2$} and the solution for any value of $\mu$ can be obtained from the
\mbox{$\mu = 1$} solution by the scale transformation:
\begin{equation}
\begin{array}{ccc}
A(p^2;\mu) = A\left(\frac{p^2}{\mu^{2}};1\right) & \;\;\;\;{\rm and} \;\;\;\;
& B(p^2;\mu) = \mu B\left(\frac{p^2}{\mu^{2}};1\right)~.
\end{array}
\end{equation}
It suffices therefore to solve the equations for \mbox{$\mu = 1$} and hence
this choice has been made in Eqs.~(\ref{a4}) and (\ref{b4}).  As remarked
above, this feature of scale invariance means that there can be no critical
coupling parameter in quenched, massless \qedt because there are no
dimensionless parameters: once a chiral symmetry breaking solution exists for
one value of $e^2$ it exists for all values of $e^2$.

The most commonly used procedure for solving this type of coupled, nonlinear
integral equations is simple iteration: a guess is made for $A(p^2)$ and
$B(p^2)$ and substituted; the result of the integration is then resubstituted
and the procedure repeated until the functions in the integrand reproduce
themselves.  This is a practical and efficacious approach.  There are
analytic methods that can be used to establish the uniqueness of the
solutions of certain nonlinear integral equations (McDaniel \etal, 1972) but
hitherto no attempt has been made to employ them in the DSE approach.  The
numerical stability of the solutions has been used as an implicit indication
of uniqueness.

This procedure can be used to solve Eqs.~(\ref{a4}) and (\ref{b4}) with the
result that for all values of the parameter $a$ in \Eq{eq:TEVtx}
\begin{equation}
A(p^2)\not\equiv 1 \mbox{ even for } \xi = 0\;.
\end{equation} 
This is a direct result of the transverse piece in \Eq{eq:TEVtx}; i.e., of
the requirement that a vertex which satisfies the Ward identity be regular in
the limit $p^2\rightarrow\,q^2$.  [A study of a ``light-cone singular''
vertex that has been used by a number of authors confirms this (Burden and
Roberts, 1991).]
\begin{figure}[tb] 
 \psrotatefirst \centering{\
\epsfig{figure=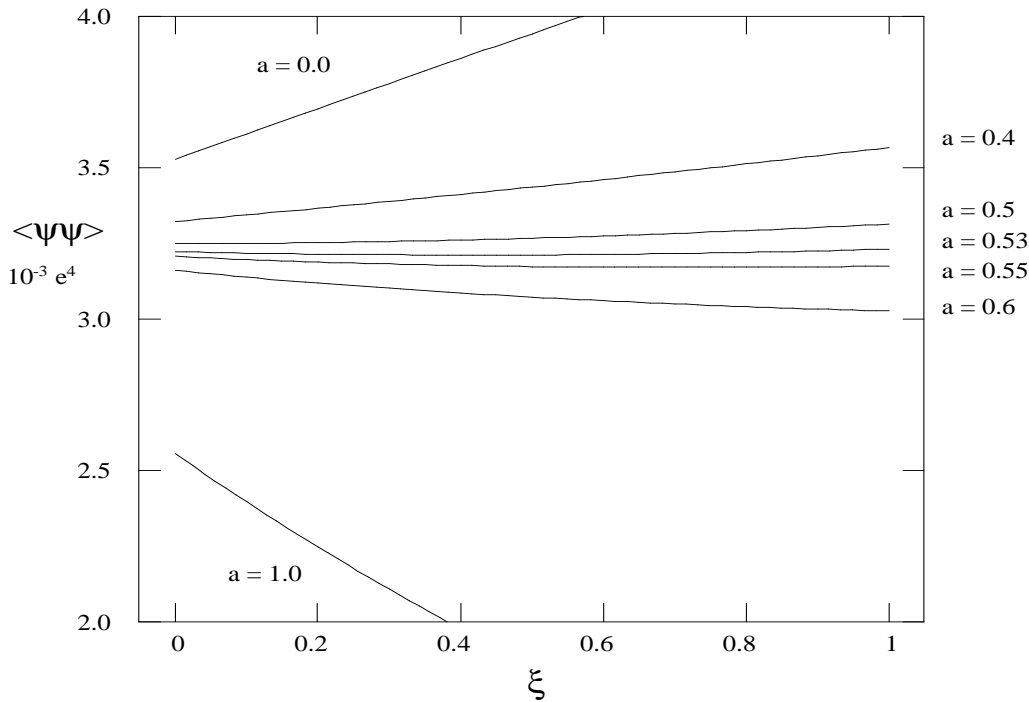,rheight=9.5cm,height=8.5cm,angle=-90} }
\parbox{130mm}{\caption{ 
Plot of \psibpsi, obtained from the solutions of Eqs.~(\protect\ref{a4}) and
(\protect\ref{b4}) as a function of the gauge parameter $\xi$ for various
values of $a$, the parameter that varies the transverse part of the vertex.
A value of \mbox{$a\approx 0.53$} minimises the sensitivity of \psibpsi to
$\xi$ (Burden and Roberts, 1991).
\label{qedtcndst}  }}
\end{figure}
The vertex of \Eq{eq:TEVtx} also has the highly desirable feature that it
leads to an [almost] gauge invariant fermion condensate.  This is illustrated
in \Fig{qedtcndst}.  It will be observed that on the domain \mbox{$0\leq
\xi\leq 1$} there is a value of  
\begin{equation}
a \approx 0.53
\end{equation}
for which the explicit gauge dependence of the photon propagator is
compensated by the implicit gauge dependence of the solution functions.  This
indicates that the transverse parts of the vertex are extremely important in
restoring gauge independence to \psibpsi.

\subsect{Comparison with Lattice Simulations}
The approach described in Sec.~\ref{beyond-rainbow-QED3} leads to an [almost]
gauge invariant condensate which invites comparison with \psibpsi obtained in
lattice simulations of \qedt.  In the lattice formulation
\begin{eqnarray}
\psibpsi^{\mbox{lattice}} & \equiv & \frac{\langle \mbox{tr} M^{-1}[U] \rangle}
                                   {V}
\end{eqnarray}
where \mbox{$\langle\cdot\rangle$} represents a Boltzmann weighted average
over the gauge field $U$, \mbox{$M^{-1}[U]$} is the inverse of the lattice
Dirac operator in a given gauge field configuration and $V$ is the number of
lattice sites. Employing the formalism developed by Burden and Burkitt (1987)
one obtains the following relationship between the lattice and continuum
condensates:
\begin{eqnarray}
-\psibpsi^{\mbox{continuum}} & = &
\frac{N_{f}}{2a^{2}}\psibpsi^{\mbox{lattice}} 
\label{eq:rltn}
\end{eqnarray}
with $N_{f}$ the number of fermion flavours which is one in the case we are
considering here.

For a proper comparison one must look to the lattice results obtained in a
scaling window where they are thought to best represent continuum physics.
One signal that a scaling window exists in the lattice theory is the
observation that
\begin{eqnarray}
\beta^{2}\psibpsi^{\mbox{lattice}} & = & {\cal K} \mbox{\hspace*{5mm}(a
constant)}
\end{eqnarray}
for large \mbox{$\displaystyle\beta = \frac{1}{e^{2}a}$}.  In the scaling
window it should be that
\begin{eqnarray}
-\frac{1}{e^{4}}\psibpsi^{\mbox{continuum}} & = & \frac{N_{f}}{2}{\cal K}~.
\label{eq:cmp}
\end{eqnarray}
The simulations of Dagotto \etal~(1990) suggest that in \qedt a scaling
window exists for \mbox{$\beta > 1$}.

The calculations of Burden and Roberts (1991) correspond to the quenched
approximation of the lattice formulation.  This corresponds to the
\mbox{$N_{f}=0$} simulations of Dagotto \etal~(1990). There exists now the
possibility for confusion because of the 
\mbox{$N_{f}$} factor in (\ref{eq:cmp}).  The quenched approximation 
actually means that the factor \mbox{$(\mbox{det}M)^{\frac{1}{2}}$}, present
in \mbox{$N_{f}=1$} simulations, is not included in the
\mbox{$N_{f}=0$} simulations. In this case the steps that led to 
(\ref{eq:rltn}) can be retraced and one finds that the proper comparison is
\begin{eqnarray}
-\frac{1}{e^{4}}\psibpsi^{\mbox{continuum}} & \mbox{with} &
\frac{1}{2}\, {\cal K}^{N_{f}=0}~.
\end{eqnarray}

\begin{figure}[tb] 
\psrotatefirst \centering{\
\epsfig{figure=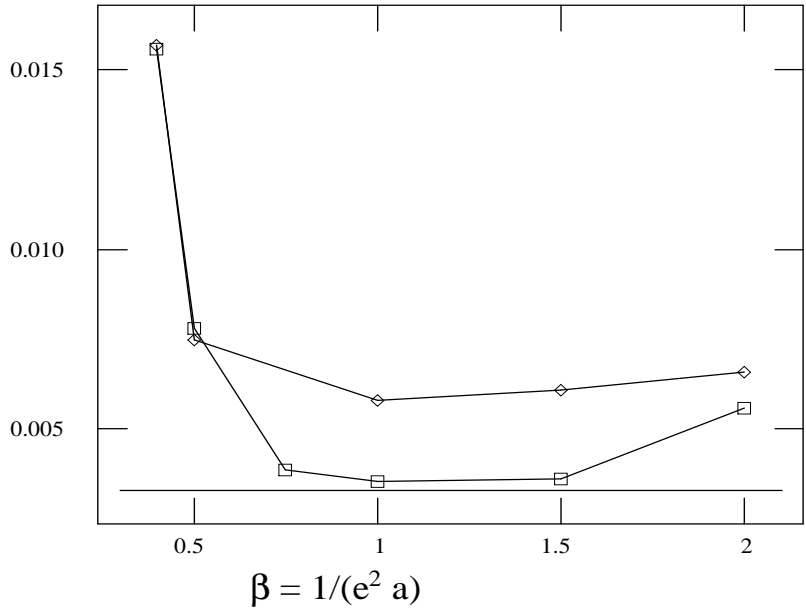,rheight=8.5cm,height=8.0cm,angle= -90} }
\parbox{130mm}{\caption{
The chiral condensate from the quenched lattice studies of Dagotto
\etal~(1990) for $8^3$ [diamonds] and $10^3$ [squares] lattices. 
The solid line is \psibpsi obtained in the DSE approach to quenched \qedt. 
The vertical axis
is \mbox{$-\psibpsi/e^4$}.  This figure is adapted from Burden (1993).
\label{DSE-Lattice-psibpsi}  }}
\end{figure}
This comparison is presented in \Fig{DSE-Lattice-psibpsi}.  It will be noted
that, within the scaling window, the agreement with the $10^3$ simulations of
Dagotto \etal~(1990) is very good.  Indeed, there is little room for
improvement which suggests that, in the quenched approximation, the Ball-Chiu
vertex,
\Eq{eq:TEVtx} with $a=0.5$, is a very good approximation to the true vertex;
or at least to that part of it which contributes to the fermion DSE.

Computationally, the DSE calculations are much simpler and quicker than the
lattice studies and \Fig{DSE-Lattice-psibpsi} illustrates that the DSE
approach to solving Abelian field theories is becoming competitive with
lattice studies.

\subsect{$1/N_f$ Expansion and a Phase Transition in \qedt}
\label{DCSBNT}
As we mentioned in association with \Eq{DqN}, including fermion loops in the
photon polarisation tensor introduces a dimensionless parameter, $N_f$, the
number of flavours of electrically active fermions, into \qedt upon which
\psibpsi may depend.  Indeed, there may be some value of $N_f=N_{\rm cr}$ for
which $\psibpsi\rightarrow\,0$.  Such behaviour has been inferred from
studies of non-compact \qedt on $8^3$ (Dagotto \etal, 1989) and $10^3$
lattices (Dagotto \etal, 1990) with $\psibpsi\neq\,0$ only for $N_f<N_{\rm
cr}=\, 3.5\pm 0.5$.  The existence of a value of $N_f=N_{\rm cr}$ above which
there is no DCSB would limit the utility of \qedt as a model for strongly
interacting systems.

The lattice studies referred to above followed a DSE study of \qedt at
leading order in an expansion in $1/N_f$ (Appelquist \etal, 1988).  In
this study, which used Landau gauge, it was argued that at lowest order in
$1/N_f$, $A(p^2)=1$; i.e., that corrections to $A$ arise at higher order in the
$1/N_f$ expansion, and that $B$ is given by the solution of
\beq
\label{ANWDSEa}
B(p)= \frac{4 \tilde{\alpha}}{N_f\pi^2 p}
\int_0^\infty\,dk\,\frac{k\,B(k)}{k^2 + B(k)^2} 
\ln\left(\frac{k+p+\tilde{\alpha}}{|k-p|+\tilde{\alpha}}\right)\;.
\eeq
We remark that \Eq{ANWDSEa} incorporates the effect of massless fermion loops
in the photon polarisation tensor; i.e., it goes beyond quenched
approximation at leading order in $1/N_f$: see \Eq{PiScal} and the associated
discussion.

This equation could, of course, be solved directly by iteration, however,
Appelquist \etal~(1988) argued that the $\ln$ term rapidly suppresses
the integrand for $p>\tilde{\alpha}$ and that
\beq
\label{ANWDSEb}
B(p)= \frac{4 }{N_f\pi^2 p}
\int_0^{\tilde{\alpha}}\,dk\,\frac{k\,B(k)}{k^2 + B(k)^2} 
\left(k+p - |k-p|\right)
\eeq
is a good approximation.  Equation (\ref{ANWDSEb}) is equivalent to the
differential equation
\beq
\label{ANWDSEc}
\frac{d}{dp}\left[ p^2\,\frac{dB(p)}{dp}\right] + 
\frac{8}{N_f\pi^2} \, \frac{p^2\,B(p)}{p^2 + B(p)^2} = 0
\eeq
with the boundary conditions
\beqn
\label{ANWDSEd}
0\leq B(0) < \infty & \;\; {\rm and} \;\; &
\left.\left[p \frac{dB(p)}{dp} + B(p)\right]\right|_{p=\tilde{\alpha}} = 0\;.
\eeqn

If one assumes that in the domain $0<p<\tilde{\alpha}$ there is a subdomain
$p_l<p<\tilde{\alpha}$ on which $B(p) \ll p$ then \Eq{ANWDSEc} can be
linearised; i.e., the $B^2$ term in the denominator neglected, and the
solution in this domain is
\beqn
B(p) \propto p^a & \;\; {\rm where}\;\; & a = -\frac{1}{2} \pm
\frac{1}{2}\sqrt{1 - \frac{32}{N_f\pi^2}} ~.
\eeqn
For $N_f> 32/\pi^2$ one has $-1<a<0$ and it is not possible to satisfy the
ultraviolet boundary condition at $p=\tilde{\alpha}$ in \Eq{ANWDSEd}.
However, for $N_f< 32/\pi^2$ the solution has the form
\beq
B(p) = \frac{1}{\sqrt{p}}\,
\sin\sqrt{\mbox{\small $\frac{1}{2}$} \left[\frac{32}{N_f\pi^2} -1\right]\,
\ln\left(\frac{p}{B(0)} + \delta\right)}
\eeq
where $\delta$ is some phase, which may depend on $N_f$, and the logarithm has
been scaled by $B(0)$.  Any dimensioned parameter would have done but all are
proportional to $B(0)$ so this is completely general.  Imposing the
ultraviolet boundary condition one finds that, for $N_f\rightarrow\,32/\pi^2$,
\beq
\label{B0behav}
B(0) = \tilde{\alpha} e^{2+\delta}\,
\exp\left(\frac{-2n\pi}{\sqrt{\frac{32}{N_f\pi^2} -1}}\right)\;.
\eeq
Since \psibpsi vanishes if $B(0)\rightarrow\,0$, the conclusion of this
analysis is that in \qedt there is a critical number of flavours of massless
fermions
\beq
\label{ANWcc}
N_{\rm cr} = \frac{32}{\pi^2}
\eeq
below which there is DCSB and above which chiral symmetry is restored. [This
assumes that $\delta$ is not singular as $N_f\rightarrow\,N_{\rm cr}$.]
Equation (\ref{ANWDSEa}) was also studied numerically by Appelquist
\etal~(1988) and the approximations associated with deriving the differential 
equation and its solution were found to be valid.  The same result has been
found by Maris (1993, Secs.~5.1 and 5.2) when $A(p^2)\equiv~1$ is imposed.

Higher order corrections in $1/N_f$, O($1/N_f^2$), have been calculated Nash
(1989).  In this study it was argued that
\beq
\label{NashA}
A(p)= 1+ 4 \, \frac{2 - 3\xi}{3\,N_f\pi^2} \,
\ln\left(\frac{p}{\tilde{\alpha}}\right)
+\,\frac{c}{N_f}~,
\eeq
where $c$ is a constant, and an integral equation was derived for
$M(p)=B(p)/A(p)$ in which the kernel was simplified;
the critical number of flavours was independent of
the gauge parameter.  The solution of the approximate equation for $M(p)$ is
of the form
\beq
M(p) = p^a
\eeq
where $a$ is the solution of
\beq
\label{Ncc}
a(a+1) = - \frac{32}{3\,N_f\pi^2} \left(1 - \frac{7.81}{N_f\pi^2}\right)~.
\eeq
The ultraviolet boundary condition of \Eq{ANWDSEd} allows a non trivial
solution for $M(p)$ if $a$ becomes complex.  Neglecting the O($1/N_f^2$) terms
in \Eq{Ncc} then this condition yields \mbox{$N_{\rm cr}=${\small
$\frac{4}{3}$}$\,32/\pi^2 \approx\,4.32$}.  [The discrepancy between this
result and that of \Eq{ANWcc} arises because Nash (1989) took account of
vertex corrections.]  Solving the full equation leads to
\beq
N_{\rm cr} \approx 3.28
\eeq
which is a $25$\% decrease from the O($1/N_f$) result.  [The solution
$N_f\approx~1.04$ was discarded by Nash (1989) as being ``unphysical''.]  This
shift in $N_{\rm cr}$ might be taken as evidence that the $1/N_f$ expansion
converges reasonably well.

The primary conclusion of this study is that the O($1/N_f^2$) terms do not
qualitatively change the nature of the solution: there is still a critical
number of massless fermion flavours above which chiral symmetry is restored.

\subsubsect{Dynamical Chiral Symmetry Breaking for all $N_f$}
\hspace*{-8pt}The existence of a critical number of massless fermi\-ons in
\qedt is not universally accepted.  Pennington and Webb (1988) and Atkinson
\etal~(1988d) studied the pair of coupled, nonlinear integral equations for
$A(p)$ and $B(p)$ obtained in \qedt including the vacuum polarisation of
\Eq{PPS}  but working in rainbow approximation, \Eq{RA}. These studies
suggest that chiral symmetry is dynamically broken for arbitrarily large
$N_f$.

The origin of this difference is the behaviour of $A(p)$ for small $p^2$.
Indeed, it has been shown (Pennington and Webb, 1988) that in perturbation
theory, when $B(p^2)$ is replaced by its value at $p=0$, \mbox{$m =
B(0)$}, in order to approximate its effect on $A(p)$:
\beq
\label{PWAa}
A(p) = 1 - \frac{4}{3 N_f \pi^2}
\left[ \frac{\tilde{\alpha}^2}{m^2 + p^2} - \frac{4}{3} + 
        2\,\frac{m^2}{p^2}\,\left(\frac{m}{p}\,
                \arctan\left(\frac{p}{m}\right) - 1\right) \right]~.
\eeq
Considering \Eq{PWAa} in the limit $p\rightarrow\,0$ one finds
\beq
A(p) = 1 + \frac{8}{3 N_f \pi^2} \left(1 + \ln\frac{m}{\tilde{\alpha}}
\right)~,
\eeq
{\it c.f.} \Eq{NashA} in Landau gauge, and with
\beq
\ln\,m \sim \,-\,\frac{1}{\sqrt{\frac{N_{\rm cr}}{N_f} - 1}}
\eeq
near the supposed critical point, see \Eq{B0behav}, then
\beq
\label{AnearNc}
A(p) \sim \,-\,\frac{1}{N_f\,\sqrt{N_{\rm cr} - N_f}};
\eeq
i.e., it is singular at the critical point.  One observes then that for
momenta such that $m\ll\,p\,<\tilde{\alpha}$, $A(p)$ does indeed receive
small corrections at ``large $N_f$''.  However, for small momenta, $p\ll m$,
which is the domain relevant to DCSB,
\beq
A(p) \neq 1 + \,{\rm O}(1/N_f)
\eeq
and the $1/N_f$ expansion is therefore not a valid tool for analysing DCSB.

There is further evidence that supports the assertion that DCSB in \qedt
cannot be properly analysed without considering $A(p)$ in detail.  A study of
the Cornwall-Jackiw-Tomboulis (1974) [CJT] effective action in \qedt using
the $1/N_f$ expansion reveals that this effective potential is always unstable
against fluctuations away from $A(p)\equiv 1$ and $B(p)\equiv 0$ (Matsuki,
1991).  This result suggests the chiral symmetry is dynamically broken in
\qedt for all $N_f$.

Going beyond perturbative arguments, the unquenched, rainbow approximation
[\Eqs{PPS} and (\ref{RA}), respectively] DSE in \qedt can be solved
numerically. This equation yields the following pair of coupled equations:
\beqn
A(p) & = & 1 +
\frac{2}{N_f\pi^2 p^3}\int_0^\infty\,dk\,\frac{k\,A(k)}{k^2 + B(k)^2} \,
\left[
\mbox{\small $\frac{1}{2}$}\,(k^2-p^2)^2\,
        \ln\left(\frac{|k-p|(k+p+1)}{(k+p)(|k-p|+1)}\right) \right. \\ & &
\left. - \mbox{\small $\frac{1}{2}$}\, \ln\left(\frac{k+p+1}{|k-p|+1}\right)
+ \mbox{\small $\frac{1}{2}$} (k+p - |k-p|)\,(1 + |p^2-k^2|) - k p \right]~,
\\
B(p) & = &
\label{PWba}
B(p)= \frac{4}{N_f\pi^2 p}
\int_0^\infty\,dk\,\frac{k\,B(k)}{k^2 + B(k)^2} 
\ln\left(\frac{k+p+1}{|k-p|+1}\right)~,
\eeqn
where the momenta have been rescaled; e.g.,
$p\rightarrow\,\tilde{\alpha}\,p$.  Analysing the numerical solution (Maris,
1993, Sec. 5.3) one observes that the result of \Eq{AnearNc} does not survive
but that instead \mbox{$A(p) \sim 1/\sqrt{N_f}$}.  This is, nevertheless, not
of the form \mbox{$1 + $O($1/N$)} and again suggests that the $1/N_f$
expansion
is not valid.  Solving these coupled equations numerically yields the result
that chiral symmetry is dynamically broken for all $N_f$.

There have been a number of other studies of this problem which go beyond the
rainbow approximation (Atkinson \etal, 1990; Walsh, 1990).  The approach in
these studies is to make an Ansatz for the fermion--gauge-boson vertex, which
satisfies certain reasonable constraints, and solve the DSE that results.
Perhaps the most sophisticated of these studies is that of Curtis
\etal~(1992) who have solved the coupled integral equations for $A(p)$ and
$B(p)$ obtained using the following Ansatz for the fermion-gauge boson vertex
which satisfies the WTI and ensures multiplicative renormalisability of the
fermion DSE in \qedf~:
\beqn
\label{VA}
\Gamma_{\mu}(k,p)
  & = & \Gamma_\mu^{\rm BC}(k,p) + \Gamma_\mu^{\rm CP}(k,p)
\end{eqnarray}
where
\begin{eqnarray}
i\Gamma_{\mu}^{\rm BC}(k,p) & = & \frac{A(p^2)+A(k^2)}{2} \,i\,\gamma_\mu
\label{VBC} \\ & & + \frac{(p+k)_{\mu}}{p^2 -k^2} \left\{ \left[
A(p^2)-A(k^2)\right] \frac{\left[ i\gamma\cdot p + i\gamma\cdot k \right]}{2}
+ \left[ B(p^2) - B(k^2) \right] \right\}\nonumber
\eeqn
is the vertex proposed by Ball and Chiu (1980) and
\beqn
i\Gamma_\mu^{\rm CP}(k,p) & = &
\frac{\gamma_\mu(k^2-p^2) - 
        (k+p)_\mu (i\gamma\cdot k-i\gamma\cdot p)}{2d(k,p)} \left[
A(k^2)-A(p^2) \right]\/, \label{CPGamT} \\ {\rm with}\;\;\;\;d(k,p) & =&
\frac{1}{(k^2+p^2)}\left( (k^2-p^2)^2 	 + \left[
\frac{B^2(k^2)}{A^2(k^2)}+\frac{B^2(p^2)}{A^2(p^2)} 	 \right]^2 \right)~,
\label{CPdkp}
\end{eqnarray}
is the correction term introduced by Curtis and Pennington (1990) in order to
restore multiplicative renormalisability in studies of the DSE in \qedf.

This study encountered numerical problems for $N_f>7.2$, simply due to the
large momentum domain on which the solution must be known very accurately at
large $N_f$.  Nevertheless, the results obtained suggested that
\beq
\ln \left( M(0) = \frac{B(0)}{A(0)}\right)  \sim \, -N_f~;
\eeq
i.e., that $M(0)$ becomes exponentially small as $N_f$ is increased but that
DCSB persists for all $N_f$.

We remark that this sort of behaviour of $M(p)$ would be very difficult to
extract from lattice simulations of \qedt and, if real, would cast doubts
upon the critical coupling inferred in those studies carried out to date
(Dagotto \etal, 1989, 1990).  It should also be noted that the
finite size of a lattice unavoidably leads to a suppression of $M(p)$ for
$N_f>1$ (Kondo and Nakatani, 1992a) which may lead to an artificial phase
transition.

There is a criticism of the vertex Ansatz approach, however, and it is that
it is not obvious how to systematically improve over a given Ansatz.  The DSE
for the vertex itself is a complicated equation and no attempt has as yet
been made to solve it directly, which would be the natural, systematic step
in the DSE approach.  Failing this, however, even applying certain, very
general constraints to restrict the form of the vertex, which we discuss in
Sec.~\ref{psi-gamma-vertex}, the Ans\"{a}tze remain {\it ad hoc}.  As a
consequence there are those who doubt the result that \qedt has dynamical
chiral symmetry breaking for arbitrary numbers of massless fermions.  Lattice
simulations of \qedt with greater accuracy might therefore be valuable.

\subsect{Photon Polarisation Tensor and Mass Generation}
Another weakness of many of the studies of DCSB in \qedt undertaken to date
is that they do not take into account the effect of the dynamically generated
fermion mass on the photon polarisation tensor.  In Abelian gauge theories,
once an Ansatz for the fermion--gauge-boson vertex is made then the DSEs for
the fermion propagator and the photon polarisation tensor form a closed
system of equations.

The DSE for the photon polarisation tensor in Euclidean momentum-space is
given in \Eq{eq:Pi} and it is a simple matter to determine whether the
dynamical fermion mass has any effect on the polarisation tensor.  At first
order one must simply evaluate the polarisation tensor using a solution of
the fermion DSE evaluated with the same vertex Ansatz but with
$\Pi(k)\equiv\,0$.  This is the first step in a self consistent solution of
the coupled fermion and photon equations and has been carried out by Burden
\etal~(1992a) using the vertex of \Eq{eq:TEVtx} with $a=0.5$; i.e., the
Ball-Chiu vertex.  Such a calculation also allows for another simple test of
gauge covariance in the DSE approach since the polarisation scalar, see
\Eq{PimnE}, should be independent of the gauge parameter; i.e.,
\beq
\label{Pixidep}
\frac{d}{d\xi}\,\Pi(k;\xi) = 0~.
\eeq

In evaluating the polarisation tensor one must first note that with the
vertex Ansatz of \Eq{eq:TEVtx}
\begin{eqnarray}
A(k\rightarrow\infty) = 1 + \frac{\xi e^2}{16 k} \; 
& \;\; {\rm and} \;\; & \; 
B(k\rightarrow \infty)  =  \frac{\lambda}{k^2} \label{eq:Basy}~,
\end{eqnarray}
with $\lambda$ a gauge parameter dependent constant (Burden and Roberts,
1991), and hence \Eq{eq:Pi} must be regularised since a direct evaluation
yields \mbox{$k_\mu\,\Pi_{\mu\nu}(k)$} as a difference of two logarithmically
divergent integrals in all but Landau gauge.  This problem can be avoided by
contracting the polarisation tensor with
\beq
{\cal P}_{\mu\nu} = \delta_{\mu\nu} - d\,\frac{k_\mu k_\nu}{k^2},
\eeq
which is orthogonal to $\delta_{\mu\nu}$ in ${\rm d}$ dimensions (Walsh, 1990,
Chap. 3).  Since the divergent part of the integral is proportional to
$\delta_{\mu\nu}$ then, after contraction, only the finite piece of the
polarisation scalar remains:
\begin{eqnarray}
\Pi(k) &=& -\frac{e^2}{2k^2}\int\dqbt\, 
\mbox{tr}\left[ \rule{0mm}{6mm}
         \gamma_\mu S(q+\hlf k)\Gamma_\mu(q+\hlf k,q-\hlf k)S(q-\hlf k)
\right. \nonumber \\ 
& & \hspace*{25mm}\left. -3\frac{\gamma\cdot k}{k^2} S(q+\hlf
k)\,k_\nu\,\Gamma_\nu(q+\hlf k,q-\hlf k)\,S(q-\hlf k)\right]~.
\label{eq:pieq}
\end{eqnarray}

At this point it is a simple matter to address the question of photon mass
generation since there are a number of lattice studies of the gauge boson
propagator in \qedf (Coddington, \etal, 1987; Mandula and Ogilvie,
1987; R. Gupta \etal, 1987; Mandula and Ogilvie, 1988) which find a
nonzero gauge boson mass after fixing a lattice Landau gauge.  To study this
one must consider $\Pi(0)$ since a $1/p^2$ pole in the polarisation scalar
signals mass generation via the Schwinger mechanism (Schwinger, 1962).  From
Eq.~(\ref{PPS}) it is clear that there is no such mass generation in
perturbation theory.

Writing
\beq
k^2 \Pi(k) = -N_f\frac{e^2}{\pi^2}f(k)
\eeq
and using the differential form of the Ward identity [\mbox{$\Gamma_\mu(q,q)
= -i\partial^{q}_{\mu}S^{-1}(q)$}] one finds
\beq
f(0) = \lim_{k^2\rightarrow 0}\frac{i}{16\pi}\int\dqbt\,\mbox{tr}\left[
\left(\gamma_\mu - 3\frac{\gamma\cdot k k_\mu}{k^2}\right) 
\partial^{q}_{\mu}S(q)\right]
\eeq
where the final limit must be taken with care after the Dirac trace is
evaluated.  Using Green's theorem:
\beq
f(0) = \lim_{k^2\rightarrow 0}
\frac{1}{4\pi}\int_{S_2}d^2S\,\left(1-3\frac{(k\cdot q)^2}{k^2 q^2}\right)
                    \frac{A(q)}{q^2A^2(q) + B^2(q)}
\eeq
and since \mbox{$\int_{0}^{\pi}d\theta\,\sin\theta[1-3\cos^2\theta] = 0$} one
has $f(0) = 0$.

Hence, as long as the Ward identity is satisfied the photon remains massless
independent of the gauge parameter and details of the transverse part of the
vertex.  This is a general result in Abelian theories, independent of the
details of the interaction and, suitably modified, it also holds in \qedf~.
One expects therefore, that in the continuum limit lattice simulations should
also yield this result.

The form of $\Pi(k)$ obtained by Burden \etal~(1992a) for a range of values
of $\xi$ is plotted in \Fig{PPS-xi}.
\begin{figure}[tb] 
 \centering{\
\epsfig{figure=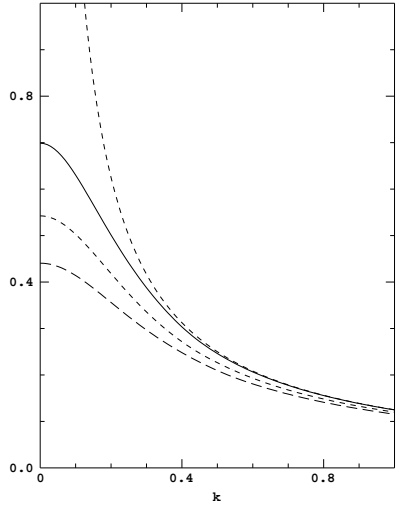,rheight=8.5cm,height=8.0cm,angle=0} }
\parbox{130mm}{\caption{
The polarisation scalar, \mbox{$\Pi(k)$}, obtained from
Eq.~(\protect\ref{eq:pieq}) using the solution of Eqs.~(\protect\ref{a4}) and
(\protect\ref{b4}) with $a=0.5$, is plotted for $\xi=0$: solid line;
$\xi=0.5$; dashed line; $\xi=1.0$; dash-dot line.  [The mass scale is fixed
by setting \mbox{$e^2 = 1$}~.]  The plot illustrates the deviation of the
result from the perturbative one at small $k$.  On the scale of this plot the
fitting function of Eq.~(\protect\ref{eq:pifit}) lies exactly on top of the
numerical results. The perturbative result of Eq.~(\protect\ref{PPS}) is the
short-dash line. [This figure is adapted from Burden \etal~(1992a)]
\label{PPS-xi}  }}
\end{figure}
The results can be summarised by the fit:
\begin{equation}
\Pi(k) =  \frac{e^2}{8(k^2+ e^4 a^2)^{\frac{1}{2}}} + be^{-c\frac{k^2}{e^4}} 
\label{eq:pifit}
\end{equation}
with $\xi$ dependent fitting parameters \mbox{$a$, $b$} and $c$ specified
below,
\mbox{$(\xi,a,b,c)$}:
\begin{eqnarray}
& & (0.0,0.2044,8.760\times 10^{-2},7.767)~, \\ & & (0.5,0.2541,5.055\times
10^{-2},11.70)~, \\ & & (1.0,0.2778,-9.349\times 10^{-3},0.5285)~.
\end{eqnarray}
[Of course, in this calculation, a trivial multiplicative factor of $N_f$
appears on the right hand side of Eq.~(\ref{eq:pifit}) if one uses $N_f$
fermions.  A complicated dependence on $N_f$ can only arise when one carries
out the simultaneous solution of Eqs.~(\ref{a4}), (\ref{b4}) and
(\ref{eq:pieq}).~]

The first thing one deduces from \Eq{eq:pifit} is that the Ball-Chiu vertex,
although yielding \psibpsi essen\-tial\-ly independent of the gauge parameter,
$\xi$, it does entail \Eq{Pixidep}; i.e., it does not lead to a gauge
independent polarisation scalar.  This is important and we will discuss it in
some detail below in Sec.~\ref{psi-gamma-vertex}.

One observes from \Eq{eq:pifit} and \Fig{PPS-xi} that, at large $k^2$, the
nonperturbative result returns to the perturbative one of Eq.~(\ref{PPS}).
For small $k^2$, \mbox{$\Pi(k)$} differs from the perturbative result which
was used in the analyses described in the preceding subsections and
predominantly in Pennington and Walsh (1991).  The difference is a
quantitative softening of the infrared behaviour of \mbox{$\Pi(k)$} and it
has been reported (Pennington and Walsh, 1991) that this has little effect on
DCSB when used simply as input to Eqs.~(\ref{a4}) and (\ref{b4}).  The
difference is not unimportant, however.  Simply assuming that $\Pi(k)$ is
bounded [in absolute value] and continuously differentiable on
\mbox{$(0,\infty)$} and that
\mbox{$\Pi(k)\sim 1/k$} for
\mbox{$k\rightarrow\infty$} is enough to ensure that, for large $r$, the
potential associated with the dressed photon propagator is
\beq
V(r) = \frac{1}{1+\Pi(0)} \frac{e^2}{2\pi} \ln e^2 r + \mbox{const.} + h(r)
\label{pot}
\eeq
where \mbox{$h(r)$} falls off at least as fast as \mbox{$1/r$} (Burden \etal,
1992a).  Hence, the large $r$ behaviour of the potential is dominated by a
confining logarithmic term.  It will be recalled that this was true of the
bare photon propagator but not of the perturbatively dressed photon
propagator.  The inclusion of a dressed fermion-photon vertex and dressed
fermion propagator [obtained with this vertex] at this level has thus
restored the confining nature of \qedt~.

The first order response of the photon polarisation scalar to dynamical
fermion mass generation is a modification on the small-$k^2$ domain which is
just that domain relevant to DCSB.  This may have an impact on the question
of the existence of a critical number of flavours in \qedt but a full
analysis of the coupled fermion-photon DSE system with a physically
reasonable vertex Ansatz would be the least necessary to address this.
Preliminary results of such a project are described by Walsh (1990) but no
firm conclusion may be drawn from this study.

\subsect{Fermion-Photon Vertex and Gauge Invariance}
\label{psi-gamma-vertex}
The issue of gauge invariance in the DSE approach to the solution of gauge
field theories is very important.  Indeed, a manifest lack of gauge
covariance in many works in this field has hindered its acceptance as an
efficacious nonperturbative tool.  [There are many contemporary works that
still fail to address this issue.]  We will discuss gauge covariance and
invariance in the context of linear, covariant gauges which is the class of
gauges in which most studies are undertaken.

The coupled fermion--gauge-boson DSE system is closed if the
fermion--gauge-boson vertex is known.  Further, it has become clear that, in
Abelian theories, the fermion-gauge-boson vertex is crucial to ensuring the
correct gauge parameter dependence of the objects one calculates in the DSE
approach.  This indicates the importance of this vertex in the DSE approach.
This presents a problem, however, because, in Abelian gauge theories, the
equation of which the vertex is the solution involves the kernel of the
fermion-antifermion BSE which cannot be expressed in a closed form; i.e., the
skeleton expansion of this kernel involves infinitely many terms.  [The
problem is even greater in non Abelian theories.]

As we have remarked, there have been no attempts to solve the DSE for the
vertex.  Instead, recognising the importance of the vertex, a number of
authors have invested some effort into enumerating constraints that this
vertex must satisfy and proposed Ans\"{a}tze which satisfy them.  The DSE for
the fermion propagator and, in some cases, the coupled fermion-photon DSEs
have then been solved using a given Ansatz.

There are six general constraints one can place on the fermion-gauge-boson
vertex (Burden and Roberts, 1993):
\begin{description}
\item[(a)] It must satisfy the WTI;
\item[(b)] It must be free of any kinematic singularities [i.e., expressing
        \mbox{$\Gamma_\mu(p,q)$} as a function of $p$ and $q$ and a
functional of the fermion propagator, $S(p)$, then $\Gamma_\mu$ should have a
unique limit as \mbox{$p^2\rightarrow q^2$}];
\item[(c)] It must reduce to the bare vertex in the free field limit in the
manner prescribed by perturbation theory;
\item[(d)] It must have the same transformation properties as the bare vertex,
        $\gamma_\mu$, under charge conjugation, $C$, and Lorentz
transformations [such as $P$ and $T$, for example];
\item[(e)] It should ensure local gauge covariance of the propagators and
vertices; and
\item[(f)] It should ensure multiplicative renormalisability of the DSE in
which it appears.
\end{description}
Criterion (b) follows from Ball and Chiu (1980) and criterion (c) is related
to this since together they are necessary to ensure that the vertex Ansatz
has the correct perturbative limit.  The charge conjugation element of
criterion (d) is essential since it constrains the properties of
\mbox{$\Gamma_\mu(p,q)$} under \mbox{$p \leftrightarrow q$}. 

Although condition (a) is a consequence of gauge invariance, it is only a
statement about the longitudinal part of the vertex, and says nothing about
the transverse part.  By itself it is insufficient to ensure condition (e)
(Burden and Roberts, 1991).  A well defined set of transformation laws, which
describe the response of the propagators and vertex in quantum
electrodynamics to an arbitrary gauge transformation are given in papers by
Landau and Khalatnikov (1956) and Fradkin (1956) [LKF] [see
Appendix~\ref{appendix_LK-Tran}].  These laws leave the DSEs and WTI
form-invariant and one can, in principle, ensure condition (e) by choosing an
Ansatz for $\Gamma$ which is covariant under the action of the LKF
transformations.  Unfortunately, however, the transformation rule for the
vertex is quite complicated, making this procedure difficult to implement.
Alternatively, since the LKF transformation rule for the fermion propagator
and polarisation scalar are straightforward, one can use the LKF
transformations to check {\it a posteriori} whether solutions for propagators
obtained using a particular vertex Ansatz transform appropriately and hence
whether the Ansatz used can possibly have been consistent with criterion (e).

The fermion-gauge-boson vertex in an Abelian gauge theory can be written in
the form (Ball and Chiu, 1980):
\beq
\Gamma_\mu(p,q) = \Gamma_{\mu}^{\rm BC}(p,q) + 
           \sum_{i=1}^{8}\; f^i(p^2,q^2,p\cdot q)\; T_{\mu}^{i}(p,q)~,
\label{gvtx}
\eeq
where $\Gamma_{\mu}^{\rm BC}(p,q)$ is given in \Eq{VBC} and $T_{\mu}^{i}$ are
the eight transverse tensors of Eq.~(3.4) in Ball and Chiu (1980) of which
those with \mbox{$i=1,2,3$} are symmetric under \mbox{$p\leftrightarrow q$}
and the remainder are antisymmetric.  The requirement (d) above implies that
all of the \mbox{$f^i$} are symmetric except for \mbox{$f^6$} which is
antisymmetric.  All vertex Ans\"{a}tze which satisfy the WTI can be written
in this form.

Criteria (a)-(d) and (f) have been used (Curtis and Pennington, 1990) in an
analysis of \qedf to argue that a minimal admissible form of the vertex has
\mbox{$ f^i \equiv 0 \:\: \forall \;i\neq 6$} and $f^6$ chosen so as to yield
\Eq{CPGamT}. 

\subsubsect{Gauge Technique and DCSB}
This is an ideal point to remark that there is an approach to solving the
DSEs that differs substantially in method to the studies described above: the
``gauge technique'' (GT)
(Salam, 1963).  This approach, based in Minkowski space,
assumes that the elements of the DSE approach [i.e., the propagators and
vertices] have spectral representations in terms of which the DSEs are
reformulated and then solved for directly.  For example, in the context of
studies of DCSB, it is assumed that a spectral density, $\rho_\psi$, exists
such that the fermion propagator can be written:
\begin{equation}
\label{GTS}
S(p) = \int_{-\infty}^{\infty}d\omega\, \frac{\rho_\psi(\omega) } {\not\!p -
\omega} \label{spP}
\end{equation}
and that the fermion--gauge-boson vertex has a similar form.  In fact, a
commonly used vertex Ansatz in this approach is (Delbourgo and West, 1977)
\begin{equation}
S(p)\Gamma_\mu^{\rm GT}(p,q)S(q) = \int_{-\infty}^{\infty}d\omega \,
\rho_\psi(\omega) \frac{1}{\not\!p - \omega} \gamma_\mu \frac{1}{\not\!q -
\omega}~, \label{spV}
\end{equation}
which has been used in an analysis of \qedt (Waites and Delbourgo, 1992).
For comparison it is useful to note that \Eq{spV} corresponds, in Euclidean
space, to the vertex
\begin{eqnarray}
\label{VmGT}
\Gamma_\mu^{\rm GT}(p,q) & = & 
i \left(\alpha_\mu S^{-1}(q) - S^{-1}(p)\alpha_\mu\right), \label{Hspec}
\end{eqnarray}
with \mbox{$\alpha_\mu = [\gamma\cdot p \gamma_\mu + \gamma_\mu\gamma\cdot
q]/[p^2-q^2]$}, or alternatively:
\begin{equation}
i\Gamma_{\mu}^{\rm GT}(p,q) = \frac{p^2A(p) - q^2A(q)}{p^2-q^2}\,i\,
\gamma_\mu + \frac{A(p)-A(q)}{p^2-q^2} \,i\,\gamma\cdot p \gamma_\mu
\gamma\cdot q + \,
\frac{B(p)-B(q)}{p^2-q^2}(\gamma\cdot p \gamma_\mu + \gamma_\mu\gamma\cdot q).
                          \label{Ha}
\end{equation}
The vertex, $\Gamma_{\mu}^{\rm GT}$, is easily seen to satisfy criteria (a)
to (d) and must therefore be of the form Eq.~(\ref{gvtx}).

Inserting \Eqs{spP} and (\ref{spV}) into the DSE for the fermion propagator
one obtains a linear equation for the spectral density.  This is a feature of
the gauge technique: it can reduce the fermion DSE to a linear equation.  In
fact, in practice, Ans\"{a}tze for the propagators and vertices are almost
always chosen so that the DSEs under consideration are linear integral
equations for the spectral density.

In a manner similar to that in which the Ball-Chiu Ansatz can be improved
using criterion (f) it is also possible to improve \Eq{spV} by adding a
correction term which ensures that the fermion DSE is multiplicatively
renormalisable in this approach (King, 1983) and ensures gauge covariance of
the fermion propagator for small- and large-$p^2$ [but not for intermediate
momenta].

A property of this approach which may prove undesirable, however, is that the
equations for the Dirac-vector and -scalar parts of the spectral density
decouple in the limit of zero Lagrangian- and renormalised-bare-mass.  In
this case one always has the chiral symmetry preserving solution
\begin{equation}
S^{\rm W}(p) = -i\gamma\cdot p \sigma_{V}^{\rm W}(p) \label{Csym}
\end{equation}
and, in addition, it is also probable that the equation admits a dynamical
chiral symmetry breaking solution which would have the form
\begin{equation}
S^{\rm NG}(p) = -i\gamma\cdot p \sigma_{V}^{\rm W}(p) + \sigma_{S}^{\rm
NG}(p).
\label{Casym} 
\end{equation}
Here the notation indicates the Weyl (W) and Nambu-Goldstone (NG) modes
respectively.
[This was the case, for example, in the phenomenological QCD studies of Haeri
and Haeri (1991).]  We remark that in \Eq{Csym} and \Eq{Casym} the vector
part of the propagator is necessarily the same.

The fermion DSE is the stationary point equation for the effective action
discussed in in Corwnall, Jackiw and Tomboulis (1974) which, evaluated at
this stationary point is (Stam, 1985; see also
Sec.~\ref{subsect-eff-actions}):
\begin{equation}
V[S] = \int\frac{d^{\rm d}p}{(2\pi)^{\rm d}}
\left[{\rm tr}\ln[1-\Sigma(p) S(p)] + \frac{1}{2}{\rm tr}[\Sigma(p)S(p)]\right].
\end{equation}
One might measure the relative stability of these extremals by evaluating the
difference \mbox{$V[S^{\rm NG}] - V[S^{\rm W}]$}.  For an Abelian gauge
theory with $N_f$ flavours of fermion one finds [for ${\rm d}=3$ or $4$
since we use $4$ component spinors] that
\begin{equation}
V[S^{\rm NG}] - V[S^{\rm W}] = 2 N_f \int\frac{d^{\rm d}p}{(2\pi)^{\rm d}}
\ln \left[ 1 + \frac{1}{p^2}
\frac{\sigma_{S}^{2}(p)}{\sigma_{V}^{2}(p)}\right]  >  0,
\end{equation}
since it is reasonable to assume that $\sigma_S$ and $\sigma_V$ are real for
real Euclidean $p^2$.  [Since the equation for $\sigma_S$ is homogeneous,
this difference can, in fact, be made arbitrarily large:
\mbox{$\sigma_S\rightarrow \lambda\sigma_S$}.] Hence, based on this effective
action [which is the same as the auxiliary field effective action at the
stationary point] one finds that the gauge technique cannot support dynamical
chiral symmetry breaking (Burden and Roberts, 1993).  This result is true of
any approach in which the equations for the Dirac-vector and -scalar pieces
of the fermion propagator decouple.  [We note that an extension of the CJT
effective action to include a functional dependence on the
fermion-gauge-boson vertex {\it may} allow for DCSB even when the equations
decouple (Haeri, 1993).]

\subsubsect{Gauge Covariance in the Quenched Approximation}$\!$
Criterion (c) has an important consequence.  The unrenormalised fermion DSE
in \qedt and \qedf can be written
\begin{equation}
1 = (i\gamma\cdot p + m)S(p) + e^2 \int \frac{d^{\rm d} q}{(2\pi)^{\rm d}}
D_{\mu \nu}(p-q) \gamma_\mu S(q)\Gamma_\nu(q,p) S(p). \label{SDE}
\end{equation}
If one works in the quenched approximation, \Eq{QuenchedA}, then since
\begin{equation}
\int d\Omega_{\rm d} \frac{1}{(p-q)^2}
\left( ({\rm d}-3)\,p\cdot q + 2\frac{p\cdot(p-q)\, (p-q)\cdot q}{(p-q)^2}\right)
\equiv 0 \label{angi}
\end{equation}
it follows that, in Landau gauge, \Eq{SDE} admits the free propagator
solution
\begin{equation}
S(p) = \frac{1}{i \gamma\cdot p}
\end{equation}
for any vertex that satisfies criterion (c).  It follows directly from this
that if a given vertex Ansatz is to satisfy the gauge covariance criterion,
(e), then, for arbitrary $\xi$, the associated fermion DSE must have the LKF
transform of the free field propagator as its solution [\Eq{LKF}].

We can elucidate this further by considering the massless fermion DSE in
configuration space:
\begin{eqnarray}
\lefteqn{\delta^{\rm d}(x-y) = \gamma\cdot\partial^x S(x-y) +} \nonumber\\
 & & \!\!\!\!\!\!\!\!\!\!\!\!\!\!\!\!  e^2 \int d^{\rm d}z d^{\rm d}x' d^{\rm
d}y'
\gamma_\mu \left( D_{\mu\nu}^{\rm T}(x-z)+\partial^z\partial^z\Delta(x-z)\right)
 S(x-x') \Gamma_\nu(z;x',y')S(y'-y)
\end{eqnarray}
where we have explicitly divided the gauge-boson propagator into the sum of a
transverse, gauge independent piece, \mbox{$D_{\mu\nu}^{\rm T}$}, and
longitudinal, gauge dependent piece, $\Delta$.  Making use of the WTI
\begin{equation}
\partial_\mu \Gamma_\mu(z;x',y') = 
S^{-1}(z-y')\delta^{\rm d}(x'-z) - \delta^{\rm d}(z-y')S^{-1}(x'-z)
\end{equation}
and the identity
\mbox{$  \int_{x'} \gamma\cdot\partial^x S(x,x')S^{-1}(x',z)  = 
        \gamma\cdot\partial^x \delta^{\rm d}(x-z)~,$} one obtains the
massless DSE in the following form:
\begin{eqnarray}
\delta^{\rm d}(x-y) & = & \gamma\cdot\partial^x S(x-y) \nonumber \\
& - & e^2 \left\{
\int d^{\rm d}z [\gamma\cdot\partial^x \Delta(x-z)]\delta^{\rm d}(x-z)
        -[\gamma\cdot\partial^x \Delta(x-y)]\right\}S(x-y) \nonumber \\ & + &
e^2 \int d^{\rm d}z d^{\rm d}x' d^{\rm d}y'
\gamma_\mu  D_{\mu\nu}^{\rm T}(x-z) S(x-x') \Gamma_\nu(z;x',y')S(y'-y). 
\label{SDEx}
\end{eqnarray}
Now it is clear by inspection that if
\begin{equation}
\int d^{\rm d}z d^{\rm d}x' d^{\rm d}y' 
\gamma_\mu  D_{\mu\nu}^{\rm T}(x-z) S(x-x') \Gamma_\nu(z;x',y')S(y'-y) = 0~;
\label{SDExc}
\end{equation}
then Eq.~(\ref{LKF}), with \mbox{$S(x;\xi=0)$} given in Eq.~(\ref{Fxfree}),
is a solution of the massless DSE; i.e., it is a solution if the last term on
the right hand side of Eq.~(\ref{SDE}) is identically zero in Landau gauge.
[We remark that \Eq{angi} is only relevant in quenched approximation: beyond
quenched approximation the integrand is modified and the integral is nonzero
so that, even given (c), the free propagator is not a solution and gauge
covariance plus \Eq{SDEx} doesn't require \Eq{SDExc}.]

As we have seen, most studies of the DSEs are undertaken in momentum space
and it is a simple matter to transcribe Eqs.~(\ref{SDEx}) and (\ref{SDExc}).
We see that the solution of the DSE is LKF covariant if
\begin{equation}
\int \frac{d^{\rm d}q}{(2\pi)^{\rm d}}\, D_{\mu \nu}^{\rm T}(p-q) \gamma_\mu
S(q) \Gamma_\nu(q,p) =0,
\label{Ceprime}
\end{equation}
where $D_{\mu \nu}^{\rm T}(k) = (\delta_{\mu \nu} - k_\mu k_\nu/k^2)/k^2$ in
the quenched theory, in which case the propagator satisfies:
\begin{equation}
1= i\gamma\cdot p S(p) + \xi e^2 \int \frac{d^{\rm d}q}{(2\pi)^{\rm d}}
\frac{i\gamma\cdot (p-q)}{(p-q)^4}\left[ S(p) - S(q)\right]
\label{SDEcov}
\end{equation}
in the covariant gauge fixing procedure.

Equation~(\ref{Ceprime}) provides a much needed additional constraint upon
the vertex function which, while not a full implementation of criterion (e),
is nevertheless a restriction on the form of the transverse part of the
vertex:

\mbox{\bf (e$^\prime$)} In the absence of dynamical chiral symmetry
breaking; i.e., for $\sigma_S \equiv 0$, the vertex must be such that
Eq.~(\ref{Ceprime}) is satisfied,

where \mbox{$D_{\mu\nu}^{\rm T}(k)$} is the transverse part of the quenched
photon propagator.  This is a necessary condition which must be satisfied by
any vertex Ansatz if it is to confer gauge covariance on the quenched QED$_3$
and QED$_4$ DSEs.

We can apply this check in the case of the Curtis-Pennington vertex, \Eq{VA},
for which the fermion DSE in \qedt is, in the absence of DCSB,
\begin{equation}
A(p) - 1 = \frac{-e^2\xi}{4\pi^2p^2} \int_0^\infty dq\, \frac{1}{A(q)}
\left(\frac{p^2A(p) - q^2A(q)}{p^2-q^2} - \frac{p^2A(p) + q^2A(q)}{2pq}
\ln\left|\frac{p+q}{p-q}\right|\right), \label{CPE}
\end{equation}
in which the right hand side is clearly zero in Landau gauge.  Hence this
vertex, or at least that part of it which contributes to the DSE, has the
form necessary to ensure gauge covariance of the chirally symmetric fermion
propagator.  [We remark that here an infrared regulator, $\lambda_{\rm IR}$,
is used to define $d(k,p)$ in \Eq{CPdkp} when $B(k)\equiv\,0$ and the limit
\mbox{$\lambda_{\rm IR}\rightarrow 0 $} taken before evaluating the integral
because it is finite.]

It is possible to solve this equation analytically (Burden and Roberts,
1993).  The solution, for $\xi>0$, is
\begin{equation}
\frac{1}{A(p)} = 
  1 - \frac{e^2\xi}{8\pi p} \arctan \left(\frac{8\pi p}{e^2\xi}\right),
\label{Vee}
\end{equation}
as it should be since this corresponds to the LKF transform of the massless,
free fermion propagator in QED$_3$ [see Appendix~\ref{appendix_LK-Tran}].

The fermion DSE for \qedf using the Curtis-Pennington vertex is given in
Curtis and Pennington (1991) and can be written formally as
\begin{equation}
\label{CPQEDF}
A(p) - 1 = \frac{\xi \alpha_0}{4\pi p^2}
\int_{0}^{\infty} dq^2 \left[ \theta(p^2-q^2) \frac{q^2}{p^2} 
+\theta(q^2-p^2)\frac{p^2}{q^2}\frac{A(p)}{A(q)} \right].\label{CPqed4}
\end{equation}
In Curtis and Pennington (1991) this equation was solved by introducing an
upper bound on the $q^2$ integral with the result 
\mbox{$A_{\rm R}(p^2;\xi) = A_{\rm R}(\mu^2;\xi) (p^2/\mu^2)^\gamma$}, 
\mbox{$ \gamma = \alpha_0 \xi/(4\pi)/[1 + \alpha_0 \xi/(8\pi)]$}, which
illustrates that \Eq{VA} ensures multiplicative renormalisability of the
chirally symmetric fermion DSE for all $p^2$. However, this solution only
agrees with the LKF transform of the free propagator at O$(\xi$).  This defect
arises because a hard cutoff was used, which violates Poincar\'e invariance
and generates a spurious term in the equation.  A detailed discussion of this
can be found in Dong, \etal~(1994), where the actual solution is shown to be
\mbox{$A_{\rm R}(p^2;\xi) = A_{\rm R}(\mu^2;\xi) (p^2/\mu^2)^\nu$} with 
$\nu= \alpha_0 \xi/(4\pi)$, which agrees with the LKF transform at all $\xi$. 

It is interesting to note that the Ball-Chiu vertex alone, \Eq{VBC}, does not
satisfy criterion (e') (Burden and Roberts, 1993).  This is also a defect of
the minimal gauge-technique vertex, \Eq{VmGT}, however, the addition
suggested by King (1983) goes some way to eliminating this problem.

\sect{Four-dimensional Quantum Electrodynamics}
\label{sect-QED4}
Many of the remarks made and studies described in Sec.~\ref{sect-QED3} are
directly relevant to \qedf.  In the absence of a Lagrangian bare mass for the
electron, the action for \qedf has the same chiral symmetry as massless QCD.
This being the case, one of the main goals of DSE studies in \qedf over the
last twenty years has been to answer the question of whether there is a phase
in which chiral symmetry is dynamically broken and, if so, what is the order
of the phase transition.  [As we will see, this is related to the question of
whether \qedf is trivial or not.]  Naturally, the tool used to address this
question is the DSE for the renormalised electron self energy in \qedf which,
in Euclidean space, is
\begin{eqnarray}
\Sigma (p) & = & i \gamma\cdot p \,(Z_1 - 1) - (m - Z_1\,m_0)
+ Z_1 e^{2}\int \dqbf
\Gamma_{\mu}(p,q) 
                D_{\mu\nu}(p-q;\xi)S_{F}(q)\gamma_{\nu}~;
\label{eq:ERDSEqedf}
\end{eqnarray}
a minor modification of \Eq{eq:ERDSE} with significant consequences.

The first of these we have mentioned above.  In Landau gauge and at O($e^2$)
one has
\beq
\label{Zisone}
Z_1\equiv\,1~.
\eeq
This result has been used in many studies to simplify this equation.
Further, it has been used as an approximation in many model DSE studies in
QCD.

\subsect{Dynamical Chiral Symmetry Breaking}
The second consequence is of particular use in the rainbow, quenched
approximation to \qedf.  In this approximation, it follows from \Eq{angi}
that the angular integral for the Dirac-vector component of the electron
self-energy vanishes in Landau gauge so that, subject to
\Eq{Zisone},  \Eq{eq:ERDSEqedf} admits the solution \mbox{$A(p^2) \equiv 1$}
with $B(p^2)$ obtained as the solution of the following nonlinear integral
equation:
\begin{eqnarray}
B(p^2) & = & m_0 + (3+\xi)\,e^2\,\int^{\Lambda_{\rm UV}}\,\dqbf\,
\frac{1}{(p-q)^2}\,
\frac{B(q^2)}{q^2 +B^{2}(q^2)}~,
\label{eq:SDSEF}
\end{eqnarray}
where $\Lambda_{\rm UV}$ is a cutoff which is necessary if $m_0\neq\,0$.

In addition, one may make use of the result:
\beq
\label{AngRes}
\int\,d\Omega_{\rm 4}\, \frac{e^2}{(p-q)^2} =
\theta(p^2-q^2)\,\frac{e^2}{p^2} + \theta(q^2-p^2)\,\frac{e^2}{q^2},
\eeq
where
\mbox{$
\int\,d\Omega_{\rm 4} \; \equiv \; \frac{1}{2\pi^2}\,
\int_0^\pi\,d\beta\,\sin^2\beta\,
\int_0^\pi\,d\theta\,\sin\theta\,
\int_0^{2\pi}\,d\phi
$}, 
to reduce \Eq{eq:SDSEF} to the following differential equation:
\beq
\label{QEDFde}
x\,B''(x) + 2\,B'(x) + \frac{3\alpha}{4\pi}\,\frac{B(x)} {x + B^2(x)} = 0,
\eeq
where $x=p^2$, with the boundary conditions
\beqn
\label{QEDFbc}
\lim_{x\rightarrow 0}\left( \frac{d}{dx}[x^2\,B(x)]\right) = 0\;\; & \;\;{\rm
and}\;\; & \;\;
\lim_{x\rightarrow\Lambda^2_{\rm UV}}\left(
\frac{d}{dx}[x\,B(x)]\right) = m_0~.
\eeqn
The problem of studying DCSB in the rainbow, quenched approximation to \qedf
is thus reduced to a study of this differential equation.

This is an opportune point to remark that \Eq{AngRes} has been used as the
basis for an extensively used approximation in phenomenological DSE studies
of QCD and \qedf beyond quenched approximation.  Many authors have assumed
that it is reasonable to write [i.e., the angle approximation]
\beq
\label{AngApprox}
\int\,d\Omega_{\rm 4}\, \frac{\alpha((p-q)^2)}{(p-q)^2} \approx
\theta(p^2-q^2)\,\frac{\alpha(p^2)}{p^2} 
        + \theta(q^2-p^2)\,\frac{\alpha(q^2)}{q^2}~,
\eeq
with the function $\alpha(p^2)$ assumed to be monotonically decreasing on
\mbox{$[0,\infty)$}.  This approximation has been critically analysed
(Roberts and McKellar, 1990) and found to be reasonable only if $\alpha(0)$
is not large; i.e., of O($1$) or less.

\subsubsect{Miransky Scaling}
Equation~(\ref{QEDFde}) has been studied by Fukuda and Kugo (1976) and the
results have a direct analogue in those of Sec.~\ref{DCSBNT}.  If one sets
$m_0 = 0$ then \Eq{QEDFde} is obviously scale invariant.  In this case the
scale is set by choosing $B(0)= 1$ and the solution then has the following
power series expansion around $x=0$:
\beq
B(x) = 1 - \frac{3\alpha}{8\pi}\,x -
\frac{\alpha}{64\pi}\,\left(\frac{3\alpha}{\pi} - 8\right)\,x^2 + \ldots~.
\eeq
The asymptotic behaviour of the solution for large-$x$ is easily found to be
\beq
\label{BasymptQED}
B(x) \sim \left\{
\begin{array}{c}
\displaystyle
x^{-\frac{1}{2} + \frac{1}{2}\sqrt{1-3\alpha/\pi}}, \;\;
\alpha\leq\frac{\pi}{3}\\
\displaystyle
\frac{1}{\sqrt{x}}\cos\left[\frac{1}{2}\sqrt{3\alpha/\pi -1}\,\ln\,x\right],
        \;\;\alpha\geq\frac{\pi}{3}
\end{array}
\right.~.
\eeq

If one demands that the ultraviolet boundary condition of \Eq{QEDFbc} be
satisfied at some finite $\Lambda_{\rm UV}$ then it is clear that a nonzero
solution for $B(p^2)$ is admitted only if
\beq
\alpha > \alpha_c = \frac{\pi}{3} \;\; (\approx 1.047)
\eeq
since the form of the solution valid for $\alpha<\alpha_c$ cannot satisfy
this condition.  In this case one has [see Sec.~\ref{DCSBNT}]
\beq
\label{MirScal}
B(0) = \Lambda_{\rm UV} \,
\exp\left(\,-\,\frac{\pi}{\sqrt{\alpha/\alpha_c - 1}} + \delta\right)~,
\eeq
where $\delta$ is some phase, and the conclusion is that chiral symmetry may
be dynamically broken in rainbow-quenched \qedf if the coupling exceeds
$\alpha_c$.  A study of the stability of the nonzero solution for $B$, the
solution which corresponds to a Nambu-Goldstone mode realisation of chiral
symmetry, relative to the solution $B=0$, which corresponds to a realisation
of chiral symmetry in the Wigner-Weyl mode, using the auxiliary field
effective action (Roberts and Cahill, 1986) shows that the Nambu-Goldstone
mode is indeed dynamically favoured for $\alpha>\alpha_c$.

As the critical coupling is approached, $B(0)$ and \psibpsi are
interchangeable as order parameters of DCSB. Equation~(\ref{MirScal}) shows
then that, in rainbow-quenched approximation, \qedf has an infinite order
phase transition.  In models such as this one \Eq{MirScal} is often referred
to as ``Miransky Scaling'' (Miransky, 1985a).

It has been pointed out by Roberts and Cahill (1986), however, that in the
limit $m_0=0$ the integral equation admits a solution even when the cutoff is
removed; i.e., when $\Lambda_{\rm UV} = \infty$.  In this case the ultraviolet
boundary condition can be satisfied for any value of $\alpha$ and hence
rainbow-quenched \qedf admits DCSB for all values of the coupling.  [See
also, Atkinson and Johnson (1987).]  This problem has also been studied using
the vertex Ansatz of \Eq{VA} (Curtis and Pennington, 1992); i.e., beyond
rainbow approximation.  In this case it is convenient to solve the integral
equations directly because of the complicated structure of the vertex but the
same result is obtained; i.e., a dynamical mass is generated for all values
of the coupling.

In this light the existence of a critical coupling in massless \qedf might be
seen as the result of the imposition of a cutoff which can be viewed as an
auxiliary condition, not intrinsic to the model, but relevant to the
consideration of whether the model is a reasonable representation of \qedf.
For example, it might be argued that the appearance of the Landau ghost in
perturbation theory suggests that the quenched approximation cannot possibly
be valid at all momentum scales and hence that introducing a cutoff is a
simple manner in which to model the effect of vacuum polarisation
contributions.

However, dynamical chiral symmetry breaking in theories with ultraviolet
divergences is a subtle problem (see, for example, Fomin \etal, 1983, pp.
9-10 and pp. 64-65; Miransky, 1985a and 1985b).  In a recent reanalysis of
this problem (Curtis and Pennington, 1993) it is argued that dynamical chiral
symmetry breaking in quenched \qedf cannot be studied simply by setting
$m_0=0$ in \Eq{eq:ERDSEqedf} and allowing the momentum integration to range
over \mbox{$[0,\infty)$} but rather that renormalisation plays an important
role.  The vertex Ansatz of \Eq{VA}, which ensures multiplicative
renormalisability of the fermion DSE in \qedf, is therefore well suited to
this study.  These authors suggest that in order to ensure conservation of
the axial vector current in the renormalised theory, which can be interpreted
as a definition of masslessness in the theory, one must take the cutoff to
infinity and the bare mass to zero in such a way that their product
approaches zero.  This ensures that, since the bare composite operator
\mbox{$(\overline{\psi}\gamma_5\psi)_{\Lambda_{\rm UV}}$} depends on the
cutoff,
\beq
\partial_\mu\,J^5_\mu = 2\,m_0\,(\overline{\psi}\gamma_5\psi)_
{\Lambda_{\rm UV}} 
        = 2 \,(Z_2^{-1}\,m_0)\,(\overline{\psi}\gamma_5\psi)_{\rm R}\;
\stackrel{\Lambda_{\rm UV}\rightarrow\infty}{\rightarrow} \;0~,
\eeq
where $Z_2$ is the mass renormalisation constant, and can be implemented
through a definition of the renormalised theory with $m_0= 0$ and with a
cutoff \mbox{$\Lambda_{\rm UV} < \infty$}.  In this way a Nambu-Goldstone
realisation of chiral symmetry; i.e., DCSB, was found to be possible only for
\beq
\alpha > \alpha_c \approx 0.93
\eeq
and Miransky scaling was observed at the critical coupling.  This result was
only weakly dependent on the gauge parameter, a result that one can attribute
to the efficacy of the vertex Ansatz [see Sec.~\ref{psi-gamma-vertex}].

\subsubsect{Mean Field Scaling - Vacuum Polarisation Ansatz}
A question that arises is whether the existence and order of the phase
transition in \qedf is sensitive to the nature of the approximations.  The
studies described above suggest that going beyond rainbow approximation does
not qualitatively affect the results but what is the effect of the quenched
approximation?  Given the results discussed in Sec.~\ref{DCSBNT} it is clear
that going beyond quenched approximation can introduce another dimensionless
parameter into the theory, the number of electrically active fermions, with
respect to which there may exist critical behaviour.  A number of authors
have addressed these questions.

The first step is to choose a form for the vacuum polarisation.  Motivated by
the large-$k^2$ form of the vacuum polarisation in perturbation theory at one
loop order, the usual choice is
\beq
\label{PiOLQEDF}
\omega(k^2) \equiv 1+ \Pi(k^2) = 1+ \frac{N_f\,\alpha}{3\pi}\,
        \ln\left[\frac{\Lambda_{\rm UV}^2}{k^2 + \mu^2}\right]
\eeq
where $N_f$ is the number of active fermions and $\mu^2$ is an infrared
regulator.  In some studies $\mu^2$ is set to $0$ (Kondo, 1990; Gusynin,
1990; Kondo, 1992; Kondo and Nakatani, 1992b; Maris, 1993, Sec. 4.3) while in
others it is related to $B(0)$, e.g., $\mu^2=B(0)^2$ (Oliensis and Johnson,
1990), in order to incorporate some information from the fermion DSE in the
expression for the vacuum polarisation.  In any event it has little effect on
the results.  [Equation~(\ref{PiOLQEDF}) can be obtained from \Eq{eq:Pi} with
\mbox{${\rm d}=3\rightarrow\,4$} using bare fermion propagators and a bare
fermion-photon vertex.]  Such a form has been used in almost all studies to
date in concert with the approximation of \Eq{AngApprox} in which case one
obtains $A(p^2)\equiv\,1$ and the following differential equation for
$B(x=p^2)$:
\beq
x\,B''(x) + B'(x)\,\left[ 1 + \frac{\omega +\omega'}{\omega}
\       \left(1 + \frac{\omega'}{\omega + \omega'}
                        - \frac{\omega\,[\omega' + \omega'']} {[\omega +
\omega']^2}\right)\right] + \,\frac{3\alpha}{4\pi}\,\frac{\omega +
\omega'}{\omega^2} \,\frac{B(x)}{x + B(x)^2} =0~,
\eeq
where $\omega'$ and $\omega''$ represent derivatives of $\omega$ with respect
to the variable \mbox{$t = \ln\,x$}, with boundary conditions
\beqn
\left[x^2\,B'(x)\right]_{x=0} = 0\;\; & \;\; {\rm and} \;\;
& \left[ (\omega' + \omega)\,B(x) + \omega\,x\,B'(x) \right]_{x =
\Lambda^2_{\rm UV}} = 0~.
\eeqn

This differential equation can be solved numerically (Oliensis and Johnson,
1990) and one finds that, with $N_f=1$, it admits a nonzero solution for
\beq
\label{OJacN}
\alpha > \alpha_c \approx 2.00
\eeq
and that the phase transition is of finite order.  Further analysis (Kondo,
1990; Gusynin, 1990; Kondo, 1992; Kondo and Nakatani, 1992b; Maris, 1993,
Sec. 4.3) shows that in this case, in contrast to \Eq{MirScal},
\beq
B(0) \sim \Lambda_{\rm UV}\, \left(\alpha - \alpha_c\right)^{\frac{1}{2}}~;
\eeq
i.e., one has a second order phase transition with the classical mean field
value for the critical exponent. Both \psibpsi and $B(0)$ behave in the same
way near the critical coupling (Kondo and Nakatani, 1992b).

It is obvious (Kondo and Nakatani, 1992b; Maris, 1993, Sec. 4.3) that if
\Eq{AngApprox} is not used in order to obtain an approximate differential
equation but instead \Eq{PiOLQEDF} is used directly in the integral equation
form of the fermion DSE in \qedf, then one encounters a problem with the
Landau ghost: the kernel of the integral equation, which involves
\beq
1 + \Pi(k^2 + p^2 - 2\,k\,p\,\cos\theta)
\eeq
where the argument of $\Pi$ can attain a maximum value of
\mbox{$4\,\Lambda_{\rm UV}^2$}, has a nonintegrable divergence unless
\beq
\alpha\,N_f \, < \frac{3\,\pi}{2\,\ln2} \approx \,6.80
\eeq
which restricts the number of flavours that one may consider.  A first
estimate based on \Eq{OJacN} is that \Eq{PiOLQEDF} will not be a viable
Ansatz for $N_f > 3$.

Numerical studies of the coupled integral equations for $A(p^2)$ and $B(p^2)$
using \Eq{PiOLQEDF} in rainbow approximation (Kondo and Nakatani, 1992b;
Maris, 1993, Sec. 4.3) find nontrivial solutions only for \mbox{$N_f = 1,
2$} with critical couplings (Maris, 1993, Sec. 4.3)
\beqn
\alpha_c(N_f=1) \approx 2.07 \;\; & \;\; {\rm and} \;\; &
\alpha_c(N_f=2) \approx 2.82~. 
\eeqn
Kondo and Nakatani (1992b) used an Ansatz for the vacuum polarisation that
differs slightly from \Eq{PiOLQEDF}, one which reflects the possible
regularisation ambiguities, and in this case the value of the critical
coupling depends on the choice of the additional parameters.  Indeed, it is
possible to choose the additional parameters and the number of active
fermions, $N_f^C$, such that one does not have DCSB irrespective of the
strength of the coupling.  Avoiding such a complication, the numerical
solution of the coupled integral equations leads to the same mean field
scaling behaviour as the approximate differential equation.

\subsubsect{Mean Field Scaling - Solving for the Vacuum Polarisation}
Given these last remarks, an obvious improvement would be to solve the
coupled fermion and photon DSEs in Landau gauge using rainbow approximation:
\beqn
\lefteqn{A(p^2)  =  1 }\\
& +& \alpha_0\,\int_0^{\Lambda^2_{\rm UV}}\,
\frac{dk^2\,k^2}{4\pi}\,\frac{A(k^2)}{k^2\,A(k^2)^2+B(k^2)^2}
\int_0^\pi\,\frac{d\theta}{\pi}\,2\,\sin^2\theta\, \frac{k\,\left[
3\,(k-p)^2\,\cos\theta -2\,k\,p\,\sin^2\theta\right]}
{p\,(k-p)^4\,[1+\Pi((k-p)^2)]}~, \nonumber \\ 
B(p^2) & = & \!\! m_0 + \!\!
\alpha_0\,\int_0^{\Lambda^2_{\rm UV}}\,
\frac{dk^2\,k^2}{4\pi}\,\frac{B(k^2)}{k^2\,A(k^2)^2+B(k^2)^2}
\int_0^\pi\,\frac{d\theta}{\pi}\,2\,\sin^2\theta\,
\frac{3}{(k-p)^2\,[1+\Pi((k-p)^2)]},
\label{BRakow}\\
\Pi(p^2) & = & -\,\frac{4}{3}\,\alpha_0\,N_f\,\int_0^{\Lambda^2_{\rm UV}}\,
        \frac{dk^2\,k^2}{4\pi}\,
\int_0^\pi\,\frac{d\theta}{\pi}\,2\,\sin^2\theta\, \frac{A(k_+^2)\,A(k_-^2)\,
\left(\frac{k^2}{p^2}\,[8\,\cos^2\theta - 2] - \frac{3}{2}\right)} {[k_+^2
A(k_+^2)^2+ B(k_+^2)^2]\,[k_-^2 A(k_-^2)^2+ B(k_-^2)^2]}
\eeqn
where $\alpha_0$ is the bare coupling, $m_0$ is the bare electron mass and
$k_\pm^2 = k^2 + \mbox{\small $\frac{1}{4}$}p^2 \pm \,k\,p\,\cos\theta$.
Such a study has been undertaken by Rakow (1991).

The main body of the study concentrates on $N_f = 1$ and it is assumed that
\mbox{$Z_1 \equiv 1$}, \Eq{Zisone}, while
\beqn
Z_2 \equiv \frac{1}{A(0)} \;\; & {\rm and} \;\; & \;\; Z_3 \equiv
\frac{1}{1+\Pi(0)}~,
\eeqn
which follows from $S^f=Z^f_2\tilde S^f$, $D_{\mu\nu}=Z_3\tilde D_{\mu\nu}$
and \Eq{QED_renorm_BCs} in Sec.~\ref{QEDRen}, with renormalisation at $p^2=0$
rather than on the mass shell. Clearly, deviations from $A(p^2)\equiv\,1$
measure the violation of the Ward Identity in rainbow approximation,
\Eq{Ward_Z}.  The results are then analysed in terms of a renormalised mass
and coupling:
\beqn
m_{\rm R} = \frac{B(0)}{A(0)} = Z_2 \, B(0) \; & \;\; {\rm and} \;\;&
\alpha_{\rm R} = {\alpha_0\over A(0)^2\,[1+ \Pi(0)]} 
        = Z_2^2\,Z_3 \,\alpha_0~,
\eeqn
respectively, with $\alpha_{\rm R}$ measuring the effect of charge screening
due to fermion loops.

It is found that with $m_0=0$ and for $N_f=1$ there is a critical coupling
\beq
\alpha_c \approx 2.25
\eeq
such that
\beq
\label{cndstbeta}
\langle\overline{\psi}\psi\rangle_{\Lambda_{\rm UV}} = 
\Lambda^3_{\rm UV}\,\left(\alpha_0-\alpha_c\right)^{\frac{1}{2}}~;
\eeq
i.e., that one has mean field scaling behaviour in the model, but that the
critical exponent approaches the mean field value of $\frac{1}{2}$ very
slowly.  [See Creswick \etal~(1992) for a pedagogical discussion of phase
transitions and renormalisation group methods in physics.]  It has been shown
that
\beq
\psibpsi \sim \Lambda^2_{\rm UV}\,m_{\rm R}
\eeq
and hence, identifying $\alpha_0$ as $\alpha(\Lambda_{\rm UV})$,
\Eq{cndstbeta} leads to the following $\beta$-function:
\beq
\beta(\alpha) \equiv 
\Lambda_{\rm UV}\,\frac{\partial}{\partial\Lambda_{\rm UV}}
        \ln\,\alpha(\Lambda_{\rm UV}) = \, -2 \left[\alpha(\Lambda_{\rm UV})
-\alpha_c\right]
\eeq
which suggests that unquenched, rainbow approximation \qedf has an
ultraviolet stable fixed point at \mbox{$\alpha_0 = \alpha_c$}; a result
supported by lattice simulations (Kogut \etal, 1988a, 1988b).

In the limit \mbox{$\alpha_0 \rightarrow \alpha_c$} from above it is found
that \mbox{$\alpha_{\rm R} \rightarrow 0$} logarithmically; i.e., recall that
$m_{\rm R}$ vanishes as $\alpha_0 \rightarrow \alpha_c$,
\beq
\alpha_{\rm R} \propto\, -\,\frac{1}{\ln\,m_{\rm R}}
        \propto \frac{1}{\ln\xi} \label{triv}
\eeq
where \mbox{$\xi = \Lambda_{\rm UV}/m_{\rm R}$} is the correlation length in
\qedf.  The continuum limit of the theory is defined as the limit
$\xi\rightarrow\infty$.  It follows from \Eq{triv} that the continuum limit
of unquenched, rainbow approximation \qedf is trivial in the sense that it
does not involve photons interacting with charged fermions, since it is not
possible to obtain an infinite correlation length at nonzero coupling.

It is clear from these observations that solving for the vacuum polarisation
yields the same results as the Ansatz of \Eq{PiOLQEDF} with only minor
quantitative differences.  The problem of the ``Landau-ghost'' is not removed
by a simultaneous solution of the fermion and photon DSEs in rainbow
approximation.

\bigskip\underline{Interacting Continuum Limit?}\ipar
The existence of a phase in which chiral symmetry is dynamically broken
suggests that at strong coupling [$\alpha_0 \geq \alpha_c$] in \qedf there
are Goldstone modes; i.e., massless, pseudoscalar $e^+$-$e^-$ bound states.
In this connection, one notes that in addition to studying the DSEs for the
fermion and photon propagators in rainbow approximation, Rakow (1991) solved
the following approximate form of the integral equations for the pseudoscalar
and scalar electron-positron scattering amplitudes:
\beqn
\lefteqn{S_5(p^2,r^2) =
\alpha_0\,\int_0^\pi\,\frac{d\theta}{\pi}\,2\,\sin^2\theta 
                \frac{3\pi}{t^2\,[1+ \Pi(t^2)]} }
\label{SPRakow}\\
& & + \alpha_0\,\int_0^{\Lambda^2_{\rm UV}}\,\frac{dk^2\,k^2}{4\pi}\,
\frac{S_5(k^2,r^2)}{k^2\,A(k^2)^2+B(k^2)^2}
\int_0^\pi\,\frac{d\theta}{\pi}\,2\,\sin^2\theta \frac{3}{q^2\,[1+
\Pi(q^2)]}~, \nonumber \\
\lefteqn{S_1(p^2,r^2) = 
\alpha_0\,\int_0^\pi\,\frac{d\theta}{\pi}\,2\,\sin^2\theta 
                \frac{3\pi}{t^2\,[1+ \Pi(t^2)]} } \\ & & +
\alpha_0\,\int_0^{\Lambda^2_{\rm UV}}\,\frac{dk^2\,k^2}{4\pi}\,
S_1(k^2,r^2)\, \frac{k^2A(k^2)^2 - B(k^2)^2}{[k^2\,A(k^2)^2+B(k^2)^2]^2}
\int_0^\pi\,\frac{d\theta}{\pi}\,2\,\sin^2\theta \frac{3}{q^2\,[1+ \Pi(q^2)]}
\nonumber \\
\eeqn
where \mbox{$\;t^2 = p^2+r^2-2\,p\,r\,\cos\theta\;$} and
\mbox{$\;q^2 = k^2+p^2-2\,k\,p\,\cos\theta$}.  These equations have been
averaged over \mbox{$p\cdot r$} and describe
\mbox{$[\psi(p);\overline{\psi}(-p)]$}$\rightarrow$
\mbox{$[\psi(r);\overline{\psi}(-r)]$} scattering.

In studying the solution of these equations it was observed that, at fixed
values of \mbox{$\alpha_{\rm R}$}, these amplitudes did not scale; i.e.,
there was a dependence on $p/m_{\rm R}$.  Indeed, the pseudoscalar amplitude
manifested a contribution that diverged in the limit $m_0\rightarrow\,0$
obviously as a result of the formation of the Goldstone mode.  This lack of
scaling at constant \mbox{$\alpha_{\rm R}$} suggests that another operator is
relevant in the continuum limit.

The fact that another operator is relevant in the continuum limit can also be
inferred from \Eq{BasymptQED}.  In the absence of interactions the mass
operator, \mbox{$\overline{\psi}\psi$}, diverges as $\Lambda^2_{\rm UV}$:
\beq
\overline{\psi}\psi \sim \int^{\Lambda^2_{\rm UV}}\,ds\,s\,
        {\rm tr}\,[S(p)] \;
\stackrel{\Lambda^2_{\rm UV}\rightarrow\infty}{\propto} \;
        \Lambda^2_{\rm UV}~.
\eeq
In the interacting case, using \Eq{BasymptQED}, it diverges as
\mbox{$\Lambda_{\rm UV}^{1+\surd[1-3\alpha/\pi]}$} from which one can  infer that
its interaction-induced anomalous dimension is
\beq
\gamma_{\overline{\psi}\psi} = \sqrt{1-\frac{3\alpha}{\pi}} - 1,
\eeq
in quenched approximation and hence that the net dimension of the operator is
\beq
d_{\overline{\psi}\psi} = 2 + \sqrt{1-\frac{3\alpha}{\pi}}~.
\eeq
In quenched, rainbow approximation [also called the planar approximation] the
dimension of \mbox{$[\overline{\psi}\psi]^2$} is just
$2\,d_{\overline{\psi}\psi}$ and one is therefore lead to infer that, in the
continuum limit [obtained, in this approximation, as
\mbox{$\alpha\rightarrow\,\frac{\pi}{3}$}]
\mbox{$ d_{[\overline{\psi}\psi]^2}\rightarrow\,4$} and hence that it becomes a
relevant operator.  This inference is supported by an analysis of quenched
\qedf which goes beyond the rainbow approximation, using the dressed
electron-photon vertex of \Eq{VA} (Atkinson \etal, 1993).

Rakow (1991) studied \qedf with the additional chirally-invariant,
four-fermion interaction term
\beq
\label{AIRakow}
\int\,d^4x\,
\frac{G_0}{2}\,\left[ \overline{\psi}(x)\psi(x)\overline{\psi}(x)\psi(x)
      - \overline{\psi}(x)\gamma_5\psi(x)\overline{\psi}(x)\gamma_5\psi(x)
\right]
\eeq
which is just the interaction of the Nambu--Jona-Lasinio model [NJLM] (Nambu
and Jona-Lasinio, 1961).  With this term, \Eqs{BRakow} and (\ref{SPRakow})
receive the additive corrections:
\beqn 
\frac{G_0}{\pi}\,\int_0^{\Lambda^2_{\rm UV}}\,
\frac{dk^2\,k^2}{4\pi}\,\frac{B(k^2)}{k^2\,A(k^2)^2+B(k^2)^2}
\; & \;\; {\rm  and} \;\; & \; 
G_0\,+\, \int_0^{\Lambda^2_{\rm UV}}\,
\frac{dk^2\,k^2}{4\pi}\,\frac{S_5(k^2,r^2)}{k^2\,A(k^2)^2+B(k^2)^2}~,
\eeqn
respectively, and the other equations remain unchanged.  [This model was
first considered by Leung \etal~(1986).]  With this additional
interaction, \Eq{AIRakow}, all of the functions scale with $p/m_{\rm R}$, the
fermion and photon propagators and the pseudoscalar and scalar scattering
amplitudes, showing that all of the interactions that are relevant in the
continuum limit are now included.

With $\alpha_0=0$ one simply has the NJLM in which chiral symmetry is
dynamically broken for $G_0\Lambda^2_{\rm UV}\geq 4\pi^2$.  In
general the model has a two-dimensional phase diagram in the
\mbox{$(G_0\Lambda^2_{\rm UV},\alpha_0)$}-plane.  From Fig.~14 of Rakow (1991)
one infers that the model has DCSB on a set which, in the first quadrant, can
be represented approximately as:
\beq
{\cal S}_1= \left\{ (G_0\Lambda^2_{\rm UV},\alpha_0)\,:\; G_0\Lambda^2_{\rm
UV}\geq 0, \;\alpha_0\geq 0,\; \frac{G_0\Lambda^2_{\rm UV}}{4\pi^2} +
\frac{\alpha_0}{\alpha_c} \geq 1
\right\}~.
\eeq
This is only a part of the picture, however, and for a given value of
$\alpha_0\in[0,\infty)$ there is always a value of
\mbox{$G_0=G_0^c\in\,(-\infty,4\pi^2]$} such that the model exhibits DCSB for
$G_0 > G_0^c$: as $\alpha_0$ is increased $G_0^c$ moves towards $-\infty$.
The domain in the \mbox{$(G_0\Lambda^2_{\rm UV},\alpha_0)$}-plane on which
chiral symmetry is dynamically broken is similar to that found in quenched,
rainbow approximation (Leung \etal, 1986).

One may now return to the question of triviality.

One notes that, for a given value of $\alpha_{\rm R}$, the correlation length
increases as $G_0$ decreases and takes its maximum value at $G_0=\,-\infty$
but that at any point not on the phase boundary this maximum value is finite.
On the phase boundary the renormalised coupling is always zero and hence,
even with the four-fermion interaction of \Eq{AIRakow}, the theory doesn't
have a continuum limit of interacting fermions.

There is another possibility, however.  There has been speculation for some
time (Miransky, 1985a; Kogut \etal, 1988a, 1988b) that \qedf has an
interesting continuum limit of interacting, strongly-bound $e^+$-$e^-$
states.  One can convince oneself of this possibility simply by looking at
the effective action for \qedf as obtained in Roberts and Cahill (1986), for
example.  Expanding this action, following the procedure of Roberts
\etal~(1988), for example, one finds interactions between the $e^+$-$e^-$ bound 
states whose strength is characterised by
\beq
g_{\rm Y} = \frac{{\rm const.}<\infty}{f}
\eeq
where
\beq
f^2 = \frac{1}{2\pi^2}\,\int_0^{\Lambda^2_{\rm UV}}\,ds\,s\,B(s)\, \frac{B(s)
- \frac{1}{2}\,s\,B'(s)}{(s+B(s)^2)^2}~.
\eeq
It is clear then that as one takes the cutoff to infinity, in order to
recover the continuum limit, the theory will be one of interacting bosons if
$f<\infty$ and free bosons if it is not.

It is quite clear from \Eq{BasymptQED} that in quenched approximation $f$ is
indeed finite.  In going beyond quenched approximation, however, one must use
numerical solutions, or some approximation to them, in which case one finds
that $f=\infty$ (Gusynin, 1990; Kondo, 1990; Maris 1993); a result which
suggests that the continuum limit is one of noninteracting bosons.

\subsect{Analytic Structure of the Electron Propagator}

The differential equation, \Eq{QEDFde}, has been studied by Atkinson and
Blatt (1979).  This study revealed a ``pathology'' of the ladder
approximation, used in all of the studies discussed in the previous section
[except Kondo (1992)], that has recently attracted a good deal of attention.
Solving \Eq{QEDFde} one finds that instead of an electron propagator with a
real branch cut on the timelike $p^2$ axis, as expected on physical grounds,
the electron propagator in rainbow-quenched \qedf has a pair of complex
conjugate branch points in the complex $p^2$ plane whose position depends on
$\alpha$.  The recent studies of rainbow approximation model DSEs for QCD
(Stainsby and Cahill, 1990; Maris and Holties, 1992; Stainsby and Cahill,
1992; Stainsby, 1993; Maris, 1993) suggest that this pathology is an artifact
of the rainbow approximation; a conclusion supported by the extended analysis
of rainbow approximation \qedf by Maris (1993, Sec.~4) whose analysis also
suggests that this ``pathology'' survives the inclusion of the one-loop
vacuum polarisation, itself calculated in rainbow approximation.

It may be that the resolution of this problem lies simply in using a dressed
fermion-gauge boson vertex; the possibilities and constraints on which we
discussed in Sec.~\ref{psi-gamma-vertex}.  However, there have been no
studies to date that test this hypothesis directly in \qedf. It should be
noted, however, that, in a QCD-based model, Burden \etal~(1992b) demonstrated
that the structure of the vertex does significantly affect the singularity
structure of the fermion propagator.

In Sec.~\ref{psi-gamma-vertex} we briefly discussed the gauge technique.
This has also been applied to \qedf by Delbourgo and West (1977) who advocate
an iterative solution of the DSEs for the spectral densities.

At lowest order, as it is defined in this approach, the only equation of
interest is
\beq
\label{DWDSE}
Z_2^{-1} = (\pslash - m_0)\,S^{(0)}(p) - ie^2\,\int\dqbf\,S^{(0)}(p) \,
\Gamma^{(0)}_\mu(p,p-q)\,S^{(0)}(p-q)\, D^{\mu\nu(0)}(q)\,\gamma_\nu
\eeq
where $S^{(0)}$ is given by the expression in \Eq{GTS}, $\Gamma^{(0)}$ by
\Eq{spV} and 
\beq
\label{DGTZ}
D^{\mu\nu(0)}(q) = \left( -g^{\mu\nu} + (1-\xi)\frac{q^\mu q^\nu}{q^2}\right)
{1\over q^2}
\eeq
with $\xi$ the gauge parameter.

Making use of these definitions \Eq{DWDSE} becomes
\beq
\label{GTDSEF}
Z_2^{-1} = \int_{-\infty}^{\infty}d\omega\, \frac{\rho_\psi(\omega) }
{\not\!p - \omega} \left(\pslash - m_0 + \Sigma(p,\omega)\right)
\eeq
with $\Sigma(p,\omega)$ obtained in the lowest order of perturbation theory:
\beq
{\rm Im}\,\Sigma(p,\omega) = \frac{e^2\,(p^2-\omega^2)}{16\,\pi\, p^3}
\left[ \xi\, (p^2 + \omega^2) - (\xi + 3)\, p\,\omega \right]~.
\eeq
Making use of the definitions
\beqn
Z_2^{-1} = \int_{-\infty}^{\infty}d\omega\,\rho_\psi(\omega) \;\; & \;\; {\rm
and} \;\; & m_0 = Z_2
\int_{-\infty}^{\infty}d\omega\,\omega\,\rho_\psi(\omega)
\eeqn
then \Eq{GTDSEF} yields the following result in Landau gauge:
\beqn
\label{GTSZ}
\lefteqn{S^{(0)}(p) = \frac{1}{\pslash - m} }\\
& & - \left(\frac{m^2}{\mu^2}\right)^{2\eta}
\,\Gamma(1-\eta)^2\,\Gamma(1+2\xi)\, \left[
\frac{\pslash}{p^2}\left[ F\left(1-\eta,1-\eta;1;\frac{p^2}{m^2}\right)
                        - 1 \right] +
\frac{1-\eta}{m}\,F\left(1-\eta,2-\eta;2;\frac{p^2}{m^2}\right) \right]
\nonumber
\eeqn
where $\mu^2$ is an infrared cutoff and $\eta = -3\alpha/(4\pi)$.  The next
iteration in the procedure uses this result coupled with \Eq{spV} and
\Eq{DGTZ} in the DSEs for the photon propagator and electron-photon vertex. 

The interesting feature of \Eq{GTSZ} is that the fermion propagator has a
branch cut at $p^2 = m^2$; i.e., it has the singularity structure expected on
physical grounds.  This is a feature that it has in common with the
propagator obtained with the solution of a linearised form of \Eq{QEDFde}
(Maris, 1993, Sec. 4.2):
\beq
x\,(x+m^2)\,B''(x) + 2\,(x+m^2)\,B'(x) + \frac{3\alpha}{4\pi} \,B(x) =0~,
\eeq
with $m\equiv B(0)$, which is a hypergeometric differential equation whose
solution is
\beq
B(x) = m\,F\left(\frac{1}{2}+\frac{1}{2}\sqrt{1 - \frac{3\alpha}{\pi}}\, , \,
\frac{1}{2}-\frac{1}{2}\sqrt{1 - \frac{3\alpha}{\pi}}\,;\,2\,;
\,-\,\frac{x}{m^2}\right)~.
\eeq
This is, perhaps, not surprising since the operational procedure in the
gauge-technique is to ensure that one has linear equations in the spectral
density.  The study by Maris (1993, Sec. 4.2.3) emphasises that the
singularity structure of the solution of the nonlinear differential equation
is quite different from that of the solution of the linearised equation: it
has complex conjugate branch points.

The application of the gauge technique to \qedf has not been pursued further.
It is not known whether the procedure, outlined briefly here, converges nor
whether the desirable singularity structure of \Eq{GTSZ} survives further
iteration.  Given the results of Stainsby and Cahill (1990), Maris and Holties
(1992), Stainsby and Cahill (1992), Stainsby (1993, Sec. 3) and Maris (1993,
Sec. 4) it would be interesting to answer these questions.  In addition to
the limitations mentioned in Sec.~\ref{psi-gamma-vertex}, however, one must
add that the procedure advocated is quite difficult to implement.

At present, in the study of Abelian gauge theories the ingredients that are
necessary and/or sufficient to ensure physical singularity structure in the
propagators remain unknown.

\sect{Gauge Boson Sector of QCD}
\label{sect-QCD-gluon}
We have briefly discussed the DSEs for QCD in Sec.~\ref{subsect-QCD} with the
equation for the vacuum polarisation and gluon propagator represented
diagrammatically in \Fig{gluon_dse_fig}.  In this section we will review the
attempts that have been made to solve these equations, the approximations
used and the results obtained.  [See also, H\"{a}dicke (1991).]  This is
important because in the application of DSEs to hadronic physics a physically
reasonable form for the quark propagator is needed and, in practice, this is
obtained by solving the quark DSE with Ans\"{a}tze for the gluon propagator
and quark-gluon vertex, with the former often being motivated by studies of
the DSE for the gluon propagator.

Before proceeding, however, this is an opportune point to make a number of
observations.  We use our Euclidean space conventions throughout this
section, see Sec.~\ref{subsect-Euclidean}.

In an arbitrary covariant gauge, specified by $\xi$, the gluon propagator can
be written as
\beq
D_{\mu\nu}(q) = \left\{\delta_{\mu\nu} - \frac{q_\mu q_\nu}{q^2}\right\}
\frac{1}{q^2[1+\Pi(q^2)]} + \xi \frac{q_\mu q_\nu}{q^4}
\eeq
where, as usual, $\Pi(q^2)$ is the vacuum polarisation.  It will be observed
that only the transverse piece of the gluon propagator is modified by
interactions.  This is a result of BRS invariance which yields, in the
covariant gauge fixing scheme, the STI (Pascual and Tarrach, 1984, pp. 42-45)
\beq
q_\mu\,D_{\mu\nu}(q) = \xi \frac{q_\nu}{q^2}~.
\eeq
This is an important constraint.

The least ambiguity and/or uncertainty exists in the form of the gluon
propagator at large spacelike-$q^2$ since this is the domain on which
perturbation theory and the renormalisation group can be easily applied.
Indeed, the evaluation of the asymptotic form of $\Pi(q^2)$ is a textbook
exercise (for example, Pascual and Tarrach, 1984, pp. 70-76).  Determining
the form of the propagator at smaller spacelike-$q^2$ is, however, a
difficult problem.

In this connection, one can employ the renormalisation group in QCD to obtain
an approximate relation between the vacuum polarisation and the running
coupling constant in QCD.  In Landau gauge, the renormalisation group
equation for
\beq
\Delta(q^2;\alpha_0) \equiv \frac{\alpha_0}{1 + \Pi(q^2;\alpha_0)}
\eeq
is, with $Z_\Delta = (Z_1^{gh})^{-2}\,(Z_3^{gh})^2$ where $Z_1^{gh}$ and
$Z_3^{gh}$ are renormalisation constants for the ghost-gluon vertex and ghost
wave function, respectively,
\beq
\label{RGED}
\left[ \mu\, \frac{\partial}{\partial \mu} 
        +\beta(\alpha(\mu))\,\alpha(\mu)\,\frac{\partial}{\partial \alpha} -
\mu\,\frac{d}{d \mu} \ln\,Z_\Delta\,\right]\, \Delta_{\rm R}(q^2;\alpha(\mu))
=0
\eeq
where $\beta(\alpha)$ is the QCD $\beta$-function:
\beq
\beta(\alpha(\mu)) = \frac{\alpha(\mu)}{\pi}\,\frac{2\,N_f -33}{6}
                + {\rm O}((\alpha(\mu)/\pi)^2)~.
\eeq
If one assumes that
\beq
Z_1^{gh} =Z_3^{gh}~,
\eeq
which is sometimes called the ``Abelian approximation'', then the last term
on the left-hand-side of \Eq{RGED} vanishes and $\Delta_{R}$ satisfies the
same renormalisation group equation as $\alpha_{\rm R}$.  Since at large
spacelike momenta one has
\mbox{$\Delta(q^2\rightarrow \infty) = \alpha$} then the equivalence
of the renormalisation group equations leads to
\beq
\Delta(q^2) = \alpha\left(\frac{q^2}{\Lambda_{\rm QCD}^2}\right)~;
\eeq
i.e., to an identification of $\Delta_{\rm R}$ with the running coupling
constant in QCD and to
\beq
\label{AbApprox}
\left(g^2\,D_{\mu\nu}(q)\right)_{\rm R} \approx 
\left\{\delta_{\mu\nu} - \frac{q_\mu q_\nu}{q^2}\right\}
                \frac{4\pi\,\alpha(q^2)}{q^2} + \xi \frac{q_\mu q_\nu}{q^4}~.
\eeq
[This argument is a minor modification of that presented in Bar-Gadda (1980);
see also, Itzykson and Zuber (1980, Chap. 13).] Equation~(\ref{AbApprox})
provides a prescription for an extrapolation of the known form of the
quark-quark scattering kernel at large spacelike-$q^2$ (Marciano and Pagels,
1978) to the entire $q^2$-domain on which the running coupling constant is
known.

At two-loop order
\beq
\label{alphaTL}
\alpha_2(q^2)= \alpha_1(q^2)\left[ 1 - 
        \frac{1}{2}\,\frac{153 - 19 N_f}{33- 2 N_f}\,
\frac{\alpha_1(q^2)}{\pi}\, \ln\left(\frac{1}{2}\, \ln\frac{q^2}{\Lambda_{\rm
QCD}^2}\right)\right]
\eeq
where
\beq
\alpha_1(q^2) = \frac{d_M\,\pi}{\ln(q^2/\Lambda_{\rm QCD}^2)}
\eeq
with $[d_M = 12/(33 - 2 N_f)$] [see, for example, Pascual and Tarrach, 1984,
pp. 128-131].  The renormalisation group invariant mass scale in QCD,
$\Lambda_{\rm QCD}$, is determined by fitting data in a number of high energy
experiments and, in the $\overline{MS}$ scheme and with four quark flavours,
\beq
\Lambda_{\rm QCD} = 0.20\,\pm\,
\begin{array}{l}
0.15 \\ 0.080
\end{array}
\;{\rm GeV}
\eeq
(Particle Data Group, 1990).  It is worth noting that at
\mbox{$q^2/ \Lambda_{\rm QCD}^2 = 100$}; i.e., $\sqrt{q^2}\sim 2$~GeV, 
\beq
\frac{\alpha_2 - \alpha_1}{\alpha_2} \approx -0.3~:
\eeq
$\alpha_2 \approx 0.25$.  This suggests that the three-loop contribution may,
already at this $q^2$, not be negligible and that the strength of the
interaction extracted from this expression for the running coupling constant
is unreliable for $q^2$ less than this.

\subsect{Infrared Behaviour of the Gluon Propagator}
\label{subsect:IRGP}
The discussion above illustrates that the tools of perturbation theory cannot
provide information about the structure of the gluon propagator at small
spacelike-$q^2$.  This region is very important as there is an expectation
that the structure of the gluon propagator at \mbox{$q^2\simeq 0$} has
important implications for quark confinement: in an imprecise way one might
say that the behaviour of the \mbox{$q-\overline{q}$} interaction in this
region determines the long range properties of the \mbox{$q-\overline{q}$}
potential and hence incorporates the physics of confinement.  [The discussion
of confinement in QED$_3$, Sec.~\ref{sect-QED3}, illustrates this.]  On this
domain nonperturbative techniques are necessary and studying the DSE for the
gluon propagator is one such approach.

\subsubsect{Singular in the Infrared? Axial gauge studies}
Some often quoted studies of the DSE for the gluon propagator (Baker \etal,
1980a, 1980b, 1983) adopt the axial-gauge formalism which implements the
constraint \mbox{$n\cdot A^a = 0$}, with $a$ the gluon-field's colour index
and $n_\mu$ an arbitrary four-vector: $n\cdot n > 0$, and in which ghost
fields are absent, Sec.~\ref{subsect-QCD}. [With our Euclidean conventions,
temporal gauge is specified by $n\cdot n < 0$.]  In this approach the gluon
propagator has the form
\beqn
\label{Daxg}
D_{\mu\nu}(q,\gamma) &=& F_1(q,\gamma)M_{\mu\nu}(q,n) +
F_2(q,\gamma)N_{\mu\nu}(q,n)
\eeqn
with
\beqn
M_{\mu\nu}(q,n) = \delta_{\mu\nu}-\frac{q_\mu n_\nu+q_\nu n_\mu}{q\cdot n}
+n^2 \frac{q_\mu q_\nu}{(q\cdot n)^2}~
\;\; & \;\; {\rm and} \;\; & \;\; 
N_{\mu\nu}(q,n) = \delta_{\mu\nu} - \frac{n_\mu n_\nu}{n^2}
\eeqn
and where \mbox{$\gamma =[q\cdot n]^2/[q^2 n^2]$} is the ``gauge parameter''.
The free propagator in axial gauge has $F_1 = -q^{-2}$ and $F_2=0$.  As
remarked in Sec.~\ref{subsect-QCD}, the benefit of axial gauge is the absence
of ghosts; the drawbacks are the loss of Lorentz invariance and the
gauge-dependent singularities in the propagator at $q\cdot n = 0$.

In axial gauges the DSE for the gluon vacuum polarisation, given
diagrammatically in \Fig{gluon_dse_fig}, does not have a ghost-loop term.
Further, because of the structure of the bare 4-gluon vertex, the term
involving the dressed 4-gluon vertex vanishes in the contraction
\mbox{$n_\mu \,\Pi_{\mu \nu}$}. Hence one can obtain a simplified integral
equation for \mbox{$n_\mu\Pi_{\mu\nu}n_\nu$} which only receives
contributions from the first two diagrams and the last diagram on the
right-hand-side of \Fig{gluon_dse_fig}.  This is the approach commonly taken
to simplify the analysis of the gluon DSE.  Further, fermion loop
contributions are often neglected leaving only the second and last diagrams
and thus a closed equation for the gluon vacuum polarisation once the dressed
3-gluon vertex is known.

The STI for the 3-gluon vertex, which takes the following simple form in
axial gauges:
\beq
\label{STIGGGAG}
p_\lambda\,\Gamma^3_{\lambda\mu\nu}(p,q,r) \,
D_{\mu\rho}(q)\,D_{\nu\sigma}(r)= D_{\rho\sigma}(r) - D_{\rho\sigma}(q),
\eeq
fixes its longitudinal part, however, the transverse part remains unknown.
[The colour structure of the vertex is, of course, $f^{abc}$.]  One way to
proceed is to neglect the transverse piece entirely and simply solve the STI
which leads to the Ansatz (Baker \etal, 1980a)
\beq
\Gamma^3_{\lambda\mu\nu}(p,q,r) = \Xi_{\lambda\mu\nu}(p,q,r) 
+ \Xi_{\mu\nu\lambda}(q,r,p) + \Xi_{\nu\lambda\mu}(r,p,q)
\eeq
where
\beqn
\lefteqn{\Xi_{\lambda\mu\nu}(p,q,r) = } \\
& & \delta_{\lambda\mu}\left[\frac{p_\nu}{p^2\,F_1(p^2,\gamma_p)}
-\frac{q_\nu}{q^2\,F_1(q^2,\gamma_q)}\right]+
\left[\frac{1}{p^2\,F_1(p^2,\gamma_p)}
                        -\frac{1}{q^2\,F_1(q^2,\gamma_q)}\right]\, \left[
p\cdot q\,\delta_{\lambda\mu} - q_\lambda\,p_\mu\right]\,
\frac{q_\nu-p_\nu}{p^2 - q^2} \nonumber
\eeqn
which is free of kinematic singularities.  At this point the further
assumption $F_2 \equiv 0$ has been made; i.e., that the dressed gluon
propagator has the same tensor structure as the free one.  Substituting this
into the DSE and regularising to remove ultraviolet divergences one obtains
an equation of the form
\beq
\frac{1}{F_1(p^2,\gamma_p)} = 
\frac{p^2 + \int\dqbf\,K(p,q,n)\,F_1(q,\gamma_q)}
        {1 - \int\dqbf\,L(p,q,n)\,F(q,\gamma_q)\,F(r,\gamma_r)}
\eeq
where $K$ and $L$ are complicated functions of their arguments.  Baker
\etal~(1981b) constructed a $\gamma$ independent, few-parameter form of $F_1$
which satisfied this integral equation to an accuracy of a few percent.  This
form has the asymptotic behaviour:
\beqn
F_1(p^2,\gamma_p) \;\stackrel{p^2 \simeq 0}{\sim}\; \frac{1}{p^4} & \;\; {\rm
and} \;\; & F_1(p^2,\gamma_p) \;\stackrel{p^2 \sim\infty}{\sim}\;
\frac{1}{p^2\,\left[\ln p^2\right]^{-11/16}}~.
\eeqn

The possibility that the gluon propagator has a double-pole at the origin is
appealing to many because it implies area law behaviour of the Wilson loop
(West, 1982) which is often regarded as a signal of confinement.  In this
connection, the ``potential'' obtained from a propagator with this form in
the infrared, using the four-dimensional analogue of \Eq{Vclass}, has a
linearly rising piece at large distances.  [This Fourier transform can be
defined as the solution of the differential equation obtained by operating on
it with $\bigtriangledown^2$.]

The gauge technique can also be employed to solve the gluon DSE (Delbourgo,
1979a, 1979b, 1981; Atkinson \etal, 1983; Delbourgo and Zhang, 1984).  Using
an Ansatz for $\Gamma^3_{\lambda\mu\nu}$ that is consistent with the STI,
Atkinson \etal~(1983) obtained the following equation;
\beq
\label{AetalGT}
n^2\,(1-\gamma)\,\left[ F_1(p^2)^{-1} - p^2 \right] =
\int_0^\infty\,ds\,\hat{\rho}(s)\,R(p,s,n)
\eeq
where $R(p,s,n)$ is a complicated function of its arguments and
\beq
F_1(p^2) = \int_0^\infty\,ds\,\frac{\hat{\rho}(s)}{(p^2 + s)^3}~.
\eeq
Equation~(\ref{AetalGT}) is a nonlinear integral equation for the spectral
density.  The nonlinearity arises because of the demand that the Ansatz for
the 3-gluon vertex satisfy the STI and distinguishes this analysis from the
other gauge technique analyses referred to above.  Analysing this equation,
Atkinson \etal~(1983) concluded that the axial-gauge gluon propagator is
likely to have a $q^{-4}$ singularity in the infrared but that $F_1$ is
unlikely to be independent of $\gamma$, as claimed by Baker
\etal~(1981b). 

There is a caveat to be borne in mind when considering these axial-gauge
studies.  It has been argued (West, 1983), from a general study of the
properties of the spectral representation of the gluon propagator in axial
gauge, that the coefficient of the $\delta_{\mu\nu}$ term in \Eq{Daxg} cannot
be more singular than $q^{-2}$.  This suggests that neglecting the $F_2$
term, as was done in the studies described above, is a poor approximation
because there are cancellations in the infrared between $F_1$ and $F_2$.

\subsubsect{Singular in the Infrared? Landau gauge studies}
There have also been similar studies of the DSE for the gluon vacuum
polarisation in Landau gauge (Mandelstam, 1979; Bar-Gadda, 1980; Brown and
Pennington, 1988a, 1988b, 1989).  In such approaches the caveat mentioned
above is circumvented because of the presence of ghost fields, however, this
approach has its own problems: 1) There is no way to eliminate the diagram
involving the 4-gluon vertex as there was in axial gauge.  In all of the
studies to date this contribution has simply been neglected in the hope that
the 3-gluon vertex alone will contain the essence of the infrared behaviour
of QCD if not the details; and 2) The ghost contribution is present.  Indeed,
it is necessary to ensure that the vacuum polarisation is transverse.
However, in a one-loop, perturbative calculation the ghost-loop only makes a
numerically small contribution to the vacuum polarisation in Landau gauge.
For this reason the diagram is neglected, again in the hope that this will
not remove the dominant infrared contributions in QCD.

Subject to these approximations the DSE for the gluon vacuum polarisation
contains only the first two terms on the right-hand-side of
\Fig{gluon_dse_fig}.  A final step of neglecting the fermion-loop
contribution yields, as in the axial gauge studies, a closed equation for the
vacuum polarisation once the dressed 3-gluon vertex is known.

In a general covariant-gauge the STI for the 3-gluon vertex involves the
ghost self-energy, $b(q^2)$, and the proper ghost-gluon vertex function
\mbox{$G_\mu(k;q,r) = \,r_\nu G_{\mu\nu}(k;q,r)$}, where $k_\mu$ is the
incoming gluon momentum.  In Landau gauge the gluon propagator is transverse
which entails that, in the limit of vanishing ghost momentum,
\beqn
G_{\mu\lambda}(k;q,r)\,D_{\lambda\nu}^{-1}(r) \approx D_{\mu\nu}^{-1}(r) \; &
\;\; 
{\rm and} \;\; & G_{\mu\lambda}(k;q,r)\,D_{\lambda\nu}^{-1}(q) \approx
D_{\mu\nu}^{-1}(q)
\eeqn
(Marciano and Pagels, 1978, pp. 171-172).  This result is commonly used to
justify a simplification of the STI that is assumed, again, to capture the
essence of the infrared behaviour of QCD.  A final commonly used assumption
is simply to set $b(q^2)=0$, which reduces the STI to the one given in
\Eq{STIGGGAG}; i.e., to the axial-gauge identity.  This is solved by
\beq
\label{GSTIQCD}
\Gamma^3_{\lambda\mu\nu}(p,q,r) = \Theta_{\lambda\mu\nu}(p,q,r) 
+ \Theta_{\mu\nu\lambda}(q,r,p) + \Theta_{\nu\lambda\mu}(r,p,q)
\eeq
with
\beq
\Theta_{\lambda\mu\nu}(p,q,r) = 
\delta_{\lambda\mu}\left[\frac{p_\nu}{D^\Pi(p^2)}
                        -\frac{q_\nu}{D^\Pi(q^2)}\right]+
\frac{1}{p^2 - q^2}\left[\frac{1}{D^\Pi(p^2)}
                        -\frac{1}{D^\Pi(q^2)}\right]\,
\left( p_\mu\,q_\lambda - p\cdot q\,\delta_{\lambda\mu}\right)
        \,\left(p_\nu - q_\nu\right)
\eeq
where $D^\Pi(p^2) = [1+\Pi(p^2)]^{-1}$.  As usual, the STI has fixed the
vertex up to the addition of a term which vanishes when contracted with any
of the external momenta and also when any of the external momenta become
zero.  This last fact suggests that the 3-gluon vertex in
\Eq{GSTIQCD} may alone be sufficient to capture the essence of the infrared
behaviour of QCD.

Subject to the approximations discussed above, one now has a single integral
equation for the gluon vacuum polarisation given by \Fig{gluon_dse_fig} with
only the second diagram on the right-hand-side closed by using \Eq{GSTIQCD}
for the 3-gluon vertex and this is the equation studied by Brown and
Pennington (1989).  This integral equation has the usual ultraviolet
divergences but infrared divergences are also possible, especially if
\mbox{$D^\Pi(p^2)\sim p^{-2}$}.  The ultraviolet divergences were handled in
the usual way and an integral equation, involving the QCD running coupling
constant, obtained for the renormalised function $D^\Pi_{\rm R}(p^2)$.

The possibility that $D^\Pi_{\rm R}$ has a term of the form $p^{-2}$ was
handled by using the ``plus'' definition in the theory of distributions:
\beq
\label{pluspres}
\int_0^\infty\,dx\,\left(\frac{1}{x}\right)_+\,\phi(x,y)
= \int_0^\infty\,dx\,\frac{1}{x}
\left(\phi(x,y) - \theta(y-x)\,\phi(0,y)\right)~.
\eeq
This enabled Brown and Pennington (1989) to find a solution of the
renormalised equation of the form
\beq
D^\Pi_{\rm R}(p^2) = \frac{A \mu^2}{p^2} 
        + D^\Pi_\infty(p^2)\,\sum_{n=1}^{N_f}\,a_n\,
\left[\frac{p^2}{p^2+p_0^2}\right]^{nb}~,
\eeq
where
\beq
D^\Pi_\infty(p^2) = \left( 1 + \frac{\alpha(\mu)}{d_M\pi} \ln\left[1 +
\frac{p^2}{\mu^2}\right]^{-28/53} \right)^{-1}
\eeq
with $\mu$ the renormalisation point, and where $A$, $p_0$, $a_n$ and $b$
were determined by requiring self-consistency of this from under iteration in
the integral equation; i.e., in Landau gauge, subject to the [severe?]
approximations described above, the gluon propagator has a double pole at
small-$p^2$.  This result is in accord with all of the other Landau-gauge
studies.

The effect of fermion loops on the form of the gluon propagator in the
infrared has been studied (Brown and Pennington, 1988b) using a simpler
Ansatz for the 3-gluon vertex, first proposed by Mandelstam (1979):
\beq
\Gamma^3_{\lambda\mu\nu}(p,q,r) = \Gamma^{3(0)}_{\lambda\mu\nu}(p,q,r)
        \frac{1}{D^\Pi(p^2)}
\eeq
where $\Gamma^{3(0)}$ is the bare 3-gluon vertex.  Based on the fact that the
results obtained with this Ansatz neglecting the fermion loop are
qualitatively the same as those obtained using \Eq{GSTIQCD} (Brown and
Pennington, 1989), this was judged to be a reasonable simplification.  This
study is not a self-consistent solution of the full coupled DSEs for the
fermion and gluon propagators and it has a number of flaws introduced by the
approximations, however, it is a first step.  The results suggest that
fermion loops act to suppress the infrared singularity in the gluon
propagator; i.e., that the double-pole found in the pure-gauge sector may be
removed by nonperturbative fermion-loop corrections.

\bigskip\underline{Vanishing at $q^2=0$?}\ipar
In QCD there are seven superficially divergent proper vertices [quark-:
$\Gamma_{FF}$; gluon-: $\Gamma_{VV}$; and ghost-: $\Gamma_{GG}$;
self-energies; quark-gluon-: $\Gamma_{FVF}$; three-gluon-: $\Gamma_{VVV}$;
four-gluon-: $\Gamma_{VVVV}$; and gluon-ghost-: $\Gamma_{GVG}$; vertices].
The set of DSEs that couple these vertices is the starting point for an
alternative approach to determining the form of the gluon [and quark]
propagator (Stingl, 1986; H\"{a}bel \etal, 1990a, 1990b; Stingl, 1992).  The
main difference between this approach and those discussed above is in the
method employed to solve the DSEs but behind this there is a different
philosophy.

The DSEs for the vertices in QCD can be written in the general form
\beq
\Gamma^i = g^2\,\Phi[\{\Gamma\}] 
\eeq
where \mbox{$\{\Gamma\} = \{ \Gamma_{FF}, \Gamma_{VV}, \Gamma_{GG},
\Gamma_{FVF}, \Gamma_{VVV}, \Gamma_{VVVV}, \Gamma_{GVG}\}$}.  Standard
perturbation theory solves this system of equations by iteration, starting
with $\Gamma^{(0)pert}$ and generating a power series:
\beq
\Gamma^{i\,pert} = \Gamma^{i\,(0)pert} + 
        \sum_{n=1}^{\infty}\,g^{2n} \Gamma^{n\,(0)pert}
\eeq
where the first iteration is obtained via
\beq
g^2\,\Phi[\{\Gamma^{(0)pert}\}] \equiv g^2 \Gamma^{(1)pert} + {\rm O}(g^4).
\eeq
The philosophy of this alternative approach is not to abandon the organising
principle of perturbation theory altogether; the solution of the DSEs is
still obtained as a power series in the coupling, but to allow each of the
vertices at each order to have an essentially nonanalytic dependence on the
coupling.  In employing this approach one assumes a form for the
nonperturbative part of the zeroth order vertices, \mbox{$\Gamma^{(0)nonp}=
\Gamma^{(0)} - \Gamma^{(0)pert}$}, in practice a rational polynomial Ansatz
is chosen, and requires that this reproduce itself under iteration, which
places self-consistency constraints on the coefficients in the
parametrisation via
\beq
g^2\,\Phi[\Gamma^{(0)}] = \Gamma^{(0)nonp} + g^2 \, \Gamma^{(1)} + {\rm
O}(g^4).
\eeq
The outcome of this procedure, however, is simply a set of
truncated/approximate DSEs which must be solved self-consistently.

Using this approach H\"{a}bel \etal~(1990a) studied approximate
coupled DSEs for the ghost self energy, $b(k^2)$ and proper ghost-gluon
vertex function \mbox{$G_{\mu\nu}(k;q,r)$}, using rational polynomial
Ans\"{a}tze for these functions, the 3-gluon vertex and gluon propagator.  A
consistent solution was found to be
\beqn
b^{(0)}(k^2) = 0 \;&\;\;{\rm and} \;\; & \; G_{\mu\nu}^{(0)}(k;q,r) =
\delta_{\mu\nu}~;
\eeqn
i.e., no zeroth-order nonperturbative parts is self-consistent at the
one-loop level.  The reason for this is that the loops in the integral
equations are either convergent or purely perturbative and thus cannot
generate a nonanalytic dependence on the coupling.

This result was used by H\"{a}bel \etal~(1990b) in a study of the
coupled DSEs for the gluon propagator and 3-gluon vertex.  In this study the
only diagrams that survived in the DSE for the gluon vacuum polarisation,
\Fig{gluon_dse_fig}, were the second and fourth on the right-hand-side
because all 4-gluon terms are of higher order in $g^2$ and fermions were
neglected.  The equation for the 3-gluon vertex was reduced to one involving
itself, the gluon propagator and ghost loops, which are perturbative
(H\"{a}bel \etal, 1990a).  With a rational polynomial Ansatz for the
zeroth-order, nonperturbative elements: the gluon vacuum polarisation is
assumed to be of the form
\beq
q^2\,\Pi^{(0)}(q^2) = c^2 + 2 a^2 + \frac{b^4}{q^2 + c^2}
\eeq
where, in this particular calculation the simplifying assumption $a=0=c$ was
made, and the 3-gluon vertex, $\Gamma^{3\,(0)}_{\lambda\mu\nu}(p,q,r)$,
depends on $9$ parameters, each of which multiplies a ratio of $p^2$ and/or
$q^2$ and/or $r^2$, the coupled DSEs yield $11$ equations for the $10$
variables.  The DSE for the 3-gluon vertex yields $9$ algebraic, coupled,
cubic equations and the DSE for the vacuum polarisation yields two linear
equations.  [The fact that the system is overdetermined is an artifact of the
simplifying assumption $a=0=c$.]  One solution is found which, although
giving a 3-gluon vertex which is complex, is judged to be ``reasonable'' by
the authors on the grounds that the fact that a solution exists at all, given
the crudity of the assumptions, is very encouraging.  This solution yields
\beq
\label{DStingl}
\frac{D^{\Pi\,(0)}(q^2)}{q^2} = \frac{1}{q^2[1+ \Pi^{(0)}(q^2)]} =
         \frac{q^2}{q^4+b^4}~;
\eeq
i.e., a gluon propagator that vanishes at $q^2=0$.

It has been argued (Stingl, 1986; H\"{a}bel \etal, 1990a, 1990b;
Zwanziger, 1991; Stingl, 1992) that \Eq{DStingl} is a propagator which
represents confined gluons because there are no poles on the timelike real
axis in the complex-$q^2$ plane and it allows an interpretation of the gluon
as an unstable excitation which fragments into hadrons before observation [in
a time of the order of $1/b$].  It is also argued (H\"{a}bel \etal,
1990a, 1990b; Stingl, 1992) that in this framework such a gluon propagator
should lead to a quark propagator with similar structure in the complex
plane, and hence a similar interpretation, but this result has not been
proven.

It should be remarked that a possible flaw in these studies is that the
Ansatz for the 3-gluon vertex, \mbox{$\Gamma^{3\,(0)}(p_1,p_2,p_3)$}, has
kinematic light-cone singularities of the form $1/(p_i^2)$.  Such
singularities cannot arise in perturbation theory (Ball and Chiu, 1980) and
hence such an Ansatz cannot reduce to the renormalisation group improved
3-gluon vertex in the deep spacelike region.

It is interesting that the form in \Eq{DStingl} is suggested by a number of
other studies.  It has been argued (Zwanziger, 1991) that in order to
completely eliminate Gribov copies (Gribov, 1979) and hence to fix Landau or
Coulomb gauge uniquely in lattice studies, one must introduce new ghost
fields into QCD in addition to those associated with the Faddeev-Popov
determinant in the continuum.  Analysing the lattice action thus obtained
suggests that the gluon propagator vanishes as $(q^2)^\gamma$, with
$\gamma>0$ not determined.  Subsequent analysis of a simplified model yields
$\gamma=1$ and, in fact, a gluon propagator of the form in \Eq{DStingl} with
$b$ a finite constant in Landau gauge.  Similar considerations lead Gribov
(1979) to the same result.

\subsubsect{Lattice Simulations}
There have been a number of lattice simulations of the gluon propagator
(Mandula and Ogilvie, 1987a, 1987b, 1988; Bernard \etal, 1993).  All of these
studies have fixed a lattice Landau gauge, which is plagued by Gribov copies,
but did not make use of the modifications suggested by Zwanziger (1991).
Using $16^3\times 40$ and $24^3\times 40$ lattices at $\beta=6.0$, Bernard
\etal~(1993) obtained a gluon propagator which allowed a fit of
the form in \Eq{DStingl} at small $q^2$ but which could not rule out a fit
using a standard massive particle propagator.  Other lattice sizes and values
of $\beta$ were also studied.  The results at $\beta=6.3$ on a lattice of
dimension $24^4$ were not inconsistent with these results but in this case
the small physical size of the lattice was a problem.  On a lattice of
dimension $16^3\times 24$ at $\beta=5.7$ it was found that the gluon
propagator was best fit with a standard massive vector boson propagator with
mass $\sim 600$~MeV.

These studies represent an improvement in both technique and lattice sizes
over earlier lattice studies of the gluon propagator (Mandula and Ogilvie,
1987a, 1987b, 1988) but the conclusions are not markedly different.  Mandula
and Ogilvie (1987a, 1987b) using $\beta= 5.6, 6.0$ on a $4^3\times 8$ lattice
and $\beta= 5.8$ on a $4^3\times 10$ lattice, obtained results that were
consistent with a free massive boson propagator with mass $\sim 600$~MeV.

Viewed as a whole, these studies appear to suggest that the Landau-gauge
gluon propagator is finite and nonzero at $q^2=0$.  This is consistent with
an analysis of an approximate DSE for the gluon vacuum polarisation using
the gauge technique (Cornwall, 1982).

\subsect{Summary}
A number of observations regarding the studies described above are in order.
It is clear that the behaviour of the gluon propagator in the infrared is
poorly understood with the results obtained depending on the
approximations/truncations used to obtain tractable DSEs and/or the
Ans\"{a}tze for the propagators and vertex functions.  From the point of view
of a DSE based phenomenological approach to QCD, there remains a great deal
of freedom in parametrising the infrared behaviour of the gluon propagator.

If one chooses to believe that those studies which yield $q^{-4}$ behaviour
are the most reliable then a caveat must be borne in mind.  This structure
leads to an infrared divergence in the DSEs for both the gluon and quark.  It
is therefore only defined with respect to some regularisation procedure.  One
such procedure is given in \Eq{pluspres} but this is not the only one
possible; for example, another commonly used regularisation is to replace
$q^{-4}$ by $\delta^{4}(q)$, which is a distribution that is integrable on
any domain containing the origin.  It is not known whether such a
prescription can reproduce itself under an iterative solution of the DSE for
the gluon vacuum polarisation.  The inference that $q^{-4}$ behaviour
corresponds to a linearly rising potential in configuration space is also
subject to this caveat since the Fourier transform is not defined until a
regularisation prescription is specified.

The lattice studies of the gluon propagator are at an early stage; gauge
fixing is a difficult problem in lattice simulations.  However, these studies
are of interest because they may be able to support or undermine one or
another of the approaches to simplifying the gluon DSE.  It seems reasonable
to doubt, however, whether the finite lattice size and associated infrared
cutoff will permit the identification of an infrared singular gluon
propagator.

Finally, as with \qedt and \qedf, the higher $n$-point functions have not
really been studied at all.  The studies described above show that they
clearly have an important bearing on the structure of the quark-quark
interaction in the infrared and, therefore, that they deserve further
attention.

\sect{Fermion Sector of QCD}
\label{sect-QCD-quark}
In this section we work primarily in Minkowski space using Minkowski-space
conventions unless explicitly specified otherwise.  Many elements of the
detailed discussions of QED in the preceding sections are relevant here and
we will draw on them as needed, introducing only those aspects of QCD which
differentiate it from QED in order to proceed.

\subsect{Dynamical Chiral Symmetry Breaking}
\label{subsect-quark-DCSB}
The starting point for studies of DCSB in QCD is the DSE for the quark
self-energy given in Eqs.~(\ref{Squark}) and (\ref{Sigmaprime_QCD}).  We will
discuss various models and approximation schemes, which attempt to
incorporate, as far as possible, the symmetries and leading-log perturbative
QCD results discussed in the previous section.  Of course, it is of great
interest to see how confinement can arise in such a scheme and we discuss
this in Sec.~\ref{subsect-quark-conf}.

\subsubsect{Proper Quark-Gluon Vertex}
\label{subsubsect-QCDvertex}
A common first approximation is to neglect the effect of ghosts in the quark
sector and assume that this can be compensated for by fine-tuning the
phenomenology in the gluon sector.  The justification of this, such as it is,
and its inherent flaws is described in Sec.~\ref{subsect:IRGP}.  It is
undesirable and raises the question of how important the resulting violation
of QCD gauge invariance is.  Indeed, it can be argued that this approximation
may compromise confinement.  Nevertheless, it does allow good use to be made
of what has already been learnt form QED studies.  The neglect of ghosts in
the quark sector implies a theory of quarks with some Abelian-like
characteristics, with QCD gauge invariance reducing to the requirement that
the colour-current be conserved at the quark-gluon vertex.  For this purpose
it is sufficient to require that \Eq{quark_gluon_STI} [without ghosts] is
satisfied for the renormalised quantities, which has then the same form as
\Eq{WTI_fermion}.  [The discussion associated with \Eq{GSTIQCD} is relevant
here.]  Making use of the QED studies discussed previously [e.g., Ball and
Chiu (1980), Curtis and Pennington (1990, 1991, 1992), and Burden and Roberts
(1993)], we can write then
\beq
\Gamma^\mu(p',p) =\Gamma_{\rm BC}^\mu(p',p) +
\Gamma_{\rm CP}^{{\rm T}\mu}(p',p) +
{\rm transverse\;\;parts}
\equiv \gamma^\mu + \Lambda^\mu(p',p)~,
\label{Gamma_noghosts}
\eeq
where we have indicated that only the longitudinal behaviour is specified by
\Eq{Gamma_noghosts}.  
The two specified parts of the vertex are the so-called Ball-Chiu vertex
$\Gamma_{\rm BC}$ and the transverse addition $\Gamma^{\rm T}_{\rm CP}$
suggested by Curtis and Pennington on the basis of requirements of
multiplicative renormalisability.  The transverse pieces are proportional to
$[g^{\nu\sigma}-(k^\nu k^\sigma/k^2)]$ where $k^\nu\equiv(p'-p)^\nu$ and so
we have then $k_\mu\Gamma^\mu(p',p)=k_\mu\Gamma_{\rm BC}^\mu(p',p)$.  These
are defined by
\beqn
\Gamma^{\mu}_{\rm BC}(p',p)
  & \equiv&\frac{A(p^2)+A(p'^2)}{2} \gamma^\mu 
+ \frac{(p+p')^{\mu}}{p^2 -p'^2} \left\{ \left[ A(p^2)-A(p'^2)\right]
\frac{\left[ \not\!{p} + \not\!{p'} \right]}{2} - \left[ B(p^2) - B(p'^2)
\right] \right\}\/, \label{Gamma_BC} \\
\Gamma^{{\rm T}\mu}_{\rm CP}(p',p) & \equiv &
\frac{\gamma^\nu(p'^2-p^2) - (p'+p)^\nu (\not\!{p'}-\not\!{p})}{2d(p',p)}
      \left[ A(p'^2)-A(p^2) \right]\/, \label{Gamma_trCP} \\
{\rm with}\;\; & & 
d(p',p) \equiv  
\frac{(p'^2-p^2)^2 + \left[ M(p'^2)^2 + M(p^2)^2\right]^2}
        {(p'^2+p^2)} \label{Gamma_CPd}
\eeqn
where $M(p^2) = B(p^2)/A(p^2)$.  The QED vertex proposed by Curtis and
Pennington has the unspecified transverse parts in \Eq{Gamma_noghosts} set to
zero; i.e., it is defined as
\beq
\Gamma_{\rm CP}^\mu(p',p)\equiv
\Gamma_{\rm BC}^\mu(p',p) + \Gamma_{\rm CP}^{{\rm T}\mu}(p',p)~.
\label{Gamma_CP}
\eeq
It is straightforward to verify that $(p'-p)_\mu\Gamma^\mu_{\rm
BC}=S^{-1}(p')-S^{-1}(p)$ and that $(p'-p)_\mu\Gamma^\mu_{\rm CP}=0$ as
required by \Eq{WTI_fermion}.  Since \Eq{QCD_renorm_BCs} implies that
$A(-\mu^2)\simeq 1$, then in the limit that $p'^2\simeq p^2\simeq -\mu^2$ the
RHS of \Eq{Gamma_noghosts} becomes $\gamma^\mu$ plus terms proportional to
$dA/dp^2$ and $dB/dp^2$ respectively, which for large $\mu^2$ will be very
small.  Thus in this limit we have $\Gamma^\sigma
\simeq\gamma^\sigma$ which is the usual perturbative vertex.  The only
way of determining the nonperturbative transverse pieces and ghost
contributions in
\Eq{Gamma_noghosts}
is to attempt to simultaneously solve the DSE for the quark-gluon vertex.
While $\Gamma_{\rm CP}$ does not include ghost effects it does give the
correct perturbative limit, reduces to the free vertex in the absence of
interactions, is free of unphysical kinematic singularities, transforms
correctly under parity and charge conjugation, is Lorentz covariant, and
preserves multiplicative renormalisability.

It is unfortunate that there are difficulties in using ghost-free gauges,
such as the axial gauge.  Any approximation made in a noncovariant gauge
tends to destroy the covariance of physical observables.  In addition, in
axial gauge additional singularities of the form $n\cdot p$ enter into
propagators.  The biggest difficulty, however, is that since one now has two
four-vectors [$p^\nu$ and $n^\nu$] the most general Lorentz structure of the
inverse quark propagator is $S^{-1}=(A\rlap/p-B)+(C\rlap/p-D)\rlap/n$, where
the functions $A, B, C,$ and $D$ depend on each of the Lorentz scalars $p^2$,
$p\cdot n$, and $n^2=\pm 1$. The fact that the scalar functions $A, B, C$,
and $D$ can depend explicitly on $n\cdot p$ and $n^{2}$ significantly
complicates the DSE analysis.  For these reasons studies of DSEs where
Lorentz covariance is important are most frequently carried out in a
covariant gauge.

\subsubsect{Asymptotic Quark Self-Energy Behaviour}
\label{subsubsect-QuarkAsympt}
Recall that the renormalised DSEs relate quantities at the renormalisation
point $\mu$ and that the appearance of the regularisation parameter $\Lambda$
is through the renormalisation constants $Z_S(\Lambda,\mu)$,
$Z_D(\Lambda,\mu)$, $Z_\Gamma(\Lambda,\mu)$, $Z_m(\Lambda,\mu)$,
$Z_g(\Lambda,\mu)$ and hence also through the bare quantities: the bare
coupling constant $g_0(\Lambda)$, gauge parameter $\xi_0(\Lambda)$, and quark
mass $m_0(\Lambda)$.  So for the quark self-energy in Eqs.~(\ref{Squark}) and
(\ref{Sigmaprime_QCD}) we understand that
\beq
-i\Sigma'(p) = {4\over 3}Z_\Gamma(\Lambda,\mu)
\;g^2\int^\Lambda {d^{\rm d}\ell\over (2\pi)^{\rm d}}(i\gamma_\mu)
(iS(\ell))(iD^{\mu\nu}(p-\ell))(i\Gamma_\nu(\ell,p))\;.
\label{new_Sigmaprime_QCD}
\eeq
In practice, numerical calculations of DSE solutions are carried out in
Euclidean space where it is convenient to take the regularisation parameter
$\Lambda$ to be a momentum cut-off, $\Lambda_{\rm UV}$.  In this case, it is
further convenient to choose $\mu$=$\Lambda_{\rm UV}$ since then all of the
$Z$-factors are unity; i.e., $Z_S(\Lambda_{\rm UV},\Lambda_{\rm UV})=1$, etc.  Thus for
$\Lambda_{\rm UV}$ and $\mu$ sufficiently large we can use the known asymptotic
behaviour to obtain information about quantities for $\mu\neq\Lambda_{\rm UV}$.  It is
of course then a standard check on numerical solutions that physical results
so obtained at fixed renormalisation point $\mu$ are independent of $\Lambda_{\rm UV}$
for $\Lambda_{\rm UV}\to\infty$.  Discussions of these issues can be found in Fomin
{\it et al.} (1983) and references therein.  An added advantage of choosing
$\mu=\Lambda_{\rm UV}$ with $\Lambda_{\rm UV}\gg\Lambda_{QCD}$ is that $\Sigma'(p)\simeq 0$ at
$p^2=-\mu^2$ and so this can be essentially neglected, which gives
$\Sigma(p)\simeq \Sigma'(p)$.  Using the above arguments we have arrived at
the commonly used expression for the quark self-energy DSE [with
$\mu=\Lambda_{\rm UV}$]
\beq
\Sigma(p) = i\;{4\over 3}\;4\pi\int^{\Lambda_{\rm UV}}
{d^4k\over (2\pi)^4}\;\gamma_\sigma\;S(k) {g^2\over
4\pi}\;D^{\sigma\nu}(k-p)\Gamma_\nu(k,p)\;.
\label{Sigma_q_exact}
\eeq

The asymptotic behaviour of the quark-quark scattering kernel $(K)$ is known
from renormalisation group analysis and to leading order [in Landau gauge] is
$[Q^2\equiv -(q'-q)^2]$
\beq
K_{nm;n'm'}(q',q;P)\simeq -i\;g^2(Q^2)\gamma^\sigma_{nm}D_{0\sigma\nu}
(q'-q)\gamma^\nu_{n'm'},
\label{qq_kernel}
\eeq
where to this order $K$ is independent of the center-of-mass momentum, $P$, 
and the renormalisation point, $\mu$, and where $n,m,n'$, and $m'$ are spinor
indices, [see, e.g., Nakanishi (1969)].  $D_0$ is the perturbative gluon
propagator; i.e., $D$ with $\Pi=0$.  In the asymptotic region $K$, as
expected, is dominated by one-gluon exchange ($K\simeq -i\;g^2\Gamma
D\Gamma)$ which leads to \Eq{qq_kernel}.  Since in this region
$\Gamma\simeq\gamma$ up to logarithmic corrections [which are unimportant for
generating the leading behaviour of the self-energy], we can use
\Eq{qq_kernel} for $g^2\gamma D\Gamma$ in \Eq{Squark}.  It is relatively
straightforward to verify that this produces the correct asymptotic behaviour
for the running mass [for a detailed discussion see, e.g., Fomin
\etal~(1983)].

As we showed in Sec.~\ref{sect-QCD-gluon}, if one makes the assumption
\mbox{$Z_1^{gh} =Z_3^{gh}$} then one may write  
$\alpha_s(Q^2) D_0^{{\rm T}\nu\sigma}(q)$ in place of the transverse part of
$(g^2/4\pi) D^{\nu\sigma}(q)$ in \Eq{Sigma_q} and so, in this approximation,
the nonperturbative part of the gluon propagator is absorbed into the
nonperturbative structure of $\alpha_s(Q^2)$.  In addition, it can argued
that [at least part] of the ghost effects omitted from $\Gamma_{\rm CP}$
and/or $\Gamma_{\rm BC}$ can be absorbed into the nonperturbative behaviour
of $\alpha_s(Q^2)$.  This then gives rise to the semi-phenomenological
expression for the quark self-energy [in Landau gauge]
\beq
\Sigma(p) = i\;{4\over 3}\;4\pi\int^{\Lambda_{\rm UV}}
{d^4k\over (2\pi)^4}\;\gamma_\sigma\;S(k)
\alpha_s(-(k-p)^2)\;D_0^{\sigma\nu}(k-p)\Gamma_\nu(k,p)\;,
\label{Sigma_q}
\eeq
where $\Gamma^\nu$ is then to be chosen consistent with the WTI and where
nonperturbative gluon and ghost effects have been [as far as possible]
absorbed into $\alpha_s$.  Hence, the problem has been reduced to motivating
particular nonperturbative gluon and ghost modifications and exploring the
resulting forms for the nonperturbative behaviour of the function $\alpha_s$.
Equation~(\ref{Sigma_q}) is a form of the quark self-energy DSE in Landau gauge
that we will consider in some detail here.  The advantages of Landau gauge
are that the gluon propagator is purely transverse and that the gauge
parameter does not run in Landau gauge $\xi(Q^2)=0$, [$\xi=0$ is an
ultraviolet fixed point, albeit not a stable one in general.  See Marciano
and Pagels (1978) and Pascual and Tarrach (1984)].

\subsubsect{Quark Dyson-Schwinger Equation}
\label{Quark_DSE}
Using \Eq{Squark} in \Eq{Sigma_q} and using $\Gamma^\nu=\Gamma^\nu_{\rm BC}$
we obtain in Euclidean space in Landau gauge [$\xi=0$]:
\begin{eqnarray}
A(p^2) & = & 1 + \frac{16\pi}{3}\int^{\Lambda_{\rm UV}} \frac{d^4k}{(2\pi)^4}
\frac{\alpha_s((p-k)^2)}{(p-k)^2} \frac{1}{A^2(k^2)k^2+B^2(k^2)} \times
\nonumber\\ 
& & \left\{ A(k^2)\frac{A(k^2)+A(p^2)}{2} \frac{1}{p^2}
\left[3p\cdot k - h(p,k)\right] \right.  \nonumber \\ 
& & \left. -
A(k^2)\Delta A(k^2,p^2) \left[k^2-\frac{(k\cdot p)^2}{p^2} + \frac{k\cdot
p}{p^2}h(p,k)\right] - B(k^2)\Delta B(k^2,p^2) \frac{h(p,k)}{p^2 }
\rule{0mm}{7mm}\right\}\/, \label{Aeqn} \\
B(p^2) & = & m_{\Lambda_{\rm UV}}+
\frac{16\pi}{3} \int^{\Lambda_{\rm UV}} \frac{d^4k}{(2\pi)^4}
\frac{\alpha_s((p-k)^2)}{(p-k)^2} \frac{1}{A^2(k^2)k^2+B^2(k^2)} \times
\nonumber\\ 
& & \left\{3B(k^2)\frac{A(k^2)+A(p^2)}{2} + \left[B(k^2)\Delta
A(k^2,p^2) -A(k^2)\Delta B(k^2,p^2)\right]h(p,k)\rule{0mm}{7mm}\right\}\/,
\label{Beqn}
\end{eqnarray}
where $h(p,k)=2\left[k^2p^2-(k\cdot p)^2\right]/(k-p)^2$ and
\mbox{$\Delta F(k,p) = [F(k^2) - F(p^2)]/[k^2 - p^2]$} 
and where $m_{\Lambda_{\rm UV}}$ is the running quark mass when
$\mu=\Lambda_{\rm UV}$.  Calculations to be reported later typically have no
ECSB [i.e., $m_{\Lambda_{\rm UV}}=0$].  For a calculation that explicitly
includes ECSB quark masses see Williams \etal~(1991).  Including the
additional Curtis-Pennington term in the vertex, [i.e., using
$\Gamma^\nu=\Gamma^\nu_{\rm CP}$] means that these equations are modified as
follows:
\begin{eqnarray}
A(p^2) & = & {\rm RHS~of~(\ref{Aeqn})} \nonumber\\ &+& \frac{16\pi}{3}
\int^{\Lambda_{\rm UV}}
\frac{d^4k}{(2\pi)^4} \frac{\alpha_s((p-k)^2)}{(p-k)^2}
\frac{A(k^2)\Delta A(k^2,p^2)}{A^2(k^2)k^2+B^2(k^2)}
\frac{(k^2-p^2)}{2d(k,p)} \frac{3(k^2-p^2)k\cdot p}{p^2}\/,\label{AT}\\
B(p^2) & = & {\rm RHS~of~(\ref{Beqn})} \nonumber\\ &+&
\frac{16\pi}{3}\int^{\Lambda_{\rm UV}} \frac{d^4k}{(2\pi)^4}
\frac{\alpha_s((p-k)^2)}{(p-k)^2} \frac{B(k^2)\Delta A(k^2,p^2)
}{A^2(k^2)k^2+B^2(k^2)} \frac{(k^2-p^2)}{2d(k,p)} 3(k^2-p^2)~.\label{BT}
\end{eqnarray}
These Euclidean space equations are obtained by a simple transcription from
the corresponding Min\-kow\-ski space equations as discussed in
Sec.~\ref{subsect-Euclidean}.  In view of those discussions of rotations to
Euclidean space, it is understood that the semi-phenomenological
nonperturbative part of $\alpha_s$ is a Euclidean space Ansatz.  The results
obtained are independent of the ultraviolet cut-off [$\Lambda_{\rm UV}$]
provided it is chosen sufficiently large since the integrals in
Eqs.~(\ref{Aeqn}) and (\ref{Beqn}) are convergent.  Note that this would not
be the case were we to take $\alpha_s(Q^2)$ constant.  If we were to make the
the approximation $\alpha_s((p-k)^2)\simeq\alpha_s(p^2_>)$ where $p^2_>\equiv
{\rm max}(p^2,k^2)$, then we would find, after an angle integration, that
$A(p^2)$ = 1.  This unnecessary but simplifying approximation
[already discussed in \Eq{AngApprox}] has sometimes
been used in the literature and we refer to it as the angle
approximation, [see, e.g., Fomin \etal~(1983) and references therein].

\subsubsect{Chiral Symmetry Considerations}
\label{Chiral_symm}
When there is no ECSB renormalised quark mass [$m=0$], we have exact chiral
symmetry and conservation of the axial-vector current leads to [in Landau
gauge]
\beq
M(-Q^2) \ \mathrel{\mathop=_{Q^2\to \infty}}\ {c\over Q^2}\;\left[
\ln(Q^2/\Lambda^2_{QCD})\right]^{d_M-1}\;,
\label{chiral_q_mass}
\eeq
where $c$ is some constant independent of $\mu$.  This is in contrast with
the form of the running mass given in \Eq{MQ2}.  Then the asymptotic form for
the quark mass can be written in convenient shorthand as
\beq
M(-Q^2) \ \mathrel{\mathop=_{Q^2\to\infty}}\ {c\over Q^2}\left[\ln(Q^2/
\Lambda^2_{QCD}\right]^{d_M-1} + m\left[{\ln(\mu^2/\Lambda^2_{QCD})\over
\ln(Q^2/\Lambda^2_{QCD})}\right]^{d_M}\;,
\label{asympt_q_mass}
\eeq
where again the $\mu$-dependence is not explicitly indicated.  Note that
\Eq{asympt_q_mass} is to be understood in the sense that for exact chiral
symmetry the second term on the RHS is zero and the first term is the
dominant one, while in the presence of ECSB [i.e., $m\neq 0$] the second term
is the dominant asymptotic behaviour.  With ECSB there will be many terms
which are suppressed by powers of $\ln (Q^2)$ with respect to the second term
but which dominate the first as $Q^2\to\infty$.  The [renormalised] quark
condensate $\langle\bar qq\rangle$ is a measure of the degree of DCSB and can
be defined as $\langle\bar qq\rangle\equiv
\langle {\rm vac}|:\bar q(0)q(0):|{\rm vac}\rangle$,
where $|{\rm vac}\rangle$ refers to the nonperturbative vacuum and the normal
ordering of the [renormalised] operators is with respect to the perturbative
vacuum.  Denoting $S^{-1}(p) \equiv Z^{-1}(p^2)[\pslash-M(p^2)]$ as the
nonperturbative quark propagator we have, in the absence of ECSB,
\beq
<\!\bar qq\!> = - \lim_{x\to 0^+}\;{\rm tr}\left\{S(x,0)\right\} = -
12i\int^{\Lambda_{\rm UV}}
{d^4p\over (2\pi)^4}{Z(p^2)M(p^2)\over p^2-M^2(p^2)} = -
12i\int^{\Lambda_{\rm UV}}
{d^4p\over (2\pi)^4}{B(p^2)\over A^2(p^2)p^2-B^2(p^2)}\;,
\label{q_condensate}
\eeq
where $S(x,y)$ is the coordinate space quark propagator and where the trace
over spinor [4] and colour [3] indices gives the factor 12.  In Appendix
\ref{appendix_condensate} we show that $c$, as defined in \Eq{chiral_q_mass},
satisfies, in the limit of exact chiral symmetry [$m$ = 0] in Landau gauge,
\beq
c\simeq - {4\pi^2 d_M\over 3}\;{<\!\bar qq\!>\over \left[\ln(\mu^2/\Lambda^2
_{QCD})\right]^{d_M}}\;,
\label{c_condensate}
\eeq
which since $c$ is a constant independent of the renormalisation point $\mu$
also implies that $<\!\bar
qq\!>\sim\left[\ln(\mu^2/\Lambda^2_{QCD})\right]^{d_M}$.
\Eq{chiral_q_mass} is a result also obtained in
discussions of the operator product expansion [OPE] and QCD sum rules [see,
e.g., Politzer (1976,1982), Gasser and Leutwyler (1982), and Reinders
\etal~(1985)].  In the OPE the quark condensate is taken to be a coefficient
with the above $\mu$-dependence irrespective of whether or not ECSB is
present.  From the discussions of Appendix \ref{appendix_condensate} it is
clear that the OPE quark condensate only corresponds to the ``natural''
definition of \Eq{q_condensate} in the limit of exact chiral symmetry.  Thus
in the limit of exact chiral symmetry there are two ways of obtaining the
quark condensate: 1) directly from the integral of \Eq{q_condensate} and 2)
from the asymptotic behaviour of the quark self-mass as follows from
\Eq{asympt_q_mass} and \Eq{c_condensate}.

In the limit of exact chiral symmetry we have the WTI of
\Eq{axial_vector_WTI} which gives
\beqn
k_\sigma\vec\Gamma^{\;\sigma}_5(p',p) &=& {\vec\tau\over 2}\left[S^{-1}(p')
\gamma_5 + \gamma_5\;S^{-1}(p)\right]
=\frac{\vec\tau}{2}\left[A(p'^2)\pslash'-A(p^2)\pslash-
\left(B(p'^2)+B(p^2)\right)\right]\gamma_5 \nonumber\\
&\mapright {k\to 0}& -\vec\tau\gamma_5 B(p^2)~.
\label{soft-pion-limit}
\eeqn
Note that the usual perturbative axial-vector vertex is
$(\vec\tau/2)\;\gamma^\sigma\gamma_5$.

In the chiral limit as $k\to 0$ the axial-vector quark proper vertex [see
\Fig{gamma5_etc_fig}(b)] becomes completely dominated by the pseudoscalar
coupling of a massless pion to the quark [$\vec\Gamma_5$ in
\Fig{gamma5_etc_fig}(a)] and the subsequent weak decay of the pion into an
axial-vector current, which is depicted in \Fig{gamma5_etc_fig}(c).
\begin{figure}[tb] 
 \centering{\ \epsfig{figure=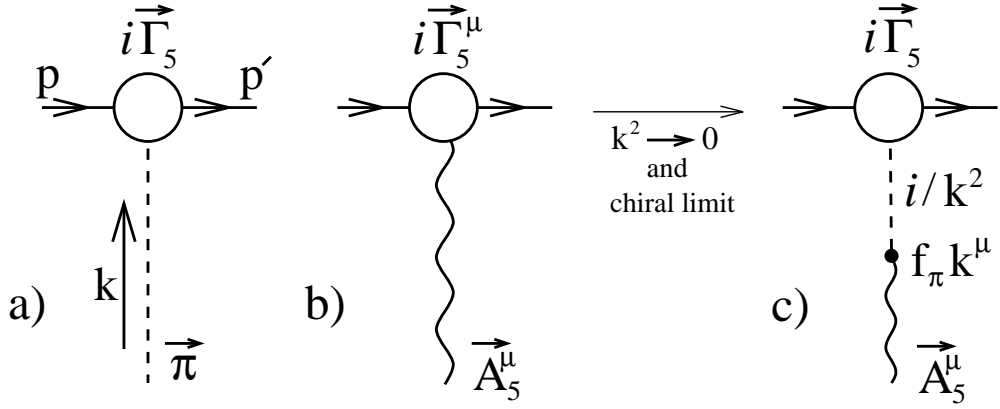,height=5.5cm} } \parbox{130mm}{
\caption{Shown are (a) the proper quark-pseudoscalar vertex coupling to a pion,
(b) the proper quark-axial-vector vertex, and (c) the pion decay component of
the quark axial-vector vertex which dominates in the soft-pion, chiral limit.
\label{gamma5_etc_fig}  }}
\end{figure}
The pion decay constant $f_\pi\simeq 93$MeV is defined by the axial-vector
transition amplitude for an on-shell pion [$k^2=m_\pi^2$ and $m,n$ are
isospin indices]
\beq
\langle 0|A^{m\sigma}_5(0)|\pi^n(k)\rangle = i\;f_\pi\;k^\sigma\;
\delta^{mn}\;.
\label{fpi_defn}
\eeq
{}From \Fig{gamma5_etc_fig}(c) and
\Eq{fpi_defn} we find then that as $k\to 0$
\beq
i\vec\Gamma^\sigma_5(p',p)\ \mapright {k\to 0}\ \left[
i\Gamma^m_5(p',p)\right]\left[{i\over k^2}\right]\left[i\langle 0|
\vec A^\sigma_5|\pi^m(-k)\rangle\right]
\mapright {k\to 0}\ -\vec\Gamma_5(p',p) f_\pi (k^\sigma/k^2)\;.
\label{kto0}
\eeq
Acting on both sides of \Eq{kto0} with $k_\sigma$, we find the that $k_\sigma
\vec\Gamma^\sigma_5(p',p)\to i f_\pi \vec\Gamma_5(p',p)$ as $k\to 0$, which
when combined with \Eq{soft-pion-limit} gives the Goldberger-Treiman relation
for the quark-pseudoscalar vertex:
\beq
\vec\Gamma_5(p,p) = i\vec\tau\gamma_5 \frac{B(p^2)}{f_\pi}
= i\vec\tau\gamma_5 \frac{Z^{-1}(p^2)M(p^2)}{f_\pi}\;.
\label{GT_relation}
\eeq

The simplest generalisation of \Eq{GT_relation} to $k\not= 0$ is perhaps
\beq
\vec\Gamma_5(p',p) = i\;\vec\tau\;\gamma_5\;{1\over 2}[B(p^{\prime 2}) + B
(p^2)]/f_\pi\;.
\eeq
However, to properly determine the pseudoscalar vertex away from the
soft-pion limit obviously requires solving the pseudoscalar vertex DSE.  This
is equivalent to the solution of the pion BSE, where the pion Bethe-Salpeter
amplitude is defined as
\begin{equation}
\chi^n_{\alpha\beta}(p,P)
\equiv \int d^4z\; e^{i[p-(P/2)]
\cdot z}\langle 0|T\psi_\alpha(0)
\bar\psi_\beta(z) |\pi^n(P)\rangle\;,
\label{Phi-PT}
\end{equation}
where $|\pi^n(P)\rangle$ denotes a pion state with momentum $P$, $n$ is the
isospin label, and $\vec\chi$ is related to $\vec\Gamma_5$ by
\beq
\vec\chi(p,P)=\left[iS\left(p+\frac{P}{2}\right)\right]
\left[i\vec\Gamma_5\left(p+\frac{P}{2},p-\frac{P}{2}\right)\right]
\left[iS\left(p-\frac{P}{2}\right)\right]\;.
\label{pion_BS_amplitude}
\eeq
In Eq.~(\ref{Phi-PT}) all colour indices are suppressed.  In the amplitude
$\chi$ the incoming pion has 4-momentum $P$, the quark has 4-momentum
$p+(P/2)$, and the antiquark has 4-momentum $p-(P/2)$.  The parity and
time-reversal invariance properties of a pseudoscalar imply that we can write
\begin{equation}
\chi(p,P) = \gamma_5 \chi_P(p,P) + \Pslash \gamma_5 \chi_A(p,P) +
\pslash\gamma_5\chi_{A'}q(p,P)
+ [\pslash,\Pslash]\gamma_5\chi_T(p,P)\;. \label{Phi}
\end{equation}
Here $\chi_i(p,P)$ for $i=P,A,A'$, and $T$ are scalar functions of
$p^2,(p\cdot P)$, and $P^2$.  The subscripts P, A, and T denote the spinor
matrix structure of the particular component of $\chi$.  The three functions
$\chi_P$, $\chi_A$, and $\chi_T$ are even functions of $p\cdot P$, whereas
$\chi_{A'}$ must be an odd function of $p\cdot P$.  A similar expansion can
also be made for the proper quark-pseudoscalar vertex $\Gamma_5$.  In the
chiral limit where the pion is massless we obviously have for an on-shell
pion $P^2=m^2_\pi=0$.

Using the above it is possible to obtain an expression for $f^2_\pi$ from the
integral equation for pion decay illustrated in \Fig{f_pi_fig}.
\begin{figure}[tb] 
 \centering{\ \epsfig{figure=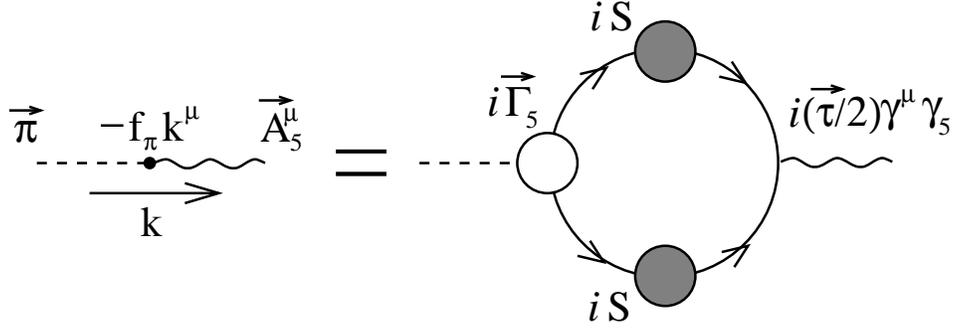,height=4.5cm} } \parbox{130mm}{
\caption{ The integral equation for the pion decay constant $(f_\pi)$.
\label{f_pi_fig} }}
\end{figure}
Our discussion here is a relatively straightforward extension of the
arguments presented by Pagels and Stokar (1979,1980) and Cornwall (1980) to
the case where $A(p^2)\neq 1$.  From \Fig{f_pi_fig} we find that
\beqn
i\langle 0|A^{m\sigma}_5(0)|\pi^n(k)\rangle &=& i(ik^\sigma f_\pi\delta^{mn})
= -k^\sigma f_\pi\delta^{mn} \nonumber\\ &=& (-1)\int {d^4p\over
(2\pi)^4}\;{\rm tr}\biggl\{
[iS(p+k)][i\Gamma^m_5(p+k,p)][iS(p)][i(\tau^n/2)\gamma^\sigma\gamma_5]
\biggr\}\nonumber\\
&=& (-1)\int {d^4p\over (2\pi)^4}\;{\rm tr}\biggl\{
[\chi^m_5(k,p)][i(\tau^n/2)\gamma^\sigma\gamma_5]
\biggr\}\;,
\label{f_pi}
\eeqn
where, as usual, the factor [-1] on the RHS arises from the fermion [i.e.,
quark] loop.  In order to avoid double counting we need to take the
axial-vector vertex as a perturbative vertex for the same reason as this is
done in the photon self-energy diagram in \Fig{photon_dse_fig}.  [The
argument for $f_\pi$ in Williams \etal~(1991) is erroneous for this reason,
although the error actually has very little effect on the numerical results
reported therein.]  If, in the limit $k\to 0$ in \Eq{f_pi}, we equate
coefficients of $k^\sigma$, using the approximation:
$\Gamma_5(p',p)\approx\Gamma_5(p,p)+O(k)\gamma_5$; i.e., assuming the Dirac
structure of the $O(k)$ correction is purely $\gamma_5$, and use the
Goldberger-Treiman relation \Eq{GT_relation} for $\vec\Gamma_5$, we obtain,
in the limit of exact chiral symmetry, [see Appendix
\ref{appendix_fpi} for details]
\beq
f^2_\pi = -12i\int {d^4p\over (2\pi)^4}\;{Z(p^2)M(p^2)\over [p^2-M^2(p^2)]^2}
\left[M(p^2)-{p^2\over 2}{dM\over dp^2}\right]
\label{f_pi2}
\eeq
which as a physical observable must be independent of the choice of
renormalisation point, as can be verified numerically for solutions.  It
should be emphasised that, while this derivation is quite general, the
assumed form of $\Gamma_5(p',p)$ for $p'\neq p$ influences the result, which
is therefore model dependent.  [Another form appears in \Eq{Fpi}.  Both forms
reduce to the Pagels and Stokar (1979) result when $Z(p^2)=A^{-1}(p^2)=1$
which indicates that the O$(k)$ corrections to $\Gamma_5(p',p)$ measure
deviations from $A(p^2)\equiv 1$, consistent with the result of Delbourgo and
Scadron (1979).]  It is straightforward to write down the Euclidean space
versions of \Eq{q_condensate} and \Eq{f_pi2} by inspection and for
completeness these have been given in Appendices \ref{appendix_condensate}
and \ref{appendix_fpi} respectively.

\subsubsect{Numerical Results}
\label{QCD_DCSB_results}
As a specific example of the above arguments for studying DCSB in QCD we
describe numerical results from one model DSE for the quark propagator (Hawes
\etal~, 1993).  [Other models and results will be compared and summarised
below.]  This study uses the Ball-Chiu and Curtis-Pennington vertices
together with an Ansatz for the gluon propagator in Landau gauge of the form
\beq
\label{Gprop}
(g^2/4\pi) D^{\nu\sigma}(q)\to \alpha_s(Q^2) D_0^{\nu\sigma}(q)
\eeq
with
\beq
\label{PiA}
\alpha_s(Q^2)= \alpha_s(\tau;Q^2) \frac{(Q^2)^2}{(Q^2)^2 + b^4}~
\;\;\;\; {\rm and}\;\;\;\;
\alpha_s(\tau; Q^2) = \frac{d\pi}
{\ln\left[\displaystyle\tau + \left(Q^2/\Lambda_{\rm QCD}^2\right)\right]}~,
\label{alphas_hrw}
\eeq
where $\ln\tau>0$ [i.e., $\tau>1$] and $b$ are real constants to be
determined.  This is to be understood in terms of the arguments which lead to
\Eq{Sigma_q}; i.e., it is understood that the ghost and nonperturbative
gluon propagator behaviour have been absorbed into the nonperturbative
structure of $\alpha_s(Q^2)$.  It is clear that this Ansatz has the
asymptotic form of \Eq{alpha_s} as $Q^2\to\infty$.  In the infrared, we see
that the effective gluon propagator vanishes as $Q^2\to 0$; i.e., \mbox{$D(q)
\sim q^2/[(q^2)^2 + b^4]$}.  [The origins of this form of gluon propagator
are discussed in detail in Sec.~\ref{subsect:IRGP}.]  A closely related
study, where the infrared gluon propagator has a Gaussian form and where the
Ball-Chiu vertex was used, is described in Hawes and Williams (1991).

We see that the parameter $\tau>1$ plays the dual role of regulating the
infrared behaviour of the logarithmic term and of determining the strength of
the nonperturbative infrared interaction.  Using the quark condensate as an
order parameter, we find that there is a critical value of $b=b_c$ such that
the model does not support dynamical chiral symmetry breaking for $b>b_c$.

In Sec.~\ref{subsect-quark-conf} we discuss confinement and illustrate a
confinement test by applying it to this model.  The results suggest that, for
all values of $b$, the quark propagator in the model is not confining.  Two
properties of a particle's propagator which together are at least sufficient to
ensure confinement of the particle are: 1) the absence of a
K\"{a}llen-Lehmann representation and 2) no singularity on the timelike $q^2$
axis.  This will be discussed in more detail in
Sec.~\ref{subsect-quark-conf}.  The gluon propagator implied by
\Eq{alphas_hrw} satisfies both of these requirements.  Although it can be
argued that it describes a confined gluon, it may seem that quark confinement
is counterintuitive, since it would appear to provide a weak interaction
between quarks at small $p^2$, corresponding to large distances.
Nevertheless, as discussed in Sec.~\ref{subsect:IRGP}, quark confinement is
not implausible in this model.

We note too that there has been an attempt to employ a gluon propagator
similar to \Eq{alphas_hrw} in a study of quarkonium spectra by Becker
\etal~(1991).  A Blankenbecler-Sugar reduction of a ladder-like approximation
to the BSE is used and it is argued that, in the bound state equation that
results, one can approximate the effect of \Eq{alphas_hrw} by a Coulomb
potential for all $r$.

We can now study the implications of \Eq{Sigma_q} and \Eq{alphas_hrw} for the
structure of the quark propagator.  A similar study has been carried out by
Alkofer and Bender (1993).  Using Landau gauge [$\xi=0$] since, as discussed
earlier, the form of gluon propagator we consider here is most realistic in
this gauge, we find that in both the Ball-Chiu and Curtis-Pennington cases
there are regions of DCSB and unbroken chiral symmetry characterised by a
two-dimensional phase diagram in $(b^2,\ln\tau)$ space, where $b$ and $\tau$
appear in \Eq{alphas_hrw}. The phase transition is second order.

For the most part in this study $\Gamma^\mu_{\rm CP}(k,p)$ was neglected;
i.e., a minimal Ball-Chiu Ansatz was used.  This has the virtue of
simplifying the integral equations.  This term was included for a single
value of $\tau$ and a number of values of $b$ and found to generate a small
quantitative change in some of the characteristic quantities calculated in
the model but not to alter its qualitative features.  The integral equations
were solved numerically by iteration on a logarithmic grid of
\mbox{$x= p^2/\Lambda_{\rm QCD}^2$} and
\mbox{$y=k^2/\Lambda_{\rm QCD}^2$} points.   This ensured that the results
were independent of the seed-solution and grid choice.  The results were
independent of the UV cutoff, which was \mbox{$\Lambda_{\rm UV}^2 = 5\times
10^8\;\Lambda_{\rm QCD}^2$}, and this value was also sufficient to ensure
that the leading-log behaviour of the mass function had become evident.

We are interested in determining whether the model gluon propagator specified
by \Eq{alphas_hrw} can support DCSB - a crucial feature of QCD.  The quark
condensate, which is gauge-invariant, was studied earlier and is an order
parameter for DCSB.  In cases for which our iterative solution procedure for
the DSE converged quickly, with relative errors of less than
\mbox{$1\times10^{-6}$}, the condensate could be obtained easily.  However,
for values of $b^2$\ near a phase transition the convergence could be
extremely slow.  In those cases, the numerical solution was examined at
constant intervals through the run [say, every 50th cycle], and the
condensate evaluated in each case.  Aitken extrapolation (Wimp, 1981) was
then used to find the ``infinite-cycles'' limit.  In several cases the
program was subsequently run until the solutions had converged to within
\mbox{$1\times10^{-6}$} and the extrapolated result always matched the actual
result to within a few parts in $10^{-6}$\/.

\begin{figure}[tb] 
  \centering{\ \epsfig{figure=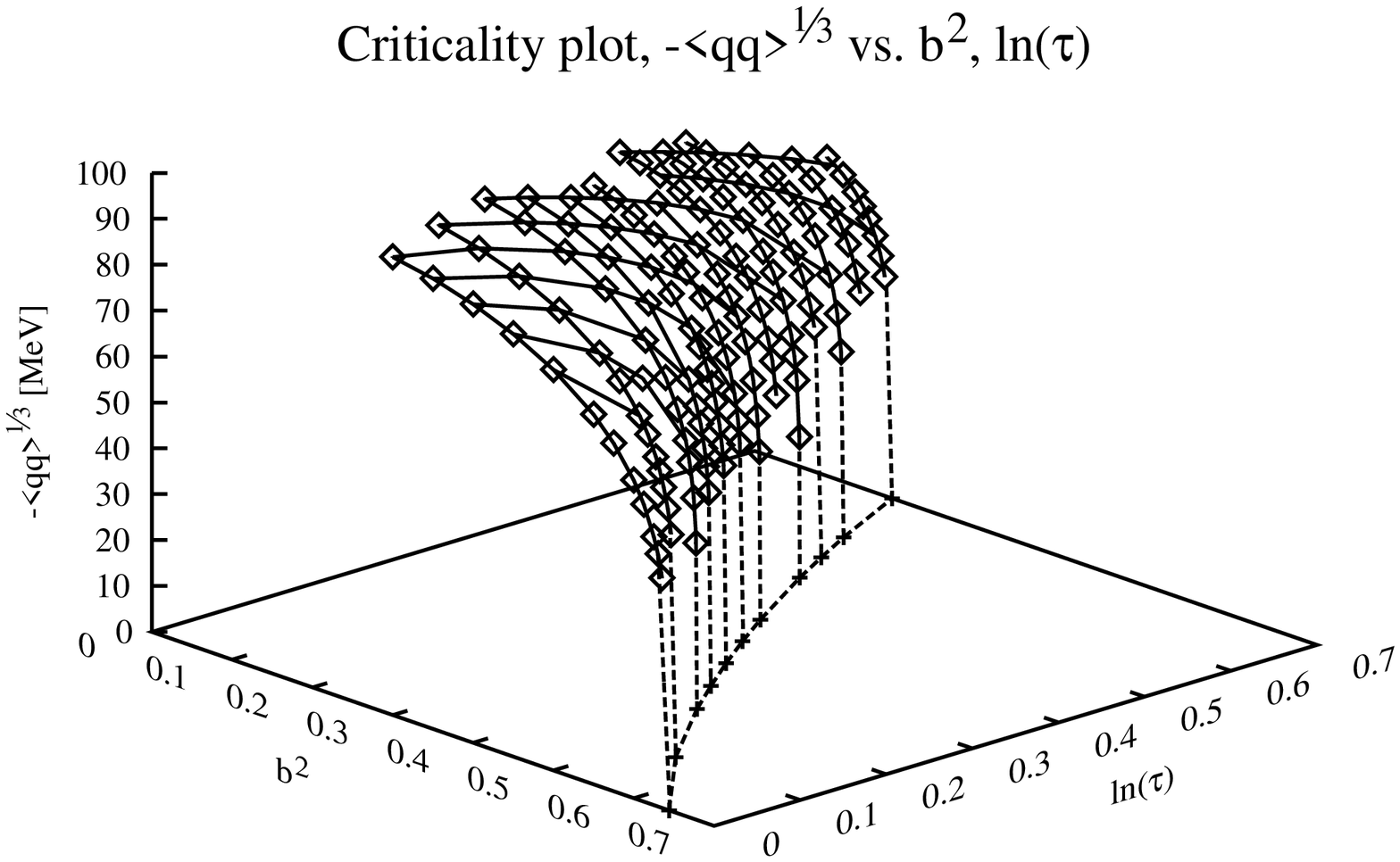,height=10.0cm} }
\parbox{130mm}{\caption{ Criticality plot for
  $(-\langle \bar{q}q \rangle_\mu)^{\frac{1}{3}}$\ as a function of $\ln \tau$\
and $b^2$\/.  The condensate, $(-\langle \bar{q}q \rangle_\mu)^{\frac{1}{3}}$,
is in units of MeV, scaled to $\mu^2 = 1 {\rm GeV}$\/, and $b^2$\ is in units
$\Lambda_{QCD}^2$\/; the gluon regulator $\tau$\ is dimensionless.
\label{dse_hrw_fig1}  }}
\end{figure}
We solved Eqs.~(\ref{Aeqn}) and (\ref{Beqn}) for values of $\ln\tau$ in the
domain \mbox{$[0.0,0.7]$} and $b^2$ in \mbox{$[0.1,1.0]$} using the minimal
Ball-Chiu vertex and we plot the condensate obtained from our solutions in
\Fig{dse_hrw_fig1}.  This figure shows regions of unbroken and dynamically
broken chiral symmetry.  Our numerical results suggest that the condensate
rises continuously from the transition boundary and hence that the transition
is second order.  As a consequence we assumed that the order parameter,
\mbox{$\langle\overline{q}q\rangle_\mu$}, behaves as
\begin{equation}
      \langle\overline{q}q\rangle_\mu(z) \approx C
\left(1-\frac{z}{z_c}\right)^{\beta}
\end{equation}
for $z \rightarrow z_c^-$\ [for z equal to either $\ln\tau$\ or $b^2$\/] and
extracted the critical points, $z_c$, and critical exponents, $\beta$, using
ratio-of-logs methods.  We list these quantities in \Table{betas} and, in
\Fig{dse_hrw_fig2}, plot the critical curve in the $(b^2,\ln\tau)$ plane.
%
\begin{table}[tbh]
\begin{center}
  \parbox{130mm}{\caption{The critical points and exponents extracted for
various values of $\ln\tau$\/; the cumulative result is 
$\beta_{\rm BC}=0.575$\/, with $\sigma_\beta=0.024$\/; excluding the
point with $\ln\tau=0.0$\/, $\beta_{\rm BC}=0.572$\ with
$\sigma_\beta=0.020$\/.}} 
\begin{tabular}{|c|c|c|c|}
\hline $\ln\tau$&$b^2_{\rm C}$ - Critical $b^2$
value &$\beta$ - Critical Exponent &$\sigma_\beta$ - standard deviation in
$\beta$\\ \hline\hline 0.00 & 0.6439 & 0.609 & 0.030 \\ \hline 0.10 & 0.5448
& 0.579 & 0.021 \\ \hline 0.20 & 0.4642 & 0.570 & 0.021 \\ \hline
      0.30 & 0.3932 & 0.573 & 0.021 \\ \hline
      0.40 & 0.3289 & 0.567 & 0.021 \\ \hline 0.50 & 0.2706 & 0.567 & 0.021
\\ \hline
      0.60 & 0.2180 & 0.561 & 0.021 \\ \hline 0.70 & 0.1710 & 0.579 & 0.021
\\ \hline \end{tabular} \label{betas} 
\end{center}
\end{table}
\begin{figure}[tb] 
  \centering{\ \epsfig{figure=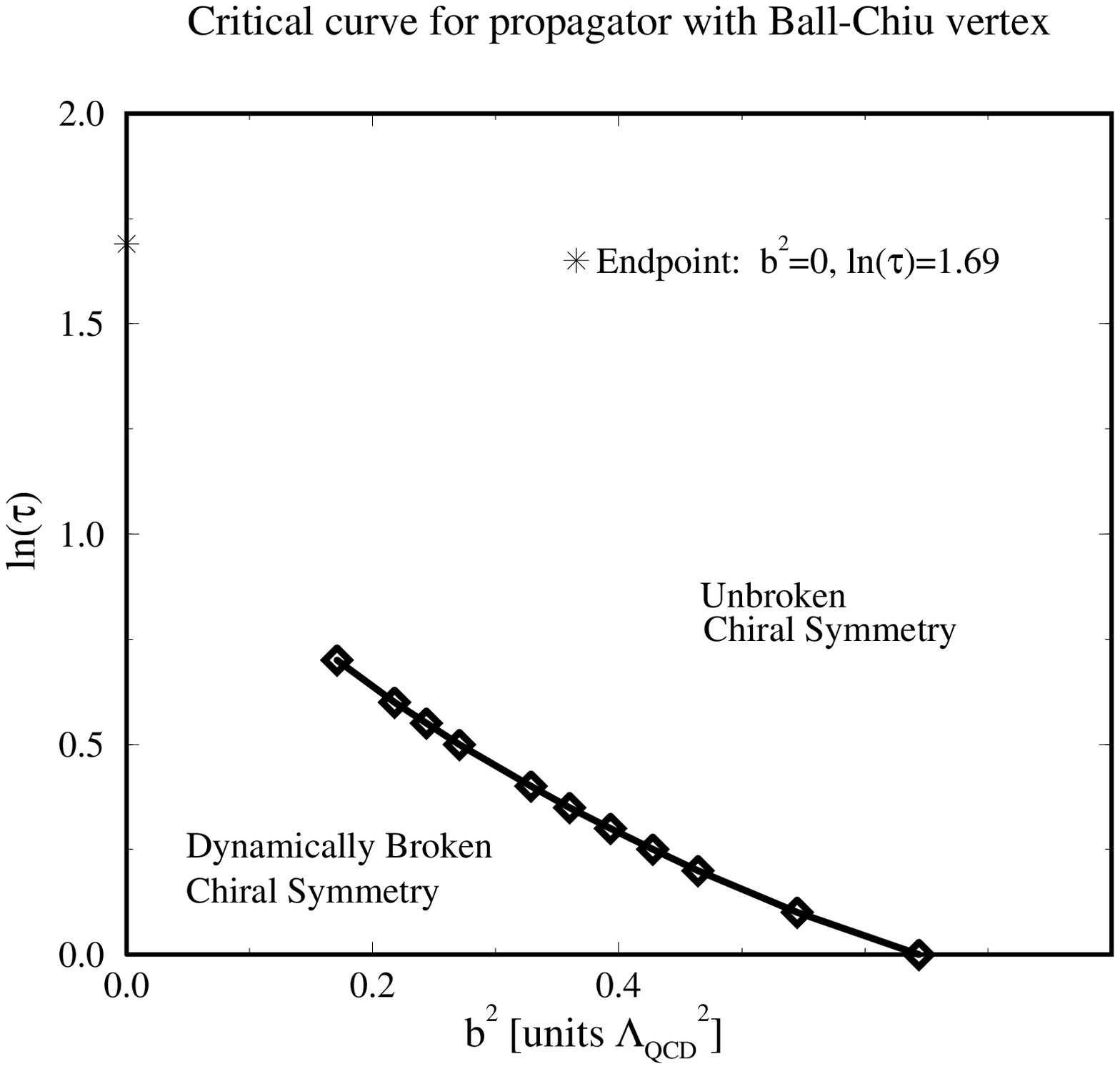,height=9.0cm} }
\parbox{130mm}{\caption{ Critical curve for the phase transition in the
  $(\ln\tau, b^2)$\ plane.  The asterisk is the result extracted from Roberts
and McKellar (1990).
\label{dse_hrw_fig2} }}
\end{figure}
\begin{figure}[tb] 
  \centering{\ \epsfig{figure=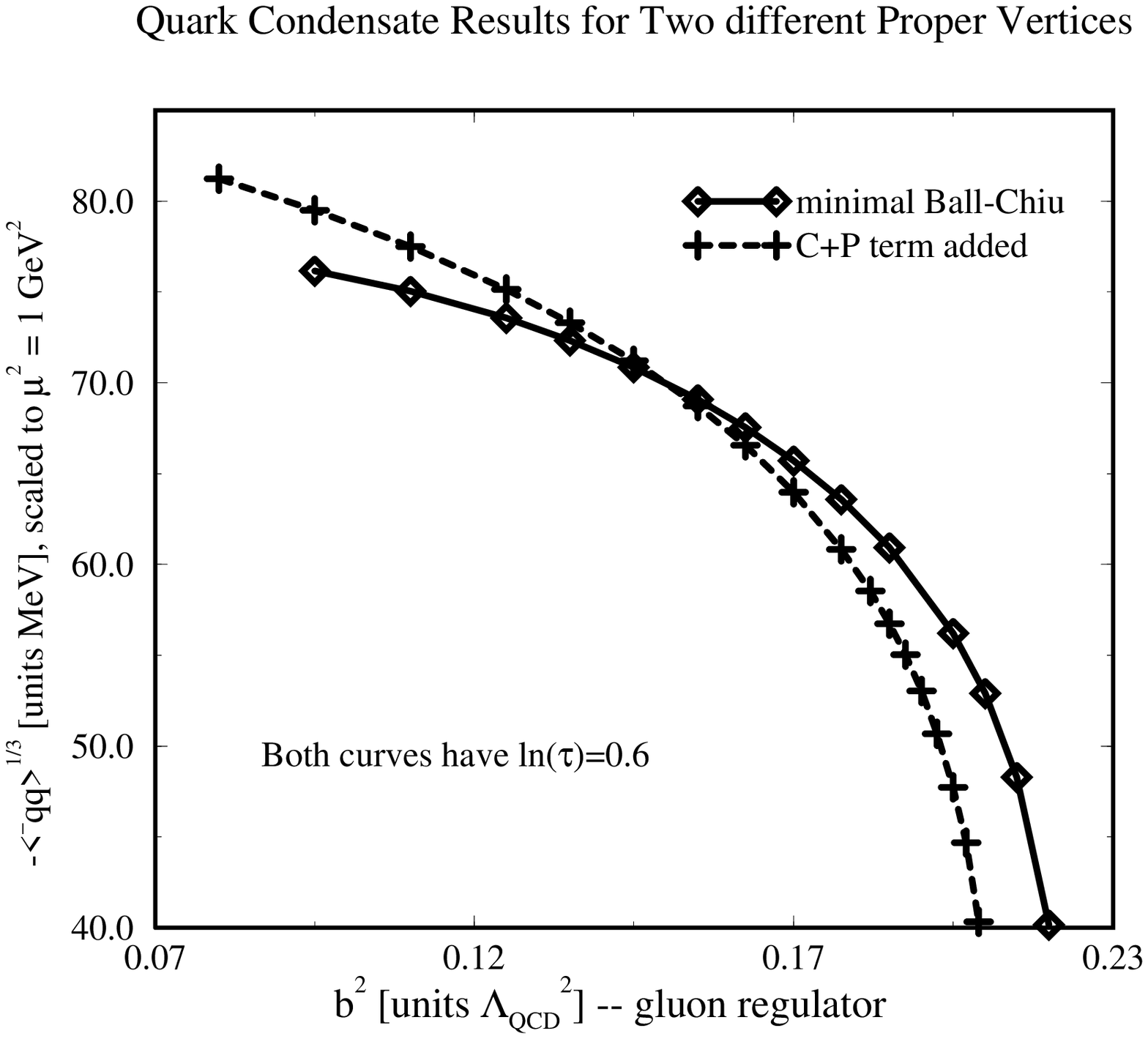,height=9.0cm} }
\parbox{130mm}{\caption{ Comparison of the
  $(-\langle \bar{q}q \rangle_\mu)^{\frac{1}{3}}$\ condensate curves for the
minimal Ball-Chiu and Curtis-Pennington Ans\"{a}tze for the proper vertex.
Both curves have $\ln\tau=0.6$\/.  Diamonds, $\diamond$, connected with solid
lines are the results from the B-C vertex; plus-signs, $+$, connected with
dashed lines are results from the C-P vertex.
\label{dse_hrw_fig3} }}
\end{figure}

We also solved Eqs.~(\ref{Aeqn}) and (\ref{Beqn}) with the Curtis-Pennington
additions, Eqs.~(\ref{AT}) and (\ref{BT}), using \mbox{$\ln\tau=0.6$}.  The
critical curve [in $b^2$] in this case is shown in \Fig{dse_hrw_fig3} along
with the minimal Ball-Chiu results for the same value of $\ln\tau$.  The
effect of the Curtis-Pennington addition is to lower the critical value of
$b^2$ but, as we show below, the critical exponent is unchanged.  This curve
illustrates the point that the qualitative features of the model are not
affected by this modification of the model quark-gluon vertex.

{}From Table~\ref{betas} we find:
\begin{eqnarray}
\label{betaa}
\beta_{\rm BC} = 0.575\; & \;\; {\rm and} \;\; &\; \sigma_{\beta} = 0.024~.
\end{eqnarray}
We note that the critical exponent obtained with \mbox{$\ln\tau=0$} is quite
different from the others.  This is a special case since for this value the
propagator does not vanish in the infrared:
\begin{equation}
\frac{g^2}{4\pi} D(Q^2=0) = d_{\rm M}\pi\,\frac{\Lambda_{\rm QCD}^2}{b^4}~.
\end{equation}
If we neglect this point in our analysis then we find
\begin{eqnarray}
\label{betab}
\beta_{\rm BC} = 0.572 \; & \;\; {\rm and}\;\; & \;\sigma_{\beta} = 0.020~.
\end{eqnarray}

The results in Eqs.~(\ref{betaa}) and (\ref{betab}) are in agreement with
those of Roberts and McKellar (1990) where it is argued that \mbox{$\beta =
0.589 \pm 0.031$}.  That study used $b^2=0$ and found a critical value of
\mbox{$\ln\tau = 1.69$} which
complements the results reported here.

We also calculated the critical exponent using our numerical DSE solutions
obtained with the Curtis-Pennington addition to the vertex at
\mbox{$\ln\tau=0.6$}:
\begin{eqnarray}
\label{betac}
\beta_{\rm CP} = 0.579 & \;\; ,\;\; & \sigma_{\beta} = 0.015~.
\end{eqnarray}
This suggests that the vertex modification does not alter the critical
exponent of the transition; a conclusion that is also supported by the
observation that the vertex used in Roberts and McKellar (1990) was not of
either of the above forms but was, effectively, a simple modified rainbow
approximation: \mbox{$\Gamma_\mu(k,p) = A(k^2) \gamma_\mu$}.

\subsubsect{Other Studies and Results}
\label{QCD_DCSB_other}
There have been numerous studies of DCSB in QCD using the quark DSE and it is
unfortunately not possible to give an exhaustive summary of all of them
herein.  However, all of the studies are well represented by the above
exemplary study: each necessarily assumes or argues for one or another form
of effective gluon propagator and makes some assumption about the quark-gluon
vertex.  In addition, many of the earlier studies used the approximation of
\Eq{AngApprox} in order to obtain an approximate differential equation for
the quark propagator.  Many also used rainbow approximation, \Eq{RA}, in Landau
gauge.

There are some interesting early studies by Higashijima (1983, 1984) in
Landau gauge and using rainbow approximation with the following Ansatz for
the running coupling:
\beqn 
\alpha_s(Q^2< Q_0^2) = 
\frac{d_M\pi}{\ln\left({Q_0^2/\Lambda_{\rm QCD}^2}\right)}
\; & \;\; {\rm and} \;\; & \;
\alpha_s(Q^2> Q_0^2) = 
\frac{d_M\pi}{\ln\left({Q^2/\Lambda_{\rm QCD}^2}\right)}~.
\eeqn
Employing \Eq{AngApprox} in order to obtain a differential equation it was
shown (Higashijima, 1984) that for $\ln[ Q_0/\Lambda_{\rm QCD}] < 0.88$,
corresponding to $\alpha_s(0) \sim 1$, the model admits DCSB.  Importantly,
it was shown that in order to obtain a differential equation for the mass
function whose solution agreed with the one-loop renormalisation group
result, \Eq{chiral_q_mass}, one must not neglect derivatives of
$\alpha_s(Q^2)$.

Subsequently there were a series of studies by Atkinson and Johnson, (1987,
1988a, 1988b).  It was found that in Landau gauge, with
\beq
\alpha_s(Q^2) = \frac{d_M \pi}{\ln \left( \tau + 
                \frac{Q^2}{\Lambda_{\rm QCD}^2} \right) }~,
\eeq
with a vertex Ansatz which satisfied the WTI, but which involved kinematic
singularities, and using \Eq{AngApprox}, one obtains a differential equation
for the mass quark mass function whose solution is such that the model
manifests DCSB for $\alpha_s(0)>\alpha_s^c(0) \approx 0.9$ (Atkinson and
Johnson, 1988b).

This model was reanalysed by Roberts and McKellar (1990) who solved the
integral equations for $A(p^2)$ and $B(p^2)$ directly and analysed the
reliability of \Eq{AngApprox}.  It was found that using \Eq{AngApprox} the
model manifests DCSB for $\alpha_s(0)>\alpha_s^c(0)\approx 0.78$.  The
difference between this result and that of Atkinson and Johnson (1988b) being
due to a more detailed analysis by Roberts and McKellar (1990) of the
behaviour of the mass function near the critical coupling.  Upon solving the
integral directly it was found that the critical coupling increased to
$\alpha_s^c(0) \approx 0.89$: a $12$\% rise.  The approximation of
\Eq{AngApprox} is thus seen to be qualitatively reliable in this case.  This
is the best possible application of the approximation, however.  It was shown
that \Eq{AngApprox} becomes unreliable as $\alpha_s(0)$ is increased making
it suspect in any study in which the model gluon propagator is large or
singular at $Q^2=0$.  An important result of this study was that, even in
Landau gauge, $A(p^2)\neq 1$ when the integral equations are solved directly
without further approximation.  [The same it true in the \qedt studies
described in Sec.~\ref{sect-QED3}.]

Equation~(\ref{AngApprox}) was used by Krein \etal~(1988, 1990) in a study of
a $1/Q^4$ infrared form for the effective gluon propagator, which lead to a
confined quark.  However, these results were later seen to be completely
dependent on the use of the angle approximation to regulate the divergent
infrared gluon behaviour.  This emphasises the limitations of this
approximation discussed by Roberts and McKellar (1990).

The difficulties associated with using a model gluon propagator with a
nonintegrable singularity in the infrared in a covariant quark-DSE were are
also illustrated in a study by von Smekal \etal~(1991).  In this study
\beq
\alpha_s(Q^2) = \frac{d_M \pi}
        {\ln\left(1 + \frac{Q^2}{\Lambda_{\rm QCD}^2} \right)}
        + d_M \pi f(Q^2)~,
\eeq
with $f(Q^2)$ regular as $Q^2 \rightarrow 0$, and an Ansatz for the
quark-gluon vertex was chosen such that it satisfied the Ward identity.  This
was subsequently approximated in order to simplify the analysis.  As in all
cases with a $Q^{-4}$ singularity in the model gluon propagator a
regularisation procedure was necessary and these authors chose to make the
replacement $Q^2 \rightarrow [Q^2 + \mu^2]$ in the singular part of the
running coupling.  [A value of $\mu = 10^{-4}\Lambda_{\rm QCD}$ was found to
reduce the sensitivity of the results to this parameter to less than 1\%.
This might be compared with the prescription employed by Brown and Pennington
(1989), \Eq{pluspres}.]  The coupled, nonlinear integral equations for
$A(p^2)$ and $B(p^2)$ obtained with $f\equiv 0$ were solved and the model
shown to manifest DCSB.  However, this version yielded a rather poor
phenomenology in the sense that the quark condensates and pion decay
constant, $f_\pi$, calculated from the solution for the quark propagator, did
not agree well with the accepted values.  It was found that this could be
remedied by using $f(Q^2) = \gamma^2/[Q^2 + m_g^2]$ with $\gamma = 2.5
\Lambda_{\rm QCD}$, $m_g = 0.2 \Lambda_{\rm QCD}$ and $\Lambda_{\rm QCD} =
0.5$~GeV.  [The value of the effective gluon mass implied by this parameter
choice is smaller than the value of $\sim 0.5$~GeV suggested by other
studies; for example, those described in Sec.~\ref{subsect:IRGP}.]

In many respects this study is qualitatively and quantitatively similar to
that of Williams \etal~(1991) in which the $Q^{-4}$ singularity was
regularised by replacing it with an integrable distribution:
\beq
\alpha_s(Q^2) = C\,\Lambda_{\rm QCD}^2 \,Q^2\,\delta^4(Q) + 
        \frac{d_M  \pi}{\ln \left( \tau + 
                \frac{Q^2}{\Lambda_{\rm QCD}^2} \right)}~.
\eeq
[The first term alone was studied in rainbow approximation by Munczek and
Nemirovsky (1983) and Cahill and Roberts (1985).]  Williams \etal~(1991) used
Landau gauge, a simplified vertex:
$\Gamma^\nu=(1/2)[A(p'^2)+A(p^2)]\gamma^\nu$, and solved the coupled,
nonlinear integral equations for $A(p^2)$ and $B(p^2)$.  It was found that
once $f_\pi$ was fitted by adjusting the infrared coupling strength, $C$,
then the quark condensate assumed a value in the typical range,
$(-\langle \bar q q\rangle)^{1/3} \simeq 225\pm 25$MeV
at $\mu=1$GeV.  A similar study
with the Ball-Chiu vertex and a Gaussian form for the gluon propagator in the
infrared reached similar conclusions (Hawes and Williams, 1992).

An effective long-distance quark-quark interaction was derived from dual QCD,
which is a long-distance treatment of QCD where the vacuum has the properties
of a dual superconductor.  The roles of the electric and magnetic fields are
interchanged in this picture [for a detailed review see Baker \etal~(1991)].
This effective quark interaction was then used as input into studies of DCSB
[see Baker\etal~(1988) and Krein and Williams (1991)] with some success.  It
is unfortunate, that a sign error was recently detected in the original
derivation by S. Kamizawa [Baker \etal~(1988, erratum 1991)], which leaves
open the question of whether or not DCSB occurs naturally in dual QCD.
Studies of DCSB have also been generalised to a chromodielectric model of QCD
where a mean field treatment of the colour-dielectric constant and DCSB
arguments leads to a chiral, nontopological soliton model (Krein
\etal, 1988, 1991).

A model gluon propagator that is not singular in the infrared has also been
studied (Haeri and Haeri, 1991; Papavassiliou and Cornwall, 1991).  Haeri and
Haeri (1991) use
\beq
\label{Cornalpha}
\frac{\alpha_s(Q^2)}{Q^2} \sim \frac{1}{Q^2 + m^2} 
        \frac{1}{\ln\left(\frac{Q^2 + 4 m^2}{\Lambda_{\rm QCD}^2}\right)}~,
\eeq
based on the analysis by Cornwall (1982), where $m$ is an effective gluon
mass, and a vertex Ansatz constructed so as to eliminate overlapping
divergences in the quark DSE.  As one would expect in such a model, there is
a critical value of the ratio $m/\Lambda_{\rm QCD} = R_m^c$ such that only
for $m/\Lambda_{\rm QCD} < R_m^c$ does the model manifest DCSB.  In this
particular model, however, such a result is phenomenologically inconsistent
because it means that DCSB requires an effective gluon mass which is
significantly less than the value of $\sim 0.5$~GeV suggested by other
studies [see Sec.~\ref{subsect:IRGP}].  A similar internal inconsistency was
found by Papavassiliou and Cornwall (1991) in their study of approximate
coupled DSEs for the 3-gluon and quark-gluon vertices and subsequent analysis
of the DSE for the quark propagator.  The conclusion in both of these studies
is that confinement is not incorporated in the model specified by
\Eq{Cornalpha} whose infrared behaviour therefore requires modification if a
reasonable phenomenological model of DCSB is to be obtained.

There have also been studies of the non-covariant, Coulomb-gauge, quark DSE
(Finger and Mandula, 1982; Govaerts \etal, 1983, 1984; Alkofer and Amundsen
1988; Langfeld \etal, 1989).  As in the covariant studies, it is found that
one has DCSB if the coupling at zero momentum transfer exceeds a critical
value.  We remark that Alkofer and Amundsen (1988) used a model interaction
with $\alpha(|\vec{Q}|^2) \sim |\vec{Q}|^{-2}$ which was regularised using a
cutoff mass: $|\vec{Q}|^2
\rightarrow [|\vec{Q}|^2 + \mu^2]$.  In all cases a not unreasonable
phenomenological description of DCSB was obtained.

\subsubsect{Summary}
The body of work exemplified and summarised in this section clearly
demonstrates that a very good phenomenological description and understanding
of DCSB in QCD can be obtained using the DSE for the quark self energy.
Given a gluon propagator whose behaviour at large spacelike-$q^2$; i.e., in
the ultraviolet, is given by the renormalisation group in QCD and whose
behaviour at small spacelike-$q^2$; i.e., in the infrared, yields an
effective coupling strength that is $> 1$, then one will have
$\langle\overline{q}q\rangle \neq 0$.  With just one or two parameters to
describe the transition from the known behaviour of the gluon propagator in
the ultraviolet to the unknown behaviour in the infrared [see
Sec.~\ref{subsect:IRGP}], one can obtain a good phenomenology of DCSB; i.e.,
one can obtain values for quantities such as the quark condensates and the
pion decay constant that agree with those extracted from experiment.  This
will be discussed in more detail in Sec.~\ref{sect-hadrons}.

\subsect{Quark Confinement}
\label{subsect-quark-conf}
\subsubsect{General Observations and Remarks}
It is now generally accepted that QCD is the appropriate theory for
describing the strong interaction.  The lack of observation of free quarks
and gluons [and ghosts] leads then to the assumption of colour confinement;
i.e., that only quantities which transform as colour-singlets can be physical
observables.  Theoretical studies also lend credence to this expectation,
where, for example, lattice gauge theory studies are consistent with a string
tension or linearly rising potential between coloured objects.  The fact that
QCD is a gauge theory also implies that physical observables must be gauge
invariant.  The BRS invariance properties of QCD can be used to show that the
$S$-matrix is gauge invariant, [see Itzykson and Zuber (1980, pp.~604--605)].
At the observational level, the confinement of colour is the statement that
the cross-section for the production of any combination of asymptotic
coloured states from a colour-singlet initial state must be zero.  A more
extensive treatment of some of the following discussion can be found in
Roberts \etal~(1992).

The failure of the cluster decomposition property [CDP] of vacuum expectation
values of the coloured fields in QCD [the QCD Wightman functions] is
certainly sufficient to ensure confinement [Marciano and Pagels (1978),
Greenberg (1978), Strocchi (1976,1978); also see Streater and Wightman (1980)
for a discussion of the CDP].  The CDP can be associated with charge
screening and so, in simple terms, the failure of the CDP means that,
irrespective of how large one makes the spacelike separation between two
coloured objects, the interaction between them never becomes negligible. This
is the common intuitive basis for understanding confinement that underlies
the idea of a string tension in QCD and provides a basis for the linear and
harmonic-oscillator potential models.  In QED the locality property of the
field theory is sufficient to ensure that the CDP is satisfied (Strocchi,
1976, 1978), however, the proof used in QED to demonstrate this property
fails in QCD.  Crucial to the failure of this proof is the fact that an
indefinite metric is unavoidable in local gauge theories and further that QCD
is asymptotically free. Of course, the failure to prove the CDP is not
equivalent to proving its failure.  It is interesting that in QED$_2$, which
has a linear potential, the failure of the CDP can be explicitly demonstrated
(Strocchi, 1976).

The $n$-point Green's functions of a quantum field theory are the
time-ordered vacuum expectation values $G^{(n)}(x_1,\dots,x_n)\equiv\langle
{\rm vac}|T(\phi(x_1)\dots\phi(x_n))| {\rm vac}\rangle$, where here the field
operators $\phi$ generically represent gluon, quark, and ghost field
operators in QCD.  The Wightman functions are simply Green's functions
without time-ordering; i.e., $W^{(n)}\equiv\langle {\rm
vac}|\phi(x_1)\dots\phi(x_n)|{\rm vac}\rangle$.  The CDP is that as $\lambda
\to +\infty$
\begin{eqnarray}
W^{(n)}(x_1,\dots,x_j,x_{j+1}+\lambda a,\dots,x_n+\lambda a) &
\longrightarrow & W^{(j)}(x_1,\dots,x_j) \,W^{(n-j)}(x_{j+1},\dots,x_n)\;,
\label{eq:CDP}
\end{eqnarray}
where $a$ is an arbitrary spacelike vector.  The actual CDP requirement is
more than this. In theories with a mass gap [no zero mass states] the
difference between the LHS and RHS of Eq.~(\ref{eq:CDP}) must decrease
exponentially, whereas for theories without a mass gap [e.g., QED$_4$] the
decrease must be at least as fast as $1/\lambda^2$, (Strocchi, 1976, 1978).
No explicit proof that the CDP fails for coloured Wightman functions in QCD
is currently known, although the fact that the usual QED proof of this
property does not apply in QCD is certainly suggestive.  Clearly, if all the
$W$ in Eq.~(\ref{eq:CDP}) are colour singlets then the CDP is expected to
hold.  The failure of the CDP property does not {\em necessarily} entail the
absence of free coloured states in the spectrum of the theory, e.g., in
Greenberg (1978) an effort was made to construct an explicit confining,
nonrelativistic, harmonic-oscillator model in which the CDP fails.

A different perspective is discussed in Nishijima (1986) where the
non-Abelian nature of QCD and the necessity of an indefinite metric give rise
to confinement as a result of the complete cancellation of colour states from
the S-Matrix, in a manner analogous to that in which the longitudinal and
scalar photons decouple in QED.  The mechanism is associated with the
extended gauge transformations manifest in the BRS invariance of QCD.  In
simple terms, this picture of confinement corresponds to the idea that after
averaging over gauge field configurations only the colour singlet pieces of
operator expectation values remain.  This can also be related to the
breakdown of the CDP for coloured states.

Another sufficient condition for confinement is if QCD has only
colour-singlet free [i.e., on-shell] states.  Hence confinement might be
equivalent to the statement that the propagator of a coloured state should
have no singularities on the real, positive $\psq$-axis, [see, e.g., Cornwall
(1980), Krein \etal~(1988,1990), Gogokhia and Magradze (1989)].  Note that
any singularities on the negative [i.e., spacelike] $\psq$-axis would imply
tachyonic behaviour and hence are not allowed by causality considerations.
To discuss this possible form of confinement we will concentrate on the quark
propagator, the most general form of which [in an arbitrary covariant gauge]
is given in \Eq{Squark}.  This statement means that the quarks have no
mass-pole \mbox{$M^2(\psq)\not=\psq~\mbox{for any } \psq\geq 0$}, and so
cannot go on mass shell.  It might be argued that if the propagator has a
timelike singularity, then it should have a mass-shell and hence there
appears to be nothing to prevent us from defining the corresponding free
state.  [It is always possible to imagine $M(p^2)$ having some very unusual
behaviour at the singular point of the propagator, which would complicate
this argument.]  This suggests that confinement might be synonymous with the
requirement that coloured particles and bound states have no poles in their
propagators.  [As a counter-example one might consider QCD$_2$ in which, at
leading order in a $1/N_c$ expansion, confinement is ensured in a
conspiratorial fashion: the quark propagator has a mass pole but the vertex
in the BSE has zeros at the momenta which correspond to one of the quarks
going on-shell and hence there is no scattering out of the bound state
(Einhorn, 1976).]

We remark here that we are discussing the gauge-fixed quark propagator which
is a well defined and nontrivial element of field theory and is an important
input to any covariant calculation of scattering or bound state amplitudes.
This propagator is gauge-dependent but does contain gauge invariant
information; e.g., the quark condensate.  Of course, a nonzero value of
\cndst\ does not imply confinement.  Our remarks here are meant simply to
emphasise that there is gauge invariant content in propagators.  A separation
between the scales for confinement and dynamical chiral symmetry breaking is,
in general, to be expected and DCSB is certainly possible without
confinement.

It might be argued that the analytic structure of the gauge-fixed quark
propagator would be a genuine signal of confinement in QCD only if the
presence or absence of a pole [or branch cut] was also a gauge invariant
property. In this connection the Nielsen Identities (Nielsen, 1975) of QCD
are important.  These identities, derived using the BRS invariance of QCD,
ensure that, for example, the quark mass value obtained from the QCD
effective action is gauge invariant.  This gauge invariance is a result of
the cancellation between the gauge dependence of the effective action and
that of the vacuum configuration.  Hence, one can infer from this that the
absence of a mass pole [or branch point] is also a gauge invariant property
of the theory.  Practically, however, there is a problem with the effective
action approach: a non-perturbative calculation of the QCD effective action
is prohibitively difficult.

Given the possible implications for confinement it is of interest to examine
the consequences of assuming that the quark propagator has no singularities
on the real $\psq$-axis.  We see that the quark propagator will have a mass
pole when $M^2(\psq)=\psq$ provided that $Z(\psq)$ is not zero in the
neighborhood of this value of $\psq$.  Alternatively, we might have that $Z$
goes to zero in just the right way to cancel a singularity of the type
$1/[\psq-M^2(\psq)]$ or that $Z$ should be zero over a finite region
containing just this value of $\psq$.  In the absence of compensating zeros
in $Z$ then we would be left with the requirement that $M^2(\psq)\neq \psq$
for {\em any} real $\psq$.  On perturbative grounds we expect the same
behaviour in the asymptotic timelike and spacelike regimes.  The only way
that $M(\psq)$ can be finite in the deep timelike and spacelike regions of
$\psq$ and yet never have a pole at $M^2(\psq)=\psq$, for any real $\psq$, is
if $M(\psq)$ is discontinuous across the $\psq$ line or if $Z(\psq)$ has a
compensating zero.  The possible implications of complex poles in the
propagators are discussed in Sec.~\ref{sect-QCD-gluon}.

As a further observation regarding the absence of mass poles for coloured
states we remark that this would imply, for example, that the quark and gluon
propagators have no singularities on the real axis.  Consider then
\Eq{Sigma_q} and assume that $\Gamma^\nu$ has no poles [e.g.,
$\Gamma^\nu\to\gamma^\nu$ as for the rainbow or ladder approximation] and
that the integral exists.  We see that the RHS of this equation will be pure
imaginary if $\Sigma$ is initially pure real, and hence in general this DSE
will give rise to a complex self-energy.  This represents an additional
complication if this picture of confinement is pursued.  We note, however,
that if we assume that it is reasonable to begin with the Euclidean
transcription of \Eq{Sigma_q} then the quark self-energy will be real.
Apparently then, in this confinement scenario the naive Wick rotation is not
valid, which should not be surprising.

There is one further point that it is interesting to consider here.  Marciano
and Pagels (1978) have discussed the possibility of confinement occurring
through the power counting of divergences, which were presumed to arise in a
full solution for the various propagators and proper vertices of QCD.  If
confinement results from the non-Abelian nature of QCD then these divergences
would also be expected to have non-Abelian origins.  At the simplest level
one could imagine, for example, associating with each coloured leg of an
$n$-point Green's function a factor $\sqrt{\epsilon}$, while for
colour-singlet bound-state legs the factor would be unity.  Similarly, for
proper [i.e., one-particle irreducible] vertices coloured legs would carry a
factor $1/\sqrt{\epsilon}$ and colour-singlet legs again would have the
factor 1.  It follows that $S$ and $D$~$\sim\epsilon$,
$\Gamma^\nu\sim1/\epsilon^{(3/2)}$ for the quark-gluon vertex,
$\overrightarrow{\Gamma_5}\sim 1/\epsilon$ for the quark-pion vertex, etc.
It is straightforward to see that any process with only colour-singlet
external legs is of ${\cal O}(1)$, while any process with $m$ coloured
external legs will be of ${\cal O}(\epsilon^{(m/2)})$ and so will be
completely suppressed as $\epsilon\to 0$.  In any perturbative QCD
calculations the selection of only those soft final state interactions which
give rise to colour singlet asymptotic states is thus ensured.  [Many other
power counting schemes are possible and may incorporate ghost fields etc.
but this illustration exemplifies the general idea behind them.]  This is an
example of the failure of the CDP for non-colour-singlet asymptotic states,
whereas it would not fail for allowed colour-singlet processes.  It can
simultaneously be considered as an example of the absence of coloured states
since free quarks cannot propagate as $\epsilon\to 0$.  We do not argue that
this is a realistic picture of confinement, rather we take it as a lesson
that confinement might arise from subtly cancelling and possibly
momentum-dependent divergences in the Green's functions and proper vertices
of QCD.

\underline{To summarise}: The failure of the cluster decomposition property
is sufficient to ensure confinement and forms the basis for many intuitive
arguments and simple nonrelativistic models of confinement.  One can think of
the CDP as a measure of charge screening: having the CDP in a theory means
that the interaction is screened and falls off with separation.  Failure of
\Eq{eq:CDP}, when $W^{(j)}$ and $W^{(n-j)}$ are not colour singlets, is
consistent with the notion that colour-charge is anti-screened.  There is as
yet, however, no proof that the CDP fails in QCD.  In simple models the
failure of the CDP appears linked to this infrared behaviour.  The absence of
a pole at timelike momenta in a particle's propagator is also sufficient to
ensure confinement [although not necessary, as the QCD$_2$ example shows
(Einhorn, 1976)].  It is certainly conceivable that the failure of the CDP
and the absence of a pole in the propagator of confined particles are related
and that one is not possible without the other.  This interesting speculation
receives some support in the simple model described in the next section.
Explicit demonstrations of confinement in more realistic models are still
wanting, however.

\subsubsect{Examples of Confining Models}
Let us consider two simple models of the quark DSE which indeed lead to a
confined quark and gluon.  More details of this work can be found in Burden
\etal~(1992b).  We work in Euclidean space and use the notation of Burden
\etal~(1992b), where we write for the quark self-energy
\mbox{$\Sigma(k) = S^{-1}(k)-S_0^{-1}(k)$} with the renormalised
quark propagator
\beqn
S(k) & = &  \frac{1}{i\gamma\cdot k A(k^2) + B(k^2)} 
       = \frac{Z(k^2)}{i\gamma\cdot k + M(k^2)} \label{repa} \\
      & = &  -i\gamma\cdot k \sigma_V(k^2) + \sigma_S(k^2)
\label{skab} 
\eeqn
and the perturbative quark propagator
\mbox{$S_0(k)=1/(i\gamma\cdot p + m)$}.

The first model of interest is obtained with the choice
(Munczek and Nemirovsky, 1983)
\beq
g^2 D_{\mu\nu}(k)  =  D^{0}_{\mu\nu}(k) \; 8\pi^4 D\, k^2\,\delta^4(k)
\label{gprop}
\eeq
where \mbox{$D_{\mu\nu}^{0}(k) = (\delta_{\mu\nu} - k_\mu k_\nu/k^2)/k^2$}
and 
\beq
\Gamma_\mu(k,p)=\gamma_\mu~.
\label{bva}
\eeq
This is an infrared dominant model which underestimates the strength of the
interaction in QCD at large-$k^2$.  In a study whose focus is confinement
this is not unreasonable since confinement is expected to be a predominantly
infrared effect.  The quark-quark interaction resulting from \Eq{gprop} is a
constant in configuration-space and so the usual proof of the CDP will fail.
Using
\Eq{gprop} and (\ref{bva}), we find that \Eq{Sigma_q_exact} reduces to the
following pair of coupled, algebraic equations for \mbox{$Z(p^2)$} and
\mbox{$M(p^2)$}~\footnotemark[1]:
\footnotetext[1]{In deriving these equations we have made use of the fact that
\[ 
\int d^4q\,f((q+p)^2)\frac{(q\cdot p)^2}{q^2p^2}\delta^4(q) = 
\frac{1}{4}f(p^2).
\]
We remark  that Eq.~(\ref{gprop}) represents a Landau gauge propagator in
contrast to the work of  Munczek and Nemirovsky (1983) where a Feynman gauge
propagator was used.  To reproduce these earlier results simply write 
\mbox{$D = \eta^2/2$}.}
\beqn
\label{AEab}
Z  = \frac{M}{2M-m}~, & & 
0  =  2M^4 - 3mM^3 + M^2(2p^2 -D  + m^2) 
                    - 3mp^2M + p^2m^2~.
\eeqn
For $m=0$ there is a dynamical chiral symmetry breaking solution of these
equations (\mbox{$s=p^2$})
\beqn
M(s) & = & \left\{\begin{array}{ll}
                  \sqrt{\hlf D - s}~, & s < \hlf D \\
                  0~,           & s > \hlf D
                \end{array} \right. \label{Mz}\\
Z(s) & = & \left\{\begin{array}{ll}
                   \frac{1}{2}~, & s< \hlf D \\
                   \displaystyle
                   \frac{s}{2D}\left(\sqrt{1+\frac{4D}{s}}-1\right)~, & 
                   s>\hlf D
                   \end{array} \label{Zz}\right.
\eeqn
The CJT effective action [Cornwall \etal~(1974)] evaluated at its extremals
is (Stam, 1985; Sec.~\ref{subsect-eff-actions}) when evaluated in Euclidean
space:
\beq
V[S] = \int \frac{d^4q}{(2\pi)^4} \, 
\left\{\mbox{trln}[S_0^{-1}(q)S(q)]
        +\hlfsm\mbox{tr} [\Sigma(k)S(k)]\right\} 
\label{cjteff}
\eeq
and is the same as the auxiliary field effective action in this case (McKay
and Munczek, 1989).  To determine whether chiral symmetry is dynamically
broken for \mbox{m=0} one must evaluate the difference
\beqn
\beta & = & V[Z=(s/2D)\left(\sqrt{1+4D/s}-1\right),~M=0]
-V[Z = \mbox{Eq.~(\ref{Zz})},~M=\mbox{Eq.~(\ref{Mz})}]\nonumber\\
& = & N_c N_f \frac{D^2}{32\pi^2}\left(4\ln 2 -\frac{11}{4}\right) >0
\eeqn
which reveals that the ground state with a nonzero fermion mass function is
dynamically favoured and hence chiral symmetry is dynamically broken in this
approximation.  The dynamical chiral symmetry breaking solution is also a
confining solution since $S(p)$ constructed from these functions has no pole
on the real $s$ axis.  This feature survives for $m>0$ with $M$ and $Z$
smooth on the real $s$ axis: in the spacelike UV limit\footnotemark[2]
\footnotetext[2]{
Deriving~these~results~is~easiest~if~one~uses~the~equations:\\
\hspace*{\fill}\( (1-Z) = (DZ^2)/(s+M^2)\; \; \; \mbox{ and } \;\;\;(M-mZ)
=(2DMZ^2)/(s+M^2)\)\hspace*{\fill}\\
which are equivalent to Eqs.~(\ref{AEab}).}
\beq
\begin{array}{lcr}
\displaystyle
M(s) = m\left(1+\frac{D}{s}\right) &
\; \; \mbox{and} \; \; &
\displaystyle
Z(s) = 1-\frac{D}{s}~;
\end{array}
\label{AEbv}
\eeq
while in the timelike UV limit
\beq
\begin{array}{lcr}
\displaystyle
M(s) = M_{0}(s) + m\frac{3D}{8}\frac{1}{M_{0}^2(s)} \; & 
\; \; \mbox{and} \; \; & \;
\displaystyle
Z(s) = Z_{0}(s) + m \frac{1}{4}\frac{1}{M_{0}(s)}~,
\end{array}
\label{AEtbv}
\eeq
where \mbox{$M_0$} and \mbox{$Z_0$} are the $m=0$ solutions of
Eqs.~(\ref{Mz}) and (\ref{Zz}).  In this case the integrable $\delta^4(k)$
singularity in the model gluon propagator of Eq.~(\ref{gprop}) lies behind
the absence of singularities in the quark propagator and has been used as the
starting point for confining, phenomenological models of nonperturbative QCD
[see, e.g., Praschifka \etal~(1989) and Frank \etal~(1991)].

The second model has the same form for the gluon propagator as
\Eq{gprop}, but uses the Curtis-Pennington vertex of \Eq{Gamma_CP}.
This represents a test of whether the absence of a pole in the quark
propagator is sensitive to the details of the quark-gluon vertex.
Substituting Eqs.~(\ref{Gamma_CP}) and (\ref{gprop}) into
(\ref{Sigma_q_exact}) then, instead of Eqs.~(\ref{AEab}), we obtain the
following pair of coupled, nonlinear, differential equations:
\beqn
\mu'(x) & = & \frac{2\mu}{x} + 2\frac{x+\mu^2(x)}{x\zeta(x)}
(\zeta(x)\mbar -\mu(x))~, \label{eqf}\\
\zeta'(x) & = & \frac{\mu(x)\mu'(x) -1}{x+\mu^2(x)}\,\zeta(x) + 
2(1-\zeta(x))\label{eqg}
\eeqn
where \mbox{$x=s/(2D)$}, \mbox{$\mbar = m/\sqrt{2D}$},
\mbox{$M(s)=\sqrt{2D}\mu(x)$} and \mbox{$Z(s) = \zeta(x)$}.  [Note
that the transverse addition \Eq{Gamma_trCP} does not survive
when \Eq{gprop} is used and so the Ball-Chiu vertex of
\Eq{Gamma_BC} gives the same results.] The boundary conditions for
these equations are, of course:
\beq
\begin{array}{lcr}
\mu(x\rightarrow +\infty)  =  \mbar \; & \; \; \mbox{and} \; \; 
& \zeta(x\rightarrow +\infty)  =  1 
\label{fgbc}
\end{array}
\eeq
consistent with Eq.~(\ref{AEbv}).  The pragmatic benefit of Eq.~(\ref{gprop})
is again obvious: we are confronted with differential equations instead of
integral equations [e.g., Roberts and McKellar (1990), Haeri and Haeri
(1991), Williams \etal~(1991), Hawes and Williams (1991), H\"adicke (1991)].
Hence we can obtain solutions for both spacelike and timelike momenta, which
enables a study of the singularity structure of the quark propagator.  [The
infrared dominant \mbox{$k^{-4}$} model also has this feature in axial gauge
as discussed in Ball and Zachariasen (1981).] These differential equations
take a particularly simple form when expressed in terms of
\mbox{$\sgmvb(x)=\zeta(x)/(x+\mu^2(x))$} and
\mbox{$\sgmsb(x)=\sgmvb(x)\mu(x)$}~ (\mbox{$x=k^2/(2D)$}, 
\mbox{$\sigma_V(k^2)=\sgmvb(x)/(2D)$},
\mbox{$\sigma_S(k^2)=\sgmsb(x)/\sqrt{2D}$}):  
\beqn
\sgmsb'(x)  =  2[\mbar\,\sgmvb(x)-\sgmsb(x)]~,
& & 
\sgmvb'(x)  =  -\frac{2}{x}[\sgmvb(x)\,(x+1)+\mbar\,\sgmsb(x) - 1]~.
\label{ldeab}
\eeqn
We begin by considering the $\mbar=0$ case [in which case Eqs.~(\ref{ldeab})
decouple].  Obviously there is the trivial \mbox{$\mu\equiv 0$} solution that
corresponds to a Wigner-Weyl realisation of chiral symmetry. In this case
\beq
\zeta(x) = 1- \frac{1}{2x}+C\frac{\mbox{e}^{-2x}}{x}
\label{guv}
\eeq
with $C$ an arbitrary constant.  The choice \mbox{$C=1/2$} yields a
propagator with no singularity on the real $x$ axis and is required if this
solution of the differential equation is to be consistent with the integral
equation, so that there is no singularity within the Euclidean integration
domain [in agreement with footnote 3 in Munczek (1986)].  The solution can
therefore be written as:
\beqn
\sgmvb(x)  =  \frac{2x-1+\mbox{e}^{-2x}}{2x^2}~, \label{svmzro}
& & 
\sgmsb(x)  =  0~.
\eeqn
There is also a nontrivial solution.  In searching for this solution one finds
that for $\mbar=0$ Eqs.~(\ref{eqf}) and (\ref{eqg}) are scale invariant;
i.e., given $\mu$ and $\zeta$ as solutions then
\beq
\begin{array}{lcr}
\mu_\lambda (x)= \lambda \mu(x) \; & \; \; \mbox{and} \; \; & \; 
\displaystyle
\zeta_\lambda(x) = \frac{x+\lambda^2 \mu^2(x)}{x+\mu^2(x)} \, \zeta(x)
\end{array}
\eeq
are also solutions.  It is therefore sufficient to find the solution for which
\mbox{$\mu(0) = 1$}.  These scale transformations correspond to:
\beq
\begin{array}{lcr}
\overline{\sigma}_{S}^{\lambda} (x)= \lambda \sgmsb(x) \; & \; \; \mbox{and} \;
\; & \;  
\displaystyle
\overline{\sigma}_{V}^{\lambda}(x) = \sigma_{V}(x)
\end{array}
\eeq
and the scale invariance of Eqs.~(\ref{eqf}) and (\ref{eqg}) is a result of
the decoupling of Eqs~(\ref{ldeab}).  It is, of course, easiest to find the
solution from Eqs~(\ref{ldeab}) which yield $\sgmvb$ of
Eq.~(\ref{svmzro}) and
\beq
\sgmsb(x) = \mbox{e}^{-2x}
\label{ssmzro}
\eeq
which correspond to:
\beqn
\mu(x)  =  \frac{2x^2}{1+(2x-1)\,\mbox{e}^{2x}}~,
& & 
\zeta(x)  = \frac{4x^3+\left[1+(2x-1)\,\mbox{e}^{2x}\right]^2}
                {2x\,\mbox{e}^{2x}\,[1+(2x-1)\,\mbox{e}^{2x}]}
\eeqn
To determine whether chiral symmetry is dynamically broken in this model one
must evaluate the difference
\beqn
\lefteqn{V[\sgmvb=\mbox{Eq.~(\ref{svmzro})},\sgmsb=0] 
- V[\sgmvb=\mbox{Eq.~(\ref{svmzro})},\sgmsb=\mbox{Eq.~(\ref{ssmzro})}] }
\nonumber\\ 
& & \; \; \; \; \; \; \; \; \; \; \; \; \; \; \; \; 
\; \; \; \; \; \; \; \; \; \; \; \; \; \; \; \; 
= -2N_cN_f\int\dq\, \ln\left(1+\frac{\mu^2(q^2)}{q^2}\right) <0
\eeqn
and hence the chirally symmetric solution is dynamically favoured in this
case, as pointed out by Munczek (1986).  The underlying reason for this is
the decoupling of the equations for $\sgmvb$ and $\sgmsb$.  It is reasonable
to expect that with a model gluon propagator that incorporates asymptotic
freedom [i.e., one which behaves as \mbox{$\sim [q^2\ln(q^2)]^{-1}$} in the
deep spacelike region] these equations will not decouple and hence chiral
symmetry will be dynamically broken.

Comparing these two confining models with $\mbar=0$ we see that the choice of
vertex has a quite dramatic qualitative and quantitative effect on the nature
of the solution and the realisation of chiral symmetry.  Qualitatively, the
dressed vertex has ensured that the quark propagator is an entire function,
whereas in the bare vertex case it had a branch point at \mbox{$x=1/4$}, and
it has also eliminated dynamical chiral symmetry breaking from the model.
Quantitatively, there is a dramatic enhancement of the propagator at large
$(-x)$; i.e., the deep timelike region.  This is easily seen since, using
Eq.~(\ref{bva}), one finds that in the bare vertex case
\mbox{$\sgmvb(x=k^2/(2D)) \rightarrow 2$} and
\mbox{$\sgmsb(x) \rightarrow 2\sqrt{-x}$} as \mbox{$x\rightarrow -\infty$}
which are to be compared with Eqs.~(\ref{svmzro}) and (\ref{ssmzro}).

In the second model with \mbox{$\mbar \neq 0$} there is no trivial
[\mbox{$\mu\equiv 0$}] solution and solving Eqs.~(\ref{eqf}) and (\ref{eqg})
[or, equivalently, Eqs.~(\ref{ldeab})] is a little more difficult.  These
equations can be combined to yield:
\beqn
\frac{d^2}{dx^2}\sgmsb(x)
+4\left(1+\frac{1}{2x}\right)\frac{d}{dx}\sgmsb(x)
+4\left(1+\frac{1+\mbar^2}{x}\right) \sgmsb(x) & = & \frac{4\mbar}{x}
\eeqn
from which one obtains (\mbox{$x=y^2$}): 
\beq
\sgmsb(y) = \frac{c_1}{2\mbar y}\,\exp(-2y^2) \,J_1(4\mbar y) + 
        \frac{\mbar^2}{y}\int_{0}^{\infty}d\xi\,
     \xi K_{1}(\mbar\xi)J_1(y\xi)\mbox{e}^{-\frac{\xi^2}{8}} 
\label{ssy}
\eeq
with $c_1$ an undetermined constant and $J_1$ and $K_1$ Bessel and modified
Bessel functions of order one, respectively.\footnotemark[4] [In the limit
$\mbar\rightarrow 0$ \Eq{ssy} reduces to \Eq{ssmzro}; the dynamical chiral
symmetry breaking solution.  In the following we set $c_1=0$ and concentrate
on the behaviour of that part of the solution associated with explicit
chiral symmetry breaking.]
\footnotetext[4]{
In this expression the additional solution of the associated homogeneous
equation:\\
\hspace*{\fill}\(\displaystyle\frac{c_2}
{2\overline{m} y}\exp(-2y^2) Y_1(4\mbar y)~,\)
\hspace*{\fill}\\
\mbox{$x=y^2$}; $Y_1$ a Bessel functions of order one, has been discarded
because it is irregular at $y=0$.} 
The form of \mbox{$\sgmvb$} follows from Eq.~(\ref{ldeab}) ($y^2=x$):
\beq
\sgmvb(y) = \frac{1}{\mbar}\left(\sgmsb(y)+\frac{1}{4y}\frac{d}{dy}
                                        \sgmsb(y)\right)~.
\label{svy}
\eeq
It is clear that \mbox{$\sgmvb(y)$} and \mbox{$\sgmsb(y)$} are even under
\mbox{$y\rightarrow\,-y$} and, hence,
are functions of \mbox{$x=y^2$}.  Further,
after a little thought it becomes clear that \mbox{$\sgmsb(x)$} and
\mbox{$\sgmvb(x)$} are entire functions in the complex $x$ plane.  
A combination of numerical and analytic methods establishes that both
$\sgmsb$ and $\sgmvb$ are positive definite on the real $x$-axis.
It is of interest to determine the asymptotic behaviour of these functions at
large spacelike and timelike \mbox{$x=k^2/(2D)$}.  In the deep spacelike region
[\mbox{$x\sim +\infty$}] one finds easily that \mbox{$\sgmsb(x)=\mbar/x$} and
\mbox{$\sgmvb(x)=1/x$}.  In the deep timelike region (\mbox{$x\sim -\infty$})
the analysis is a little more complicated.  Using Laplace's method
one finds that at large \mbox{$E=\sqrt{-x}$}:
\beq
\sgmsb(E) = \sqrt{\frac{\pi\mbar^3}{2E^3}}\exp\left(2[E-\mbar]^2\right)
              \left[1+\mbox{erf}\left(\sqrt{2}[E-\mbar]\right)\right]
\eeq
which yields the following asymptotic timelike behaviour:
\beqn
\sgmsb(E) & = & \sqrt{\frac{2\pi\mbar^3}{E^3}}\exp\left(2[E-\mbar]^2\right)~,
\\ 
\sgmvb(E) & = & \sqrt{\frac{8\pi\mbar}{E^3}}\exp\left(2[E-\mbar]^2\right)
                    \left(1-\frac{\mbar}{2E}-\frac{3}{16E^2}\right)~.
\eeqn
This again is a dramatic enhancement compared with the bare vertex solutions:
\mbox{$\sgmvb \simeq 2(1-\mbar/\sqrt{\frac{1}{4}-x})$} and
\mbox{$\sgmsb \simeq 2(\sqrt{\frac{1}{4}-x}- \mbar)$}. 

Hidden in this discussion of $\sgmsb$ and $\sgmvb$ is the behaviour of the
functions $\mu$ and $\zeta$ [which are related to the mass function, $M$, and
wave function renormalisation, $Z$: see after Eq.~(\ref{eqg})] that can be
expressed [\mbox{$K(x)=1+\sgmsb'(x)/(2\sgmsb(x)$}]:
\beqn
\mu(x)  =  \frac{\sgmsb(x)}{\sgmvb(x)} \equiv \frac{\mbar}{K(x)} \;
& \;\; {\rm and} \;\; & 
\zeta(x) = 
\frac{1}{\mbar}\left[x+\mu^2(x)\right]\sgmsb(x) K(x)~.\label{cang}
\eeqn
The propagator is very often discussed and modelled in terms of these functions
[equivalent to $M$ and $Z$].  Our analysis leads us to the conclusion that 
\mbox{$K(x)>0$} for all finite, real $x$ and hence so is $\mu$; i.e., the mass
function has no zeros or singularities for real $x$.
\begin{figure}[tb] 
 \centering{\
  \epsfig{figure=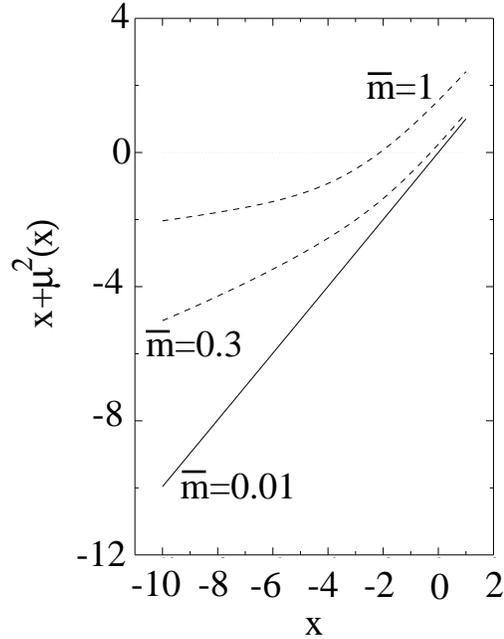,height=9.0cm} }
  \parbox{130mm}{
\caption{ A plot illustrating the $\overline{m}$-dependence of the zero
of the propagator denominator in the \{$Z,M$\} representation for $S(p)$.
\label{brw_conf_fig} }}
\end{figure}
In \Fig{brw_conf_fig} we have plotted
\mbox{$[x+\mu^2(x)]$} for a range of values of \mbar.
The striking observation is that this has a \mbar\ dependent zero.  A pole in
the propagator is only avoided, therefore, because $\zeta(x)$ also has a zero
at this point, Eq.~(\ref{cang}); i.e., $\zeta$ and $x+\mu^2$ have coincident
zeros.  [We remark that a zero in $\zeta$ with $\mu$ finite entails a pole in
the functions \mbox{$A(x)$} and \mbox{$B(x)$} of Eq.~(\ref{repa}).]

\underline{To summarise}: In the first model [the bare vertex] we see that
$M(p^2)$ diverges in the timelike region [Euclidean $p^2<0$] so that a pole
in the quark propagator is avoided and hence the theory is confining.  In the
second model [the CP or BC vertex] for $m=0$ there is confinement [the quark
propagator is entire] and the phase with no DCSB [i.e., with $M(p^2)=0$] is
energetically favoured.  When $m\neq 0$ in the second model we find DCSB and
that confinement occurs through cancelling zeros in the numerator and
denominator of the quark propagator.  It is interesting to speculate that the
effective gluon propagator is primarily responsible for confinement and that
the vertex modifies the particular manifestation for the quark propagator.
However, other models need to be studied before any such conclusion can be
drawn.

\subsubsect{Application of a Confinement Test}
The confining models discussed above could be studied in the timelike region
because their simple form allowed them to be written as coupled
differential equations.  In the general case an integral equation must
be solved and the extension into the timelike region is highly nontrivial.
It is still possible to probe the timelike behaviour of the quark propagator
obtained by numerical solution of a model quark DSE.
In order to determine whether the quark propagator
obtained as a solution to our DSE can represent a confined particle we follow
Hollenberg \etal~(1992) and adapt a method commonly used in lattice QCD to
estimate bound state masses.
We write as before
\begin{equation}
\sigma_S(p^2) = \frac{B(p^2)}{p^2 A(p^2)^2 + B(p^2)^2}
\end{equation}
and define 
\begin{equation}
\Delta_S(T,\vec{x}) = \int\,\frac{d^4p}{(2\pi)^4} 
{\rm e}^{i(p_4T+\vec{p}\cdot\vec{x})} \sigma_S(p^2)~.
\end{equation}
This is the scalar part of the Schwinger function of the model quark
propagator.  If we now define
\begin{equation}
\Delta_S(T) = \int\, d^3x\,\Delta_S(T,\vec{x})
\end{equation}
and, for notational convenience, 
\mbox{$E(T) = -  \ln\left[\Delta_S(T)\right]$}, then it follows that if there
is a stable {\it asymptotic state} with the quantum numbers of this Schwinger
function then
\begin{equation}
\label{ConfT}
\lim_{T\rightarrow\infty} 
 \frac{dE(T)}{dT} \, = m~;
\end{equation}
where $m \geq 0$ is the asymptotic [on-shell] mass of this excitation; i.e.,
this limit yields the dynamically generated quark mass evaluated at the mass
pole.  If the limit in Eq.~(\ref{ConfT}) exists for a given propagator then
we can conclude that the propagator has a timelike mass pole, which suggests
that the associated excitation is not confined.

For example, one can consider the model of Burden \etal~(1992b)
discussed above and one finds
\begin{equation}
\frac{dE(T)}{dT} \stackrel{T\rightarrow\infty}{\sim} \kappa T
\end{equation}
where $\kappa$ is a constant and hence the limit in Eq.~(\ref{ConfT}) does
not exist.  In this case the Schwinger function does not satisfy the cluster
decomposition property and hence the quarks are confined (Roberts \etal,
1992).  Alternatively one may say that through self interaction the quark
acquires an infinite dynamical-mass.  This provides another way of
understanding the claim that the model of Burden \etal~(1992b) is confining.

Another example of this approach is to apply it to the IR vanishing gluon
propagator of \Eq{alphas_hrw}.  We set $\alpha_s$ to a constant here for
simplicity and then, for the boson analogue of $\Delta_S(T)$, one finds:
\begin{equation}
\label{BosonD}
\Delta(T) \propto \frac{1}{ b \surd 2}
        \exp\left(-\frac{b\, T}{\surd 2}\right)
        \left(\cos\left(\frac{b\, T}{\surd 2}\right)
                - \sin\left(\frac{b\, T}{\surd 2}\right)\right)~.
\end{equation}
Note that the Schwinger function in this case is not positive definite, which
is an easily identifiable signal in $\Delta(T)$ due to the pair of complex
conjugate poles.  This violates the axiom of reflection positivity.  It
follows from this that Eq.~(\ref{BosonD}) describes a field with a complex
mass spectrum and/or residues that are not positive.  This is appropriate for
particles that decay and forms the basis of the argument that such a
propagator allows coloured states to exist only for a finite time [of the
order of $1/b$] before hadronising; i.e., that the propagator describes
confined gluons (Stingl, 1986; H\"abel \etal~1990a, 1990b; Zwanziger, 1991).

To illustrate this procedure for an explicit numerical solution we return to
the example of the model quark DSE discussed in Sec.~\ref{QCD_DCSB_results}
and attempt to determine whether the model gluon propagator specified by
Eqs.~(\ref{Gprop}) and (\ref{PiA}) leads to quark confinement; i.e., the
absence of free quarks in the QCD spectrum (Hawes \etal~(1993).  In applying
this method here it is obvious that numerical evaluation of the Fourier
transforms required in using Eq.~(\ref{ConfT}) will be hindered by numerical
noise as $T$ is increased.  In order to minimise the effect of this noise on
the derivative, we fitted $\Delta_S(T)$ to a form
\begin{equation}
\label{expfit}
C \exp\left(-m T\right)
\end{equation}
and extracted the derivative from this fit.  [Importantly, there is no
indication in the results of the structure suggested by Eq.~(\ref{BosonD}).]
This was particularly useful with the propagators obtained using small values
of $b$ which had large dynamical masses [as one would expect since the
condensate is large in this case] and hence a rapid decline with $T$.
Results were found for the confinement test in the following cases: 1) The
propagators obtained with \mbox{$\ln\tau=0.1$} and $b^2$ in the range
\mbox{$[0.1,1.0]$}; 2) The propagator obtained with \mbox{$\ln\tau = 0$} and
\mbox{$b^2 = 0.35$} which yields the largest value of
\mbox{$-\langle\overline{q}q\rangle_\mu$} on the \mbox{$(b^2,\ln\tau)$} domain
considered; 3) Two propagators obtained with 
\mbox{$(b^2,\ln\tau) = (0.1,0.6)$} - one using the Ball-Chiu vertex and
another using the Curtis-Pennington addition.  The results obtained by
fitting the form Eq.~(\ref{expfit}) to our numerical output are presented in
Table.~\ref{TPMs}. 
\begin{table}[tbh]
\begin{center}\parbox{130mm}{\caption{Asymptotic dressed-quark-mass
           values for the family of
	   propagators with $\ln\tau=0.1$\/,
	   the propagator which showed maximal DCSB\ 
	   (at $\ln\tau=0$\ and $b^2 = 0.35$\/),
	   and for $\ln\tau=0.6$\/, $b^2=0.1$\ with
	   both the Ball-Chiu and Curtis-Pennington vertices.
	   These values were arrived at by exponential fit and
	   match the limits of the asymptotic mass curves.
	   }}
  \begin{tabular}{|c|c|c|c|c|}\hline 
  $\ln\tau$&$b^2$&$m_{\rm free}$&$C$&Comments\\
  \hline\hline
     .1 & .25    & 0.410   & 0.664 & B-C vertex\\ \hline
     .1 & .35    & 0.296   & 0.633 & ''\\ \hline
     .1 & .45    & 0.176   & 0.619 & ''\\ \hline
     .1 & .54    & 0.0275  & 0.619 & ''\\ \hline
     .0 & .35    & 0.406   & 0.667 & ''\\ \hline
     .6 & 0.1    & 0.210   & 0.648 & B-C vertex\\ \hline
     .6 & 0.1    & 0.210   & 0.507 & C-P vertex\\ \hline
  \end{tabular}
  \label{TPMs} 
\end{center}
\end{table}
\begin{figure}[tb] 
 \centering{\
  \epsfig{figure=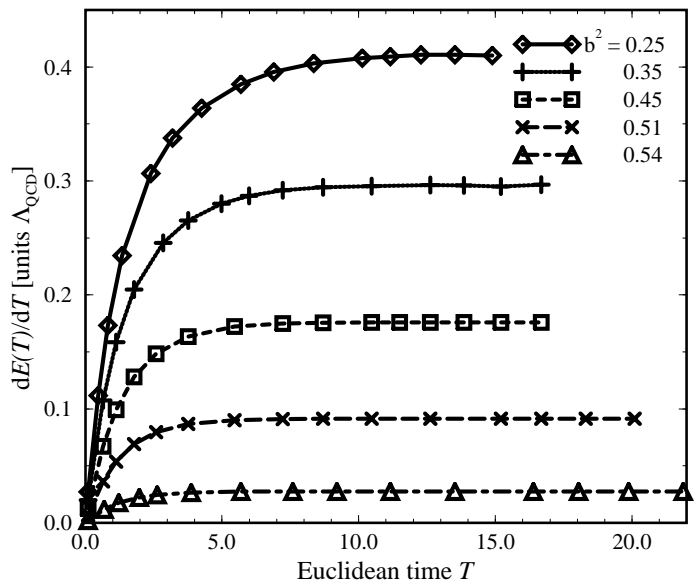,height=9.0cm} }
  \parbox{130mm}{
\caption{ Dressed-quark-mass curves for the family of propagators with
the minimal Ball-Chiu vertex and $\ln\tau=0.1$. The masses are in units
of $\Lambda_{\rm QCD}.$
\label{dse_hrw_fig4} }}
\end{figure}

In \Fig{dse_hrw_fig4} we present plots of \mbox{$E'(T)$} for the family of
propagators obtained with \mbox{$\ln\tau = 0.1$} and this clearly illustrates
that an unambiguous determination of the dressed-quark-mass is possible.  It
will be observed that the mass decreases with increasing $b^2$.  This is
easily understood in terms of the chiral phase transition: as $b$ increases
beyond $b_c$ there is no DCSB and massless current quarks remain massless.
Since the behaviour of all the other solutions we obtained was qualitatively
the same as that described by the results presented in
\Table{TPMs} and \Fig{dse_hrw_fig4} we infer that the model considered here
does not yield a confining quark propagator.  We also note that the rainbow
approximation studies of the fermion DSE in Alkofer and Bender (1993),
which address the question of confinement by a direct
continuation to Minkowski momentum space, found a quark propagator with a
pole at timelike $p^2$; i.e., a non-confining propagator.
We remark that, within numerical noise, the Curtis-Pennington addition made
no difference to the dressed-quark-mass value extracted in the cases
considered and only slightly reduced the normalisation constant $C$. Clearly,
the Curtis-Pennington addition leads only to a minor quantitative effect in
this part of our study too.

It is worthwhile to recall the suggestion [H\"abel \etal~(1990a,b)] that a
model gluon propagator with the infrared form of
\Eq{alphas_hrw} would lead to a confining quark propagator of the form
\begin{equation}
\label{FStingl}
\frac{i\gamma\cdot p + c_0}{
(i\gamma\cdot p + c_1)\,(i\gamma\cdot p + c_1^{\ast})}
\end{equation}
where $c_0$ and $c_1$ are real and complex constants, respectively; i.e., to
a fermion propagator with complex conjugate poles just as the gluon
propagator has; confinement being realised through the absence of poles on
the real timelike axis.  As Eq.~(\ref{BosonD}) shows, such a propagator has a
characteristic signature in \mbox{$\Delta_S(T)$} which we do not see in
\Fig{dse_hrw_fig4}.  In this model then it is clear that a fermion propagator
of the type in Eq.~(\ref{FStingl}) does not arise.  This does not eliminate
the possibility that it can arise in the approach of H\"abel \etal~(1990a,
1990b), however, since the rational-polynomial {\it Ans\"{a}tze} employed for
the vertex functions therein leads to a completely different quark-gluon
vertex to that used here.  The above results do however
suggest that a fermion
propagator of the type in Eq.~(\ref{FStingl}) cannot arise if the quark-gluon
vertex is free of kinematic singularities.


\subsect{Euclidean $\leftrightarrow$ Minkowski Continuation}
\label{subsect-E-to-M}
We have discussed this issue in Sec.~\ref{subsect-Euclidean} and here we
simply put these earlier remarks into context by way of a few examples.

The model of Eqs.~(\ref{gprop}) and (\ref{bva}) yields a quark propagator
which has complex conjugate branch points with spacelike real parts.
Clearly, therefore, the naive transcription rules of
Sec.~\ref{subsect-Euclidean} will not be suitable in this case. In the
modification of this model, Eqs.~(\ref{Gamma_CP}) and (\ref{gprop}), one
obtains a quark propagator that is an entire function with an essential
singularity at timelike infinity.  In this case it is not possible to close
the integration contour and hence a ``Wick rotation'' is simply not possible.
This latter case illustrates that the inability to perform a Wick rotation
appears even in models which go beyond the rainbow approximation.

Detailed analyses of the analyticity properties of quark propagators obtained
with other model gluon propagators in rainbow approximation have been carried
out (Stainsby and Cahill, 1992; Maris and Holties, 1992; Stainsby, 1993, Sec.
3; Maris, 1993, Sec. 3.3).  In the studies by Stainsby and Cahill (1992) and
Stainsby (1993) three forms of gluon propagator were considered:
\beq
D_{\mu\nu}(q) = \delta_{\mu\nu}\,\frac{4\pi \alpha(q^2)}{q^2}
\eeq
with [$s=q^2$]
\beqn
\label{GCMalpha}
\alpha_1(s) & = &\frac{3\pi \chi^2}{4 \Delta^2}\,
        \exp\left(\,-\,\frac{s}{\Delta}\right) + 
        \frac{4\pi}{11\ln\left(1 + \epsilon 
        + \frac{s}{\Lambda^2_{\rm QCD}}\right)}~,
\eeqn
where $\chi= 1.14$~GeV, $\Delta=0.002$~GeV$^2$, $\Lambda_{\rm QCD}=0.19$~GeV
and $\epsilon= 2.0$, taken from Praschifka \etal~(1989) where
it was shown that such a gluon propagator is well suited to a description of
low energy hadronic physics;
\beqn
\alpha_2(s)  =  \frac{C_1}{s+s_0}
& \;\; {\rm and} \;\; &
\alpha_3(s)  =  C_2 \exp\left(\,-\,\frac{s}{s_0'}\right)
\eeqn
with $C_1 \approx 1$, $s_0 \approx (0.0707)^2$~GeV$^2$ and $C_2 \approx 50.0$
and $s_0' \approx (0.245)^2$~GeV$^2$ chosen so that the solutions obtained
for $A(p^2)$ and $B(p^2)$, with $p^2$ on the real spacelike axis, were very
similar to those found with $\alpha_1$.  Solving the integral equations
directly, with no approximation of the kernel, it was found that in each case
there were pairs of complex-conjugate, logarithmic branch points with
spacelike real part in the functions $A(p^2)$ and $B(p^2)$ and hence in the
quark propagator.  The position of the branch point depended on the form of
the propagator but not its nature or existence:
\beq
\begin{array}{ccc}
\alpha_1: s_B \approx 0.22 + 0.25\,i~, &
\alpha_2:  s_B \approx 0.015 + 0.33\,i~, &
\alpha_3: s_B \approx 0.06 + 0.41 \, i~.
\end{array}
\eeq

Maris (1993) considered two model forms of gluon propagator
\beq
D_{\mu\nu}(q) = \left(\delta_{\mu\nu}- \frac{q_\mu q_\nu}{q^2}\right) 
        \,\frac{4\pi \alpha(q^2)}{q^2}
\eeq
with
\beqn
\alpha_1(q^2) = \frac{d_M  \pi}
        {\ln\left(\tau + \frac{q^2}{\Lambda_{\rm QCD}^2}\right)}
\; & \;\; {\rm and} \;\; & \;
\frac{4\pi \alpha_2(q^2)}{q^2} = \eta^2 \delta^4(q) 
                + \frac{4 \pi \alpha_1(q^2)}{q^2}~.
\eeqn
The first form is a simple extension of the one loop renormalisation group
result for the running coupling in QCD to small-$q^2$ and the second form
involves an additional ``confining term'', $\delta^4(q)$.  The DSEs obtained
were analysed using the approximation of \Eq{AngApprox} to obtain
differential equations.

In the case of $\alpha_1$ there are five singularities, two of which are
simply those of the running coupling; i.e., at $x = -\tau$ and $x= -\tau +
1$, where $x= p^2/\Lambda_{\rm QCD}^2$, and can therefore be discarded as
artifacts of the model.  Of the three not trivial singularities, two are
complex conjugate branch points and one is a branch point on the timelike
$p^2$ axis with \mbox{$-\tau + 1 < x_B < 0$}.  This latter singularity might
be used to define the quark mass shell. 

The singularity structure of the $\alpha_2$ model is interesting and
complicated.  The most interesting feature is that the singularity on the
timelike axis bifurcates into a pair of complex conjugate branch points in
the presence of the ``confining'' term thereby eliminating the quark mass
shell.  In addition, the branch points that were present with $\alpha_1$
alone also bifurcate and move as $\eta^2$ is changed.

In each of these examples the naive Wick rotation, given by the transcription
rules of \Eqs{TMEpa}, (\ref{TMEpb}) and (\ref{TMEpc}), does not yield the
correct continuation of the Euclidean space DSE to Minkowski space.  This
serves to emphasise the points we made in the last subsection of
Sec.~\ref{subsect-Euclidean}.


\sect{Applications to Hadron Structure}
\label{sect-hadrons}
In this section we will apply the knowledge harvested from the studies of the
DSEs in QCD, all of which is necessary, to the study of the kinematic and
dynamical properties of hadrons.  We will work for the most part in Euclidean
space.
\subsect{Coupled Dyson-Schwinger and Bethe-Salpeter equation Phenomenology}
\label{subsect-DSE-BSE}
One direct application of the tower of DSEs to hadron phenomenology is to
solve the bound state equations using a kernel constructed from the dressed
quark and gluon propagators discussed above.  In \Fig{bse_fig} we illustrate
the BSE for mesons.  We are assuming non-flavour-singlet mesons and so only
the homogeneous BSE is shown.  In this figure we employ the convention that
$i\chi=S\Gamma_{\rm meson} S$ and the BSE quark-anti-quark kernel is
$K=(i\Gamma)(iD)(i\Gamma)+\cdots$.  Retaining only the first term shown for
$K$ yields the dressed-ladder approximation for the BSE.
\begin{figure}[tb] 
 \centering{\
  \epsfig{figure=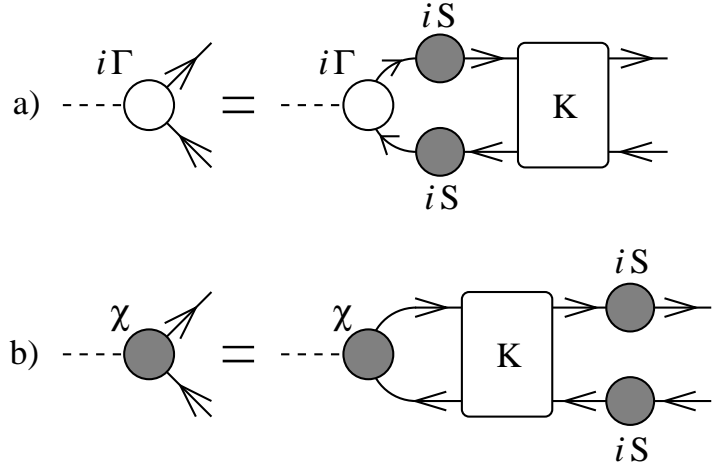,height=6.5cm} }
  \parbox{130mm}{
\caption{ Shown are two equivalent forms of the BSE; i.e., (a) the proper
meson-quark vertex [$\Gamma_{\rm meson}$] and (b) the Bethe-Salpeter
amplitude for the meson [$i\chi=S\Gamma_{\rm meson}S$].
\label{bse_fig}  }}
\end{figure}

There have been a number of studies of the covariant BSE for mesons using a
kernel constructed from a model dressed gluon propagator, whose ultraviolet
behaviour is fixed by the QCD renormalisation group and whose infrared
behaviour is based on the studies described in Sec.~\ref{subsect:IRGP}, and a
dressed quark propagator obtained as a solution of the rainbow approximation
DSE with the same model gluon propagator.  In a series of studies (Cahill
\etal, 1987; Praschifka, 1988; Praschifka \etal, 1989;
Stainsby, 1993, Sec. 4; Stainsby and Cahill, 1993) it was shown that this
system of equations provides a very good description of the $u$- and
$d$-quark sector of the mesonic spectrum with the choice
\begin{equation}
\label{GCMD}
D_{\mu\nu}(q) = \delta_{\mu\nu}
\left( C\,\Lambda_{\rm QCD}^2\,\delta^4(q) 
        + \frac{d_M  \pi}
        {q^2\,\ln\left(\tau + \frac{q^2}{\Lambda_{\rm QCD}^2}\right)}\right)
\end{equation}
with $C= (3\pi)^3$ and $\tau = 3$.  These model studies incorporated quark
confinement in the sense discussed in Sec.~\ref{subsect-quark-conf} and will
be discussed further in Sec.~\ref{subsect-eff-actions}.

In addition, there have been phenomenologically successful studies of this
coupled system using an infrared-finite form of the model gluon propagator
subject to the approximation of \Eq{AngApprox} (Dai \etal, 1991; Aoki
\etal, 1991) and noncovariant, Coulomb-gauge studies of a pion BSE
whose kernel is constructed using a dressed quark propagator obtained as a
solution of the Coulomb-gauge quark-DSE with a bare gluon propagator
(Govaerts \etal, 1984) and a model dressed gluon propagator which is
singular in the infrared (Alkofer and Amundsen, 1988).

We cannot summarise all of these studies and choose instead to discuss, as a
particular example, recent studies by Munczek and Jain (1992) and Jain and
Munczek (1993) which consider pseudoscalar- and vector-$q$-$\overline{q}$
bound states of $u$-, $d$-, $s$-, $c$- and $b$-quarks.  The treatment
necessarily used significant approximations and was carried out in Euclidean
space, which required extrapolation into the timelike region in order to
extract the meson masses.  The second of these articles is somewhat simpler
and more systematic and we summarise its arguments and results here.  It is
important to note that solving the BSE with a kernel that is consistent with
the solution of the quark DSE is necessary to guarantee the Goldstone-boson
nature of the pion in the chiral limit (Delbourgo and Scadron, 1979).

As in the studies summarised briefly above, the principal initial
approximation in this model is that the quark-gluon vertex is simply given by
$\Gamma_\mu=\gamma_\mu$ and the $q-\bar q$ scattering kernel is given by an
effective single-gluon propagator.  This is a ladder-type approximation with
a modified gluon propagator $D_{\mu\nu}$.  In this approximation, the BSE for
a $q\bar q$ bound state of mass $m^2_{fg} = p^2$ can be written in Minkowski
space as,
\beq
(S^f)^{-1}(q+\alpha p)\chi(p,q)(S^g)^{-1}(q-\beta p) = 
{16\pi\over 3}i\int {d^4k\over (2\pi)^4} \gamma_\mu
\chi(p,k)\gamma_\nu D^{\mu\nu}(k-q)\ ,
\eeq
with $\alpha+\beta = 1$.  The quark propagators are obtained by solving the
DSE for a fermion flavor $f$, in the rainbow approximation,
\beq
(S^f)^{-1}(q) = \qslash - m_f -
i{16\pi\over 3}\int {d^4k\over (2\pi)^4} \gamma_\mu S^f(k)
\gamma_\nu D^{\mu\nu}(k-q)\; .
\eeq
Note that for convenience $\alpha_s(-(k-q)^2)$ has been absorbed into the
definition of the effective gluon propagator here [as in Jain and Munczek
(1993)].  The Euclidean space form of these equations is [see
Secs.~\ref{subsect-Euclidean} and \ref{subsect-E-to-M}]:
\beqn
(S^f)^{-1}(q+\alpha p)\chi(p,q)(S^g)^{-1}(q-\beta p) = 
 - {16\pi\over 3}\int {d^4k\over (2\pi)^4} \gamma_\mu
\chi(p,k)\gamma_\nu D_{\mu\nu}(k-q)~, \\
 (S^f)^{-1}(q) = i\gamma\cdot  q + m_f +
{16\pi\over 3}\int {d^4k\over (2\pi)^4} \gamma_\mu S^f(k)
\gamma_\nu D_{\mu\nu}(k-q)\ .
\eeqn
The ladder-rainbow approximation has been argued to be most appropriate in
Landau gauge (Fomin \etal, 1983; Atkinson \etal, 1988c) in which
case the model dressed gluon propagator can be written
\beq
D_{\mu\nu}(k) = \bigg(\delta_{\mu\nu} - {k_\mu k_\nu\over k^2}\bigg) D(k^2)~.
\eeq
Jain and Munczek (1993) choose the momentum dependence of the gluon
propagator taken to be
\beq
\label{JMD}
D(k^2) = 
\bigg[ {d_M  \pi\over k^2\ln (\tau+x)}\bigg( 1+b
{\ln\lbrack\ln(\tau+x)\rbrack\over \ln(\tau+x)}\bigg)\bigg]
+ D_{IR}(k^2) \;,
\eeq
where $x\equiv k^2/\Lambda_{\rm QCD}^2$ and $\Lambda_{\rm QCD}\simeq 200$MeV
is the \hspace*{-2pt}QCD scale parameter.  In \Eq{JMD}, \mbox{$d_M\! =
12/[33-2N_f]$},
\mbox{$b= 2\beta_2/\beta_1^2$},  \mbox{$\beta_1 =N_f/3-11/2$},
\mbox{$\beta_2 = 19N_f/12-51/4$} and $N_f$ is the number of quark flavors,
[taken as five in these calculations] and so, at large $k^2$, the expression
in parentheses agrees with the two-loop renormalisation group result for the
running coupling in QCD, \Eq{alphaTL}.  As in many of the studies described
above, the parameter $\tau$ regulates the infrared behaviour of the running
coupling. As long as its value is larger than $\sim 2$, the results of the
calculations are largely insensitive to it.  The value adopted by Jain and
Munczek is $\tau=10$. 

The infrared effects in the model are dominated by $D_{IR} (k^2)$, which for
simplicity was taken to have a modified Gaussian form
\beq
\label{JMDIR}
D_{IR}(k^2) = \pi a\ k^2\ e^{-\mu k^2}\ ,
\eeq
with $a=(0.387\ GeV)^{-4}$ and $\mu=(0.510\ GeV)^{-2}$.  Defining the
classical potential by analogue with \Eq{Vclass}, including the negative sign
associated with attraction and multiplying by the factor $16\pi/3$, this form
corresponds to
\beq
V(r) = \frac{a}{6\mu^3}\,\sqrt{\frac{\pi}{\mu}}\,(r^2 - 6\mu) 
        \exp\left(\frac{r^2}{4\mu}\right)
\eeq
which is approximately linear out to 0.4 fm. We note that \Eq{JMDIR} does not
incorporate the infrared singularity $\sim \delta^4(k)$ that was associated
with quark confinement in Sec.~\ref{subsect-quark-conf}.  This term was
included in the earlier study by Munczek and Jain (1992).

The most general decomposition for the vector, pseudoscalar and
scalar bound state wave functions can be written as
\beqn
\chi_{_V}(p,q) &=& i\gamma\cdot \epsilon \,\chi_{_{V0}} 
        - i\gamma\cdot p \, \epsilon\cdot q \,\chi_{_{V1}} 
        - i\gamma\cdot q\,\epsilon\cdot q \,\chi_{_{V2}} 
        + \epsilon\cdot q \,\chi_{_{V3}} \nonumber\\
        &+& [\gamma\cdot\epsilon,\gamma\cdot p]\,\chi_{_{V4}} 
        + [\gamma\cdot\epsilon,\gamma\cdot q]\,\chi_{_{V5}} 
        - [\gamma\cdot p,\gamma\cdot q]\,\epsilon\cdot q\,\chi_{_{V6}} 
        +i \gamma_5 \gamma\cdot t\, \chi_{_{V7}}
\eeqn
with $t_\mu = \epsilon_{\mu\nu\alpha\beta}q_\nu p_\alpha \epsilon_\beta$,
$\epsilon^2 = -1$ and $ \epsilon\cdot p = 0$;
\beq
\chi_{_P}(p,q) = \gamma_5\bigg[\chi_{_{P0}} - i\gamma\cdot p\,\chi_{_{P1}} 
      -i\gamma\cdot q\,\chi_{_{P2}} 
        - [\gamma\cdot p,\gamma\cdot q]\,\chi_{_{P3}}\bigg]~;
\eeq
\beq
\chi_{_S}(p,q) = \chi_{_{S0}} -i \gamma\cdot p\,\chi_{_{S1}} 
        -i \gamma\cdot q\,\chi_{_{S2}}
        - [\gamma\cdot p,\gamma\cdot q]\,\chi_{_{S3}}~.
\eeq
In these equations, flavor indices are implicit on $\chi_{_{Vi}}$,
$\chi_{_{Pi}}$ and $\chi_{_{Si}}$, which are functions of $p^2$, $q^2$ and
$p\cdot q$.  

An important result in this study is that the dominant contributions to the
vector and pseudoscalar wave functions is provided by $\chi_{_{J0}}^{(0)}$.
The BSEs for the dominant wave functions, $\chi_{_{J0}}$, are:
\beq
\chi_{_{V0}}(p^2,q^2,p\cdot q) =
 {1\over A_aA_bD}\lbrace (q^2-\alpha\beta p^2+m_am_b) +
q\cdot p (\alpha - \beta)\rbrace I_{V0}\ +\Delta\chi_{_{V0}},
\eeq
with
\beq
I_{V0} = \int {d^4k \over (2\pi)^4} \,D(k-q)\chi_{_{V0}}\bigg\lbrace 1
+{2\over 3}
\bigg(1-{q\cdot (k-q)^2\over q^2(k-q)^2}\bigg)\bigg\rbrace~;
\eeq
\beq
\chi_{_{P0}}(p^2,q^2,p\cdot q) =
{3\over A_aA_bD}\lbrace (q^2-\alpha\beta p^2+m_am_b) +
 (\alpha - \beta) q\cdot p\rbrace I_{P0}+\Delta\chi_{_{P0}}\ ,
\eeq
with
\beq
I_{P0} = \int {d^4k \over (2\pi)^4}\, D(k-q)\chi_{_{P0}}~;
\eeq
\beq
\chi_{_{S0}}(p^2,q^2,p\cdot q) =
{3\over A_aA_bD}\lbrace (q^2-\alpha\beta p^2-m_am_b) +
 (\alpha - \beta) q\cdot p\rbrace I_{S0}+\Delta\chi_{_{S0}}\ ,
\eeq
with
\beq
I_{S0} = \int {d^4k \over (2\pi)^4} \,D(k-q)\chi_{_{S0}}~;
\eeq
and where
\beq
\begin{array}{cccc}
A_a=A_a(q+\alpha p)~, & A_b=A_b(q-\beta p)~, & 
\displaystyle m_a = {B_a(q+\alpha p)\over A_a(q+\alpha p)}~, & 
\displaystyle m_b = {B_b(q-\beta p)\over A_b(q-\beta p)}
\end{array}
\eeq
and $ D = [(q+\alpha p)^2+m_a^2][(q-\beta p^2)+m_b^2]$.

In these equations the $\Delta\chi_{_{J0}}$ contain contributions from
several of the wave functions $\chi_{_{Ji}}$.  The technique adopted by Jain
and Munczek (1993) to solve these equations was to write $p\cdot
q=pq\cos\theta$ and expand the functions $\chi_{_{Ji}}$ in terms of
Tschebyshev polynomials, $T^{(n)}(\cos\theta)$:
\beq
\chi_{_{Ji}}\biggl(q^2,M_B^2,\cos\theta\biggr)
=\sum\limits_n\,\chi_{_{Ji}}^{(n)}\biggl(q^2,M_B^2\biggr)
        \,T^{(n)}(\cos\theta)~,
\eeq
where the subscript $J$ ranges over $V$, $P$ and $S$ and
\mbox{$i=1, 2, \ldots$} over the independent wave functions.  The equation
was then projected with the zeroth order Tschebyshev polynomial 
and the right-hand-side approximated by keeping only the leading order term
in the Tschebyshev expansion for $\chi_{_{Ji}}$, $i\geq 1$, and by retaining
the three lowest terms, $\chi_{_{J0}}^{(0)}$, $\chi_{_{J0}}^{(1)}$ and
$\chi_{_{J0}}^{(2)}$, for $\chi_{_{J0}}$.

The vector and pseudoscalar meson equations were reduced to single equations
for $\chi_{J0}^{(0)}$ by neglecting all contributions on the right-hand-side
of the equations for $\chi_{_{J0}}^{(1)}$, $\chi_{_{J0}}^{(2)}$ and
$\chi_{J\,i\geq 1}^{(0)}$ other than the one involving $\chi_{J0}^{(0)}$,
thus breaking the self-consistency of the tower of equations that couple the
Tschebyshev moments.  In the scalar meson case, however, the contribution due
to $\chi_{_{S2}}^{(0)}$ was found to be important.  Consequently the
procedure used in the vector and pseudoscalar case was adapted in order to
obtain a pair of coupled equations for $\chi_{_{S0}}^{(0)}$ and
$\chi_{_{S2}}^{(0)}$ which were solved self consistently.

In arriving at the final equations to be studied Jain and Munczek (1993) also
employed the approximation of expanding the mass functions, $m_a$ and $m_b$,
and the functions $A$ and $B$ in a Taylor series around $q^2+\alpha^2p^2$ or
$q^2+\beta^2p^2$, depending on their argument, retaining up to first
derivative terms whose contribution to the calculated meson masses was less
than 10-15\% .  Where necessary the solutions for $A$ and $B$, obtained on
the spacelike real axis from the quark DSE, were extrapolated to timelike
values of their arguments.  Electromagnetic effects were incorporated by
including the one photon exchange kernel in the BSE and, in order ensure that
Goldstone's theorem remained valid, in the quark DSE.  Isospin mass
splittings were included by using different masses for the $u$- and $d$-
quarks.  The projected BSEs were then solved numerically for bound-state
mass-eigenvalues and wave functions.  The total mass difference between
states of the type $q\bar u$ and $q\bar d$, where $q$ can be any one of the
quarks, was obtained as a sum of the isospin and electromagnetic mass
splittings except in the case of the pion for which it was assumed that the
pure isospin contribution to the mass difference between $\pi^0$ and $\pi^+$
is negligible.

\begin{table}[tbh]
\begin{center}\parbox{130mm}{\caption{Calculated pseudoscalar and vector
meson masses and pseudoscalar decay constants obtained by Jain and Munczek
(1993) using renormalisation group invariant quark masses of $\hat{m}_u
= 8.73\;{\rm MeV}\;=\hat{m}_d$, $\hat{m}_s = 203$~MeV; ``constituent-quark''
masses of $M_c = 1.54$~GeV and $M_b=4.74$~GeV, defined as the value of the
mass function at $q^2=0$, and the values of $a$ and $\mu$ given after
Eq.~(\protect\ref{JMDIR}). These tables were adapted from Jain and Munczek
(1993). }}
\begin{tabular}{|c|c|c|c|c|c|c|c|c|c|c|}
\hline
meson  & $\pi$ & $K^+$ & $s\bar s$ & $\eta_c$ &
$ D_s$ & $D_0$ & $\eta_B$ & $B_c$ &
 $B_s$  & $B_+$ \\
\hline
mass (MeV) & 135 & 494 &  & 2979 & 1969 & 1865 & & & & 5279\\
(exp.)& & & & & & & & & &\\
\hline
mass (MeV) & 135 & 494 & 703 & 2821 & 1872 & 1756
& 9322 & 6126 & 5249 & 5149\\
\hline
$f_M$ (MeV) & 93 & 114 & & 213 & 148 & 125 & 287 & 207 & 119 & 119\\
\hline
$\alpha$ & 0.5 & 0.35 & 0.5 & 0.5 & 0.25 & 0.2 & 0.5 & 0.27 & 0.10 & .09\\
\hline
\hline
meson  & $\rho$ & $K^{*+}$ & $\phi$ & J/$\psi$ & $D_s^*$ & $D_0^*$ &
$\Upsilon $ & $B_c^*$ & $B_s^*$ & $B_+^*$ \\
\hline
mass (MeV) & 770 & 892 & 1019 & 3097 & 2110 & 2007 & 9460 &  & &5325\\
(exp.)& & & & & & & & & &\\
\hline
mass (MeV) & 760 & 976 & 1184 & 3100 & 2187 &  1997 & 9460 & 6277 &
5415 & 5290\\
\hline
$\alpha$ & 0.5 & 0.65 & 0.5 & 0.5 & 0.3 & 0.3 & 0.5 & 0.2 & .1 & 0.07\\
\hline
\end{tabular}
\label{JMtable} 
\end{center}
\end{table}

There are a number of important conclusions: 1) The contribution to a given
meson mass due to $\chi_{J\,i \geq 1}$ is negligible in all cases as was the
contribution of all Tschebyshev moments of order greater than zero.  The net
contribution being less than $10$\%; 2) There was some dependence of the
meson mass on $\alpha$, the relative-momentum distribution parameter, but in
all cases it was less than $15$\%.  In quoting results the authors settled on
a value of $\alpha$ for which the response of the mass to changes in $\alpha$
was minimised; 3) The hyperfine splitting was in general too large but the
splitting between the ground state and radial excitations was in good
agreement with experiment.  Some of these observations are apparent from
Table~\ref{JMtable}.

As remarked by Jain and Munczek (1993), a number of improvements of this
study are possible.  The fact that the splitting between vector and
pseudoscalar mesons is larger than experimentally observed may indicate the
need for an effective scalar interaction in the kernel of the BSE.  This must
be introduced, however, in such a manner as to ensure the preservation of
Goldstone's theorem and hence will entail a modification of the quark DSE.  A
natural way to proceed in this direction is to go beyond ladder-rainbow
approximation, which requires a dressed vertex in the quark DSE.

Simple improvements in methodology are also possible, for example, one might
avoid the Tschebyshev expansion, Taylor expansion and extrapolation
procedures.  In a study of the pion and scalar-diquark BSEs using a quark
propagator based on the DSE studies of Burden \etal~(1992b), which are
summarised in Sec.~\ref{subsect-quark-conf}, and subject to the
approximations
\beqn
\chi_{_P}\approx \gamma_5\, \chi_{_{P0}}\; &\;\; {\rm and}\;\; &\;
\chi_{0^+}\approx \, \chi_{0^+ 0}~,
\eeqn
Stainsby and Cahill (1993) solved the integral equations directly by matrix
inversion on a discrete grid.  The fact that the arguments of the quark
propagators in the kernel of the BSEs became complex in Euclidean space,
since $P^2 = -M^2$ at the solution, did not present a problem because the
quark propagators were regular on the complete integration domain as a
consequence of the quark confinement mechanism.

This approach can also be applied to baryons using the relativistic Faddeev
equations (Cahill \etal, 1989; Burden \etal, 1989; Cahill, 1992).  Such
studies are facilitated by, but not founded upon, the observation that the
scalar-diquark, $(qq)_{0^+}$, acts effectively as a resonance in the two-body
subchannel in which case the Faddeev equation can be written in the form
\beqn
\label{RFE}
\Psi(p;P) & = &\frac{1}{f^2} \int\dqbf\,
\Gamma_{0^+}(p + \mbox{\small $\frac{1}{2}$}q + 
        \mbox{\small $\frac{2 - 3\alpha}{2}$}P)
\Gamma_{0^+}(q + \mbox{\small $\frac{1}{2}$}p + 
        \mbox{\small $\frac{2 - 3\alpha}{2}$}P)\\
& &  \times 
S([2\alpha-1]P - p - q) \, S([1-\alpha]P + q) \, D_{0^+}(\alpha P -q)\,
\Psi(q;P) \nonumber
\eeqn
where $f$ is the normalisation, $\Gamma_{0^+}$ is the scalar-diquark
Bethe-Salpeter amplitude, which, at the same level of approximation, is
obtained as the solution of (Cahill \etal, 1987)
\beq
\Gamma_{0^+}(p;P) = \int\dqbf\,
\frac{4}{3}D_{\mu\nu}(p-q)\,\gamma_\mu S(q-\hlfsm P)\Gamma_{0^+}(p;P)
S(q+\hlfsm P)\gamma_\nu~,
\eeq
$S$ is the dressed quark propagator, obtained as the solution its DSE with a
model dressed gluon propagator, and $D_{0^+}$ is the scalar-diquark
propagator, which can be extracted from the $(qq)_{0^+}\rightarrow
(qq)_{0^+}$ scattering matrix.  The first study of \Eq{RFE} (Burden \etal,
1989) used parametrisations of the quark and scalar-diquark propagators based
on DSE studies.  These lead to a quark-core contribution to the baryon mass
of $\sim 1.2$~GeV.  Various studies of the effect of the pion-cloud suggest
that this will lead to a $(-200)\,\sim\,(-300)$~MeV contribution and hence this
is a good result.  However, there was a flaw in this study: the
parametrisations of the quark and scalar-diquark propagators introduced
unphysical poles into the integration domain, poles which would not be
present in the kernel of \Eq{RFE} if confinement is taken into account.
Current studies (Cahill, 1993) are addressing this problem.

\subsect{Effective Actions and QCD}
\label{subsect-eff-actions}
Coupled BSE-DSE systems such as the ones described above arise naturally in
the effective action approach to field theories.  Indeed, the coupled DSE-BSE
approach and the effective action approach are equivalent; a given effective
action serves to specify a certain order of resummation of the Feynman
diagrams that contribute to a given DSE.  The application of effective
actions and their interplay with the DSE tower has been discussed by McKay
and Munczek (1989) and reviewed by Haymaker (1991).  There are two commonly
used effective actions.

\subsubsect{CJT Effective Action}
The CJT effective action (Cornwall \etal, 1974) which can be written in
Euclidean space as:
\beq
\label{CJTEA}
\Gamma^{\rm CJT}[S] = {\rm TrLn}\left[ S_0^{-1} S \right] 
        + {\rm Tr}\left[ 1 - S_0^{-1}\,S \right] 
                + V_2[S]
\eeq
where $S$ is the dressed fermion propagator, $S_0$ is the free fermion
propagator and $V_2[S]$ is the sum of all two-particle, irreducible vacuum
diagrams (Stam, 1985):
\beqn
V_2[S]  = 
\mbox{\small $\frac{1}{2}$}\,{\rm Tr}\left[D_{\mu\nu}\,\Pi_{\mu\nu}\right]
& = &        \,- \mbox{\small $\frac{1}{2}$} g^2\,
        \int\frac{d^4k}{(2\pi)^4}\int\frac{d^4q}{(2\pi)^4}\,
D_{\mu\nu}(k-q)\,{\rm tr}
\left[ \gamma_\mu\,S(q)\,\Gamma_\nu(q,k)\,S(k)\right]
\eeqn
where $D_{\mu\nu}$ is the dressed gluon propagator and $\Gamma_\nu$ is the
dressed quark-gluon vertex.  From \Eq{new_Sigmaprime_QCD}, one recognises
this as
\beqn
& = & \,- \mbox{\small $\frac{1}{2}$} \int\frac{d^4k}{(2\pi)^4}
        \,{\rm tr}\left[\Sigma(k)\,S(k)\right]~.
\label{VTC}
\eeqn
It is therefore obvious that the variational equation
\beqn
\frac{\delta}{\delta S} \Gamma^{\rm CJT}[S] = 0 \;
& \;\; {\rm leads~to} \;\; & 
S^{-1} = S_0^{-1} + \Sigma
\label{VTD}
\eeqn
which is the quark DSE.

One can make use of \Eq{VTD} to find that, evaluated at the stationary point,
\beq
\Gamma^{\rm CJT}[S] = {\rm TrLn}\left[ S_0^{-1} S \right] 
+ \mbox{\small $\frac{1}{2}$}{\rm Tr}\left[ 1 - S_0^{-1}\,S \,\right]~,
\eeq
which was used above in \Eq{cjteff}. 

\subsubsect{Auxiliary Field Effective Action}
The ``auxiliary field'' effective action, $\Gamma^{\rm AF}[\Sigma]$, is best
introduced by way of an example since it cannot be written in a closed form.
Consider a field theory defined by the generating functional
\beq
\label{ZAF}
{\cal Z }[J] = \frac{1}{\cal N}\int {\cal D}\overline{q}\,{\cal D}q\,
\exp\left(- {\cal A}[\overline{q},q] \, + \,
\int d^4x d^4y\,  J_{\alpha\beta}(x,y) q_\alpha(x)\overline{q}_\beta(y)
\right)
\eeq
with an action
\beq
{\cal A}[\overline{q},q] = 
\int d^4x d^4y \left[
        \overline{q}_\alpha(x)S_0^{-1}(x-y)q_\alpha(y)
  -\hlfsm  q_\alpha(y)\overline{q}_\beta(x)\, 
{\cal M}_{\beta\gamma}^{\delta\alpha}(x,y)
          \, q_\gamma(x)\overline{q}_\delta(y)\right]~,
\eeq
where 
\mbox{${\cal M}^{\delta\alpha}_{\beta\gamma}(x,y)= 
        {\cal M}_{\delta\alpha}^{\beta\gamma}(y,x)$}, and normalised such
that ${\cal Z}[0] = 1$.  Equation~(\ref{ZAF}) can be rewritten using the
Hubbard-Stratonovich transformation to introduce hermitian, bilocal auxiliary
fields, $\beta_{\alpha\beta}(x,y)$.  One multiplies ${\cal Z}$ by
\beqn
\label{IHS}
\lefteqn{\int \prod_{\alpha\beta}{\cal D}\beta_{\alpha\beta}\, 
{\rm e}^{-I_{\rm HS}[\beta] 
- \int d^4x d^4y\,J_{\alpha\beta}(x,y)\beta_{\alpha\beta}(x,y)} = }\\
& & \int \prod_{\alpha\beta}{\cal D}\beta_{\alpha\beta} \,
\exp\left(-\int d^4x d^4y\,\left[
\beta_{\alpha\beta}(y,x)\,\hlfsm 
\,{\cal M}_{\beta\gamma}^{\delta\alpha}(x,y)\,
\beta_{\gamma\delta}(x,y) + J_{\alpha\beta}(x,y)\beta_{\alpha\beta}(x,y)
\right]\right) \nonumber
\eeqn
and renormalises; interchanges the order of integration and performs a change
of functional-integration variables:
\beq
\beta_{\alpha\beta}(x,y) \rightarrow \beta_{\alpha\beta}(x,y)
                + q_\beta(x)\overline{q}_\alpha(y) 
\eeq
which yields a functional integral over the quark fields that is quadratic
and hence can be evaluated:
\beq
\label{GFGCM}
{\cal Z }[J] = \frac{1}{{\cal N}'}
\int \prod_{\alpha\beta}{\cal D}\beta_{\alpha\beta}\,
\exp \left[-\overline{\cal A}[\beta] 
- \int d^4x d^4y\,  J_{\alpha\beta}(x,y)\beta_{\alpha\beta}(x,y)
\right]
\eeq
where
\beq
\label{Abar}
\overline{\cal A}[\beta] = 
-{\rm Tr Ln}\left[ {\cal G}^{-1}_{\alpha\beta}(x,y) \right] + I_{\rm HS}[\beta]
\eeq
with 
\beq
{\cal G}^{-1}_{\alpha\beta}(x,y) \equiv 
S_0^{-1}(x-y)\delta_{\alpha\beta} + \Sigma_{\alpha\beta}(x,y) = 
S_0^{-1}(x-y)\delta_{\alpha\beta} - 
{\cal M}^{\delta\beta}_{\alpha\gamma}(x,y)\,\beta_{\gamma\delta}(x,y)~.
\eeq
The auxiliary field effective action, $\Gamma^{\rm AF}[\beta]$, is obtained
from ${\cal Z}$ via a Legendre transformation
\beq
\Gamma^{\rm AF}[\beta] 
+ \int d^4x d^4y J_{\alpha\beta}(x,y)\beta_{\alpha\beta}(x,y) = 
W^{\rm AF}[J] \equiv \,-\,\ln\left[ {\cal Z}[J]\right]~,
\eeq
with
\beqn
\frac{\delta W^{\rm AF}[J]}{\delta J_{\alpha\beta}(x,y)} 
= -\beta_{\alpha\beta}(x,y) \; & \;\; {\rm and } \;\; & \;
\frac{\delta \Gamma^{\rm AF}[\beta]}{\delta \beta_{\alpha\beta}(x,y)} 
= \,J_{\alpha\beta}(x,y)~,
\eeqn
from which it is clear that, at tree level, 
\beq
\Gamma_{\rm AF}[\beta] \approx \overline{\cal A}[\beta]~.
\eeq

The stationary point equation
\beqn
\label{CFE}
\frac{\delta\overline{\cal A}[\beta]}{\delta \beta_{\alpha\beta}(x,y)} = 0
 \; & \;\; {\rm yields} \;\; & \;
\Sigma_{\alpha\beta}(x,y) = 
{\cal M}_{\alpha\gamma}^{\delta\beta}(x,y)\,{\cal G}_{\gamma\delta}(x,y)\,
\eeqn
which is simply the Hartree approximation to the DSE for the fermion self
energy in this theory.  [As an example one might consider the cases of \qedt
and \qedf, which are obtained with
\beq
{\cal M}_{\beta\gamma}^{\delta\alpha}(x,y) = 
 e^2 (\gamma_\mu)_{\beta\gamma} 
(\gamma_\nu)_{\delta\alpha} D_{\mu\nu}(x-y)~,
\eeq
and serve to illustrate this point.]  It follows that in Hartree
approximation $\Gamma^{\rm CJT}$ and $\Gamma^{\rm AF}_{\rm tree-level}$ have
the same stationary point equation and solutions.  Indeed, when evaluated at
the stationary point, $\Gamma^{\rm AF}_{\rm tree-level}$=$\Gamma^{\rm CJT}$.
However, $\Gamma^{\rm AF}_{\rm tree-level}$ is bounded below whereas
$\Gamma^{\rm CJT}$ is not (Haymaker, 1991) and so, for many applications,
$\Gamma^{\rm AF}_{\rm tree-level}$ is more useful.

\subsubsect{QCD Phenomenology using the Auxiliary Field Effective Action}
The generating functional in equation (\ref{ZAF}) is directly applicable to
\qedt and \qedf since the gauge boson field can be integrated out to yield a
quartic  four-fermion interaction [see, for example, Roberts and Cahill
(1986)].  Here, however, we are interested in its application to QCD.  A
model field theory that has received a good deal of attention is defined by
the action:
\begin{eqnarray}
\label{GCMA}
{\cal A}[\overline{q},q] & = & \int\,d^4x\, 
\overline{q}(x)[\gamma\cdot\partial +M]q(x) 
+ \hlfsm\int d^4x\, d^4y\, j_{\mu}^a(x)\,g^2D_{\mu\nu}(x-y)j_{\nu}^a(y)
\end{eqnarray}
where $M$ is the quark mass matrix, \mbox{$j_{\mu}^a(x) =
\overline{q}(x)\frac{\lambda^a}{2}\gamma_\mu q(x)$} and 
\begin{equation}
g^2 D_{\mu\nu}(x-y) = \int\,\frac{d^4k}{(2\pi)^4} e^{ik\cdot x}
\left( \left[\delta_{\mu\nu} - \frac{k_\mu k_\nu}{k^2}\right]D(k^2)
        +\xi \frac{k_\mu k_\nu}{k^4} \right)
\label{eqD}
\end{equation}
represents a model quark-quark interaction, realistic constraints on which
are discussed in Sec.~\ref{sect-QCD-gluon}.  [We note that a form of the
Nambu--Jona-Lasinio model can be obtained by writing $D(k^2) = $~constant but
this does not satisfy the constraints discussed in
Sec.~\ref{sect-QCD-gluon}.]  With this current-current interaction one
identifies
\beq
\label{MGCM}
{\cal M}_{\beta\gamma}^{\delta\alpha}(x,y) = g^2 D_{\mu\nu}(x-y)\,
I_{\beta_F\gamma_F} \, (\gamma_\mu)_{\beta_D\gamma_D} \,
\left(\frac{\lambda^a}{2}\right)_{\beta_C\gamma_C} \,
I_{\delta_F\alpha_F} \, (\gamma_\nu)_{\delta_D\alpha_D} \,
\left(\frac{\lambda^a}{2}\right)_{\delta_C\alpha_C} \,
\eeq
where $D$, $C$, $F$ refer to Dirac-, colour- and flavour-space, respectively.
Although such a model field theory does not include explicit three- or
four-gluon vertices, non-Abelian effects can be incorporated through the
structure of $D(k^2)$, as discussed in Sec.~\ref{sect-QCD-gluon}.  Many
independent studies can be correlated within this framework; for example,
those of McKay and Munczek (1985); Cahill and Roberts (1985); Praschifka
\etal~(1987a); McKay \etal~(1988); Barducci \etal~(1988); Haymaker (1991) and
references therein.  The nature of the truncations and approximations
necessary to obtain \Eq{GCMA} from QCD is discussed by Cahill and Roberts
(1985) and McKay \etal~(1988).

As mentioned above, there is a large body of work in this area and we choose
to summarise it by way of a discussion of a series of studies using a form of
\Eq{GCMA} that has come to be called the Global Colour-symmetry Model [GCM]
(Roberts and Cahill, 1985) and which use the following form for the model
dressed gluon propagator:
\beq
\label{GCMprop}
D(k^2) = \frac{\alpha_1(k^2)}{k^2}
\eeq
with $\alpha_1$ given in \Eq{GCMalpha}.  This form of the model has been used
successfully to calculate the kinematic and dynamical properties of mesons
and has also been applied to baryons.  One ingredient of its success is that
it incorporates confinement in the sense of ensuring the absence of quark
production thresholds in any Feynman diagram associated with a physical
process, a criterion that is discussed in Sec.~\ref{subsect-quark-conf}.  From
the point of view of a connection between the GCM and QCD, Landau gauge, $\xi
= 0$, is the best choice, as we have seen in the DSE studies described above.
However, in developing an efficacious phenomenology, a Feynman-like gauge
with
\mbox{$D_{\mu\nu}(k) = \delta_{\mu\nu}\alpha_1(k^2)/k^2$} can be used without
penalty. 

The generating functional for the GCM is given by \Eqs{ZAF} and (\ref{GCMA})
from which the auxiliary-field effective action, in the form of \Eq{GFGCM},
follows.  In making contact with low energy phenomenology in QCD and the
tower of DSEs it is useful to express the generating functional in terms of a
local functional integral.  This can be achieved (Roberts and Cahill, 1987)
by writing
\beqn
\beta_{\alpha\beta}(x,y) & \equiv &
B_{\alpha\beta}\left(w=\hlfsm(x+y),z=x-y\right) \nonumber \\
& = & \tilde{\Theta}_{\alpha\beta}(z)+ 
\sum_{k}\,\left[\Phi_{\alpha\beta}^{k}\ast\Theta_{\alpha\beta}^{k}\right]
        (w,z)
\equiv  \tilde{\Theta}_{\alpha\beta}(z)+ 
\sum_{k}\,\int d^4u\,\Phi_{\alpha\beta}^{k}(w-u)
                \, \Theta_{\alpha\beta}^{k}(u,z)~,
\label{LFBoson}
\eeqn
where \mbox{$\{\Theta_{\alpha\beta}^{k}(w,z); k=0,1,\dots\}$} is a complete
set of orthogonal functions, to be specified later, and
$\{\tilde{\Theta}_{\alpha\beta}(z)\}$ are those solutions of \Eq{CFE} which
actually minimise the action: the translationally invariant ``vacuum field
configuration''.  Using \Eq{LFBoson}, the generating functional becomes
\beq
\label{ZGCML}
{\cal Z }[J^\Phi] = \frac{1}{{\cal N}''}
\int \prod_k \prod_{\alpha\beta}{\cal D}\Phi_{\alpha\beta}^{k}\,
\exp \left[-\overline{\cal A}[\Phi]
- \int d^4w \, \sum_k\, J_{\alpha\beta}^{\Phi(w)k}\Phi_{\alpha\beta}^k(w)
\right]
\eeq
where $\overline{\cal A}[\Phi]$ follows from direct substitution of
\Eq{LFBoson} in \Eq{Abar}.  It is clear that since the solutions of \Eq{CFE}
have been isolated explicitly the lowest order contribution from
$\overline{\cal A}[\Phi]$, in an expansion about $\Phi=0$, is of O$(\Phi^2)$.
The stationary point equation in the GCM is obviously the rainbow-ladder
approximation to the quark DSE.  Clearly then, of all the combinations
$(\alpha\beta)$, the only nonzero $\tilde{\Theta}_{\alpha\beta}$ are the
isoscalar, colour-singlet, Dirac-vector and Dirac-scalar combinations, which
are related to the Fourier transforms of the functions $A(p^2)$ and $B(p^2)$,
much discussed above.  It follows that, in any expansion of $\overline{\cal
A}[\Phi]$ about $\Phi=0$, the dressed quark propagator will be an important
element.

In \Eq{ZGCML} particular $(\alpha\beta)$-combinations correspond to local
fields with the quantum numbers of $\pi$-mesons, $\rho$-meson, etc., and
whose internal structure is described by the associated
$\Theta_{\alpha\beta}$-functions.  This is most easily demonstrated by way of
an example.  Considering an expansion in the $(\alpha\beta)$-combination that
corresponds to the isovector, colour-singlet, Dirac-pseudoscalar channel;
i.e., the $\pi$-meson, one obtains
\beq
\label{ATrLnGCM}
\overline{\cal A}[\vec{\pi}] = 
-\,{\rm Tr Ln}\left[S^{-1}(x-y) 
        - g^2\,D_{\mu\nu}(x-y)\gamma_\mu \frac{\lambda^a}{2}\,
\left[\pi_i^k\ast \Theta_i^k\right]\left(\hlfsm(x+y),x-y\right)
\gamma_\nu \frac{\lambda^a}{2}\right] + I_{\rm HS}[\vec{\pi}]
\eeq
where $S(x-y)$ is the dressed quark propagator and, with $i,j$ isospin labels,
\beqn
I_{\rm HS}[\vec{\pi}] & = &
 \int\frac{d^4 P}{(2\pi)^4} \,\pi_i^k(P) \, \hlfsm \,
        _{\vec{\pi}}{\cal I}_{ij}^{kl}[\Theta](P)\,\pi_j^l(-P)
\eeqn
with
\beqn
_{\vec{\pi}}{\cal I}_{ij}^{kl}[\Theta](P)& = & \int \frac{d^4 q_1}{(2\pi)^4}
\,\frac{d^4 q_2}{(2\pi)^4} \,
 g^2\,D_{\mu\nu}(q_1-q_2)\,{\rm tr}\left[
\Theta_i^k(P,q_1) \,\gamma_\mu \frac{\lambda^a}{2}
\Theta_j^l(-P,q_2) \,\gamma_\nu \frac{\lambda^a}{2}\,\right]~,
\eeqn
where $P$ is the centre-of-mass momentum, which follows from Eqs.~(\ref{IHS})
and (\ref{MGCM}).

At O$(\pi^2)$, neglecting the constant vacuum energy, one obtains
\beq
\label{PiInvProp}
\overline{\cal A}[\vec{\pi}] = 
 \int\frac{d^4 P}{(2\pi)^4} \
        \pi_i^k(P)\, \hlfsm
        \,_{\vec{\pi}}\Delta_{ij}^{kl}[\Theta](P)\, \pi_j^l(-P) \,
\eeq
where 
\beq
_{\vec{\pi}}\Delta_{ ij}^{kl}[\Theta](P) =\,
 _{\vec{\pi}}{\cal T}_{ ij}^{kl}[\Theta](P)
        +\, _{\vec{\pi}}{\cal I}_{ ij}^{kl}[\Theta](P)
\eeq
with
\beqn
_{\vec{\pi}}{\cal T}_{ ij}^{kl}[\Theta](P) & = & 
\int \frac{d^4 k}{(2\pi)^4}\,\frac{d^4 q}{(2\pi)^4}\,\frac{d^4 r}{(2\pi)^4}\,
g^2 D_{\mu\nu}(q)\,g^2 D_{\rho\sigma}(r)\\
& & \times
{\rm tr}\left[S(k+\hlfsm P)\,\gamma_\mu\frac{\lambda^a}{2}
\Theta_i^k(P,k-q)\,\gamma_\nu\frac{\lambda^a}{2}\,
S(k-\hlfsm P)\,\gamma_\rho\frac{\lambda^a}{2}\,
\Theta_j^l(-P,k-r)\,\gamma_\sigma\frac{\lambda^a}{2}\,\right]~. \nonumber
\eeqn
{}From \Eq{PiInvProp} it is natural to identify 
\mbox{$_{\vec{\pi}}\Delta_{ ij}^{kl}[\Theta](P)$} as the inverse of the
matrix propagator for the $\vec{\pi}^k$-field and it is clear that, with the
functions $\tilde{\Theta}_{\alpha\beta}$ chosen such the action is minimised,
\mbox{$_{\vec{\pi}}\Delta_{ ij}^{kl}[\Theta](P)$} is non-negative for all
$P^2 > 0$.

Until now the functions \mbox{$\{\Theta^k(w,z), k=0,1,\ldots\}$} have not been
specified.  If one chooses them such that
\beqn
\label{orth}
_{\vec{\pi}}\Delta_{ij}^{kl}[\Theta](P)\; =0~, \;\; k\neq l
\; & \;\; {\rm and } \;\; &
\frac{\delta\; _{\vec{\pi}}\Delta_{ ij}^{kk}[\Theta](P)}{\delta \Theta^k(-P,q)} 
         = 0~,
\eeqn
which entails that the convolution
\beq
\Gamma_i^k(P,p) = \int\dqbf\, g^2\,D_{\mu\nu}(p-q)\,
        \gamma_\mu\frac{\lambda^a}{2}\,
        \Theta_i^k(P,q)
        \gamma_\nu\frac{\lambda^a}{2}\,
\eeq
satisfies 
\beq
\label{BSEGCM}
\Gamma_i^k(P,q) + 
\int\dqbf\,g^2\,D_{\mu\nu}(p-q)\,
        \gamma_\mu\frac{\lambda^a}{2}\,
        S(q+\hlfsm P) \Gamma_i^k(P,q) S(q-\hlfsm P)\,
        \gamma_\nu\frac{\lambda^a}{2}\, = 0~,
\eeq
which is the ladder BSE in the GCM, then the local field variables introduced
via \Eq{LFBoson} represent meson fields whose internal structure is described
by a ladder-BSE, the kernel of which involves the model dressed gluon
propagator and the dressed quark propagator obtained with this gluon
propagator via the quark-DSE.  The amplitudes $\Gamma^k$ can be normalised in
the standard fashion (Itzykson and Zuber, 1980, pp. 482-485).  With this
interpretation of \Eq{LFBoson} then the superscript $k$ labels excited states
in a given channel so, in the present example: $k=0$ represents the
$\pi$-meson; $k=1$, its first excited state; etc.

This discussion demonstrates that the tree-level effective action for the GCM
describes non-pointlike meson-fields whose internal structure is given by the
solution of the coupled rainbow-DSE--ladder-BSE system; i.e., it is the
``quark-core'' of the hadron.  The connection between the auxiliary-field
effective action approach and the coupled DSE-BSE approach discussed in
Sec.~\ref{subsect-DSE-BSE} is therefore clear.  The higher order terms in the
tree-level effective action, O$(\pi^4)$ etc., describe couplings between
these quark-core pions and, although only the pion has been discussed
explicitly, the approach can be generalised to incorporate other mesons
(Praschifka \etal, 1987a) and baryons (Cahill \etal, 1989).

In order to study low energy properties of QCD it is common to employ a
derivative expansion of tree-level effective action, \Eq{ATrLnGCM}.  In the
GCM one obtains a complete effective action, a small portion of the real part
of which we illustrate here:
\beqn
&& \int d^4 x\,\left\{
 \hlfsm\left( \partial_\mu \vec{\pi} \cdot \partial_\mu \vec{\pi} 
        + m_\pi^2 \vec{\pi}\cdot \vec{\pi}\right) + 
 \hlfsm \left( \partial_\mu \vec{\rho}_\nu \cdot \partial_\mu \vec{\rho}_\nu
        - \partial_\mu \vec{\rho}_\mu \cdot \partial_\nu \vec{\rho}_\nu 
        + m_\rho^2 \vec{\rho}_\mu\cdot \vec{\rho}_\mu\right) \right.
\nonumber \\
&& \left. 
        - g_{\rho\pi\pi} \vec{\rho}_\mu\cdot 
                \vec{\pi}\times\partial_\mu\vec{\pi}
        + (\kappa, \eta, \omega, a_1, \ldots )
        + {\rm tr}\left[\overline{N}\left(\gamma\cdot\partial + m_N 
                - \mbox{\small $\frac{i}{\surd{2}}$}
        m_N\gamma_5 \vec{\tau}\cdot\vec{\pi}\right)N
        + \ldots \right]\right\}~.
\eeqn
Although only shown to second order in $\pi$, the constraints placed by
chiral symmetry on the nature of pion self-interactions (Roberts \etal, 1988;
Roberts \etal, 1994) and on interactions between the pion and other hadrons
(Roberts \etal, 1989) are properly represented.  The full action also has an
imaginary part which contains the Wess-Zumino term, as any realistic model of
QCD must (Praschifka \etal, 1987a).  In following this procedure one obtains
the bound state masses by solving the BSE [or Faddeev equation], as discussed
above, and all of the couplings between quark-core mesons and baryons as
integrals whose integrand contains the dressed quark propagators and
Bethe-Salpeter amplitudes.

For example, if one retains only the dominant functions in the Bethe-Salpeter
amplitude for the $\pi$- and $\rho$-mesons, as discussed in
Sec.~\ref{subsect-DSE-BSE}: $\Gamma_\rho$ the dominant vector-amplitude and
$\Gamma_\pi=B$ the dominant pseudoscalar-amplitude; then, writing the dressed
quark propagator as 
\mbox{$S(p) = -i \gamma\cdot p \sigma_V(p) + \sigma_S(p)$}, one has 
\begin{eqnarray}
\label{Fpi}
f_{\pi}^2  & = & \!\!
\frac{N_c}{8\pi^2}\int_{0}^\infty\,dx\,x\,B^2\,
\left(  \sigma_{V}^2 - 
2 \left[\sigma_S\sigma_S' + x \sigma_{V}\sigma_{V}'\right]
- x \left[\sigma_S\sigma_S''- \left(\sigma_S'\right)^2\right]
- x^2 \left[\sigma_V\sigma_V''- \left(\sigma_V'\right)^2\right]\right), 
\eeqn
which is obtained from the normalisation condition for the pion's
Bethe-Salpeter amplitude in ladder approximation (Alkofer \etal, 1993): 
\begin{eqnarray}
\label{LAN}
\lefteqn{2 f_\pi^2 P_\mu = }\\
& & 
N_c\,\int\,\frac{d^4k}{(2\pi)^4}\,{\rm tr}_D \left[ 
\overline{\Gamma}_\pi(k;P) S(k_{0-}) \Gamma_\pi(k;-P) 
        \frac{\partial\,S(k_{0+})}{\partial P_\mu}
+\overline{\Gamma}_\pi(k;P) \frac{\partial\,S(k_{0-}) }{\partial P_\mu}
        \Gamma_\pi(k;-P) S(k_{0+})
\right]\nonumber
\end{eqnarray}
with $k_{\alpha\beta} = k + \hlfsm \alpha\, p + \hlfsm \beta\, q$,  and
\beqn
\langle\overline{q}q\rangle_{\mu^2} 
& = & \frac{N_c}{4\pi^2}\int_{0}^{\mu^2}\,dx\,x\,\sigma_S
\end{eqnarray}
with the result that
\mbox{$m_\pi^2 f_\pi^2 = 
(m^u_{\mu^2} + m^d_{\mu^2})\,\langle\overline{q}q\rangle_{\mu^2} $}.  [Recall
that this expression for $f_\pi$ differs from that in \Eq{f_pi2} and that the
difference is due to neglecting contributions to the RHS of \Eq{f_pi} from
$\Gamma_5(p',p)$, $p'\neq p$.  When $Z(p^2)=1$, both forms are the same.]
One also finds (Praschifka \etal, 1987b)
\beq
g_{\rho\pi\pi} = \frac{1}{f_\rho f_\pi^2}\frac{N_c}{2\pi^2}
        \int_0^\infty\,ds\,s\,\Gamma_\rho(s)\,A(s)\,B(s)\,
                \frac{B(s) - \hlfsm\,s\,B'(s)}
                {(s\,A(s)^2 + B(s)^2)^2} + \mbox{O}(m_\rho^2)~,
\eeq
with $f_\rho$ given by an expression similar to that in \Eq{Fpi} and obtained
in the same way.  [The fact that $\Gamma_\pi=B$ is a manifestation of
Goldstone's theorem in this approach, as it also is in the coupled DSE-BSE
approach (Delbourgo and Scadron, 1979).  This can be seen heuristically by
substituting $\Gamma_k \propto i\,\tau_k\,\gamma_5\,E$ in \Eq{BSEGCM} and
noting that the equation that results is solved by $E=B$.]  Because of the
structure of the model dressed gluon propagator, \Eq{GCMprop}, all of these
integrals are finite [except for the quark condensate, which diverges as
\mbox{$[\ln\Lambda_{\rm UV}^2/\Lambda_{\rm QCD}^2]^{d_{\rm M}}$}, as it must in
a model that is consistent with the ultraviolet constraints imposed by QCD].
The phenomenological success of the GCM is illustrated in
Table~\ref{GCMTable}. 

One can proceed beyond the tree-level effective action and calculate the
effect of meson-loops.  The introduction of the auxiliary fields, and their
subsequent identification with the quark-core; i.e., ladder bound states, via
the connection with the BSE, \Eq{BSEGCM}, provides a mechanical framework for
the systematic ordering and summation of diagrams.  As examples we note that
the model has been used successfully to calculate the $\omega$-$\rho$ mass
splitting induced by $\rho\rightarrow\pi\pi\rightarrow\rho$,
$\rho\rightarrow\omega\pi\rightarrow\rho$,
$\omega\rightarrow\rho\pi\rightarrow\omega$ and
$\omega\rightarrow\pi\pi\pi\rightarrow\omega$ self energy corrections
(Hollenberg \etal, 1992) and to analyse the contribution of pion loops to the
electromagnetic charge radius of the pion (Alkofer \etal, 1993), which is
found to be an additive correction to $r_\pi^2$ of $< 15$\% at $m_\pi =
0.14$~GeV.  Furthermore, recent calculations of the $\omega$-$\rho$ mixing
component of the $N$-$N$ potential (Goldman \etal, 1992; Krein \etal, 1993;
Mitchell \etal, 1994) also fit neatly within this framework.  The important
feature of these meson-loop contributions is that they too are finite because
the Bethe-Salpeter amplitudes that describe the quark core of the hadrons
appear in every integral and provide natural momentum cutoffs, thus ensuring
convergence.

\begin{table}[ht]
\label{GCMTable}
\begin{center}\parbox{130mm}{\caption{This is an illustrative set of
calculated physical quantities, obtained in the GCM using the propagator of
Eq.~(\protect\ref{GCMprop}).  Here $m^{\rm qq}$ are effective diquark-masses
and $a_J^I$ are scattering lengths in $\pi$-$\pi$ scattering.  The
superscripts indicate the reference that a particular result is taken from:
1) Roberts \etal~(1988); 2) Roberts \etal~(1994); 3) Praschifka
\etal~(1987a); 4) Praschifka \etal~(1989); 5) Hollenberg
\etal~(1992); 6) Cahill \etal~(1987); 7) Cahill (1992). }}
\begin{tabular}{|l|l|l|} \hline
          & Calculated & Experiment \\ 
          & Massless u,d     &    (where applicable/known)     \\ \hline
 m$_{\pi} \;^{1,2} $    &  ~0  &  (0 if quarks massless) \\ \hline
  f$_{\pi} \;^{2}$    &  ~0.091 GeV &   ~0.093 GeV            \\ \hline
  m$_{\omega} \;^{3,4}$  &  ~0.745 &   ~0.783         \\ \hline
  m$_{\omega}-\ $m$_{\rho} \;^5$ 
              &  ~0.030 &   ~0.013              \\ \hline
 $\Gamma_{\rho} \;^{3,4}$ 
              &  ~0.232 &   ~0.154              \\ \hline
  m$_{{\rm f}_{1}}\; ^{3,4}$ 
              &   ~1.310 &   ~1.283              \\ \hline
 m$_{{\rm f}_{1}}-\ $m$_{\omega}\; ^{3,4}$
              &  ~0.565 &   ~0.500               \\ \hline\hline
  m$^{{\rm qq}}_{0^{+}}\; ^{4,6}$
              &  ~0.607 &                       \\ \hline
  m$^{{\rm qq}}_{1^{+}}\; ^{4,6}$
              &  ~1.170  &                       \\ \hline
  m$^{{\rm qq}}_{0^{-}}\; ^{6}$
              &  ~0.948 &                       \\ \hline
  m$^{{\rm qq}}_{1^{-}}\; ^{6}$
              &  ~1.950  &                       \\ \hline
  m$_{N}\; ^7$  & ~1.20 $\sim$ 1.30 & ~0.939  \\ \hline\hline
 $r_\pi \; ^{1,2}$ & ~0.59 fm & ~0.66 fm  \\  \hline\hline
 $a_0^0 \; ^2 $ & ~0.17 (dimensionless) & ~0.20 (dimensionless) \\ \hline
 $a_0^2 \; ^2 $ & -0.048 & -0.037 \\ \hline
 $a_1^1 \; ^2 $ & ~0.030 & ~0.038 \\ \hline
 $a_2^0 \; ^2$  & ~0.0017 & ~0.014 \\ \hline
 $a_2^2 \; ^2 $ & -0.0011 & \\ \hline
\end{tabular}
\end{center}
\end{table}

This example illustrates the usefulness of the auxiliary-field effective
action [or, equivalently, the coupled rainbow-DSE--ladder-BSE] approach to
QCD phenomenology.  A few parameters [two or three in the studies described
in this section] in the model dressed gluon propagator, chosen in order to
parametrise the unknown behaviour of the quark-quark interaction in the
infrared, are all that is necessary in order to construct a model field
theory, with confinement and dynamical chiral symmetry breaking, which
provides a good description of low energy hadronic phenomena.  Furthermore,
the comparison of the calculated results with experimental data allows one to
place constraints on the nature of the quark-quark interaction at
small-$q^2$.  This phenomenological success provides considerable incentive
to resolve the questions regarding the formulation of the DSEs in QCD.

\sect{Summary, Conclusions and Outlook}
\label{sect-summary}
We began with an introduction to the Dyson-Schwinger equation [DSE] formalism
with specific reference to QED and to the renormalisation procedure in this
framework.  It was then relatively straightforward to extend this discussion
to QCD.  Following this we presented a survey of studies of strong-coupling
QED, both because it is an interesting theory in its own right and because it
is an ideal pedagogical tool for DSE studies and their application to QCD.
We reviewed the DSE studies of the infrared behaviour of the gluon propagator
in QCD and its role in phenomenological studies of dynamical chiral symmetry
breaking and confinement.  Finally, we collected all of this information
together and illustrated how it can be used in relativistic bound state
calculations using the BSE for mesons and the Faddeev equation for baryons.
This information also provides the foundation for the phenomenologically
successful effective-action approach to QCD.  We illustrated this approach
and demonstrated its equivalence, and importance, to the coupled
Bethe-Salpeter--Dyson-Schwinger equation studies.

The strengths of the DSE approach to nonperturbative field theory are its
manifest covariance and the natural way in which dynamical chiral symmetry
breaking can be described.  It is also straightforward to build in the
correct perturbative limits for QCD because of asymptotic freedom.  From the
various model studies it is clear that once the scale of one physical
parameter; for example, $f_\pi$, is fixed by adjusting the infrared behaviour
of the quark-quark interaction, then other quantities are automatically of
approximately the right magnitude.  The fine tuning relies on the details of
the models, with very successful descriptions of a broad range of phenomena
being possible with only two or three parameters.

While much has been accomplished, it is apparent that much more needs to be
done. It is important to develop a better understanding of the detailed way
in which confinement is manifested.  Since the DSE formalism is not closed at
any finite order it will always be important to draw on information gathered
from other nonperturbative studies.  For example, it would be very helpful if
we could learn more about the infrared behaviour of the gluon propagator and
the role of ghosts in QCD from lattice gauge theory studies.  However,
because of present day limitations on lattice studies, this information is
currently unavailable but it is an area where cross-fertilisation could
provide significant benefits.  The results of recent lattice simulations have
shown that such information may be available in the not too distant future.

An important question, in field theory in general, and in the DSE approach in
particular, is the connection between Euclidean-space and physical Minkowski
space.  As we demonstrated, a straightforward transcription of the variables
in the DSE will almost never be correct.  A direct formulation and solution
of the DSEs in Minkowski space is one possibility but this is presently
prevented by the complete lack of information about the $n$-point functions
on the timelike axis at all but very large momenta.  At the present time,
both lattice and DSE studies are carried out in Euclidean space and the
axioms of field theory used to extract values of physical observables.  It is
not yet clear when or if it will be possible to perform DSE studies directly
in Minkowski space.

\section*{\sc Acknowledgments}

We are very grateful to R.J. Crewther for several helpful discussions.
The work of CDR was supported by the U.S. Department of Energy, Nuclear Physics
Division, under contract number W-31-109-ENG-38.  The work of AGW was
supported by the Australian Research Council, the U.S. Department of Energy 
through Contract No. DE-FG05-86ER40273, and by the Florida State University
Supercomputer Computations Research Institute which is partially funded by
the Department of Energy through Contract No. DE-FC05-85ER250000.


\appendix
\sect{Landau-Khalatnikov Transformations}
\label{appendix_LK-Tran}
As we discussed in Sec.~\ref{psi-gamma-vertex}, on any Ansatz for the
fermion--gauge-boson vertex in Abelian gauge theories, one would like to
impose the constraint that, when used in a DSE, it ensures local gauge
covariance of the solutions.  The gauge transformation laws which relate the
propagators and fermion-gauge-boson vertex in QED to their Landau gauge
counterparts were first given by Landau and Khalatnikov (1956) and Fradkin
(1956).  These rules are most easily specified in coordinate space and we
give below the corresponding Euclidean space transformation laws.

In an arbitrary gauge, the photon propagator is modified from its 
transverse, Landau gauge, form $D_{\mu \nu}(x;0)$ by the addition of 
a longitudinal piece parameterised by an arbitrary function $\Delta$:
\begin{equation}
D_{\mu \nu}(x;\Delta) = D_{\mu \nu}(x;0) 
        + \partial_\mu \partial_\nu\Delta(x).   \label{LKD}
\end{equation}
The corresponding rule for the fermion propagator is
\begin{equation}
S(x;\Delta) = S(x;0) {\rm e}^{e^2[\Delta(0) - \Delta(x)]}, \label{LKF}
\end{equation}
where $e$ in the exponent is the gauge coupling constant. The rule 
for the fermion-photon vertex is
\begin{eqnarray}
B_\mu(z; x,y|\Delta) & = & B_\mu(z;x,y|0) {\rm e}^{e^2[\Delta(0) -
    \Delta(x-y)]} \nonumber \\ & + & S(x-y;0) {\rm e}^{e^2[\Delta(0) -
    \Delta(x-y)]} \frac{\partial}{\partial z_\mu} [\Delta(x-z) -
    \Delta(z-y)], \label{LKV}
\end{eqnarray}
where $B_\mu$ is the non-amputated vertex defined in momentum space in
terms of the amputated vertex, $\Gamma_\mu$, by
\begin{equation}
B_\mu(p,q) = S(p)\Gamma_\nu(p,q)S(q)D_{\mu \nu}(p-q).
\end{equation}
The transformation rule for the partially amputated vertex
\begin{equation}
\Lambda_\mu(p,q) = S(p)\Gamma_\mu(p,q)S(q)
\end{equation}
follows from Eqs.~(\ref{LKD}), (\ref{LKF}) and (\ref{LKV}) and is simply:
\begin{equation}
\Lambda_\mu(z;x,y|\Delta)  = 
    \Lambda_\mu(z;x,y|0) e^{e^2[\Delta(0) - \Delta(x-y)]}. \label{LKL}
\end{equation}

In the usual covariant gauge fixing procedure the photon propagator  
takes the form
\begin{equation}
D_{\mu \nu}(k;\xi) = \frac{1}{k^2[1+\Pi(k^2)]}
             \left(\delta_{\mu \nu} - \frac{k_\mu k_\nu}{k^2} \right)
        + \xi\frac{k_\mu k_\nu}{k^4},        \label{CD}
\end{equation}
which is obtained by taking $\Delta$ in Eq.~(\ref{LKD}) to be
\begin{equation}
\Delta(x) = -\xi \int \frac{d^{\rm d}k}{(2\pi)^{\rm d}} \frac{e^{-ik\cdot
x}}{k^4}. 
\end{equation}
Within this set of gauges, one finds that in QED$_3$ the transformation rule for
the fermion propagator, Eq.~(\ref{LKF}), becomes
\begin{equation}
S(x;\xi) = S(x;0) e^{-e^2\xi\left|x\right| / 8\pi}. \label{LKFX}
\end{equation}

The free massless propagator is \mbox{$S^{-1}(p;0) = i \gamma\cdot p$}
which corresponds to the following function in configuration space:
\begin{equation}
S(x;0) = \frac{\gamma\cdot x}{4\pi |x|^3}. \label{Fxfree}
\end{equation}
Applying \Eq{LKF} one obtains the LKF transformed function in an arbitrary
covariant gauge: 
\begin{equation}
S(x;\xi) = \frac{\gamma\cdot x }{4\pi |x|^3}
              e^{-e^2\xi\left|x\right| / 8\pi}. 
\end{equation}
For $\xi > 0$ one may evaluate the Fourier amplitude directly to obtain
\begin{equation}
S(p;\xi) = \frac{-i \gamma\cdot p}{p^2}
             \left[1 - \frac{e^2\xi}{8\pi p} 
             \arctan \left(\frac{8\pi p}{e^2\xi}\right) \right].
                          \label{ATAN}
\end{equation}

\sect{Quark Condensate}
\label{appendix_condensate}

We discuss here the meaning of the quark condensate $\langle\bar qq\rangle$.
It is convenient to perform a Wick rotation 
and evaluate $\langle\bar qq\rangle$ in Euclidean space.  Then using
\beq
\int d^4\ell f(\ell^2) = \pi^2 \int_0^\infty d\ell^2 \ell^2 f(\ell^2)
\eeq
we can write \Eq{q_condensate} as
\beq
\langle\bar qq\rangle = -{3\over 4\pi^2}
\int^{\Lambda^2_{\rm UV}}_0d\ell^2 {\ell^2Z(\ell^2)M(\ell^2)\over
\ell^2+M^2(\ell^2)}\;, 
\eeq
where we have chosen $\mu =\Lambda_{\rm UV}$ as in \Eq{Sigma_q_exact} and
\Eq{Sigma_q} etc.  Now, writing an explicit subscript $\Lambda_{\rm UV}$ where
appropriate, we have
\beq
\langle\bar qq\rangle_{\Lambda_{\rm UV}} = 
        -{3\over 4\pi^2}\int^{\Lambda_{\rm UV}^2}_0 
        d\ell^2 \frac{\ell^2 Z_{\Lambda_{\rm UV}}(\ell^2)
                        M_{\Lambda_{\rm UV}}(\ell^2)}
                {\ell^2+M^2_{\Lambda_{\rm UV}}(\ell^2)}
\mathrel{\mathop\simeq_{\Lambda_{\rm UV}\to\infty}}\ 
        -{3\over 4\pi^2}\int^{\Lambda_{\rm UV}^2}_0 
        d\ell^2 M_{\Lambda_{\rm UV}}(\ell^2)~.
\label{qbarq}
\eeq
The second result follows since the UV dominates for large $\Lambda_{\rm UV}$
and since in Landau gauge $Z_{\Lambda_{\rm UV}}(\ell^2)\simeq 1$, for large
$\ell^2$.  In the case of exact chiral symmetry [i.e., $m_{\Lambda_{\rm UV}}
= 0$] \Eq{chiral_q_mass} implies that $M_{\Lambda_{\rm UV}}(\ell^2) =
(c/\ell^2)\ln (\ell^2/\Lambda^2_{QCD})^{d_M -1}$ and hence, from \Eq{qbarq},
we find
\beq
\langle\bar qq\rangle_{\Lambda_{\rm UV}}\
\mathrel{\mathop\simeq_{\Lambda_{\rm UV}\to\infty}}\ -{3\over
4\pi^2}\ {c\over d_M }\
\left[\ln(\Lambda^2_{UV}/\Lambda^2_{QCD})\right]^{d_M} \;. 
\label{qbarq_UV}
\eeq
Since $c$ is independent of the choice of $\mu$ then \Eq{chiral_q_mass}
follows immediately.

We see from \Eq{asympt_q_mass} that $M_{\Lambda_{\rm UV}}(\ell^2) \!= \!\hat
m/[{1\over 2}\ln(\ell^2/\Lambda^2_{QCD})]^{d_{\rm M}}$ in the presence of an
ECSB mass, $m_{\Lambda_{\rm UV}}\!\neq \!0$.  It is clear that
\Eq{c_condensate} will no longer result from \Eq{qbarq} and so the quark
condensate, as defined in \Eq{q_condensate}, no longer varies with the
renormalisation point as $\langle\bar qq\rangle\sim
[\ln(\mu^2/\Lambda^2_{QCD})]^{d_M}$.

\sect{Pion Decay Constant}
\label{appendix_fpi}

We derive the expression for $f^2_\pi$ given in \Eq{f_pi2}. First
note that only terms up to ${\cal O}(k)$ need be kept since we will be
taking $k\to 0$.  For any function of $p^2$, say $f(p^2)$, we have
\beq
f((p+k)^2) = f(p^2) + 2k\cdot p\ {df\over dp^2}(p^2) + {\cal O}
(k^2)\;.
\eeq
In \Eq{f_pi2} the trace is over isospin, spinor, and colour labels.  The
spinor trace simplifies things considerably since [in an obvious shorthand
notation]
\beqn
{\rm tr}\biggl\{[S(p)&+& k\cdot\partial S(p)][\Gamma^m_5(p,p) +{\cal O}
(k)\gamma_5][S(p)][(\tau^n/2)\gamma^\sigma\gamma_5]\biggr\}\nonumber\\
&=&{\rm}tr\biggl\{k\cdot\partial S(p)\Gamma^m_5(p,p)S(p)
(\tau^n/2)\gamma^\sigma\gamma_5\biggr\} +{\cal O}(k^2)\;,
\eeqn
where we have used the fact that 
\mbox{${\rm tr}\left\{S(p)\gamma_5S(p)\gamma^\sigma\gamma_5\right\} = 0$},
which follows from Dirac matrix algebra, and the assumption that
$\Gamma_5(p',p)=\Gamma_5(p,p)+{\cal O}(k)\gamma_5$, where the ${\cal
O}(k)\gamma_5$ term is assumed to have a purely $\gamma_5$ spinor structure.
Now we can use $\vec\Gamma_5(p,p)$ which is known from the Goldberger-Treiman
relation, \Eq{GT_relation}.  We can now use the above results in \Eq{f_pi},
operate on both sides with $(k_\sigma/k^2)$ and use ${\rm tr}(\tau^m\tau^n) =
2\delta^{mn}$ to give
\beq
f^2_\pi = {3i\over k^2}\int {d^4p\over (2\pi)^4}{\rm tr_{spinor}}
\biggl\{k\cdot\partial S(p)\gamma_5B(p^2)S(p)A(p^2)\kslash\gamma_5\biggr\}
\eeq
where the trace is now over spinor labels only.  The factor of three comes
from the trace over colour.  After evaluating $(\partial/\partial p^\nu)S(p)$
and taking the spinor trace we find the result of \Eq{f_pi2}.  For
completeness we also give here the Euclidean-space form of \Eq{f_pi2}, which
is
\beq
f^2_\pi = \frac{3}{4\pi^2}\int_0^\infty dp^2\;{p^2Z(p^2)M(p^2)\over
[p^2+M^2(p^2)]^2}
\left[M(p^2)-{p^2\over 2}{dM\over dp^2}\right]\;.
\label{f_pi2_Euclidean}
\eeq

\section*{\sc References}\vspace*{-\parskip}
\addcontentsline{toc}{section}{\protect\numberline{ }{\sc References}}
\label{references}
\begin{description}
\itemsep=0pt
\parskip=0pt
\item Alkofer, R. and Amundsen, P.A. (1988).  Chiral symmetry breaking in an
instantaneous approximation to Coulomb gauge QCD, {\it Nucl. Phys. B}, {\bf
306}, 305. 
\item Alkofer, R. and Bender, A. (1993). Work in progress and private
communication.
\item Alkofer, R., Bender, A. and Roberts, C.D. (1993). Pion loop
contribution to the electromagnetic pion charge radius, {\it Argonne National
Laboratory Preprint} ANL-PHY-7663-TH-93 and {\it Universit\"{a}t T\"{u}bingen
Preprint} UNITUE-THEP-13/1993. 
\item Altarelli, G. (1982). Partons in quantum chromodynamics,
{\it Phys. Rep.} {\bf 81}, 1.
\item Aoki, Ken-Ichi, Kugo, Taichiro and Mitchard, Mark K. (1991). 
      Meson properties from the ladder Bethe-Salpeter equation,
      {\it Phys. Lett. B} {\bf 266}, 467.
\item Appelquist, T.W. and Carrazzone, J. (1975).
Infrared singularities and massive fields,  {\it Phys. Rev. D}
{\bf 11}, 2856.
\item Appelquist, T. and Pisarski, R.D. (1981). High-temperature Yang-Mills
theories and three-dimensional quantum chromodynamics, {\it Phys. Rev. D}
{\bf 23}, 2305.
\item Appelquist, T.W., Bowick, M., Karabali, D. and Wijewardhana, L.C.R.
(1986). Spontaneous chiral-symmetry breaking in three-dimensional QED, {\it
Phys. Rev. D} {\bf 33}, 3704. 
\item Appelquist, T., Nash, D and Wijewardhana, L.C.R. (1988).  Critical
behaviour in (2+1)-dimensional QED, {\it Phys. Rev. Lett.} {\bf 60}, 2575. 
\item Atkinson, D. and Blatt, D.W.E. (1979). Determination of the
singularities of the electron propagator, {\it Nucl. Phys. B} {\bf 151}, 342.
\item Atkinson, D. and Johnson, P.W., Schoenmaker, W.J. and Slim, H.A.
(1983).  Infra-red behaviour of the gluon propagator in axial gauges, {\it
Nuovo Cimento A} {\bf 77}, Series 11, 197. 
\item Atkinson, D. and Johnson, P.W. (1987). Bifurcation of the quark
self-energy: Infrared and ultraviolet cutoffs, {\it Phys. Rev. D} {\bf 35},
1943. 
\item Atkinson, D. and Johnson, P.W. (1988a). Chiral-symmetry breaking in
QCD. I. The infrared domain, {\it Phys. Rev. D} {\bf 37}, 2290.
\item Atkinson, D. and Johnson, P.W. (1988b). Chiral-symmetry breaking in
QCD. II. Running coupling constant, {\it Phys. Rev. D} {\bf 37}, 2296.
\item Atkinson, D., Johnson, P.W., Stam, K. (1988c). Chiral symmetry
breaking in QCD. III. Arbitrary covariant gauge, {\it Phys. Rev. D} {\bf 37},
2996. 
\item Atkinson, D., Johnson, P. and Pennington, M.R. (1988d).  Dynamical mass
generation in three-dimensional QED, {\it Brookhaven Nat. Lab. Preprint}
BNL-41615. 
\item Atkinson, D., Johnson, P. and Maris, P. (1990).  Dynamical mass
generation in three-dimensional QED: Improved vertex function, {\it Phys.
Rev. D} {\bf 42}, 602.
\item Atkinson, D., Bloch, J.C.R., Gusynin, V.P., Pennington, M. R. and
Reenders, M. (1993).  Strong coupling QED with weak gauge dependence: 
critical coupling and anomalous dimension, {\it Univ. Durham Preprint}
DTP-93/62. 
\item Bailin, D. and Love, A. (1986). {\it Introduction to Gauge Field 
Theory.} Adam Hilger, Bristol.
\item Baker. M, Ball, J.S. and Zachariasen, F. (1981a). A
non-perturbative calculation of the infrared limit of the axial gauge gluon
propagator (I), {\it Nucl. Phys. B} {\bf 186}, 531.
\item Baker, M., Ball, J.S. and Zachariasen, F. (1981b). A
non-perturbative calculation of the infrared limit of the axial gauge gluon
propagator (II), {\it Nucl. Phys. B} {\bf 186}, 560.
\item Baker, M., Ball, J.S. and Zachariasen, F. (1983). An analytic
calculation of the weak field limit of the static color dielectric constant,
{\it Nucl. Phys. B} {\bf 226}, 455. 
\item Baker, M., Ball, J.S. and Zachariasen, F. (1988).  Chiral-symmetry
breaking in dual QCD, {\it Phys. Rev. D} {\bf 38}, 1926; {\it erratum},
{\it Phys. Rev. D} {\bf 47}, 743 (1993).
\item Baker, M., Ball, J.S. and Zachariasen, F. (1991).  Dual QCD: A review,
{\it Phys. Reports} {\bf 209}, 73.
\item Ball, J.S. and Chiu, T.-W. (1980).  Analytic properties of the vertex
function in gauge theories. I and II, {\it Phys. Rev. D},
{\bf 22}, 2542. 
\item Ball, J.S. and Zachariasen, F. (1981).  The quark propagator in axial
gauge, {\it Phys. Lett. B} {\bf 106}, 133.
\item Barducci, A., Casalbuoni, R., De Curtis, S., Dominici, D and Gatto, R.
(1988).  Dynamical chiral-symmetry breaking and determination of the quark
masses, {\it Phys. Rev. D} {\bf 38}, 238.
\item Bar-Gadda, U. (1980). Infrared behavior of the effective coupling in
quantum chromodynamics, {\it Nucl. Phys. B} {\bf 163}, 312.
\item Bernard, C., Parrinello, C. and Soni, A. (1993). A lattice study of
the gluon propagator in momentum space, {\it Washington University preprint}
HEP/93-33.
\item Bjorken, J.D. and Drell, S.D. (1965). {\it Relativistic
Quantum Fields}. McGraw-Hill, New York.
\item Brown N. and Pennington M.R. (1988a), Preludes to confinement: Infrared
properties of the gluon propagator in the Landau gauge, {\it Phys. Lett. B}
{\bf 202}, 257.
\item Brown, N. and Pennington, M. (1988b). Studies of confinement: How
quarks and gluons propagate, {\it Phys. Rev. D} {\bf 38}, 2266.
\item Brown N. and Pennington M.R. (1989). Studies of confinement: How the
gluon propagates, {\it Phys. Rev. D} {\bf 39}, 2723.
\item Burden, C.J. and Burkitt, A.N. (1987).  Lattice fermions in odd
dimensions, {\it Europhys. Lett.} {\bf 3}, 545.
\item Burden, C.J., Cahill, R.T. and Praschifka, J. (1989). Baryon
structure and QCD: Nucleon calculations, {\it Aust. J. Phys.} {\bf 42}, 147. 
\item Burden, C.J. and Roberts, C.D. (1991). Light-cone regular vertex in
three-dimensional quenched QED, {\it Phys. Rev. D} {\bf 44}, 540. 
\item Burden, C.J., Praschifka, J. and Roberts, C.D. (1992a). Photon
polarization tensor and gauge dependence in three-dimensional quantum
electrodynamics, {\it Phys. Rev. D} {\bf 46}, 2695.
\item Burden, C.J., Roberts, C.D. and Williams, A.G. (1992b). Singularity
structure of a model quark propagator, {\it Phys. Lett. B} {\bf 285}, 347. 
\item Burden, C.J. (1993). How can \qedt help us understand QCD$_4$?, in
{\it QCD Vacuum Structure}, edited by H.M. Fried and B. M\"{u}ller, World
Scientific, Singapore. 
\item Burden, C.J. and Roberts, C.D. (1993). Gauge covariance and the
fermion-photon vertex in three- and four-dimensional, massless quantum
electrodynamics, {\it Phys. Rev. D} {\bf 47}, 5581.
\item Cahill, R.T. and Roberts, C.D. (1985).  Soliton bag models of hadrons
from QCD, {\it Phys. Rev. D} {\bf 32}, 2419. 
\item Cahill, R.T., Roberts, C.D. and Praschifka, J. (1987). Calculation of
diquark masses in QCD, {\it  Phys. Rev. D} {\bf 36}, 2804. 
\item Cahill, R.T., Roberts, C.D. and Praschifka, J. (1989). Baryon
structure and QCD, {\it  Aust. J. Phys.} {\bf 42}, 129.
\item Cahill, R.T. (1992). Hadronic laws from QCD, {\it Nucl. Phys. A} {\bf
543}, 63. 
\item Cahill, R.T. (1993). Private communication.
\item Coddington, P., Hey, A., Mandula, J. and Ogilvie, M. (1987). The lattice
photon propagator, {\it Phys. Lett. B} {\bf 185}, 191. 
\item Collins, J.C. (1984). {\it Renormalization}. Cambridge University Press,
Cambridge.
\item Cornwall, J.M., Jackiw, R. and Tomboulis, E. (1974).  Effective action
for composite operators, {\it Phys. Rev. D.} {\bf 10}, 2428.
\item Cornwall, J.M. (1980).  Confinement and chiral-symmetry breakdown:
Estimates of $F_\pi$ and of effective quark masses, {\it Phys. Rev. D}
{\bf 22}, 1452.
\item Cornwall, J.M. (1982).  Dynamical mass generation in continuum quantum
chromodynamics,  {\it Phys. Rev. D} {\bf 26}, 1453.
\item Creswick, R.J., Farach, H.A. and Poole, C.P. Jr. (1992). {\it
Introduction to Renormalization Group Methods in Physics}, John Wiley and
Sons, Inc., New York.
\item Curtis, D.C. and Pennington, M.R. (1990). Truncating the
Schwinger-Dyson equations: how multiplicative renormalizability and the Ward
Identity restrict the 3-point vertex in QED,  {\it Phys. Rev. D} {\bf 42},
4165.  
\item Curtis, D.C. and Pennington, M.R. (1991).  Nonperturbative fermion
propagator in quenched QED, {\it Phys. Rev. D} {\bf 44}, 536. 
\item Curtis, D.C. and Pennington, M.R. (1992). Generating fermion mass in
four-dimensional quenched QED, {\it Phys. Rev. D} {\bf 46}, 2663;
{\it erratum}, {\it Phys. Rev. D} {\bf 47}, 1729 (1993). 
\item Curtis, D.C., Pennington, M.R. and Walsh, D. (1992). Dynamical mass
generation in \qedt and the $1/N$ expansion, {\it Phys. Lett. B.} {\bf 295},
313.
\item Curtis, D.C. and Pennington, M.R. (1993). Non-perturbative study of
the fermion propagator in quenched QED in covariant gauges using a
renormalizable truncation of the Schwinger-Dyson equation, {\it University of
Durham preprint}, DTP-93/20.
\item Dagotto, E., Koci\'{c}, A. and Kogut, J.B. (1989).  Computer simulation
of chiral symmetry breaking in (2+1)-dimensional QED with $N$ flavors, {\it
Phys. Rev. Lett.}, {\bf 62}, 1083. 
\item Dagotto, E., Koci\'{c}, A. and Kogut, J.B. (1990).  Chiral symmetry
breaking in three-dimensional QED with $N_f$ flavours, {\it Nucl. Phys. B},
{\bf 334}, 279.
\item Dai, Y.-b, Huang, C.-s and Liu, D.-s, (1991). Calculation of
chiral-symmetry and pion properties as a Goldstone boson, {\it Phys. Rev. D}
{\bf 43}, 1717. 
\item Delbourgo, R. and West, P. (1977). A gauge covariant approximation to
quantum electrodynamics, {\it J. Phys. A} {\bf 10}, 1049.
\item Delbourgo, R. (1979a). The gauge technique, {\it Nuovo Cimento A}
{\bf 49}, Series 11, 484.
\item Delbourgo, R. (1979b). The gluon propagator, {\it J. Phys. G}
{\bf 5}, 603.
\item Delbourgo R. and Scadron M.D. (1979).  Proof of the Nambu-Goldstone
realisation for vector-gluon--quark theories, {\it J. Phys. G} {\bf 5}, 1621.
\item Delbourgo, R. (1981).  Solution of the gauge identities in the axial
gauge, {\it J. Phys. A} {\bf 14}, 753.
\item Delbourgo, R. and Zhang, R.B. (1984).  Transverse vertices in
electrodynamics and the gauge technique, {\it J. Phys. A}
{\bf 17}, 3593. 
\item Dong, Z., Munczek, H.J. and Roberts, C. D. (1994). Gauge covariant
fermion propagator in quenched, chirally-symmetric quantum electrodynamics,
{\it Argonne National Laboratory Preprint}, ANL-PHY-7711-94. 
\item Dorey, N. and Mavromatos, N.E. (1992). \qedt and two-dimensional
superconductivity without parity violation, {\it Nucl. Phys. B} {\bf 386}.
\item  Dyson, F.J. (1949). The S matrix in quantum electrodynamics. {\it Phys.
Rev.} {\bf 75} 1736. 
\item Einhorn, M. B. (1976). Confinement, form factors, and deep-inelastic
scattering in two-dimensional quantum chromodynamics, {\it Phys. Rev. D} {\bf
14}, 3451. 
\item  Faddeev, L.D. and Popov, V.N. (1967). Feynman diagrams for the
Yang-Mills Field. {\it Phys. Lett. B} {\bf 25}, 29.
\item Finger, J.R. and Mandula, J.E. (1982).  Quark pair condensation
and chiral symmetry breaking in QCD, {\it Nucl. Phys. B} {\bf 199}, 168. 
\item Fomin, P.I., Gusynin, V.P., Miransky, V.A. and Sitenko, Yu.A. (1983).
Dynamical chiral symmetry breaking and particle mass generation in
gauge field theories, {\it Rivista del Nuovo Cimento} {\bf 6}, Series 3,
Number 5.
\item Fradkin, E.S. (1956).  Concerning some General Relations of Quantum
Electrodynamics, {\it Sov. Phys. JETP} {\bf 2}, 361.
\item Frank, M.R., Tandy, P.C. and Fai, G. (1991).  Chiral solitons with
quarks and composite mesons, {\it Phys. Rev. C} {\bf 43}, 2808.
\item Fukuda, R. and Kugo, T. (1976).  Schwinger-Dyson equation for
massless vector theory and the absence of a fermion pole, {\it Nucl. Phys. B}
{\bf 117}, 250. 
\item  Gasser, J. and Leutwyler, H. (1982).  Quark masses, {\it Phys. Rep.}
{\bf 87}, 77. 
\item Glimm, J. and Jaffe, A. (1987). {\it Quantum Physics}, Springer-Verlag,
New York.  
\item Gogokhia, V.Sh. and Magradze, B.A. (1989).  Infrared finite quark
propagator and chiral symmetry breaking in QCD, {\it Phys. Lett. B}
{\bf 217}, 162.
\item Goldman, T., Henderson, J.A. and Thomas, A.W. (1992).  A New
Perspective on the $\rho$-$\omega$ Contribution to the Charge-Symmetry
Violation in the $N$-$N$ Force, {\it Few Body Systems} {\bf 12}, 123.
\item G\"{o}pfert, M. and Mack, G. (1982).  Proof of confinement of static
quarks in 3-dimensional $U(1)$ lattice gauge theory for all values of the
coupling constant, {\it Commun. Math. Phys.}{\bf 82}, 545. 
\item Govaerts, J., Mandula, J.E. and Weyers, J. (1983).  Pion properties in
QCD,  {\it Phys. Lett. B} {\bf 130}, 427.
\item Govaerts, J., Mandula, J.E. and Weyers, J. (1984). A model for chiral
symmetry breaking in QCD, {\it Nucl. Phys. B} {\bf 237}, 59. 
\item Gradshteyn, I.S. and Ryzhik, I.M. (1980). {\it Table of Integrals,
Series and Products}, Academic, New York. 
\item Greenberg, O.W. (1978).  Structure of asymptotic fields
associated with permanently confined degrees of freedom in quantum field
theory, {\it Phys. Rev. D} {\bf 17}, 2576.
\item Gribov, V.N. (1979).  Quantization of nonabelian gauge fields,
{\it Nucl. Phys. B} {\bf 139}, 1.
\item Gupta. R. \etal~(1987).  Hadron spectrum on an $18^3\times 42$
lattice, {\it Phys. Rev. D} {\bf 36}, 2813. 
\item Gusynin, V.P. (1990). Vacuum polarization and dynamical chiral
symmetry breaking in quantum electrodynamics, {\it Mod. Phys. Lett. A} {\bf
5}, 133. 
\item H\"{a}bel, U., K\"{o}nning, R., Reusch, H.-G., Stingl, M.
and Wigard, S. (1990a). A nonperturbative solution to the Dyson-Schwinger
equations of QCD. I. Nonperturbative vertices and a mechanism for their self
consistency, {\it Z. Phys. A} {\bf 336}, 423.
\item H\"{a}bel, U., K\"{o}nning, R., Reusch, H.-G., Stingl, M.
and Wigard, S. (1990b). A nonperturbative solution to the Dyson-Schwinger
equations of QCD. II. Self-consistency and physical properties, {\it Z. Phys.
A} {\bf 336}, 435.
\item H\"{a}dicke, A. (1991). Nonperturbative approaches to determining the
behavior of the gluon propagator and quark propagator in quantum
chromodynamics by Schwinger-Dyson equations, {\it Intern. J. Mod. Phys. A}
{\bf 6}, 3321.
\item Haeri, B. and Haeri, M.B. (1991). Effect of the gluon mass on
dynamical chiral symmetry breaking in QCD, {\it Phys. Rev. D} {\bf 43}, 3732.
\item Haeri, B. (1993). Vertices and the CJT effective potential, {\it Purdue
University preprint}, PURD-TH-93-12. 
\item Hawes, F.T., Roberts, C.D. and Williams, A.G. (1993). Dynamical chiral
symmetry breaking and confinement with an infrared-vanishing gluon propagator,
{\it Florida State University preprint}, FSU-SCRI-93-108.  To appear in
{\it Phys. Rev. D}.
\item Hawes, F.T. and Williams, A.G. (1991). Proper vertex in studies of
dynamical chiral symmetry breaking, {\it Phys. Lett. B} {\bf 268}, 271.
\item Haymaker, R.W. (1991).  Variational methods for composite operators,
{\it Rivista del Nuovo Cimento}, {\bf 14}, Series 3, Number 8.
\item Higashijima, K. (1983).  Effective potential for $\bar q$-$q$
condensation, {\it Phys. Lett. B} {\bf124}, 257.
\item Higashijima, K. (1984).  Dynamical chiral-symmetry breaking, {\it Phys.
Rev. D} {\bf 29}, 1228.
\item Hollenberg, L.C.L., Roberts, C.D. and McKellar, B.H.J. (1992). Two loop
calculation of the $ \omega$-$\rho$ mass splitting, {\it Phys. Rev. C} {\bf
46}, 2057.
\item  Itzykson C. and Zuber, J.-B. (1980). {\it Quantum Field Theory}.
McGraw-Hill, New York.
%
%
\item Jain, P. and Munczek, H.J. (1993).  $q\bar q$ bound states in the
Bethe-Salpeter formalism, {\it Phys. Rev. D} {\bf 48}, 5403. 
\item King, J.E. (1983).  Transverse vertex and gauge technique in quantum
electrodynamics, {\it Phys. Rev. D} {\bf 27}, 1821.
\item Kogut, J.B., Dagotto, E. and Koci\'{c}, A. (1988a).  On the existence
of quantum electrodynamics, {\it Phys. Rev. Lett.}, {\bf 61}, 2416.
\item Kogut, J.B., Dagotto, E. and Koci\'{c}, A. (1988b).  A supercomputer
study of strongly coupled QED, {\it Nucl. Phys. B},
{\bf 317}, 271.
\item Kondo, K.-I. (1990).  Critical exponents, scaling law, universality
and renormalization group flow in strong coupling QED, {\it Intern. J. Mod.
Phys. A} {\bf 6}, 5447. 
\item Kondo, K.-I. (1992).  Recovery of gauge invariance in strong-coupling
QED, {\it Intern. J. Mod. Phys. A} {\bf 7}, 7239.
\item Kondo, K. and Nakatani, H. (1992a).  Cut-off dependence of
self-consistent solutions in unquenched QED$_3$, {\it Prog. Theor.
Phys.} {\bf 87}, 193. 
\item Kondo, K. and Nakatani, H. (1992b).  Strong coupling unquenched QED.
II, {\it Prog. Theor.  Phys.} {\bf 88}, 737.
\item Krein, G., Tang, P., Wilets, L. and Williams, A.G. (1988).
Confinement, chiral symmetry breaking, and the pion in a chromo-dielectric
model of quantum chromodynamics, {\it Phys. Lett B} {\bf 212}, 362. 
\item Krein, G., Tang, P. and Williams, A.G. (1988).
Dynamical chiral symmetry breaking in quantum chromodynamics with confinement
and asymptotic freedom, {\it Phys. Lett B} {\bf 215}, 145.
\item Krein, G. and Williams, A.G. (1990).  Covariant model of chiral
symmetry breaking in QCD, {\it Mod. Phys. Lett. A} {\bf 5}, 399.
\item Krein, G. and Williams, A.G. (1991).  Dynamical chiral symmetry
breaking in dual QCD, {\it Phys. Rev. D} {\bf 43}, 3541.
\item Krein, G., Tang, P., Wilets, L. and Williams, A.G. (1991).
The chromodielectric model;  Confinement, chiral symmetry breaking,
and the pion, {\it Nucl. Phys. A} {\bf 523}, 548. 
\item Krein, G., Thomas, A.W. and Williams, A. G. (1993).  Charge-Symmetry
Breaking, Rho-Omega Mixing, and the Quark Propagator, 
{\it Phys. Lett. B} {\bf 317}, 293.
\item Landau, L.D. and Khalatnikov, I.M. (1956). The Gauge Transformations of
the Green's Function for Charged Particles, {\it Sov.  Phys. JETP} {\bf 2},
69.
\item Langfeld, K.,  Alkofer, R. and Amundsen, P.A. (1989).  Pion
electromagnetic properties in Coulomb gauge QCD, {\it Z. Phys. C} {\bf 42},
159. 
\item Leung, C.N., Love, S.T. and Bardeen, W.A. (1986). Spontaneous
symmetry breaking in scale invariant quantum electrodynamics, {\it Nucl.
Phys. B} {\bf 273}, 649. 
\item Mandelstam, S. (1979). Approximation scheme for quantum chromodynamics,
{\it Phys. Rev. D} {\bf 20}, 3223.
\item Mandula, J.E. and Ogilvie, M. (1987a). The Landau gauge gluon
propagator in lattice QCD, {\it Phys. Lett. B} {\bf 185}, 127; in {\it
Nonperturbative Methods in Field Theory}, Proceedings of the Conference,
Irvine, California, edited by H. W. Amber [{\it Nucl.  Phys. B (Pro.
Supple.)}, {\bf 1A}, 117].
\item Mandula, J.E. and Ogilvie, M. (1987b). The gluon is massive: A lattice
calculation of the gluon propagator in the Landau gauge, {\it Phys. Lett. B}
{\bf 201} 127.
\item Mandula, J.E. and Ogilvie, M. (1988). The gluon propagator at finite
temperature, {\it Phys. Lett. B} {\bf 201} 117.
\item  Marciano, W. and Pagels, H. (1977). Quantum chromodynamics. {\it Phys.
Rep.} {\bf 36} 137.
\item Maris, P. and Holties H. (1992). Determination of the singularities of
the Dyson-Schwinger equation for the quark propagator, {\it Intern. J. Mod. Phys
A} {\bf 7}, 5369.
\item Maris, P.  (1993). {\it Nonperturbative Analysis of the Fermion
Propagator: Complex Singularities and Dynamical Mass Generation}, PhD Thesis, 
Rijksuniversiteit Groningen.
\item Matsuki, T. (1991). Instability at the origin in (2+1)-dimensional QED,
{\it Z. Phys. C.} {\bf 51}, 429.
\item McDaniel, H., Uretsky, J. and Warnock, R.L. (1972).  Solutions of a
nonlinear integral equation for high-energy scattering. I. An existence
theorem, {\it Phys. Rev. D} {\bf 6}, 1588.
\item McDaniel, H., Uretsky, J. and Warnock, R.L. (1972).  Solutions of a
nonlinear integral equation for high-energy scattering. II. Numerical
solutions, {\it Phys. Rev. D} {\bf 6}, 1600.
\item McKay, D. and Munczek, H.J. (1985).  Anomalous chiral Lagrangians of
pseudoscalar, vector, and axial-vector mesons generated from quark loops,
{\it Phys. Rev. D} {\bf 32}, 266.
\item McKay, D., Munczek, H.J. and Young, B.-L., (1989).  From QCD to the
low-energy effective action through composite fields: Goldstone's theorem and
$f_\pi$, {\it Phys. Rev. D} {\bf 37}, 195.
\item McKay, D. and Munczek, H.J. (1989).  Composite-operator
effective-action considerations on bound states and corresponding $S$-Matrix
elements, {\it Phys. Rev. D} {\bf 40}, 4151.
\item Miransky, V.A. (1985a). Dynamics of spontaneous chiral symmetry
breaking and the continuum limit in quantum electrodynamics, {\it Nuovo
Cimento, A} {\bf 90}, Series 11, 149. 
\item Miransky, V.A. (1985b).  On dynamical chiral symmetry breaking, {\it
Phys. Lett. B} {\bf 165}, 401. 
\item Mitchell, K., Tandy, P. C., Roberts, C. D. and Cahill, R. T. (1994).
Work in progress. 
\item Munczek, H.J. and Nemirovsky, A.M. (1983). Ground-state $q\bar q$
mass spectrum in quantum chromodynamics, {\it Phys. Rev. D}
{\bf 28}, 181.
\item Munczek, H.J. and McKay, D.W. (1990).  The Schwinger-Dyson equation
in QCD: Comparison of some approximations, {\it Phys. Rev. D}, {\bf 42},
3548. 
\item Munczek, H.J. and Jain. P. (1992).  Relativistic pseudoscalar
$q\bar q$ bound states: Results on Bethe-Salpeter wavefunctions and
decay constants, {\it Phys. Rev. D} {\bf 46}, 438.
\item  Muta, T. (1987). {\it Foundations of Quantum Chromodynamics: An
Introduction to Perturbative Methods in Gauge Theories}.  World Scientific,
Singapore.
\item Nakanishi, N. (1969).  A general survey of the theory of the
Bethe-Salpeter equation, {\it Suppl. Prog. Theor. Phys.}
{\bf 43}, 1.
\item Nambu, Y. and Jona-Lasinio, G. (1961). Dynamical model of elementary
particles based on an analogy with superconductivity. I., {\it Phys. Rev.}
{\bf 122}, 345.
\item Nash, D. (1989).  Higher-order corrections in (2+1)-dimensional QED,
{\it Phys. Rev. Lett.} {\bf 62}, 3024.
\item Nielsen, N.K. (1975).  On the gauge dependence of spontaneous
symmetry breaking in gauge theories, {\it Nucl. Phys. B} {\bf 101}, 173.
\item Nishijima, K. (1986).  {\it Colour confinement based on BRS algebra},
in Progress in Quantum Field Theory, edited by H. Ezawa and S. Kamefuchi,
Elsevier, Amsterdam.
\item Oliensis, J. and Johnson, P.W. (1990).  Possible second-order phase
transition in strongly coupled unquenched planar four-dimensional QED, {\it
Phys. Rev. D} {\bf 42}, 656. 
\item Pagels, H. and Stokar, S. (1979).  Pion decay constant, electromagnetic
form factors, and quark electromagnetic self-energy in quantum
chromodynamics, {\it Phys. Rev. D} {\bf 20}, 2947.
\item Pagels, H. and Stokar, S. (1980).  Magnitude of the light current quark
masses, {\it Phys. Rev. D} {\bf 22}, 2876.
\item Particle Data Group (1990). Review of Particle Properties, {\it Phys.
Lett. B} {\bf 239}, 1. 
\item  Pascual, P. and Tarrach, R. (1984). {\it QCD: Renormalization for the
Practitioner}. Springer-Verlag, Berlin. 
\item  Pennington, M.R. and Webb, S.P. (1988).  Hierarchy of scales in
three-dimensional QED, {\it Brookhaven Nat. Lab. preprint}, BNL-40886. 
\item Pennington, M.R. and Walsh, D. (1991).  Masses from nothing.
A nonperturbative study of QED$_3$, {\it Phys. Lett. B} {\bf 253}, 246. 
\item  Pisarski, R.D.  (1984).  Chiral symmetry breaking in three-dimensional
electrodynamics, {\it Phys. Rev. D} {\bf 29}, 2423.  
\item  Pokorski (1987).  {\it Gauge Field Theories}.  Cambridge University
Press, Cambridge - New York.
\item  Politzer, H.D. (1976).  Effective quark masses in the chiral limit,
{\it Nucl. Phys. B} {\bf 117}, 397.
\item  Politzer, H.D. (1982).  Chiral symmetry, quenched fermions, 
nonrelativistic quarks, $\eta$'s and glueballs, {\it Phys. Lett. B}
{\bf 116}, 171.
\item  Popov, V.N. (1983). {\it Functional Integrals in Quantum Field Theory
and Statistical Mechanics}. Reidel, Dordrecht.
\item Praschifka, J., Roberts, C.D. and Cahill, R.T. (1987a).
QCD bosonization and the meson effective action, {\it Phys. Rev. D} {\bf 36},
209.
\item Praschifka, J., Roberts, C.D. and Cahill, R.T. (1987b).
A Study of $\rho\rightarrow\pi\pi$ Decay in a Global Colour Model for QCD,
{\it Intern. J. Mod. Phys. A} {\bf 2}, 1797.
\item Praschifka, J., (1988). {\it Mesons and diquarks in Bosonised QCD}, PhD
Thesis, Flinders University of South Australia. 
\item Praschifka, J., Cahill, R.T. and Roberts, C.D. (1989).
Mesons and diquarks in chiral QCD: Generation of constituent quark masses,
{\it Intern. J. Mod. Phys. A} {\bf 4}, 4929.
\item Rakow, P.E.L. (1991).  Renormalization group flow in QED - An
investigation of the Schwinger-Dyson equations, {\it Nucl. Phys. B} {\bf
356}, 27. 
\item Reinders, L.J., Rubinstein, H. and Yazaki, S. (1985).  Hadron
properties from QCD sum rules, {\it Phys. Rep.} {\bf 127}, 1.
\item Reya, E. (1981). Perturbative quantum chromodynamics, {\it Phys. Rep.}
{\bf 69}, 195.
\item  Rivers, R.J. (1987). {\it Path integral methods in quantum
field theory}, 1990 printing. Cambridge University Press, Cambridge (UK).
\item Roberts, C.D. and Cahill, R.T. (1986).  Dynamically selected vacuum
field configuration in massless QED, {\it Phys. Rev. D} {\bf 33}, 1755. 
\item Roberts, C.D. and Cahill, R.T. (1987).  A Bosonisation of QCD and
Realisations of Chiral Symmetry, {\it Aust. J. Phys.} {\bf 40}, 499.
\item Roberts, C.D., Cahill, R.T. and Praschifka, J. (1988).  The
effective action for the Goldstone modes in a global colour symmetry model of
QCD, {\it Ann. Phys. (NY)} {\bf 188}, 20.
\item Roberts, C.D., Praschifka, J.\ and Cahill, R.T. (1989). A chirally
symmetric effective action for vector and axial vector fields in a global
colour symmetry model of QCD, {\it Intern.\ J.\ Mod.\ Phys.\ A} {\bf 4},
1681.
\item Roberts, C.D. and McKellar, B.H.J. (1990). Critical coupling for
dynamical chiral symmetry breaking, {\it Phys. Rev. D} {\bf 41}, 672. 
\item Roberts, C.D., Williams, A.G. and Krein, G. (1992).  On the
implications of confinement, {\it Intern. Journal Mod. Phys. A} {\bf 7}, 5607.
\item Roberts, C.D., Cahill, R.T., Sevior, M.E., Iannella, N. (1994).
$\pi$-$\pi$ scattering in a QCD based model field theory, {\it Phys. Rev. D}
{\bf 49}, 125. 
\item  Rothe, H.J. (1992). {\it Lattice Gauge Theories: An Introduction},
World Scientific Lecture Notes in Physics - Vol.~{\bf 43}. World Scientific,
Singapore.
\item Salam, A. (1963).  Renormalizable electrodynamics of vector mesons,
{\it Phys. Rev.} {\bf 130}, 1287. 
\item  Schwinger, J.S. (1951).  On the Green's functions of quantized fields.
I and II, {\it Proc. Nat. Acad. Sc.} {\bf 37} 452, 455.
\item  Schwinger, J.S. (1962). Gauge invariance and mass. I and II,
{\it Phys. Rev.} {\bf 125}, 397; {\bf 128}, 2425. 
\item  Seiler, E. (1982).  {\it Gauge Theories as a Problem of Constructive
Quantum Theory and Statistical Mechanics}. Springer-Verlag, New York.
\item Smekal, L.v., Amundsen, P. and Alkofer, R. (1991). A covariant model
for dynamical chiral-symmetry breaking in QCD, {\it Nucl. Phys. A} {\bf 529},
663.
\item Stainsby, S.J. and Cahill, R.T. (1990). Is space-time Euclidean
``inside'' hadrons?, {\it Phys. Lett. A} {\bf 146}, 467. 
\item Stainsby, S.J. and Cahill, R.T. (1992). The analytic structure of
quark propagators, {\it Intern. J. Mod. Phys. A}, {\bf 7}, 7541. 
\item Stainsby, S.J. (1993).  {\it Quark Propagator Singularities in QCD},
PhD Thesis, Flinders University of South Australia. 
\item Stainsby, S.J. and Cahill, R.T. (1993). Quark-gluon vertex and
structure of mesons and diquarks, {\it Flinders University of South Australia
Preprint}.
\item Stam, K. (1985). Dynamical chiral symmetry breaking, {\it Phys. Lett. B}
{\bf 152}, 238.
\item Stingl, M. (1986).  Propagation properties and condensate formation of
the confined Yang-Mills field, {\it Phys. Rev. D} {\bf 34}, 3863; {\it
erratum} (1987), {\it Phys. Rev. D} {\bf 36}, 651.
\item Stingl, M. (1992). An extended perturbation theory for QCD, {\it
Universit\"{a}t M\"{u}nster preprint}, MS-TPI-92-25. 
\item Streater, R.F. and Wightman, A.S. (1980). {\it PCT, Spin and
Statistics, and All That}, 3rd ed., Addison-Wesley, Reading, Mass. 
\item Strocchi, (1976).  Locality, charges, and quark confinement,
{\it Phys. Lett. B} {\bf 62}, 60.
\item Strocchi, (1978).  Local and covariant gauge quantum field theories.
Cluster property, superselection rules, and the infrared problem,
{\it Phys. Rev. D} {\bf 17}, 2010.
\item Symanzik, K. (1969). in {\it Local Quantum Theory}, edited
by R. Jost, Academic, New York.  
\item Symanzik, K. (1973).  Infrared singularities and small-distance
behaviour analysis, {\it Comm. Math. Phys.} {\bf 34}, 7.
\item Takahashi, Y. (1957). On the generalised Ward identity. {\it Nuovo
Cimento} {\bf 6}, 370.
\item Waites, A.B. and Delbourgo, R. (1992). Non-perturbative behaviour in
three dimensional QED, {\it Intern. J. Mod. Phys. A} {\bf 7}, 6857. 
\item Walsh, D. (1990). {\it A Nonperturbative Study of Three-Dimensional
Quantum Electrodynamics with $N$ Flavours of Fermion}, PhD Thesis, University
of Durham.
\item Ward, J.C. (1950). An identity in quantum electrodynamics. {\it Phys.
Rev.} {\bf 78}, 182.
\item West, G.B. (1982).  Confinement, the Wilson loop, and the gluon
propagator, {\it Phys. Lett. B} {\bf 115}, 468.
\item West, G.B. (1983).  General infrared and ultraviolet properties of the
gluon propagator, {\it Phys. Rev. D} {\bf 27}, 1878.
\item Wick, G.C. (1954). Properties of the Bethe-Salpeter wavefunctions,
{\it Phys. Rev.} {\bf 96}, 1954.
\item Williams, A.G., Krein, G. and Roberts, C.D. (1991). Modelling
the quark propagator, {\it Annals of Physics} {\bf 210}, 464.
\item Wimp, J. (1981). {\it Sequence Transformations and their
Applications}, New York, Academic Press.
\item Yndur\'ain, F.J. (1993). {\it The Theory of Quark and Gluon
Interactions.}  Springer-Verlag, Berlin - Heidelberg - New York. 
\item Zwanziger, D. (1991). Vanishing of zero-momentum lattice gluon
propagator and color confinement, {\it Nucl. Phys. B} {\bf 364}, 127.
\end{description}
%
%
\end{document}